\documentclass[aip,cha,reprint,author-year,nofootinbib,floatfix,dvipsnames]{revtex4-2}

\usepackage{physjour_aas_macros}
\usepackage{orcidlink}
\usepackage{lmodern}
\usepackage{graphicx}
\usepackage{amsmath,amssymb}
\usepackage{stmaryrd}
\usepackage[utf8]{inputenc}
\usepackage[T1]{fontenc}
\usepackage{color}
\usepackage[raggedright]{titlesec}
\usepackage[normalem]{ulem}
\definecolor{darkblue}{rgb}{0.0,0.0,0.4}
\definecolor{darkred}{rgb}{0.7,0.0,0.0}
\definecolor{darkgreen}{rgb}{0.0,0.5,0.0}
\definecolor{C0}{HTML}{1f77b4}
\definecolor{C1}{HTML}{ff7f0e}
\definecolor{C2}{HTML}{2ca02c}
\definecolor{C3}{HTML}{d62728}
\definecolor{C4}{HTML}{9467bd}
\definecolor{C5}{HTML}{8c564b}
\definecolor{C6}{HTML}{e377c2}
\definecolor{C7}{HTML}{7f7f7f}
\definecolor{C8}{HTML}{bcbd22}
\definecolor{C9}{HTML}{17becf}

\usepackage{tikz}
\usetikzlibrary{shapes,arrows,shadows,decorations.pathreplacing,calligraphy,calc}
\usepackage[export]{adjustbox}

\usepackage{scalerel}
\usetikzlibrary{svg.path}
\definecolor{orcidlogocol}{HTML}{A6CE39}
\tikzset{
  orcidlogo/.pic={
    \fill[orcidlogocol] svg{M256,128c0,70.7-57.3,128-128,128C57.3,256,0,198.7,0,128C0,57.3,57.3,0,128,0C198.7,0,256,57.3,256,128z};
    \fill[white] svg{M86.3,186.2H70.9V79.1h15.4v48.4V186.2z}
                 svg{M108.9,79.1h41.6c39.6,0,57,28.3,57,53.6c0,27.5-21.5,53.6-56.8,53.6h-41.8V79.1z M124.3,172.4h24.5c34.9,0,42.9-26.5,42.9-39.7c0-21.5-13.7-39.7-43.7-39.7h-23.7V172.4z}
                 svg{M88.7,56.8c0,5.5-4.5,10.1-10.1,10.1c-5.6,0-10.1-4.6-10.1-10.1c0-5.6,4.5-10.1,10.1-10.1C84.2,46.7,88.7,51.3,88.7,56.8z};
  }
}
\newcommand\orcid[1]{\href{https://orcid.org/#1}{\mbox{\scalerel*{
\begin{tikzpicture}[yscale=-1,transform shape]
\pic{orcidlogo};
\end{tikzpicture}
}{|}}}}

\usepackage{hyperref}
\hypersetup{colorlinks,breaklinks,
            linkcolor=darkblue,urlcolor=darkblue,
            anchorcolor=darkblue,citecolor=RoyalBlue}
\usepackage{textcomp, gensymb}
\usepackage[mathscr]{euscript}
\usepackage{upgreek}
\DeclareFontFamily{OT1}{pzc}{}
\DeclareFontShape{OT1}{pzc}{m}{it}{<-> s * [1.10] pzcmi7t}{}
\DeclareMathAlphabet{\mathpzc}{OT1}{pzc}{m}{it}

\usepackage{tikz}
\usetikzlibrary{calc}
\usetikzlibrary{positioning, shapes.geometric}

\titleformat{\section}{\selectfont \normalfont\raggedright\sffamily\small\bfseries\uppercase}{\thesection.}{1em}{}{}
\titleformat{\subsection}{\selectfont \normalfont\raggedright\sffamily\small\bfseries}{\thesubsection.}{1em}{}{}
\titleformat{\subsubsection}{\selectfont \normalfont\sffamily\small\bfseries}{\thesubsection.\thesubsubsection}{1em}{}{}
\titleformat{\paragraph}[runin]{\selectfont \normalfont\sffamily\small\bfseries}{\thesubsection.\thesubsubsection.\theparagraph}{1em}{}[.]

\usepackage{tikz}
\usetikzlibrary{shapes,arrows,shadows}
\usepackage[export]{adjustbox}
\usepackage{makecell}

\usepackage{stmaryrd}

\usepackage[caption=false]{subfig}

\usepackage{cleveref}

\usepackage{siunitx}
\usepackage{acro}

\DeclareAcronym{selfi}{
    short = {\selfi}, 
    long  = {Simulator Expansion for Likelihood-Free Inference},
}

\DeclareAcronym{ili}{ 
    short = {ILI}, 
    long  = {Implicit Likelihood Inference},
}

\DeclareAcronym{abc}{ 
    short = {ABC}, 
    long  = {Approximate Bayesian Computation},
}

\DeclareAcronym{pmc}{ 
    short = {ABC-PMC}, 
    long  = {Population Monte Carlo},
}

\DeclareAcronym{bhm}{ 
    short = {BHM}, 
    long  = {Bayesian Hierarchical Model},
    short-plural = {s},
    long-plural = {s},
}

\DeclareAcronym{planck}{ 
    short = {Planck 2018}, 
    long  = {Planck 2018},
    first-style = short,
    cite = [][]{aghanim2020planck}
}

\DeclareAcronym{bolfi}{ 
    short = {\bolfi}, 
    long  = {Bayesian Optimization for Likelihood-Free Inference},
}

\DeclareAcronym{baos}{ 
    short = {BAOs}, 
    long  = {Baryon Acoustic Oscillations},
}

\DeclareAcronym{pdf}{ 
    short = {pdf}, 
    long  = {probability distribution function},
    short-plural = {s},
    long-plural = {s},
}

\DeclareAcronym{lrgs}{ 
    short = {LRGs}, 
    long  = {Luminous Red Galaxies},
}

\DeclareAcronym{limd}{ 
    short = {LIMD}, 
    long  = {Local-In-Matter Density},
}

\DeclareAcronym{hmc}{ 
    short = {HMC}, 
    long  = {Hamiltonian Monte Carlo},
}

\DeclareAcronym{cola}{
    short = {\textsc{COLA}},
    long = {COmoving Lagrangian Acceleration},
}

\DeclareAcronym{class}{
    short = {\class},
    long = {the Cosmic Linear Anisotropy Solving System},
}

\DeclareAcronym{2lpt}{
    short = {2LPT},
    long = {second order Lagrangian Perturbation Theory},
}

\newcommand{\eqdef}{\equiv}

\renewcommand{\i}{\mathrm{i}}

\newcommand{\p}{\mathpzc{P}}

\newcommand{\power}{initial matter power spectrum}
\newcommand{\powers}{initial matter power spectra}
\newcommand{\psa}{power spectra}
\newcommand{\selfi}{\textsc{selfi}}
\newcommand{\bolfi}{\textsc{bolfi}}
\newcommand{\class}{\textsc{class}}
\newcommand{\camb}{\textsc{camb}}

\definecolor{airforceblue}{rgb}{0.36, 0.54, 0.66}
\definecolor{babyblue}{rgb}{0.54, 0.81, 0.94}
\definecolor{babyblueeyes}{rgb}{0.63, 0.79, 0.95}
\definecolor{beaublue}{rgb}{0.74, 0.83, 0.9}
\definecolor{blizzardblue}{rgb}{0.67, 0.9, 0.93}
\definecolor{bleudefrance}{rgb}{0.19, 0.55, 0.91}
\definecolor{bondiblue}{rgb}{0.0, 0.58, 0.71}
\definecolor{brightcerulean}{rgb}{0.11, 0.67, 0.84}
\definecolor{columbiablue}{rgb}{0.61, 0.87, 1.0}
\definecolor{darkturquoise}{rgb}{0.0, 0.81, 0.82}
\definecolor{deepskyblue}{rgb}{0.0, 0.75, 1.0}
\definecolor{icblue}{rgb}{0.49, 0.98, 1.0}
\definecolor{electriccyan}{rgb}{0.0, 1.0, 1.0}
\definecolor{glaucous}{rgb}{0.38, 0.51, 0.71}
\definecolor{lightcyan}{rgb}{0.88, 1.0, 1.0}

\begin{document}

\title{Diagnosing systematic effects using the inferred initial power spectrum}



\author{Tristan~Hoellinger~\orcidlink{0000-0003-0217-8542},}
\email[]{tristan.hoellinger@iap.fr}
\homepage{https://hoellin.github.io/}
\affiliation{\mbox{Sorbonne Université, CNRS, UMR 7095, Institut d’Astrophysique de Paris, 98 bis bd Arago, 75014 Paris, France}}

\author{Florent~Leclercq~\orcidlink{0000-0002-9339-1404},}
\email[]{florent.leclercq@iap.fr}
\homepage{https://www.florent-leclercq.eu/}
\affiliation{\mbox{Sorbonne Université, CNRS, UMR 7095, Institut d’Astrophysique de Paris, 98 bis bd Arago, 75014 Paris, France}}


\date{\today}

\begin{abstract}
\noindent
{The next generation of galaxy surveys has the potential to substantially deepen our understanding of the Universe. This potential hinges on our ability to rigorously address systematic uncertainties. Until now, diagnosing systematic effects prior to inferring cosmological parameters has been out of reach in field-based implicit likelihood cosmological inference frameworks.}
{As a solution, we aim to diagnose a variety of systematic effects in galaxy surveys prior to inferring cosmological parameters, using the inferred {\power}.}
{Our approach is built upon a two-step framework. First, we employed the \ac{selfi} algorithm to infer the {\power}, which we utilised to thoroughly investigate the impact of systematic effects. This investigation relies on a single set of $N$-body simulations. Second, we obtained a posterior on cosmological parameters via implicit likelihood inference, recycling the simulations from the first step for data compression. As a demonstration, we relied on a model of large-scale spectroscopic galaxy surveys that incorporates fully non-linear gravitational evolution with \ac{cola} and simulates multiple systematic effects encountered in real surveys.}
{We provide a practical guide on how the \acs{selfi} posterior can be used to assess the impact of misspecified galaxy bias parameters, selection functions, survey masks, inaccurate redshifts, and approximate gravity models on the inferred {\power}. We show that a subtly misspecified model can lead to a bias exceeding $2\sigma$ in the $(\Omega_\mathrm{m}, \sigma_8)$ plane, which we are able to detect and avoid prior to inferring cosmological parameters.}
{This framework has the potential to significantly enhance the robustness of physical information extraction from full forward models of large-scale galaxy surveys such as DESI, \textit{Euclid}, and LSST.}
\end{abstract}

\maketitle

\section{Introduction}

Current and forthcoming large-scale galaxy surveys have the statistical capability to unveil the nature of dark energy, shed light on the mechanisms driving cosmic inflation, and constrain neutrino masses with unprecedented precision \citep{mishra2018neutrino,euclid2020euclid,goldstein2023beyond}. The vast volumes of data collected by Stage-IV surveys \citep{albrecht2006report} such as DESI \citep{aghamousa2016desi}, \textit{Euclid} \citep{mellier2024euclid}, and LSST \citep{lsst2012large,ivezic2019lsst} will be dominated by systematic rather than statistical uncertainties \citep[][]{salvati2020impact}. Extracting the subtle signals relating to the cosmological structure and content of our Universe therefore requires careful treatment of astrophysical and observational nuisances, along with their associated systematic effects \citep[e.g.][]{Meiksin1999,Desjacques_2018}. Such effects are pervasive and constitute one of the foremost challenges in modern cosmology \citep[e.g.][]{kim2004effects, kitching2008systematic,davis2011effect, kitching2016discrepancies,glanville2021effect,ayccoberry2023unions}. For instance, substantial effort is being directed towards accurately measuring the growth rate of structure from peculiar velocities and galaxy surveys, whilst properly accounting for biases and selection effects \citep[e.g.][]{said2020joint, carreres2023growth}. Crucially, the risk of mistaking systematic biases for new physics must be rigorously addressed.

To match the ever-growing precision of observations from stage-IV astrophysical surveys, Bayesian cosmological inference pipelines rely on increasingly complex probabilistic simulators. In particular, field-based forward models of galaxy surveys have become a cornerstone of modern precision cosmology \citep[e.g.][]{Jasche2013BORG,Leclercq2015ST,ramanah2019cosmological,porqueres2022lifting,andrews2023bayesian,kostic2023consistency,nguyen2024much,zeghal2024simulation,zhou2024accurate}. These models simulate the evolution of the entire density field from its initial conditions in the early Universe to the present epoch. Whilst these simulators incorporate a refined treatment of non-linear gravitational evolution, astrophysical nuisances, and instrumental responses, the resulting pipelines become increasingly dependent on the accuracy of the underlying models. Consequently, they exhibit little to no robustness to model misspecification: when models deviate from the actual data-generating process, posteriors tend to become biased or overly concentrated \citep{frazier2017model}. For instance, using explicit likelihood inference within an effective field theory model of large-scale structure, \citet{nguyen2021impacts} showed that misspecified models can significantly bias the inferred transfer function.

Current approaches to cosmological data analysis using field-based models can be classified into two broad categories, depending on whether the likelihood is explicitly or implicitly defined. Explicit likelihood methods \citep[e.g.][]{wandelt2004global,Jasche2013BORG,wang2014elucid,wang2016elucid,alsing2016hierarchical,jasche2019physical,loureiro2022almanac,sellentin2023almanac,zhou2024accurate} employ Markov chain Monte Carlo \citep[MCMC, ][]{metropolis1953equation} algorithms such as Hamiltonian Monte Carlo \citep[HMC,][]{duane1987hybrid} to sample from the target distribution. To make the likelihood tractable, they rely on approximations of the astronomical observables’ models. In contrast, implicit likelihood approaches \citep[see ][for reviews and applications to cosmology]{weyant2013likelihood,ishida2015cosmoabc,lintusaari2017fundamentals,alsing2018delfi,Leclercq2018BOLFI,Leclercq2019SELFI,taylor2019cosmic,alsing2019fast,brehmer2020mining,cranmer2020frontier,Makinen_2021,nguyen2024much,tucci2024eftoflss,ho2024ltu} can accommodate arbitrarily complex simulator-based models, in which the internal mechanisms are unknown or difficult to incorporate into likelihood-based analyses. We refer to these simulators as hidden-box models. They may include a physical treatment of structure formation through $N$-body simulations as well as the intricacies of observational processes within the survey.

Both explicit and implicit approaches are susceptible to model misspecification. To address this issue within explicit likelihood cosmological inference frameworks, several solutions have been proposed: developing flexible data models \citep{jasche2017bayesian,jasche2019physical,nguyen2021impacts,porqueres2022lifting}, marginalising over unknown systematic effects \citep{porqueres2019explicit}, heating up the likelihood \citep{jasche2019physical}, or informing the likelihood with explicit knowledge of where the model is likely to underperform \citep{doeser2023bayesian}.  Some statistical literature has recently been focused on addressing the issue of model misspecification in \acl{ili} \citep[\acs{ili}, e.g.][]{thomas2020misspecification}. Yet, until now, no method had been designed to systematically address model misspecification in field-based implicit likelihood cosmological inferences.

Among the \ac{ili} approaches to cosmological data analysis, \citet{Leclercq2019SELFI} introduced the \acf{selfi} algorithm, which makes it possible to infer the {\power} using arbitrarily complex hidden-box models of galaxy surveys. The {\power} (after recombination) is a theoretically well-understood object, and we expect its inference to be sensitive to most of the systematic effects encountered in the survey. Therefore, in this paper, we propose to use the {\power} to diagnose systematic effects in implicit likelihood cosmological inferences from hidden-box models of galaxy surveys.

The full pipeline that we construct is built upon an application to cosmological data analysis of the two-step procedure introduced by \citet{Leclercq_2022LV} for \ac{ili} of \acp{bhm} where a latent function carries relevant information about the target parameters to be inferred. The latent function is further required to be confined within a narrow region of its parameter space. Cosmological inferences naturally fall within this class of \acp{bhm}: the {\power} can be treated as a latent function within the physical model mapping the cosmological parameters to the observed galaxy count fields; and its shape is already strongly constrained by Cosmic Microwave Background experiments \citep{aghanim2020planck, balkenhol2023measurement}. For the first step of the procedure, we utilise the \ac{selfi} algorithm to infer the {\power}. The inference requires the mean data model at an expansion point along with its gradient, which can be estimated by finite differences as in \citet{Leclercq2019SELFI}. Alternatively, manual or automatic differentiation can be used: see \cite{hahn2024disco} for an automatically differentiable Boltzmann solver; \cite{wang2014elucid,jasche2019physical,modi2021flowpm} for differentiable $N$-body simulators; \textsc{PineTree} \citep{ding2024pinetree} for a differentiable halo model. The second step of the procedure addresses the primary objective of inferring the cosmological parameters given the observations. Any \ac{ili} technique such as \ac{abc} or more sophisticated techniques can be used.\footnote{If the likelihood of the model has an explicit form, explicit likelihood techniques may also be used for the second step.} \ac{ili} is known to be arduous when the dimensionality of the data space is high \citep[e.g.][]{cranmer2020frontier}, and therefore requires data compression. A benefit of the statistical framework introduced by \citet{Leclercq_2022LV} is that the simulations generated in the first step naturally provide a score compressor \citep{heavens2000massive,Alsing_Wandelt_2018} for the second step at no additional cost.

In this article, we illustrate the method with a forward model of spectroscopic galaxy surveys that includes fully non-linear gravitational evolution and simulates multiple systematic effects. We use a prior embedding substantial information about the {\power}, building upon previous experiments to effectively reduce the dimensionality of the parameter space, and infer the {\power} from synthetic observations using \ac{selfi}. To make things simple whilst keeping an \ac{ili} framework, we rely on an \ac{abc} procedure using a \ac{pmc} sampler \citep{beaumont2009adaptive} to infer the posterior on cosmological parameters. Using the \ac{abc} posterior, we show that an inconspicuously misspecified model can lead to a bias greater than $2\sigma$ in the $(\Omega_\mathrm{m}, \sigma_8)$ plane. This bias can be unambiguously detected and avoided before performing the \ac{ili} of cosmological parameters. We provide a practical guide to diagnose systematic effects in implicit likelihood cosmological inferences, demonstrating how the \ac{selfi} posterior enables a comprehensive investigation of misspecified linear galaxy bias parameters, selection functions, survey masks, and inaccurate redshifts. Importantly, this process relies on a single, tractable set of $N$-body simulations. Additionally, at the cost of using distinct sets of $N$-body simulations, we are able to investigate the effect of misspecified gravity models, which is not directly apparent from the error on the galaxy count fields.

This article is structured as follows. In \Cref{sec:Method}, we introduce the \ac{bhm} used in this work and describe the two-step framework in detail, with a novel treatment of cosmological parameters within the hidden-box model compared to the original version of the \ac{selfi} algorithm. In \Cref{sec:data_model}, we present the spectroscopic galaxy survey model used in this study, which includes a non-linear physical model of structure formation and accounts for multiple observational systematic effects. In \Cref{sec:results}, we address the issue of model misspecification by investigating the impact of systematic effects on the inferred {\power}, and we discuss their impact on the posterior on cosmological parameters. Finally, we discuss our results and conclude in \Cref{sec:conclusion}.

\section{Method}
\label{sec:Method}

\subsection{Bayesian hierarchical models of galaxy surveys with the {\power} as a latent vector}
\label{subsec:BHM}

We consider a \ac{bhm} comprising the following variables: the target parameters $\boldsymbol{\upomega} \in \mathbb{R}^N$, limited to cosmological parameters in this paper, a latent vector $\boldsymbol{\uptheta}\in\mathbb{R}^S$, corresponding to the normalised values of the {\power} at $S$ wave numbers, the summary statistics $\boldsymbol{\Phi} \in \mathbb{R}^P$ of the mock galaxy populations considered in this study, and the compressed data vector $\widetilde{\boldsymbol{\upomega}} \in \mathbb{R}^N$ composed of $N$ numbers—one per target parameter. The deterministic compression step $\widetilde{\mathpzc{C}}$ linking $\boldsymbol{\Phi}$ to $\boldsymbol{\widetilde{\upomega}}$ is discussed in \Cref{subsubsec:score_compression}. A graphical representation of the generic \ac{bhm} is presented in \Cref{fig:full_BHM}; additionally, \Cref{fig:full_BHM_detailed} provides a detailed representation of the \ac{bhm} in the context of cosmological inference from probing the large-scale structure of the Universe. In applications aimed at jointly constraining cosmological and astrophysical nuisance parameters from upcoming galaxy surveys, we expect $N \sim \mathcal{O}(5-20)$ target parameters, $S \sim \mathcal{O}(10^2-10^3)$; $P$ can be as large as $\mathcal{O}(10^4)$ for complex data models exploiting information beyond $2$-point statistics, and corresponds to a first layer of compression from the $D$-dimensional full-field data, which may be as large as $D\sim\mathcal{O}(10^{11})$. \Cref{tab:variables_table} provides an overview of the different variables appearing in this section and their respective physical interpretation within the spectroscopic galaxy survey model introduced for this study.

In this work, the variables $\boldsymbol{\upomega}$ and $\boldsymbol{\uptheta}$ are linked by a deterministic Boltzmann solver, \acl{class} \citep[\class,][]{Blas2011}, which associated with the deterministic function $\mathpzc{T}$ in the \ac{bhm}. Alternative choices include the Code for Anisotropies in the Microwave Background \citep[\camb,][]{lewis2011camb}, DIfferentiable Simulations for COsmology — Done with Jax \citep[\textsc{Disco-Dj},][]{hahn2024disco}, an emulator of cosmological \psa \citep{spurio2022cosmopower,mootoovaloo2022kernel} or a fitting function such as the Eisenstein-Hu fitting function \citep{Eisenstein1998,campagne2023jax}. In either case, $\mathpzc{T}$ is theoretically well-understood and numerically inexpensive compared to the potentially misspecified part of the \ac{bhm}, which is the probabilistic simulator $\p(\boldsymbol{\Phi}|\boldsymbol{\uptheta})$ linking the {\power} $\boldsymbol{\uptheta}$ to the simulated summary statistic $\boldsymbol{\Phi}$.

\begin{table*}
\caption{\small{Main variables, their generic role in the statistical framework, and their physical interpretation in this study.}
\label{tab:variables_table}}
\centering
  \begin{tabular}{c c c}
    \hline\hline
  Symbol & Role in the BHM & Physical meaning in this work\\ \hline
  \makecell{$\boldsymbol{\upomega} \in \mathbb{R}^N$} & Target parameters & \makecell{Cosmological parameters $h$, $\Omega_\mathrm{m}$, $\Omega_\mathrm{b}$, $n_\mathrm{S}$ and $\sigma_8$, corresponding to $N=5$.}\\
  $\boldsymbol{\uptheta} \in \mathbb{R}^S$ & Latent vector & \makecell{Vectorial parametrisation of the {\power} over $S=64$\\support points.}\\
  $\boldsymbol{\uppsi} \in \mathbb{R}^T$ & Nuisance parameters & \makecell{Random numbers involved in the phase and noise realisations.}\\
  \makecell{$\boldsymbol{\mathrm{o}} \in \mathbb{R}^N$} & \makecell{Target parameters \\ treated as nuisance\\parameters} & \makecell{Cosmological parameters $h$, $\Omega_\mathrm{m}$, $\Omega_\mathrm{b}$, $n_\mathrm{S}$ and $\sigma_8$ at play inside the probabilistic\\simulator $\p(\boldsymbol{\Phi}|\boldsymbol{\uptheta})$, which is the second part of the \ac{bhm}, used for the latent\\vector inference. The symbol $\boldsymbol{\mathrm{o}}$ is introduced to avoid confusion with $\boldsymbol{\upomega}$ in the\\first part of the \ac{bhm}.}\\
  $\boldsymbol{\textbf{d}} \in \mathbb{R}^D$ & Full data vector & \makecell{Three-dimensional maps of the galaxy number counts in the survey volume for\\the three mock populations of galaxies. $D=3\cdot 512^3\simeq 4\times 10^8$.}\\
  $\boldsymbol{\Phi} \in \mathbb{R}^P$ & \makecell{Summary statistics} & \makecell{Power spectra of the galaxy number count fields. We use the concatenation of the\\three $27$-dimensional estimated \psa for each galaxy population, so $P=111$.}\\
  $\boldsymbol{\Phi}_\textrm{O} \in \mathbb{R}^P$ & \makecell{Observed summary \\ statistics} & \makecell{Power spectra of the observed galaxy number count fields. The capital letter ``O''\\stands for ``Observations''.}\\
  $\boldsymbol{\widetilde{\upomega}} \in \mathbb{R}^N$ & \makecell{Compressed \\ summaries} & \makecell{Summary statistics of the galaxy number count fields after data compression.}\\
  \hline
  \end{tabular}
\caption{\small{The variables' roles within the \ac{bhm} are defined in \Cref{subsec:BHM}. We elaborate upon their physical interpretations in \Cref{sec:data_model}, which describes the large-scale spectroscopic galaxy survey model used in this study.}}
\end{table*}

\begin{figure}
\begin{center}
\begin{tikzpicture}
	\pgfdeclarelayer{background}
	\pgfdeclarelayer{foreground}
	\pgfsetlayers{background,main,foreground}

	\tikzstyle{probability}=[draw, thick, text centered, rounded corners, minimum height=1em, minimum width=1em, fill=darkgreen!20]
	\tikzstyle{deterministic}=[draw, thick, text centered, minimum height=1.8em, minimum width=1.8em, fill=violet!20]
	\tikzstyle{variabl}=[draw, thick, text centered, circle, minimum height=1em, minimum width=1em, fill=white]
	\tikzstyle{method}=[draw=airforceblue!80, very thick, text centered, rounded corners, minimum height=1.8em, minimum width=5em, fill=babyblue!50, text=airforceblue, font=\sffamily]

	\def\blockdist{0.7}
	\def\modeldist{2.0}

    \node (omageaproba) [probability]
    {$\p(\boldsymbol{\upomega})$};
    \path (omageaproba.south)+(0,-\blockdist) node (omega) [variabl]
    {$\boldsymbol{\upomega}$};
    \path (omega.south)+(0,-\blockdist) node (thetaproba) [deterministic]
    {$\mathpzc{T}$};
    \path (thetaproba.south)+(0,-\blockdist) node (theta) [variabl]
    {$\boldsymbol{\uptheta}$};
    \path (theta.south)+(0,-\blockdist) node (phiproba) [probability]{$\p(\boldsymbol{\Phi}|\boldsymbol{\uptheta})$};
    \node [right=of phiproba, xshift=-2\blockdist] (selfi) [method]{\text{SELFI}};
    \node [right=of selfi, xshift=-2\blockdist, align=center] (sbi) [method]{Implicit\\ Likelihood\\ Inference};
    \draw[->, color=airforceblue!80, very thick] (sbi.north) |- (omega.east);
    \draw[->, color=airforceblue!80, very thick] (selfi.north) |- (theta.east);
    \path (phiproba.south)+(0,-\blockdist) node (phi) [variabl]
    {$\boldsymbol{\Phi}$};
    \draw[-, color=airforceblue!80, very thick] (phi.east) -| (selfi.south);
    \path (phi.south)+(0,-\blockdist) node (comproba) [deterministic]
    {$\widetilde{\mathpzc{C}}$};
    \path (comproba.south)+(0,-\blockdist) node (com) [variabl]
    {$\widetilde{\boldsymbol{\upomega}}$};
    \draw[-, color=airforceblue!80, very thick] (com.east) -| (sbi.south);
    \node [left=of omega, xshift=2\blockdist, align=right, color=black!80] (cosmoparam) {\text{Cosmological}\\ \text{parameters}};
    \node [left=of theta, xshift=2\blockdist, align=right, color=black!80] (power) {\text{Initial matter}\\ \text{power spectrum}};
    \node [left=of phi, xshift=2\blockdist, align=right, color=black!80] (galaxypower) {\text{Galaxy count}\\ \text{power spectra}};
    \node [left=of com, xshift=2\blockdist, align=right, color=black!80] (comsum) {\text{Compressed}\\ \text{statistics}};

	\path [draw, line width=0.7pt, arrows={-latex}] (omageaproba) -- (omega);
	\path [draw, line width=0.7pt, arrows={-latex}] (omega) -- (thetaproba);
	\path [draw, line width=0.7pt, arrows={-latex}] (thetaproba) -- (theta);
	\path [draw, line width=0.7pt, arrows={-latex}] (theta) -- (phiproba);
	\path [draw, line width=0.7pt, arrows={-latex}] (phiproba) -- (phi);
	\path [draw, line width=0.7pt, arrows={-latex}] (phi) -- (comproba);
	\path [draw, line width=0.7pt, arrows={-latex}] (comproba) -- (com);

\end{tikzpicture}
\end{center}
\caption{\small{Hierarchical representation of the Bayesian framework employed for diagnosing model misspecification and inferring the target parameters $\boldsymbol{\upomega}$. The rounded green boxes denote probability distributions, whilst the purple squares represent deterministic functions. $\p(\boldsymbol{\upomega})$ is the prior on $\boldsymbol{\upomega}$. $\mathpzc{T}$ is a deterministic function linking $\boldsymbol{\upomega}$ to $\boldsymbol{\uptheta}$. $\p(\boldsymbol{\Phi}|\boldsymbol{\uptheta})$ denotes the probabilistic data model that maps the space of latent vectors $\boldsymbol{\uptheta}$ to the survey space. The final layer, $\widetilde{\mathpzc{C}}$, is a deterministic compression step required for the \ac{ili} of the target parameters $\boldsymbol{\upomega}$. In this study, $\boldsymbol{\upomega}$ is the vector of cosmological parameters, $\mathpzc{T}$ is derived from the Boltzmann solver \ac{class}, $\boldsymbol{\uptheta}$ is the {\power} (after recombination), and $\boldsymbol{\Phi}$ corresponds to the {\psa} of multiple galaxy number count fields.
\label{fig:full_BHM}}}
\end{figure}

\begin{figure}
\begin{center}
\begin{tikzpicture}
	\pgfdeclarelayer{background}
	\pgfdeclarelayer{foreground}
	\pgfsetlayers{background,main,foreground}

	\tikzstyle{probability}=[draw, thick, text centered, rounded corners, minimum height=1em, minimum width=1em, fill=darkgreen!20]
	\tikzstyle{deterministic}=[draw, thick, text centered, minimum height=1.8em, minimum width=1.8em, fill=violet!20]
	\tikzstyle{variabl}=[draw, thick, text centered, circle, minimum height=1em, minimum width=1em, fill=white]
	\tikzstyle{plate} = [draw, thick, rectangle, rounded corners, dashed, fill=orange!10]
	\tikzstyle{comment}=[draw=airforceblue!50, thick, text centered, rounded corners, minimum height=1.8em, minimum width=5em, fill=babyblue!20, text=airforceblue, font=\sffamily]

	\def\blockdist{0.7}
	\def\modeldist{2.0}

    \path [plate] (omageaproba.north)+(-1.5*\blockdist,+0.2em) rectangle (1.5*\blockdist,-3.5*\blockdist)
    {};
    \node (omageaproba) [probability]
    {$\p(\boldsymbol{\upomega})$};
    \path (omageaproba.south)+(0,-\blockdist) node (omega) [variabl]
    {$\boldsymbol{\upomega}$};
    \path (omega.south)+(0,-\blockdist) node (T) [deterministic]
    {$\mathpzc{T}$};
    \path (T.south)+(0,-\blockdist) node (theta) [variabl]
    {$\boldsymbol{\uptheta}$};
    \path [plate] (theta.south)+(-2*\blockdist,-0.9em) rectangle (2*\blockdist,-12.5*\blockdist)
    {};
    \path (theta.south)+(-\blockdist,-1.1*\blockdist) node (omegaproba2) [probability]
    {$\p(\textbf{o})$};
    \path (theta.south)+(\blockdist,-1.1*\blockdist) node (psiproba) [probability]
    {$\p(\boldsymbol{\uppsi})$};
    \path (psiproba.south)+(0,-\blockdist) node (psi) [variabl]
    {$\boldsymbol{\uppsi}$};
    \path (omegaproba2.south)+(0,-\blockdist) node (omega2) [variabl]
    {$\textbf{o}$};
    \path (theta.south)+(0,-4*\blockdist) node (S) [deterministic]
    {$\mathpzc{S}$};
    \path (S.south)+(0,-\blockdist) node (d) [variabl]
    {$\textbf{d}$};
    \path (d.south)+(0,-\blockdist) node (C) [deterministic]
    {$\mathpzc{C}$};
    \path (C.south)+(0,-\blockdist) node (Phi) [variabl]
    {$\boldsymbol{\Phi}$};
    \path (Phi.south)+(0,-\blockdist) node (comproba) [deterministic]
    {$\widetilde{\mathpzc{C}}$};
    \path (comproba.south)+(0,-\blockdist) node (com) [variabl]
    {$\widetilde{\boldsymbol{\upomega}}$};
    \path (S.west)+(-3*\blockdist,0) node (phiproba) [probability]
    {$\p(\boldsymbol{\Phi}|\boldsymbol{\uptheta})$};
    \path (phiproba.east)+(0.7*\blockdist,0) node (phiprobaB)
    {};
    \path (omega.west)+(-2.1*\blockdist,0) node (thetaproba) [probability]
    {$\p(\boldsymbol{\uptheta})$};
    \path (thetaproba.east)+(+0.6*\blockdist,0) node (thetaprobaB)
    {};

	\path [draw, line width=0.7pt, arrows={-latex}] (omageaproba) -- (omega);
	\path [draw, line width=0.7pt, arrows={-latex}] (omega) -- (T);
	\path [draw, line width=0.7pt, arrows={-latex}] (T) -- (theta);
	\path [draw, line width=0.7pt, arrows={-latex}] (psiproba) -- (psi);
	\path [draw, line width=0.7pt, arrows={-latex}] (omegaproba2) -- (omega2);
	\path [draw, line width=0.7pt, arrows={-latex}] (theta) -- (S);
	\path [draw, line width=0.7pt, arrows={-latex}] (omega2) -- (S);
	\path [draw, line width=0.7pt, arrows={-latex}] (psi) -- (S);
	\path [draw, line width=0.7pt, arrows={-latex}] (S) -- (d);
	\path [draw, line width=0.7pt, arrows={-latex}] (d) -- (C);
	\path [draw, line width=0.7pt, arrows={-latex}] (C) -- (Phi);
	\path [draw, line width=0.7pt, dashed] (phiproba.east) -- (phiprobaB);
	\path [draw, line width=0.7pt, dashed] (thetaproba.east) -- (thetaprobaB);
	\path [draw, line width=0.7pt, arrows={-latex}] (Phi) -- (comproba);
	\path [draw, line width=0.7pt, arrows={-latex}] (comproba) -- (com);

 \draw let \p1 = (theta.south) in
       let \p2 = (thetaproba.east) in
       let \p3 = (Phi.north) in
       [airforceblue!80,
        thick,
        decorate, 
        decoration = {brace, raise=5pt, amplitude=5pt}
        ] (\x2+10em,\y1-0.9em) -- (\x2+10em,\y3+0.7em)
        node [pos=0.5,right=1.5em,black,align=center,color=airforceblue]{Probabilistic\\hidden-box\\model of the\\galaxy survey};

 \draw let \p1 = (omageaproba.north) in
       let \p2 = (thetaproba.east) in
       let \p3 = (T.south) in
       [airforceblue!80,
        thick,
        decorate, 
        decoration = {brace, raise=5pt, amplitude=5pt}
        ] (\x2+10em,\y1+0.2em) -- (\x2+10em,\y3-0.4em)
        node [pos=0.5,right=1.5em,black,align=center,color=airforceblue]{Probabilistic\\model of\\the initial\\matter power\\spectrum};
        
\end{tikzpicture}
\end{center}
\caption{\small{Detailed representation of the \ac{bhm} used in this study for the inference of cosmological parameters $\boldsymbol{\upomega}$ from galaxy surveys. The hidden-box model corresponds to the second dashed orange rectangle and defines the true (unknown) likelihood $L(\boldsymbol{\uptheta})$ when $\boldsymbol{\Phi} = \boldsymbol{\Phi}_\mathrm{O}$. The variables are $\boldsymbol{\upomega}$ (the target cosmological parameters), $\boldsymbol{\uptheta}$ (the {\power}), $\boldsymbol{\uppsi}$ (the nuisance parameters), $\textbf{o}$ (the cosmological parameters, formally treated as nuisance parameters within the hidden-box), $\textbf{d}$ (the full data vector containing the full galaxy count fields), $\boldsymbol{\Phi}$ (the summary statistics), and $\widetilde{\boldsymbol{\upomega}}$ (the compressed summaries). We provide further details about the roles and physical meanings of these variables in \Cref{tab:variables_table}.}
\label{fig:full_BHM_detailed}}
\end{figure}

\subsubsection{The top-level Bayesian problem}
\label{subsubsec:toplevel_Bayesian_problem}

From a broad perspective, the cosmological inference problem considered here consists in updating prior knowledge $\p(\boldsymbol{\upomega})$ of the cosmological parameters $\boldsymbol{\upomega}$ through the Bayes rule
\begin{equation}
\p(\boldsymbol{\upomega}|\boldsymbol{\Phi}_\mathrm{O}) = L_{\upomega}(\boldsymbol{\upomega}) \frac{\p(\boldsymbol{\upomega})}{\p(\boldsymbol{\Phi}_\mathrm{O})},
\end{equation}
based on observations $\boldsymbol{\Phi}_\mathrm{O}$ from one or multiple cosmological probes, where the true, unknown likelihood with respect to the cosmological parameters is given by
\begin{equation}
L_{\upomega}(\boldsymbol{\upomega}) \eqdef \p(\boldsymbol{\Phi}_\mathrm{O}|\boldsymbol{\upomega}),
\label{eq:true_likelihood_omega}
\end{equation}
and the prior $\p(\boldsymbol{\upomega})$ stems from previous experiments and/or encodes physical constraints derived from theoretical considerations and heuristic rules.
From this point onwards, to avoid assuming a parametric form for the likelihood $L_{\upomega}(\boldsymbol{\upomega})$, we employ \ac{ili} techniques based on a probabilistic forward model of the observable $\boldsymbol{\Phi}$. These techniques rely on comparing the simulated data, $\boldsymbol{\Phi}$, with the observations $\boldsymbol{\Phi}_\mathrm{O}$, which is difficult when the dimension $P$ of the summarised data space is high. Consequently, the observed and simulated data must undergo an additional compression step, denoted by $\widetilde{\mathpzc{C}}$ in \Cref{fig:full_BHM}.

\subsubsection{The latent vector inference}

To obtain a posterior for the {\power} $\boldsymbol{\uptheta}$ conditional on the observations $\boldsymbol{\Phi}_\mathrm{O}$, we consider the alternative inference problem defined by
\begin{equation}
\p(\boldsymbol{\uptheta}|\boldsymbol{\Phi}_\mathrm{O}) \propto L(\boldsymbol{\uptheta}) \p(\boldsymbol{\uptheta}),
\end{equation}
where
\begin{equation}
L(\boldsymbol{\uptheta}) \eqdef \p(\boldsymbol{\Phi}_\mathrm{O}|\boldsymbol{\uptheta})
\label{eq:true_likelihood_theta}
\end{equation}
is the true intermediate likelihood. We expect the inference of $\boldsymbol{\uptheta}$ to be sensitive to most of the systematic effects of interest, as the {\power} contains a wealth of information about the cosmological parameters $\boldsymbol{\upomega}$ \citep{peebles1980large}. Harnessing our theoretical understanding of $\boldsymbol{\uptheta}$, we use the posterior $\p(\boldsymbol{\uptheta}|\boldsymbol{\Phi}_\mathrm{O})$ as a diagnostic tool to examine how simulation- and observation-related systematic effects may affect the top-level Bayesian inference problem defined by \Cref{eq:true_likelihood_omega}.

\subsection{First step: Initial matter power spectrum inference}
\label{subsec:selfi}

\subsubsection{The power spectrum prior distribution}
\label{subsubsec:prior}

At wave number $k$, we define the latent vector as
\begin{equation}
    \theta_k \eqdef \frac{P(k)}{P_0(k)},
\label{eq:latent_vector}
\end{equation}
where $P(k)$ is the {\power}, and the normalisation function $P_0(k)$ is the Bardeen-Bond-Kaiser-Szalay \citep[BBKS,][]{bardeen1986statistics} power spectrum for some fiducial cosmological parameters $\boldsymbol{\upomega}_0$. The normalisation $P_0$ plays no role in the statistical framework presented here and is introduced solely for numerical stability, as initial {\psa} span several orders of magnitude across the range of scales considered.

We assume that, given a prior $\p(\boldsymbol{\upomega})$, a tight, physically-motivated prior on the {\power} $\boldsymbol{\uptheta}$ can be obtained by sampling from $\p(\boldsymbol{\upomega})$ and propagating the samples through the Boltzmann solver associated with the deterministic function $\mathpzc{T}$. That is, $\p(\boldsymbol{\uptheta}|\boldsymbol{\upomega})=\updelta^{\mathrm{D}}_{\boldsymbol{\uptheta}}(\mathpzc{T}(\boldsymbol{\upomega}))$, where $\updelta^{\mathrm{D}}_{\boldsymbol{\uptheta}}$ is the Dirac measure centred on $\boldsymbol{\uptheta}$. Marginalising over $\boldsymbol{\upomega}$ yields:
\begin{equation}
    \p(\boldsymbol{\uptheta}) = \int \updelta^\mathrm{D}_{\boldsymbol{\uptheta}}(\mathpzc{T}(\boldsymbol{\upomega})) \p(\boldsymbol{\upomega}) \, \mathrm{d}\boldsymbol{\upomega}.
\label{eq:natural_prior}
\end{equation}

To obtain an explicit form for $\p(\boldsymbol{\uptheta})$, we assume a Gaussian-distributed prior on the {\power}, whose mean and covariance matrix can be estimated by drawing from \Cref{eq:natural_prior}. We generate an ensemble of $m$ simulated \psa $\boldsymbol{\uptheta}_{\boldsymbol{\upomega}}^{(i)} = \mathpzc{T}(\boldsymbol{\upomega}^{(i)})$, $i\in\llbracket1,m\rrbracket$, where $\boldsymbol{\upomega}^{(i)}\sim\p(\boldsymbol{\upomega})$. The resulting prior reads
\begin{equation}
-2\log \p(\boldsymbol{\uptheta}) \eqdef \log\left| 2\pi \textbf{S} \right| + \left\lVert\boldsymbol{\uptheta}-\boldsymbol{\hat{\uptheta}}_{\boldsymbol{\upomega}}\right\rVert^2_{\textbf{S}^{-1}},
\label{eq:Planck_prior}
\end{equation}
where we introduced the notation $\left\lVert \boldsymbol{\mathrm{a}}\right\rVert^2_{\boldsymbol{\mathrm{B}}}\eqdef \boldsymbol{\mathrm{a}}^\intercal \boldsymbol{\mathrm{B}} \boldsymbol{\mathrm{a}}$, with
\begin{equation}
 \boldsymbol{\hat{\uptheta}}_{\boldsymbol{\upomega}} \eqdef \mathrm{E}^m \left[ \boldsymbol{\uptheta}_{\boldsymbol{\upomega}} \right] = \frac{1}{m}\sum_{i=1}^{m} \boldsymbol{\uptheta}_{\boldsymbol{\upomega}}^{(i)}
\label{eq:prior_mean}
\end{equation}
the mean of our prior, and where
\begin{align}
    \textbf{S} & \eqdef \frac{m}{m-1} \mathrm{E}^m \left[ (\boldsymbol{\uptheta}_{\boldsymbol{\upomega}} - \boldsymbol{\hat{\uptheta}}_{\boldsymbol{\upomega}})(\boldsymbol{\uptheta}_{\boldsymbol{\upomega}} - \boldsymbol{\hat{\uptheta}}_{\boldsymbol{\upomega}})^\intercal \right] \nonumber \\
   & = \frac{1}{m-1} \sum_{i=1}^m (\boldsymbol{\uptheta}_{\boldsymbol{\upomega}}^{(i)} - \boldsymbol{\hat{\uptheta}}_{\boldsymbol{\upomega}}) (\boldsymbol{\uptheta}_{\boldsymbol{\upomega}}^{(i)} - \boldsymbol{\hat{\uptheta}}_{\boldsymbol{\upomega}})^\intercal
   \label{eq:estimated_prior_covariance}
\end{align}
is the empirical prior covariance matrix. The symbol $\mathrm{E}^m$ denotes the empirical mean computed from an ensemble of $m$ samples.

Since evaluating the deterministic function $\mathpzc{T}$ is computationally inexpensive compared to the probabilistic simulator $\p(\boldsymbol{\Phi}|\boldsymbol{\uptheta})$, $m$ can be taken as large as necessary to ensure that the prior distribution $\p(\boldsymbol{\uptheta})$ is well-sampled. In this work, we use $m=10^4$.
 
\subsubsection{The power spectrum posterior distribution}
\label{subsubsec:selfi_posterior}

In the following, we adopt the approach introduced by \citet{Leclercq2019SELFI} to obtain a posterior on $\boldsymbol{\uptheta}$, with the key distinction that we formally treat the cosmological parameters $\textbf{o}$ as additional nuisance parameters within the hidden-box model, over which we marginalise. In doing so, we rigorously account for the dependence of the mapping from {\powers} to evolved dark matter density fields on the cosmological parameters.

Once realisations of $\boldsymbol{\uptheta}$, $\textbf{o}$, and $\boldsymbol{\uppsi}$ are specified, the output $\textbf{d} \in \mathbb{R}^D$ of the simulation is a deterministic function $\mathpzc{S}$; that is, $\p(\textbf{d}|\boldsymbol{\uptheta}, \textbf{o}, \boldsymbol{\uppsi}) = \updelta^{\mathrm{D}}_\textbf{d}(\mathpzc{S}(\boldsymbol{\uptheta}, \textbf{o}, \boldsymbol{\uppsi}))$, where, again, $\updelta^{\mathrm{D}}_\textbf{d}$ denotes the Dirac measure centred on $\textbf{d}$. Including the compression $\mathpzc{C}$ of the full data $\textbf{d}$ to a summary statistic $\boldsymbol{\Phi}$, the hidden-box model can therefore be expressed as $\mathpzc{B} \eqdef \mathpzc{C} \circ \mathpzc{S}$, as illustrated in \Cref{fig:full_BHM_detailed}. Marginalising over $\textbf{o}$ and $\boldsymbol{\uppsi}$, the exact likelihood given by \Cref{eq:true_likelihood_theta} reads
\begin{align}
L(\boldsymbol{\uptheta}) &= \int \p(\boldsymbol{\Phi}_\mathrm{O}|\boldsymbol{\uptheta}, \textbf{o}, \boldsymbol{\uppsi}) \p(\textbf{o}) \p(\boldsymbol{\uppsi}) \, \mathrm{d}\textbf{o}\,\mathrm{d}\boldsymbol{\uppsi} \nonumber \\
&= \int \updelta^{\mathrm{D}}_{\boldsymbol{\Phi}_\mathrm{O}}(\mathpzc{B}(\boldsymbol{\uptheta}, \textbf{o}, \boldsymbol{\uppsi})) \p(\textbf{o}) \p(\boldsymbol{\uppsi}) \, \mathrm{d}\textbf{o}\,\mathrm{d}\boldsymbol{\uppsi},
\label{eq:true_likelihood_theta_expanded}
\end{align}
which involves an intractable integral.

To overcome this difficulty, an inefficient yet insightful approach is to condition the probabilities on an ensemble of $\boldsymbol{\uptheta}$-dependent data realisations. For a given $\boldsymbol{\uptheta}$, consider an ensemble of $K$ simulated summaries $\boldsymbol{\Phi}_{\boldsymbol{\uptheta}}^{(i)} = \mathpzc{B}(\boldsymbol{\uptheta}, \textbf{o}^{(i)}, \boldsymbol{\uppsi}^{(i)})$ with $i\in\llbracket 1,\,K\rrbracket$, where $K$ is arbitrary for now. Postulating an effective Gaussian likelihood yields $\widetilde{L}^K(\boldsymbol{\uptheta}) = \exp\left[ \Tilde{\ell}^K(\boldsymbol{\uptheta}) \right]$ with the effective log-likelihood $\Tilde{\ell}^K(\boldsymbol{\uptheta})$ defined as
\begin{equation}
-2 \Tilde{\ell}^K(\boldsymbol{\uptheta}) = \log\left| 2\pi\boldsymbol{\hat{\Sigma}}_{\boldsymbol{\uptheta}}' \right| + \left\lVert\boldsymbol{\Phi}_\mathrm{O} - \boldsymbol{\hat{\Phi}}_{\boldsymbol{\uptheta}}\right\rVert^2_{\boldsymbol{\hat{\Sigma}}_{\boldsymbol{\uptheta}}'^{-1}},
\label{eq:effective_likelihood_theta}
\end{equation}
where $\boldsymbol{\hat{\Phi}}_{\boldsymbol{\uptheta}}$ is the empirical mean of the simulated summaries, defined by
\begin{equation}
\boldsymbol{\hat{\Phi}}_{\boldsymbol{\uptheta}} \eqdef \mathrm{E}^K \left[ \boldsymbol{\Phi}_{\boldsymbol{\uptheta}} \right] = \frac{1}{K}\sum_{i=1}^{K} \boldsymbol{\Phi}_{\boldsymbol{\uptheta}}^{(i)},
\label{eq:mean_phi}
\end{equation}
and the estimators of the covariance matrix of $\widetilde{L}^K(\boldsymbol{\uptheta})$ and its inverse are given by:
\begin{align}
\frac{K-1}{K+1}\boldsymbol{\hat{\Sigma}}_{\boldsymbol{\uptheta}}^\prime &\eqdef \mathrm{E}^K \left[ (\boldsymbol{\Phi}_{\boldsymbol{\uptheta}} - \boldsymbol{\hat{\Phi}}_{\boldsymbol{\uptheta}})(\boldsymbol{\Phi}_{\boldsymbol{\uptheta}} - \boldsymbol{\hat{\Phi}}_{\boldsymbol{\uptheta}})^\intercal \right] \nonumber \\
&= \frac{1}{K} \sum_{i=1}^K (\boldsymbol{\Phi}_{\boldsymbol{\uptheta}}^{(i)} - \boldsymbol{\hat{\Phi}}_{\boldsymbol{\uptheta}}) (\boldsymbol{\Phi}_{\boldsymbol{\uptheta}}^{(i)} - \boldsymbol{\hat{\Phi}}_{\boldsymbol{\uptheta}})^\intercal\,\mathrm{and} \nonumber \\
\frac{K+1}{K}\boldsymbol{\hat{\Sigma}}_{\boldsymbol{\uptheta}}^{\prime-1} &\eqdef \alpha\,\boldsymbol{\hat{\Sigma}}_{\boldsymbol{\uptheta}}^{-1},
\label{eq:estimated_covariance_phi}
\end{align}
respectively. The factor $\alpha \eqdef \dfrac{K-P-2}{K-1}$ arises from the assumption that $\boldsymbol{\Sigma}_{\boldsymbol{\uptheta}}^{-1}$ follows an inverse-Wishart distribution \citep{hartlap2007your}; the $\dfrac{K+1}{K}$ factor originates from a derivation of $\widetilde{L}^K(\boldsymbol{\uptheta})$ based on a surrogate signal proposed in \citet{Leclercq2019SELFI}, which motivates the Gaussian parametric form.

At this point, the numerical cost of computing $\widetilde{L}^K(\boldsymbol{\uptheta})$ as given by \Cref{eq:effective_likelihood_theta} is still prohibitively large when the full parameter space has to be explored: it requires $K$ simulations per $\boldsymbol{\uptheta}$ where an evaluation of the approximate likelihood is sought. The \ac{selfi} algorithm solves this issue by Taylor-expanding the full forward model at first order around an expansion point $\boldsymbol{\uptheta}_0$, leveraging prior knowledge of the {\power} so that the linearised model remains asymptotically exact around $\boldsymbol{\uptheta}_0$. We define the expansion point as
\begin{equation}
\boldsymbol{\uptheta}_0 \eqdef \mathpzc{T}(\boldsymbol{\upomega}_0),
\label{eq:expansion_point}
\end{equation}
where $\boldsymbol{\upomega}_0$ is a vector of fiducial cosmological parameters.

The aforementioned approach translates into a linearisation of the mean data model around the {\power} $\boldsymbol{\uptheta}_0$. Namely, let $\textbf{o}$ and $\boldsymbol{\uppsi}$ denote cosmological and nuisance parameters, $\delta\boldsymbol{\uptheta}$ be a small vector in the parameter space; let $\boldsymbol{\mathrm{X}}_0=\left(\boldsymbol{\uptheta}_0,\,\textbf{o},\,\boldsymbol{\uppsi}\right)$, $\boldsymbol{\mathrm{H}}=\left(\delta\boldsymbol{\uptheta},0,\,0\right)$, $\boldsymbol{\mathrm{X}}=\boldsymbol{\mathrm{X}}_0+\boldsymbol{\mathrm{H}}$. At first order in its first variable, $\mathpzc{B}\left(\boldsymbol{\mathrm{X}}\right)\eqdef \mathpzc{C} \circ \mathpzc{S}\left(\boldsymbol{\mathrm{X}}\right)$ can be approximated by
\begin{align}
    \mathpzc{B}_j\left(\boldsymbol{\mathrm{X}}\right) & \simeq \mathpzc{B}_j\left(\boldsymbol{\mathrm{X}}_0\right) + \boldsymbol{\nabla} \mathpzc{B}_j\left(\boldsymbol{\mathrm{X}}_0\right) \cdot \boldsymbol{\mathrm{H}} \nonumber \\
    & \simeq \mathpzc{B}_j\left(\boldsymbol{\mathrm{X}}_0\right) + \partial_\theta \mathpzc{B}_j\left(\boldsymbol{\mathrm{X}}_0\right) \cdot \delta\boldsymbol{\uptheta},
\end{align}
where $\mathpzc{B}_j\left(\boldsymbol{\mathrm{X}}\right)$ is the $j$-th component of $\mathpzc{B}\left(\boldsymbol{\mathrm{X}}\right)$ and $\partial_\theta$ denotes the partial derivative with respect to the {\power}. Formally treating the cosmological parameters $\textbf{o}$ intervening in the second layer of the \ac{bhm} as a nuisance and marginalising over all nuisances yields
\begin{align}
    \mathbb{E} \left[ \boldsymbol{\Phi}_{\boldsymbol{\uptheta}} \right] & \eqdef \int \mathpzc{B}\left(\boldsymbol{\mathrm{X}}\right) \p(\textbf{o}) \p(\boldsymbol{\uppsi}) \, \mathrm{d}\textbf{o}\,\mathrm{d}\boldsymbol{\uppsi} \nonumber \\
    & \simeq \int \left[\mathpzc{B}\left(\boldsymbol{\mathrm{X}}_0\right) + \partial_\theta \mathpzc{B}\left(\boldsymbol{\mathrm{X}}_0\right)\delta\boldsymbol{\uptheta}\right] \p(\textbf{o}) \p(\boldsymbol{\uppsi}) \, \mathrm{d}\textbf{o}\,\mathrm{d}\boldsymbol{\uppsi} \nonumber \\
    & \simeq \mathbb{E} \left[ \mathpzc{B}\left(\boldsymbol{\mathrm{X}}_0\right) \right] + \partial_\theta \mathbb{E}\left[ \mathpzc{B}\left(\boldsymbol{\mathrm{X}}_0\right) \right] \cdot \delta\boldsymbol{\uptheta},
    \label{eq:expectation_linear_order}
\end{align}
under the assumption that the $\mathbb{E}$ and $\partial_\theta$ operators can be permuted in \Cref{eq:expectation_linear_order}, and where the expectation is taken with respect to the joint distribution of $\textbf{o}$ and $\boldsymbol{\uppsi}$. For any $\boldsymbol{\uptheta}$, the ensemble mean $\mathbb{E} \left[ \boldsymbol{\Phi}_{\boldsymbol{\uptheta}} \right]$ can therefore be approximated by
\begin{equation}
 \textbf{f}(\boldsymbol{\uptheta}) \eqdef \textbf{f}_0 + \boldsymbol{\nabla}_{\uptheta} \textbf{f}_0 \cdot \delta\boldsymbol{\uptheta},
\label{eq:linearised_hidden_box}
\end{equation}
where $\delta\boldsymbol{\uptheta}=\boldsymbol{\uptheta}-\boldsymbol{\uptheta}_0$, $\textbf{f}_0 \eqdef \mathrm{E}^{N_0} \left[ \boldsymbol{\Phi}_{{\boldsymbol{\uptheta}}_0} \right]$ is the empirical mean of the data model at the expansion point $\boldsymbol{\uptheta}_0$ using $N_0$ simulations for different phase and other nuisance realisations, and $\boldsymbol{\nabla}_{\uptheta} \textbf{f}_0$ is the empirical gradient of $\mathpzc{B}$.

Further assuming the covariance matrix to be independent of $\boldsymbol{\uptheta}$ close to the expansion point, that is $\boldsymbol{\hat{\Sigma}}_{\boldsymbol{\uptheta}}' \simeq \boldsymbol{\hat{\Sigma}}_{\boldsymbol{\uptheta}_0}' \eqdef \textbf{C}_0$, and replacing the exact data model in \Cref{eq:effective_likelihood_theta} with the linearised data model $\textbf{f}$, the \ac{selfi} effective likelihood is given by $\widehat{L}^{N_0}(\boldsymbol{\uptheta}) = \exp\left[ \hat{\ell}^{N_0}_{\uptheta}(\boldsymbol{\uptheta}) \right]$ with
\begin{equation}
-2 \hat{\ell}^{N_0}_{\uptheta}(\boldsymbol{\uptheta}) \eqdef \log\left| 2\pi\textbf{C}_0 \right| + \left\lVert\boldsymbol{\Phi}_\mathrm{O} - \textbf{f}(\boldsymbol{\uptheta})\right\rVert^2_{\textbf{C}_0^{-1}}.
\label{eq:linearised_effective_likelihood}
\end{equation}

At the end of the day, \ac{selfi} approximates the average hidden-box model with its linearisation $\textbf{f}$ around $\boldsymbol{\uptheta}_0$ marginalised over the nuisance parameters. These assumptions, alongside the Gaussian effective likelihood assumption, solely impact the inference of the {\power} $\boldsymbol{\uptheta}$, and do not affect the inference of the target cosmological parameters $\boldsymbol{\upomega}$ discussed later, except through data compression.

The \ac{selfi} likelihood defined by \Cref{eq:linearised_effective_likelihood} is fully characterised by $\textbf{f}_0$, $\textbf{C}_0$, and $\boldsymbol{\nabla}_{\uptheta} \textbf{f}_0$, which can all be evaluated through forward simulations. The numerical computation requires $N_0$ simulations at the expansion point to evaluate $\textbf{f}_0$ and $\textbf{C}_0$. The gradient $\boldsymbol{\nabla}_{\uptheta} \textbf{f}_0$ can be evaluated using an analytical formula or auto-differentiation when available. In this work, we perform $N_s$ simulations in each direction of the parameter space to empirically estimate $\boldsymbol{\nabla}_{\uptheta} \textbf{f}_0$ via first-order forward finite differences. The total number of simulations required is therefore $N_\textrm{tot} = N_0 + N_s \times S$ simulations. $N_0$ and $N_s$ should be of the order of the dimensionality of the data space $P$, giving a total cost of $\mathcal{O}(\gtrsim P(S+1))$ model evaluations. If not chosen in advance, a precise value for $N_0$ and $N_s$ can be obtained by ensuring sufficient convergence of the covariance matrix and the gradients (see \Cref{appendix:convergence_covariance}) or by monitoring the convergence of the \ac{selfi} posterior.

To fully characterise the Bayesian problem, any prior $\p(\boldsymbol{\uptheta})$ can be used, such as the prior naturally given by \Cref{eq:natural_prior}. Any numerical technique such as MCMC can then be employed to explore the posterior using \Cref{eq:linearised_effective_likelihood}, which does not require any additional evaluation of the hidden-box model. Remarkably, if the prior is Gaussian, then the \ac{selfi} effective posterior for the {\power} is Gaussian and reads:
\begin{equation}
-2\log \p(\boldsymbol{\uptheta}|\boldsymbol{\Phi}_\mathrm{O}) \simeq \log \left| 2\pi\boldsymbol{\Gamma} \right| + \left\lVert\boldsymbol{\uptheta}-\boldsymbol{\upgamma}\right\rVert _{\boldsymbol{\Gamma}^{-1}},
\label{eq:selfi_posterior}
\end{equation}
with mean and covariance matrix given by
\begin{align}
\boldsymbol{\upgamma} &\eqdef \boldsymbol{\uptheta}_0 + \boldsymbol{\Gamma} \, (\boldsymbol{\nabla}_{\uptheta} \textbf{f}_0)^\intercal \, \textbf{C}_0^{-1} (\boldsymbol{\Phi}_\mathrm{O} - \textbf{f}_0) + \boldsymbol{\Gamma}\textbf{S}^{-1}\boldsymbol{\Delta}, \label{eq:filter_mean} \\
\boldsymbol{\Gamma} &\eqdef \left[ (\boldsymbol{\nabla}_{\uptheta} \textbf{f}_0)^\intercal \, \textbf{C}_0^{-1} \boldsymbol{\nabla}_{\uptheta} \textbf{f}_0 + \textbf{S}^{-1} \right]^{-1}.
\label{eq:filter_var}
\end{align}
$\boldsymbol{\Delta}=\boldsymbol{\hat{\uptheta}}_{\boldsymbol{\upomega}} - \boldsymbol{\uptheta}_0$ is the difference between the prior mean and the expansion point. This result was derived by \citet{Leclercq2019SELFI} for the special case where the mean of the prior is equal to the expansion point; that is, $\boldsymbol{\Delta}=0$. We provide the general derivation in \Cref{appendix:effective_posterior}. To avoid inverting $\textbf{S}$ in case it is ill-conditioned, one may equivalently opt for the alternative form:
\begin{align}
\boldsymbol{\Gamma}\textbf{S}^{-1}\boldsymbol{\Delta} &= \left[ \textbf{S}(\boldsymbol{\nabla}_{\uptheta} \textbf{f}_0)^\intercal \, \textbf{C}_0^{-1} \boldsymbol{\nabla}_{\uptheta} \textbf{f}_0 + \textbf{I} \right]^{-1}\boldsymbol{\Delta}, \\
\boldsymbol{\Gamma} &= \left[ \textbf{S}(\boldsymbol{\nabla}_{\uptheta} \textbf{f}_0)^\intercal \, \textbf{C}_0^{-1} \boldsymbol{\nabla}_{\uptheta} \textbf{f}_0 + \textbf{I} \right]^{-1}\textbf{S}.
\end{align}

\subsubsection{Check for model misspecification}

A sensitivity analysis of how systematic effects in the data model influence the posterior $\boldsymbol{\uptheta}$ can be performed by recalculating the \ac{selfi} posterior mean $\boldsymbol{\upgamma}$ using \Cref{eq:filter_mean} whilst varying the model of the systematic effect under investigation. This approach enables qualitative assessments of the impact of model misspecification on the posterior {\power}, which can be interpreted in light of theoretical considerations. For a given posterior, as a simple quantitative check for model misspecification, we compute the Mahalanobis distance between the reconstruction $\boldsymbol{\upgamma}$ and the prior distribution $\p(\boldsymbol{\uptheta})$:
\begin{equation}
d_\mathrm{M}(\boldsymbol{\upgamma}, \boldsymbol{\uptheta}_0 | \textbf{S}) \eqdef \left\lVert\boldsymbol{\upgamma}-\boldsymbol{\uptheta}_0\right\rVert_{\textbf{S}^{-1}}.
\label{eq:Mahalanobis_distance_def}
\end{equation}
We compare it to an ensemble of values of $d_\mathrm{M}(\boldsymbol{\uptheta}, \boldsymbol{\uptheta}_0 | \textbf{S})$ for simulations $\boldsymbol{\uptheta}=\mathpzc{T}(\boldsymbol{\upomega})$, where $\boldsymbol{\upomega}$ is drawn from $\p(\boldsymbol{\upomega})$.

\subsection{Second step: Implicit likelihood inference of the cosmological parameters}
\label{subsec:second_step}

\subsubsection{Score compression}
\label{subsubsec:score_compression}

The second step of the framework focuses on inferring the cosmological parameters $\boldsymbol{\upomega}$ based on the observations $\boldsymbol{\Phi}_\mathrm{O}$. As usual in \ac{ili}, the simulated and observed summaries must undergo an additional compression step to reduce their dimensionality. In essence, the compressed summaries $\widetilde{\mathpzc{C}}(\boldsymbol{\Phi})$ should closely approximate sufficient statistics of $\boldsymbol{\Phi}$, such that $\p\left(\boldsymbol{\upomega}\middle|\widetilde{\mathpzc{C}}(\boldsymbol{\Phi})\right) \simeq \p(\boldsymbol{\upomega}|\boldsymbol{\Phi})$. Following the procedure described by \citet{Leclercq_2022LV}, we assume, for compression only, that $\p(\boldsymbol{\Phi}|\boldsymbol{\upomega})$ is Gaussian-distributed: $\p(\boldsymbol{\Phi}_\mathrm{O}|\boldsymbol{\upomega}) \eqdef \exp \left[ \hat{\ell}_\omega(\boldsymbol{\upomega}) \right]$ with $\hat{\ell}_\omega(\boldsymbol{\upomega}) \eqdef \hat{\ell}(\mathpzc{T}(\boldsymbol{\upomega}))$, where
\begin{equation}
-2 \hat{\ell}_\omega(\boldsymbol{\upomega}) \eqdef \log\left| 2\pi\textbf{C}_0 \right| + \left\lVert\boldsymbol{\Phi}_\mathrm{O} - \textbf{f}\left[\mathpzc{T}(\boldsymbol{\upomega})\right]\right\rVert^2_{\textbf{C}_0^{-1}}
\label{eq:gaussian_likelihood_omega}
\end{equation}
is the effective log-likelihood defined in the first part of the framework by \Cref{eq:linearised_effective_likelihood}, which corresponds to the exact log-likelihood under the Gaussian assumption.

We consider the score function $\boldsymbol{\nabla}_{\upomega} \hat{\ell}_\omega$, which is the gradient of the log-likelihood with respect to the parameters $\boldsymbol{\upomega}$ at a fiducial point in parameter space. This function is a sufficient statistic for the log-likelihood given by Equation~\eqref{eq:gaussian_likelihood_omega} to linear order \citep{Alsing_Wandelt_2018}, making it a natural choice for data compression. Using $\boldsymbol{\upomega}_0$ as the fiducial point, a quasi maximum-likelihood estimator for the parameters is given by $\widetilde{\boldsymbol{\upomega}}_\mathrm{O} = \boldsymbol{\upomega}_0 + \textbf{F}^{-1}_0 \boldsymbol{\nabla}_{\boldsymbol{\upomega}} \hat{\ell}_\omega({\boldsymbol{\upomega}_0})$ where the Fisher matrix $\textbf{F}_0=\mathbb{E}\left[\boldsymbol{\nabla} \hat{\ell}_\omega({\boldsymbol{\upomega}_0}) \boldsymbol{\nabla}^\intercal \hat{\ell}_\omega({\boldsymbol{\upomega}_0})\right]$ represents the expected observed information, and the gradient of the log-likelihood is evaluated at $\boldsymbol{\upomega}_0$. Compression of $\boldsymbol{\Phi}_\mathrm{O}$ to $\widetilde{\boldsymbol{\upomega}}_\mathrm{O}$ yields $N$ compressed statistics, which are optimal in the sense that they saturate the Fisher information content of the data at the expansion point under lenient assumptions provided by \citet{Alsing_Wandelt_2018}.

Further assuming that the covariance $\textbf{C}_0$ does not depend on the parameters ($\boldsymbol{\nabla}_{\upomega} \textbf{C}_0 = 0$) around the expansion point, and using \Cref{eq:expansion_point}, the compressor can be approximated by
\begin{equation}
\widetilde{\mathpzc{C}}(\boldsymbol{\Phi}) \eqdef \boldsymbol{\widetilde{\upomega}} \eqdef \boldsymbol{\upomega}_0 + \textbf{F}^{-1}_0 \left[ (\boldsymbol{\nabla}_{\upomega} \textbf{f}_0)^\intercal \textbf{C}_0^{-1} (\boldsymbol{\Phi} - \textbf{f}_0) \right],
\label{eq:compression_mle}
\end{equation}
where the Fisher matrix is given by
\begin{equation}
\textbf{F}_0 \eqdef -\mathbb{E}\left[ \boldsymbol{\nabla}_{\upomega} \boldsymbol{\nabla}_{\upomega}\hat{\ell}_\omega({\boldsymbol{\upomega}_0}) \right] = (\boldsymbol{\nabla}_{\upomega} \textbf{f}_0)^\intercal \textbf{C}_0^{-1} \boldsymbol{\nabla}_{\upomega} \textbf{f}_0,
\label{eq:Fisher_matrix}
\end{equation}
with $\boldsymbol{\nabla}_{\upomega} \textbf{f}_0 = \boldsymbol{\nabla} \textbf{f}_0 \cdot \boldsymbol{\nabla}_{\upomega}\mathpzc{T}(\boldsymbol{\upomega}_0)$ \citep{heavens2000massive,Alsing_Wandelt_2018}. \citet{Alsing_Wandelt_2018} provide a more general data compression scheme which includes the case when the dependence of $\textbf{C}_0$ on the parameter cannot be neglected.

At this point, $\textbf{C}_0$ and $\boldsymbol{\nabla} \textbf{f}_0$ have already been computed for the {\power} inference with \ac{selfi}; and the gradient $\boldsymbol{\nabla}_{\upomega}\mathpzc{T}(\boldsymbol{\upomega}_0)$ derived from the Boltzmann solver can easily be computed for instance by finite difference or using auto-differentiation \citep{hahn2024disco}. We now have a \ac{bhm} that maps the $N$ target parameters, $\boldsymbol{\upomega}$, to the same number $N$ of compressed summaries, $\widetilde{\boldsymbol{\upomega}}$, as summarised in \Cref{fig:full_BHM}, which can be used to infer the target cosmological parameters using standard \ac{ili} techniques.

\subsubsection{Implicit likelihood inference with a population Monte Carlo sampler}
\label{subsubsec:PMC}

Every \ac{ili} method, such as \ac{abc}, requires a discrepancy measure to compare the compressed simulated summaries $\widetilde{\boldsymbol{\upomega}}$ with the observed summaries $\widetilde{\boldsymbol{\upomega}}_\mathrm{O}$ \citep{beaumont2002approximate,beaumont2019approximate}. For this purpose, we re-used the Fisher matrix from \Cref{eq:Fisher_matrix} to compute the Fisher-Rao distance between the compressed observed and simulated data, defined by:
\begin{equation}
d_\mathrm{FR}(\widetilde{\boldsymbol{\upomega}}, \widetilde{\boldsymbol{\upomega}}_\mathrm{O}) \eqdef \left\lVert\widetilde{\boldsymbol{\upomega}}-\widetilde{\boldsymbol{\upomega}}_\mathrm{O}\right\rVert_{\textbf{F}_0}.
\label{eq:FisherRao_distance}
\end{equation}

To infer the target parameters $\widetilde{\boldsymbol{\upomega}}$ given the observed summaries, $\widetilde{\boldsymbol{\upomega}}_\mathrm{O}$, we relied on a simple yet theoretically well-motivated particle filter method, which is a variant introduced by \citet{Simola2021} of \ac{pmc}. Unlike the basic \ac{abc} sampling strategy, known as implicit likelihood rejection sampling \citep{neumann1951various, beaumont2002approximate}, and which does not take advantage of the information acquired during the sampling process, particle filters iteratively refine pools of candidates–referred to as particles, bringing them closer and closer to the true posterior distribution. Specifically, \ac{pmc} methods gradually enhance the quality of the samples by updating the proposal distribution.

The \ac{pmc} algorithm variant introduced by \citet{Simola2021} and employed here proceeds as follows. First, we set an initial acceptance threshold $\epsilon_0$, and generated an initial pool of $J$ particles \mbox{$\Xi^{(0)} \eqdef \left\{\boldsymbol{\upomega}_i^{(0)}\right\}_{i\in\llbracket 1,\,J\rrbracket}$} by drawing cosmological parameters from the proposal distribution $\pi^{(0)}(\boldsymbol{\upomega})\eqdef\p(\boldsymbol{\upomega})$ until they satisfied \mbox{$d_\mathrm{FR}\left(\widetilde{\mathpzc{C}} \circ \mathpzc{B}\left(\boldsymbol{\upomega}_i^{(0)}\right), \widetilde{\boldsymbol{\upomega}}_\mathrm{O}\right)<\epsilon_0$}. This initial step is equivalent to an implicit likelihood rejection sampling stage with $J$ accepted samples and a tolerance $\epsilon_0$.

To construct an intermediate proposal distribution $\pi^{(1)}(\boldsymbol{\upomega})$ and the corresponding approximate posterior, the method uses a Gaussian mixture distribution. We set initial weights $w_i^{(0)}\eqdef J^{-1}$. The initial multivariate Gaussian kernel functions are centred on $\boldsymbol{\upomega}_i^{(0)}$, with a common covariance matrix defined as twice the weighted sample covariance of the particles $\Xi^{(0)}$; that is,
\begin{equation}
\boldsymbol{\Sigma}^{(0)}=2\,\sum_{i=1}^{J} w_i^{(0)} \left(\boldsymbol{\upomega}_i^{(0)} - \boldsymbol{\widehat{\boldsymbol{\upomega}}}^{(0)}\right)\left(\boldsymbol{\upomega}_i^{(0)} - \boldsymbol{\widehat{\boldsymbol{\upomega}}}^{(0)}\right)^\intercal,
\end{equation}
where $\boldsymbol{\widehat{\boldsymbol{\upomega}}}^{(0)}$ is the sample mean of the particles $\Xi^{(0)}$. The factor $2$ arises from minimising the Kullback-Leibler divergence \citep{kullback1951information} between the proposal and the target distribution \citep{beaumont2009adaptive}. This leads to the updated proposal distribution
\begin{equation}
\pi^{(1)}(\boldsymbol{\upomega})=\sum_{i=1}^{J} w_i^{(0)}\mathcal{G}\left(\boldsymbol{\upomega}\left\vert\boldsymbol{\upomega}_i^{(0)},\boldsymbol{\Sigma}^{(0)}\right.\right),
\end{equation}
where $\mathcal{G}\left(\boldsymbol{\upomega}\left\vert\boldsymbol{\upmu},\boldsymbol{\Sigma}\right.\right)$ denotes the \ac{pdf} of a Gaussian centred on $\boldsymbol{\upmu}$ with covariance matrix $\boldsymbol{\Sigma}$, evaluated at $\boldsymbol{\upomega}$.

In subsequent iterations $t$, the algorithm proceeds by drawing particles from the approximate posterior proposal distribution $\pi^{(t)}(\boldsymbol{\upomega})$ until they satisfy \mbox{$d_\mathrm{FR}\left(\widetilde{\mathpzc{C}} \circ \mathpzc{B}\left(\boldsymbol{\upomega}_i^{(t)}\right), \widetilde{\boldsymbol{\upomega}}_\mathrm{O}\right)<\epsilon_t$} for some $\epsilon_t<\epsilon_{t-1}$. This results in the updated pool of particles $\Xi^{(t)}=\left\{\boldsymbol{\upomega}_i^{(t)}\right\}_{i\in\llbracket 1,\,J\rrbracket}$, along with an updated proposal $\pi^{(t+1)}(\boldsymbol{\upomega})$ given by the Gaussian mixture distribution
\begin{equation}
\pi^{(t+1)}(\boldsymbol{\upomega})=\sum_{i=1}^{J} w_i^{(t)} \mathcal{G}\left(\boldsymbol{\upomega}\left\vert\boldsymbol{\upomega}_i^{(t)},\boldsymbol{\Sigma}^{(t)}\right.\right),
\end{equation}
where the unnormalised importance weights are
\begin{equation}
    \Tilde{w}_i^{(t)}\eqdef \dfrac{\pi^{(t)}(\boldsymbol{\upomega}_i^{(t)})}{\displaystyle\sum_{j=1}^Jw_j^{(t-1)}\mathcal{G}\left(\boldsymbol{\upomega}_i^{(t)}\left\vert\boldsymbol{\upomega}_j^{(t-1)},\boldsymbol{\Sigma}^{(t-1)}\right.\right)},
\end{equation}
so that $w_i^{(t)}=\Tilde{w}_i^{(t)}/\sum_j{\Tilde{w}_j^{(t)}}$. The covariance matrix $\boldsymbol{\Sigma}^{(t)}$ is, again, defined as twice the weighted sample covariance of the particles $\Xi^{(t)}$. The resulting approximate posterior can thus be expressed as a kernel density estimator:
\begin{equation}
\hat{\pi}_{\epsilon_t,h}\left(\boldsymbol{\upomega}\middle|\widetilde{\boldsymbol{\upomega}}_\mathrm{O}\right)=\displaystyle\sum_{i=1}^J w_i^{(t)} k_h\left(\boldsymbol{\upomega}-\boldsymbol{\upomega}_i^{(t)}\right),
\label{eq:approximate_abc_posterior}
\end{equation}
where $k_h$ is the kernel \citep[e.g.][]{rosenblatt1956remarks, parzen1962estimation}, and where we replaced the exponent $(t)$ with $\epsilon_t$ to emphasise the dependence on the acceptance threshold. In this study, we used an isotropic Gaussian kernel with a smoothing parameter $h$. The particles were drawn from sequentially improving proposal distributions, with decreasing tolerances \mbox{$\epsilon_0>\epsilon_1>\dots>\epsilon_T$}. The algorithm starts with an initial acceptance quantile $q_0$, which is fixed by the initial tolerance $\epsilon_0$, and terminates when $q_T$ falls below a predetermined target quantile $q$. The target $q$ sets the maximum allowable pointwise difference between posteriors at successive stages for the algorithm to be deemed converged. The rule for determining $T$ and the choice of the decreasing sequence of tolerances are discussed in detail in \citet{Simola2021}. The sequence is designed to ensure efficient sampling and to provide good exploration of the relevant regions of the parameter space. Starting from the initial acceptance quantile $q_0$, subsequent tolerances were determined as follows:
\begin{align}
\epsilon_t &= \displaystyle\min_{i \in \llbracket 1, J \rrbracket} \left\{ d_\mathrm{FR}\left(\widetilde{\mathpzc{C}} \circ \mathpzc{B}\left(\boldsymbol{\upomega}_i^{(t-1)}\right), \widetilde{\boldsymbol{\upomega}}_\mathrm{O}\right) \right\}, \\
q_t &= \left( \sup_{\boldsymbol{\upomega}} \dfrac{\widetilde{\pi}_{\epsilon_t}\left(\boldsymbol{\upomega}\middle|\widetilde{\boldsymbol{\upomega}}_\mathrm{O}\right)}{\widetilde{\pi}_{\epsilon_{t-1}}\left(\boldsymbol{\upomega}\middle|\widetilde{\boldsymbol{\upomega}}_\mathrm{O}\right)} \right)^{-1},
\end{align}
where $\widetilde{\pi}_{\epsilon_t}/\widetilde{\pi}_{\epsilon_{t-1}}$ is a density ratio estimate of the posterior change between steps $t-1$ and $t$. Notably, substantial differences between the approximate posteriors at stages $t-1$ and $t$ lead to a large shrinkage of the proposed tolerance $\epsilon_{t+1}$. The shrinkage in tolerance gradually decreases as the \ac{abc} posterior converges.

\section{Data model of galaxy surveys}
\label{sec:data_model}

In the following, Model A refers to the correct, well-specified model used to generate the synthetic observations, whilst Model B refers to a misspecified model.

\subsection{Initial power spectrum parametrisation}
\label{subsec:initial_power_spectrum}

Throughout this work, we defined the initial density fluctuations and galaxy overdensity fields on a cubic equidistant grid with a comoving side length of $3.6~\mathrm{Gpc}/h$ and $512^3$ voxels, spanning scales between $k_{s,\mathrm{min}} = 1.75 \times 10^{-3}$~$h$/Mpc and $k_{s,\mathrm{max}} = 7.74 \times 10^{-1}$~$h$/Mpc. We discretised the parameter space by evaluating the continuous normalised {\psa} at $S=64$ support wave numbers $k_s$, yielding the vector $\boldsymbol{\uptheta}$ defined by \Cref{eq:latent_vector}. Between two consecutive support wave numbers, we interpolated {\psa} $P(k)$ using quintic splines. The first eight support wave numbers match the largest modes in the Fourier grid of our set-up. We manually distributed the remaining support wave numbers up to $k_{s,\mathrm{max}}$ to ensure a small maximum relative error in the representation of initial matter {\psa} across all wave numbers of the Fourier grid. We verified that, for all $k$ below the grid’s Nyquist frequency, this set-up yielded a relative interpolation error below $0.14\%$ for the fiducial power spectrum.

\subsection{Gravitational evolution}
\label{subsec:gravitational_evolution}

The data model is a non-linear process meant to approximate the large variety of physical and observational phenomena at play in actual galaxy surveys. To generate dark matter overdensity fields for a given set of cosmological parameters, we essentially followed the procedure described by \citet{Leclercq2019SELFI}. We assumed a flat $\Lambda$CDM cosmology \citep{blumenthal1984formation} and used the best fit of the \ac{planck} data, given in \Cref{tab:cosmo_fiducial}, as the fiducial cosmological parameters $\boldsymbol{\upomega}_0$.
\begin{table}[ht]
\caption{\small{Fiducial cosmological parameters $\boldsymbol{\upomega}_0$.}
\label{tab:cosmo_fiducial}}
\centering
\begin{tabular}{c c c c c }
\hline\hline
$h$ & $\Omega_\mathrm{b}\,h^2$&  $\Omega_\mathrm{m}$ & $n_\mathrm{S}$& $\sigma_8$ \\ \hline
0.6766 & 0.02242  & 0.3111 & 0.9665  & 0.8102 \\
\hline
\end{tabular}
\caption{\small{Values taken from \citet{aghanim2020planck}.}}
\end{table}

Given an {\power} $P(k)$, we generated a corresponding realisation of the initial matter density contrast field $\boldsymbol{\updelta}^\mathrm{i}$. For the gravitational evolution, we used the \textsc{Simbelmyn\"e} cosmological solver \citep[][]{Leclercq2015ST}. We populated the initial grid of $512^3$ voxels with $1024^3$ dark matter particles arranged on a regular lattice. These particles were evolved via \ac{2lpt} \citep{Moutarde1991,bouchet1994perturbative} up to redshift $z_\textrm{LPT}=19$, followed by $N$-body evolution with \acs{cola} \citep[\acl{cola},][]{Tassev2013} from $z_\textrm{LPT}$ to $z_1=0.1500$, both implemented within \textsc{Simbelmyn\"e}. We performed the \ac{cola} evolution on a particle-mesh grid of $1024^3$ voxels. We used 20 non-uniform time steps, which we distributed to ensure approximately linear scaling in the scale factor $a$ while meeting the requirements imposed by our choice of mock galaxy populations. Specifically, the first 14 time steps were linearly spaced in $a$ up to redshift $z_3=0.8182$, followed by four time steps linearly spaced in $a$ up to redshift $z_2=0.4925$, and finally two additional time steps to reach $z_1=0.1500$, corresponding to a scale factor of $a=0.8696$.

\begin{figure*}
\begin{center}
\includegraphics[width=.89\textwidth]{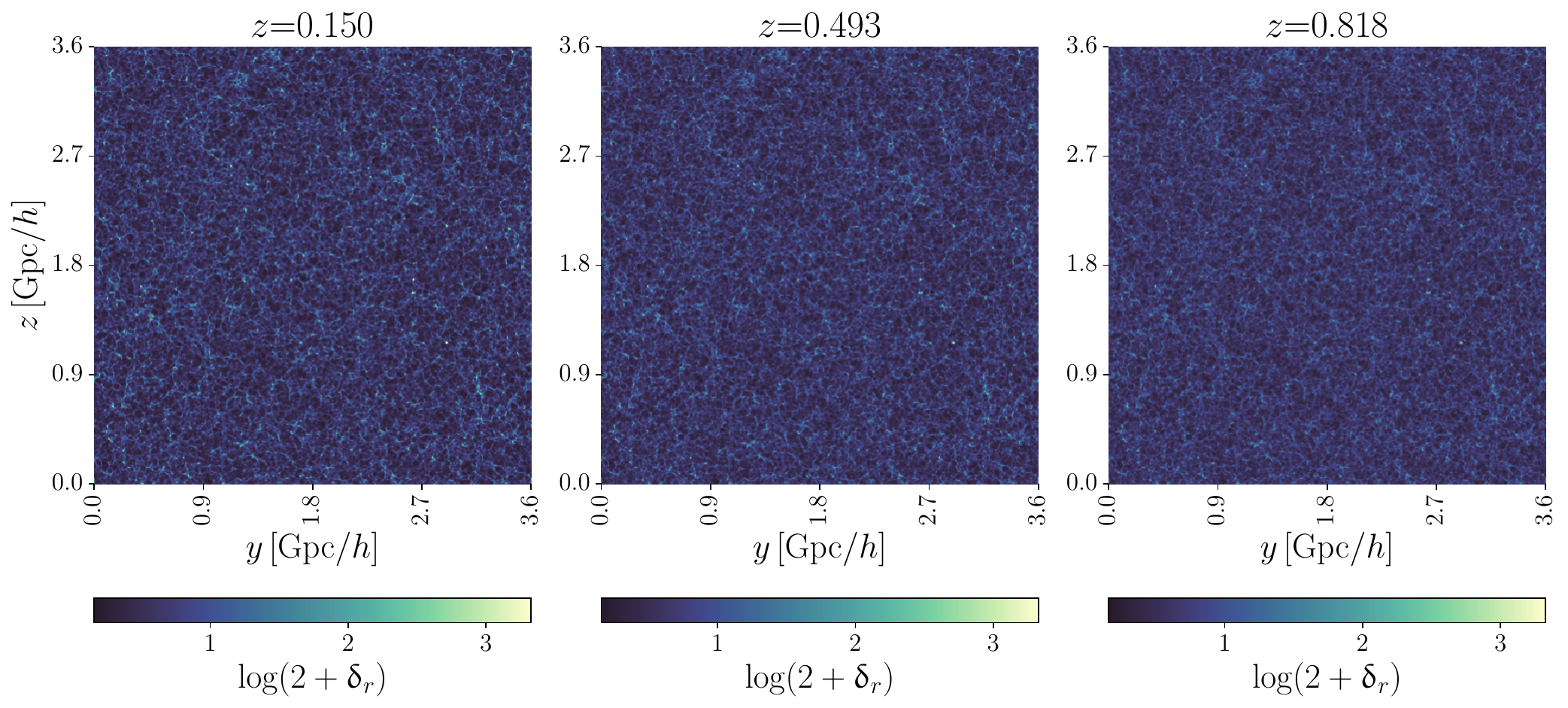}\\
\includegraphics[width=.89\textwidth]{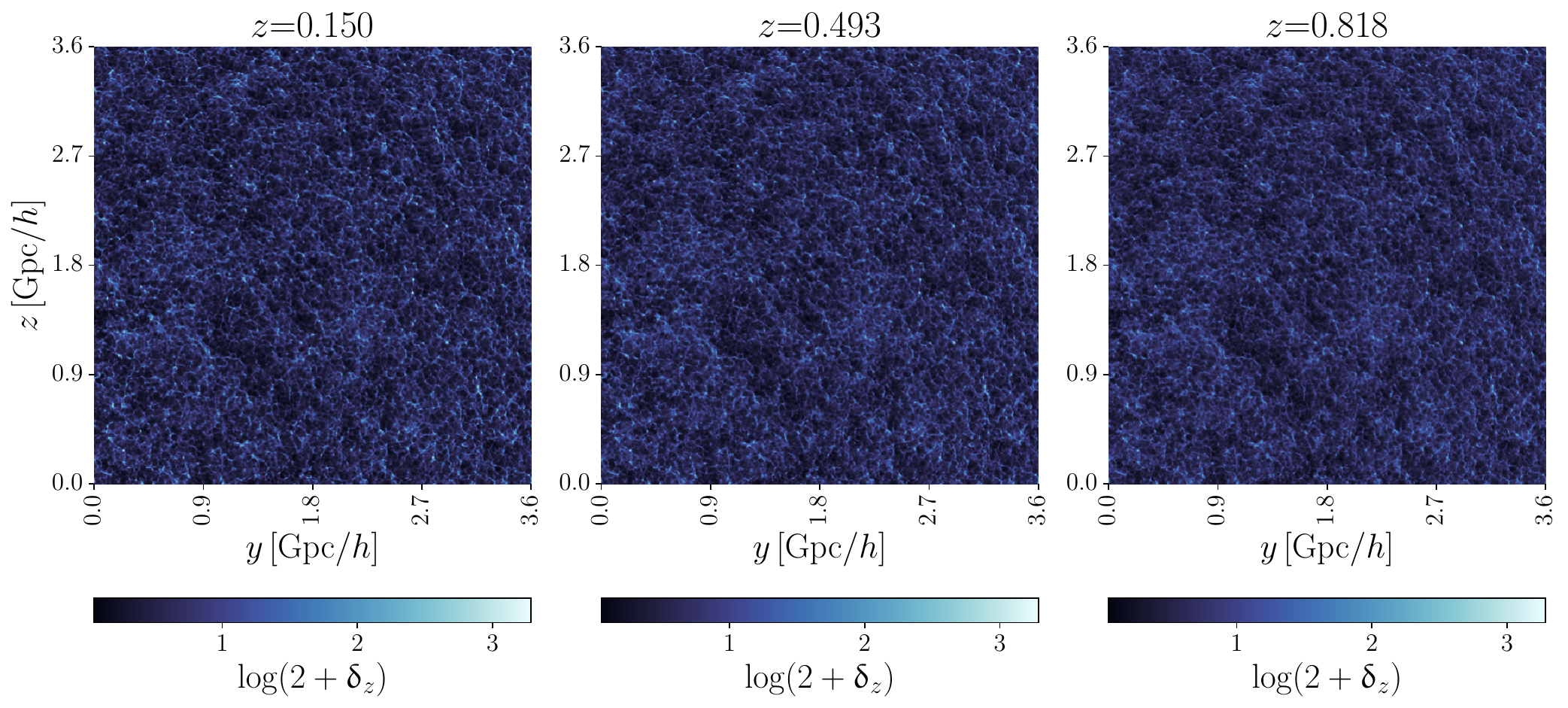}\\
\includegraphics[width=.89\textwidth]{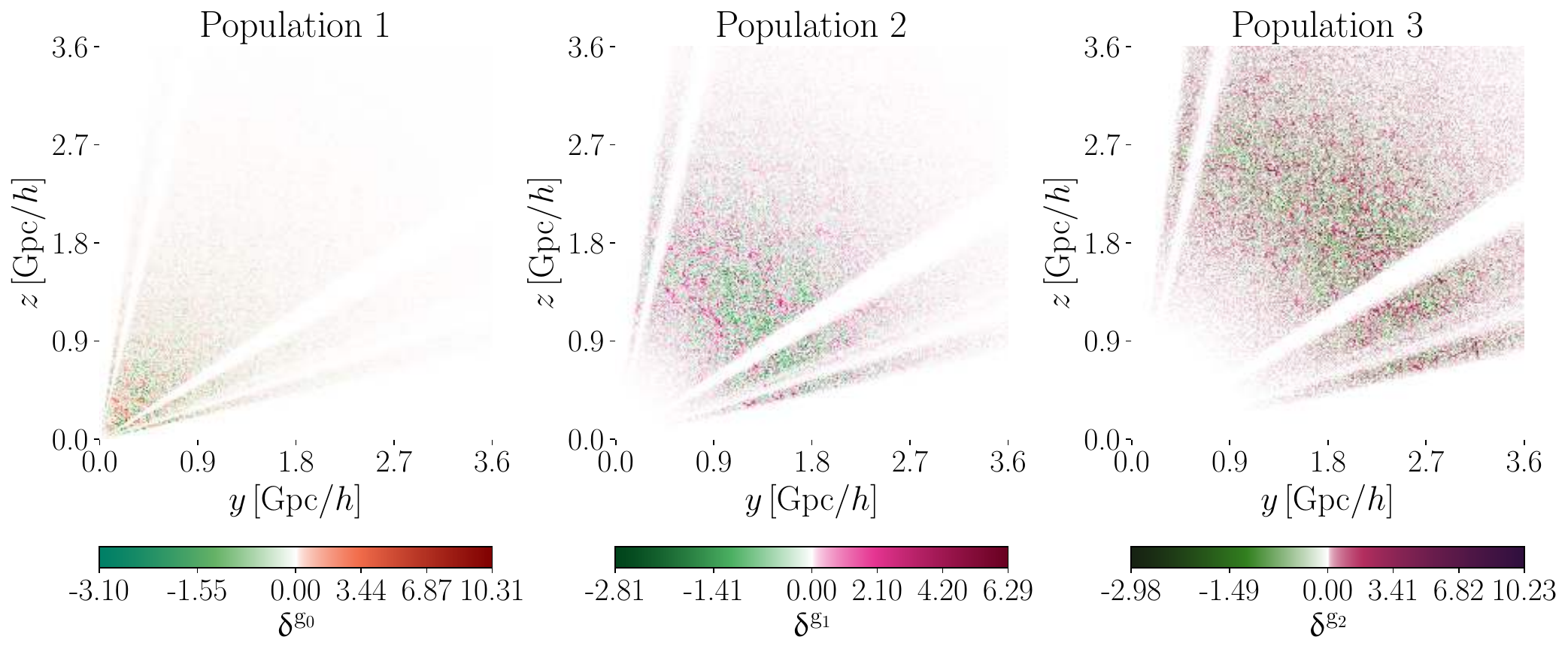}
\caption{\small{Illustration of the galaxy survey data model used in this study. (Upper and middle rows) Slices through a realisation of the dark matter overdensity fields in real space \textit{(top)} and redshift space \textit{(middle)}. The fields are evolved in time using $N$-body simulations, starting from a grid of Gaussian initial dark matter density fluctuations with power spectrum $\boldsymbol{\uptheta}_{\textrm{gt}}$, up to the redshift indicated above the corresponding panels. (Lower row) Observed galaxy number count fields for the three galaxy populations, each defined on a grid of $512^3$ cells. Light-cone effects are accounted for by using one effective redshift per galaxy population. Each slice passes through the observer located at the $(0,0,0)$ corner of the box.}
\label{fig:slices}}
\end{center}
\end{figure*}

The observer is positioned at one corner of the box, whereby they see one octant of the sky. Using the radial component  $v_r$ of their final peculiar velocities with respect to the observer, the dark matter particles are placed in redshift space according to the non-linear mapping defined for each particle by
\begin{equation}
1+z_\mathrm{obs} = (1+z_\mathrm{cosmo})(1+z_\mathrm{pec}), \enspace \enspace  z_\mathrm{pec} \eqdef -\frac{v_r}{\mathrm{c}},
\end{equation}
where $z_\mathrm{cosmo}$ is the true cosmological redshift, $z_\mathrm{pec}$ is the redshift due to the peculiar velocity, $z_\mathrm{obs}$ is the ``observed'' redshift of the particle and $\mathrm{c}$ is the speed of light. The particles are then assigned to a $512^3$-voxels grid using the cloud-in-cell scheme \citep{Hockney1981} to give the density contrast fields at redshift $z_i$, denoted  $\boldsymbol{\updelta}^{z_i}$, illustrated in the upper and middle rows of \Cref{fig:slices}.

\subsection{Galaxy biasing}

Galaxies are biased tracers of the underlying dark matter distribution \citep{kaiser1984spatial, bardeen1986statistics}. The mapping from local dark matter overdensities to the galaxy density field is a crucial element in cosmological inferences, involving numerous physical processes. Multiple questions, such as halo formation and mergers, gas cooling, and stellar feedback \citep[][]{Desjacques_2018}, have to be addressed. Despite its inherent complexity, the highly non-linear process of galaxy formation can be accurately described by a small number of bias parameters on linear and quasi-linear scales, through a bias expansion in the matter density contrast and tidal fields. The parameters of this expansion often exhibit degeneracies with cosmological parameters, making the bias model critical in cosmological inferences, with the potential to either significantly enhance or severely bias the constraints on cosmological parameters \citep[][]{Barreira_2020_2, Barreira_2020_1}. Consequently, exploiting galaxy survey data beyond \ac{baos} requires careful modelling of galaxy biases \citep[][]{Barreira_2021,krause2024parameter}, and there remains considerable scope for improving the bias models employed in cosmological analyses \citep{bartlett2024bye}.

In this study, we focus on linear galaxy bias parameters, usually denoted $b_1$ in the literature, and use our framework to assess the impact of these bias parameters on the \ac{selfi} posterior on the {\power}. The same statistical framework can be applied to study the impact of higher-order \ac{limd} bias parameters or tidal bias parameters on the {\power} reconstruction. We leave this exploration for future research. Hereafter, the first-order \ac{limd} bias parameter for the $i$-th galaxy population $\textrm{g}_i$ is denoted $b_{\mathrm{g}_i}$, to avoid confusion of $b_{\mathrm{g}_2}$ and $b_{\mathrm{g}_3}$ with higher order bias parameters.

We simulated three mock galaxy populations: one population of nearby bright galaxies and two populations of \ac{lrgs}. \ac{lrgs} are slowly evolving galaxies that have been extensively studied; due to their high spatial densities and intrinsic brightness, they are among the most valuable tracers of the dark matter distribution \citep{eisenstein2001spectroscopic} in the Universe. Notably, subsamples of the $1.5$ million spectra from the Baryon Oscillation Spectroscopic Survey \citep[][]{eisenstein2015baryon} have been prominently used to study the large-scale structure of the Universe, proving able to measure the \ac{baos} in their large-scale distribution \citep{eisenstein2005detection, anderson2014clustering}. \ac{lrgs} are notably the primary target of DESI within the redshift range $0.4<z<1.0$ \citep{zhou2023target}.

We approximated light cone effects by using snapshots from the dark matter overdensity field at three distinct redshifts $z_1,\,z_2$, and $z_3$, as input to the linear bias model to define the three populations of galaxies, where $z_i$ represents the effective redshift of the $i$-th galaxy population $\textrm{g}_i$. According to the linear bias model, the normalised galaxy density $\boldsymbol{\uprho}^{\mathrm{g}_i}$ for the galaxy population $\textrm{g}_i$ is therefore given in any cell $x$ by:
\begin{equation}
\rho_x^{\mathrm{g}_i} = \bar{N}\,(1+ b_{\mathrm{g}_i} \, \delta_x^{z_i}).
\label{eq:galaxy_density}
\end{equation}
For simplicity, we used fixed values for the $b_{\mathrm{g}_i}$ parameters, given in \Cref{tab:linear_bias}. We emphasise that, given any prior distribution for these parameters, they could be treated as additional nuisances alongside $\boldsymbol{\uppsi}$ and marginalised over, without affecting the statistical framework presented in this article. To showcase the efficiency of our method for diagnosing systematic effects with high signal-to-noise ratio data, we made the simplifying assumption that $\bar{N}=1$, effectively adopting a unit system where the mean number of galaxies per cell equals unity, which in turn affects the stochastic properties of the galaxy count fields, as detailed in \Cref{subsec:obs}. We further assumed that the contrast,
\begin{equation}
    \delta_x^{\mathrm{g}_i}=b_{\mathrm{g}_i} \, \delta_x^{z_i},
    \label{eq:density_contrast}
\end{equation}
could be observed directly, implying in particular that the mean number of galaxies per cell is perfectly known.

\begin{table}[ht]
\caption{\small{Linear galaxy bias parameters used in this study.}\label{tab:linear_bias}}
\centering
    \begin{tabular}{c c c c}
    \hline\hline
    Model & Population 1 &  Population 2 & Population 3 \\ \hline
	A & 1.47 & 1.99  & 2.32 \\
	B & 1.50 & 1.98  & 2.33 \\
    \hline
    \end{tabular}
\end{table}

To simulate LRG populations, we based our approach on the measurements reported by \citet{gil2015power}, where the authors found $b_1^{1.40}(z_\text{eff})\sigma_8(z_{\text{eff}})=1.672\pm 0.060$ and $f\sigma_8(z_{\text{eff}})=0.597$ for $z_{\text{eff}}=0.57$ (best fit of their full sample), yielding $b_1^{1.40}(z_\text{eff})=2.80\pm 0.10$. Motivated by the $(b_1,D)$ degeneracy between the linear galaxy bias and the growth factor at large scales, we assumed constant $b_1(z)D(z)$ within the relevant redshift range, yielding:
\begin{equation}
b_1(z) \simeq \frac{b_1(z_\text{eff})\times D(z_{\text{eff}})}{D(z)} \simeq \frac{2.086\times D(z_{\text{eff}})}{D(z)}.
\label{eq:biases_LRG}
\end{equation}
For the two \ac{lrgs} populations, $\textrm{g}_2$ and $\textrm{g}_3$, we applied \Cref{eq:biases_LRG} to the mean redshift of each bin, used as the effective redshift to estimate an approximate ground truth linear bias (see \Cref{fig:radial_selection,tab:linear_bias}). For nearby galaxies, we used the measurements $f\sigma_8(z_1=0.15)=0.43^{+0.15}_{-0.14}$ and $b_1\sigma_8(z_1=0.15)=1.20\pm 0.15$ provided for a sample of high-bias galaxies at $z<0.2$ by \citet{HowlettRoss2015}. Assuming $f=\Omega_\textrm{m}^{0.55}$ \citep[][]{bouchet1994perturbative}, we obtained $b_1(z_1=0.15)\simeq 1.45$. For Model A, we randomly selected the ground truth values in a $\pm 0.02$ interval around those given by the above equations. For Model B, the misspecified values were chosen manually. They are reported in \Cref{tab:linear_bias}.

\subsection{Observational processes}
\label{subsec:obs}

The last step involves a virtual observation of the galaxy fields, accounting for observational effects expected in actual surveys. One such effect is dust extinction. Dust particles in the Milky Way and the intergalactic medium, presumably ejected by stars, absorb a significant portion of UV-to-near-infrared light, contaminating observations, causing reddening that must be corrected for \citep{galametz2017sed}, and affecting the noise properties of the observed galaxies \citep{ho2015sloan}. The presence of gas and plasma in the intra- and extra-galactic medium introduces another known source of attenuation, as these media absorb, scatter, and re-emit a portion of incident radiation at longer wavelengths \citep{menard2008lensing, more2009cosmic}.

These attenuation phenomena, along with contamination from bright stars, constitute complex and often correlated systematic effects \citep{boulanger1996dust} which can significantly impact nearly all cosmological measurements, from supernova distance estimates \citep{brout2024impact} to photometric and spectroscopic galaxy surveys \citep{corasaniti2006impact, leistedt2014exploiting, ho2015sloan, bovy2016galactic}. Consequently, considerable effort is devoted to accurately modelling and accounting for dust extinction in cosmological inference pipelines \citep[e.g.][]{huterer2013calibration, jasche2017bayesian, karchev2024side}.

\begin{figure}
\begin{center}
\includegraphics[width=.5\textwidth]{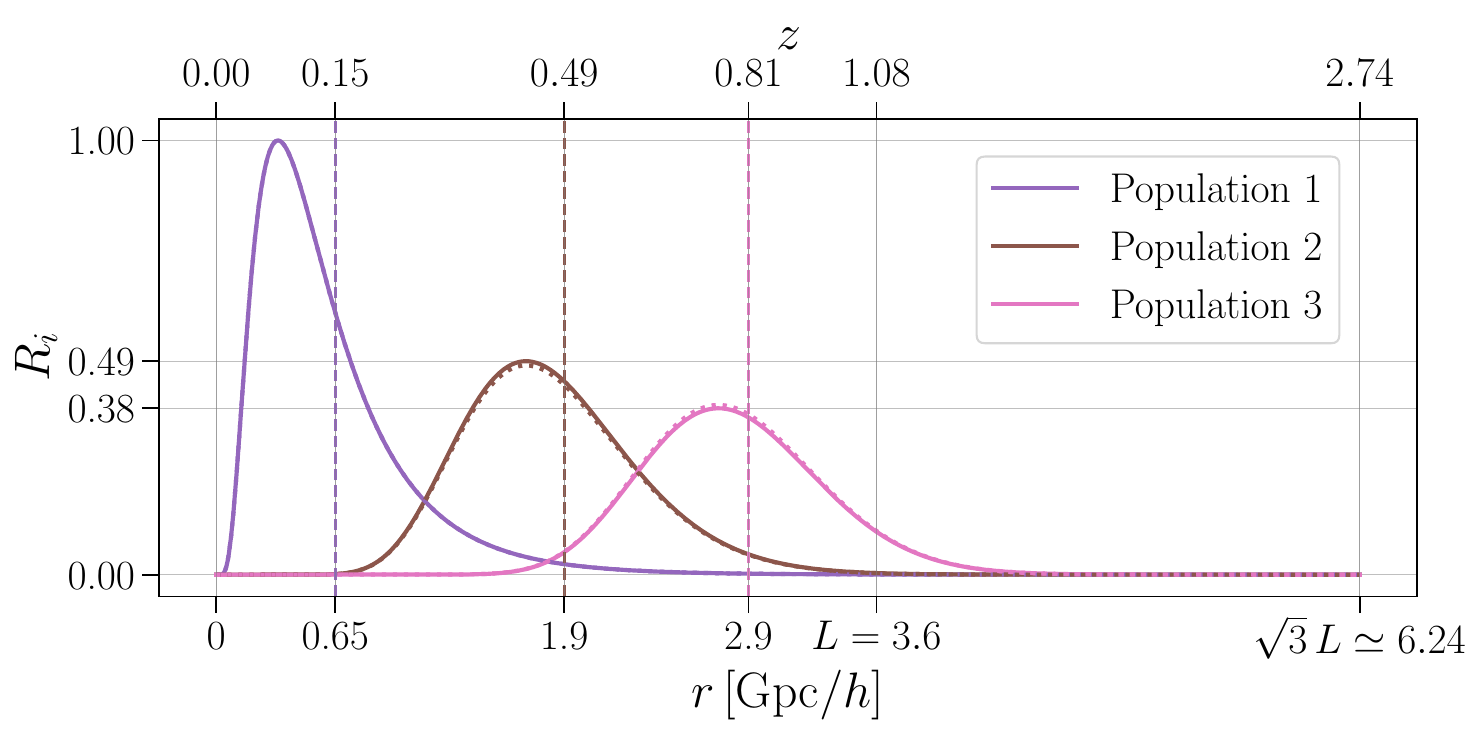}
\caption{\small{Radial selection functions modelled as log-normal distributions in redshift (upper $x$ axis) for the three mock galaxy populations. In the figure, the distributions are represented with respect to comoving distances (lower $x$ axis) using the redshift-distance relation. The continuous lines represent the selection functions for the well-specified Model A, while the dotted lines represent the misspecified Model B, both defined by the values provided in \Cref{tab:selection}. The small, percent-order difference between Models A and B renders the dotted curves nearly indistinguishable from the solid ones. The vertical dashed lines indicate the means of the log-normal distributions.}\label{fig:radial_selection}}
\end{center}
\end{figure}

\begin{figure}
\begin{center}
\includegraphics[width=.5\textwidth]{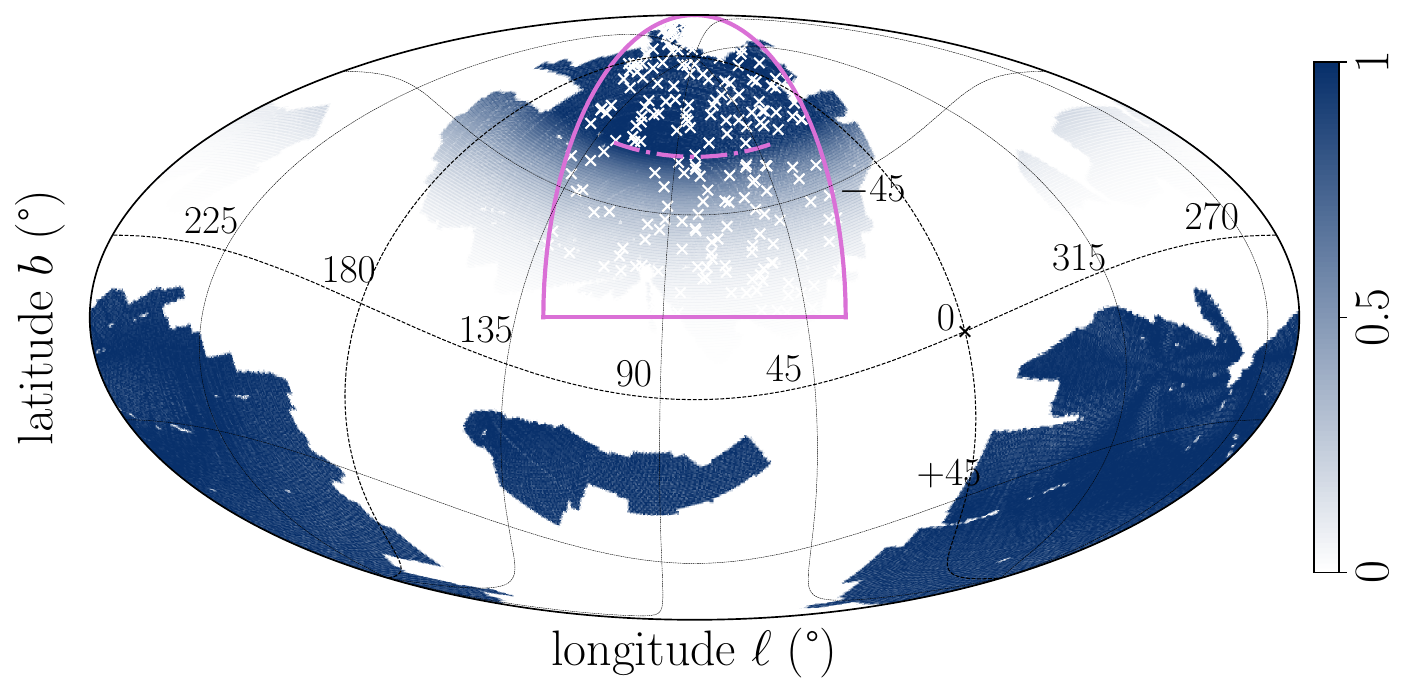}

\caption{\small{Survey mask for the well-specified Model A. We added extinction linearly decreasing from $100\%$ at the galactic plane to $0\%$ at $60\degree$ galactic latitude (dash-dotted orchid line), along with 256 holes of approximately $0.1$-degree radius (white crosses). The observed octant is delineated by the light orchid triangle. The colour scale represents the completeness function $C(\hat{\textbf{n}})$, with $0$ indicating masked regions and $1$ indicating unmasked regions, and any value in between indicating partial completeness. In the misspecified Model B, the same mask is used, except for the linear extinction which extends from $0$ to $59\degree$ galactic latitude.}\label{fig:mask}}
\end{center}
\end{figure}

To obtain the simulated data from the galaxy contrast fields, we computed the three-dimensional survey response operators $\textbf{W}_i$. For each galaxy population $\textrm{g}_i$, the corresponding operator is defined as the product of the radial selection function, $R_i(r)$, and the angular survey mask and completeness function $C(\hat{\textbf{n}})$ for any line of sight $\hat{\textbf{n}}$; that is,
\begin{equation}
    \textbf{W}_i(\hat{\textbf{n}},r) \eqdef R_i(r) \, C(\hat{\textbf{n}}).
    \label{eq:window}
\end{equation}
For the angular completeness $C(\hat{\textbf{n}})$, we used a mask mimicking the wide sky coverage of stage-IV experiments, shown in \Cref{fig:mask}. We introduced additional linear extinction near the galactic plane up to $60\degree$ galactic latitude as a simple model of dust extinction. We also introduced 256 randomly positioned holes of angular diameter $\sim 0.1\degree$, representing the masking of bright stars or other point sources contamination expected in galaxy surveys. Then, we projected the octant delineated by the light orchid triangle in \Cref{fig:mask} onto the observed $512^3$ grid to define the survey mask. We modelled the radial selection functions as log-normal distributions in redshift, which approximately captures the expected behaviour over a range of redshifts \citep[as in e.g.][]{gavazzi1986radio}. They are defined by
\begin{equation}
    R_i(z)=\dfrac{c_i}{z\sigma_i\sqrt{2\,\pi}}\left[\exp\left({-\dfrac{(\ln{z}-\mu_i)^2}{2\,\sigma_i^2}}\right)\right],
\end{equation}
where the constants $\sigma_i$ and $\mu_i$ take different values for the three galaxy populations, and $c_i$ scales the amplitude of the signal. By analogy with the scaling of redshift uncertainties in typical photometric survey models \citep[e.g.][]{laureijs2011euclid}, we defined the variance $s_{i}$ of the log-normal selection functions in our data model as
\begin{equation}
s_{i}=s\,(1+z_i),
\end{equation}
with an overall factor $s=0.1$, where $z_i$ is the effective redshift of the $i$-th galaxy population. The $R_i$ are defined in redshift using $\mu_i\eqdef\ln\left(z_i^2/\sqrt{z_i^2+s_{i}^2}\right)$ and $\sigma_i^2=\ln\left(1+s_{i}^2/z_i^2\right)$, and then mapped to distance space using tabulated values of the distance-redshift relation. The values of $z_i$, $s_i$, and $c_i$ used for Models A and B are provided in \Cref{tab:selection}; the corresponding profiles of the selection functions are illustrated in \Cref{fig:radial_selection}.

\begin{table}[ht]
\caption{\small{Values of the selection function parameters.}\label{tab:selection}}
\centering
    \begin{tabular}{c c c c}
    \hline\hline
    Parameter & $\mathrm{g}_1$ &  $\mathrm{g}_2$ & $\mathrm{g}_3$ \\ \hline
	$z_i$ & 0.1500 & 0.4925  & 0.8182 \\
	$s_{i}$ & 0.1150 & 0.1493  & 0.1818 \\
	$c_i$ (Model A) & 0.1638 & 0.1638 & 0.1638  \\ 
	$c_i$ (Model B) & 0.1638 & 0.1605 & 0.1670  \\
    \hline
    \end{tabular}
\caption{\small{The values are reported to the fourth decimal place. Only the $c_i$ values differ between Models A and B.}}
\end{table}

We emulated the survey by generating galaxy excess number counts $N_x^{\mathrm{g}_i}$ for each cell $x$ of the box across the three galaxy populations. To account for instrumental noise, the $N_x^{\mathrm{g}_i}$ were drawn from Gaussian distributions with means $\mu^{\mathrm{g}_i}_{x}$ and standard deviations $\varsigma_{i,x}$,
\begin{equation}
N_x^{\mathrm{g}_i} \curvearrowleft \mathcal{G}(\mu^{\mathrm{g}_i}_x,\varsigma_{i,x}).
\end{equation}
The mean $\mu^{\mathrm{g}_i}_x \eqdef W_{i,x} b_{\mathrm{g}_i}\, \delta^{z_i}_x$ characterises the expected number of galaxies based on \Cref{eq:density_contrast}, where $W_{i,x}$ is the value in the cell $x$ of the three-dimensional response operator defined by \Cref{eq:window}. The standard deviation is defined as $\varsigma_x \eqdef \varsigma \sqrt{W_{i,x}}$, where $\varsigma = 1$ is the overall shot noise level.
Setting $\bar{N}=1$ in \Cref{eq:galaxy_density} yields a mean galaxy number density of approximately $6\times 10^{-3}$~$h^3\,\textrm{Mpc}^{-3}$ for the mock populations of \ac{lrgs}, which is about an order of magnitude higher than the expected galaxy number density in actual surveys, for instance $5\times 10^{-4}$~$h^3\,\textrm{Mpc}^{-3}$ in \citet{berti2023galaxy}. A slice through one realisation of the galaxy counts fields $\textbf{N}^{\mathrm{g}_i}$ for each population $\textrm{g}_i$ is shown in \Cref{fig:slices}.

\subsection{Summary statistics}
\label{subsec:summary_statistics}

The full data vectors $\boldsymbol{\mathrm{d}}$, consisting of the three $512^3$-dimensional galaxy count fields, are compressed into a summary statistic believed to contain relevant information about the cosmological parameters. In this study, we use the estimated {\psa} of the observed and simulated fields, which is a standard choice in cosmological data analysis that has proven to be a powerful tool for constraining cosmological parameters \citep[e.g.][]{tegmark2004cosmological, ross2013clustering}. The transformation from the full data space to the space of observed summaries corresponds to the data compressor $\mathpzc{C}$ in the \ac{bhm} of \Cref{fig:full_BHM_detailed}.

For each galaxy population $\textrm{g}_i$, following \citet{Leclercq2019SELFI}, we define the binned data power spectrum as
\begin{equation}
P^{\mathrm{g}_i}(k_r) = \sum_{|\textbf{k}| \in \left[k_r\pm\delta r\right]} \frac{\vert\widehat{N}_k^{\mathrm{g}_i}\vert^2}{N_{k_r}-2},
\end{equation}
for $k_r-\delta r>0$, where $\widehat{\textbf{N}}^{\mathrm{g}_i}$ denotes the discrete Fourier transform of $\textbf{N}^{\mathrm{g}_i}$. $N_{k_r}$ represents the total number of Fourier modes within the wave number shell around $k_r$. The $-2$ term arises from the assumption that the data power spectrum follows an inverse-$\Gamma$ distribution with shape parameter $N_{k_r}/2$ and scale parameter $\left| \widehat{N}^{\mathrm{g}_i}_k \right|^2/2$, corresponding to the Jeffreys prior for {\psa} \citep{Jasche2010b}. The summaries $\boldsymbol{\Phi}$ are then defined as the concatenation of the three vectors $P^{\mathrm{g}_i}$, $i=1,2,3$, corresponding to the three galaxy populations.

\subsection{Synthetic observations}

The ground truth cosmological parameters $\boldsymbol{\upomega}_\mathrm{gt}$ are drawn from the \ac{planck} prior \citep{aghanim2020planck}. They are given up to the fourth decimal place in \Cref{tab:cosmo_gt}. We generate synthetic observations $\boldsymbol{\Phi}_\mathrm{O}$ with the well-specified Model A, using
\begin{equation}
\boldsymbol{\uptheta}_\mathrm{gt}=\mathpzc{T}(\boldsymbol{\upomega}_\mathrm{gt}),
\label{eq:theta_gt}
\end{equation}
as the input {\power}, with $\boldsymbol{\upomega}_\mathrm{gt}$ given in \Cref{tab:cosmo_gt}, whilst fixing $\textbf{o}=\boldsymbol{\upomega}_\mathrm{gt}$ in the simulator. The synthetic observations are therefore obtained as
\begin{equation}
\boldsymbol{\Phi}_\mathrm{O}\vert\boldsymbol{\upomega}_\mathrm{gt} \curvearrowleft \int \updelta^{\mathrm{D}}(\mathpzc{B}(\boldsymbol{\uptheta}_\mathrm{gt}, \boldsymbol{\upomega}_\mathrm{gt}, \boldsymbol{\uppsi}))\p(\boldsymbol{\uppsi}) \, \,\mathrm{d}\boldsymbol{\uppsi}.
\end{equation}
\begin{table}[ht]
\caption{\small{Ground truth cosmological parameters.}\label{tab:cosmo_gt}}
\centering
    \begin{tabular}{c c c c c }
    \hline\hline
    $h$ & $\Omega_\mathrm{b}\,h^2$&  $\Omega_\mathrm{m}$ & $n_\mathrm{S}$& $\sigma_8$ \\ \hline
	0.6792 & 0.02247  & 0.3054 & 0.9638  & 0.8210 \\
    \hline
    \end{tabular}
\end{table}

\section{Results}
\label{sec:results}
\subsection{First step: Check for model misspecification}
\label{sec:res_misspecification}

\subsubsection{SELFI posteriors with the well- and misspecified models}
\label{subsec:res_A_B}

\begin{figure*}
\begin{center}
\includegraphics[width=.53\textwidth]{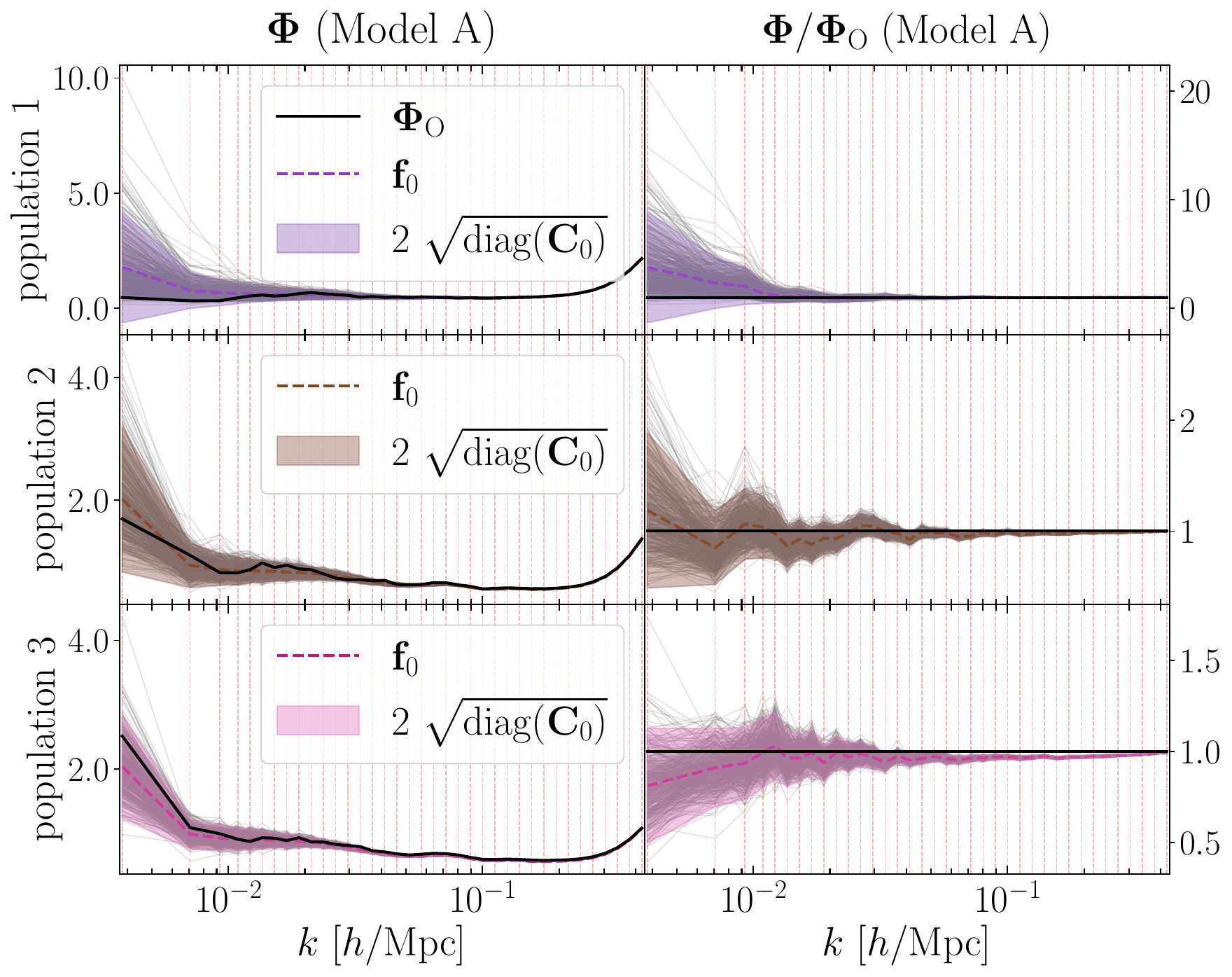}
\includegraphics[width=.451\textwidth]{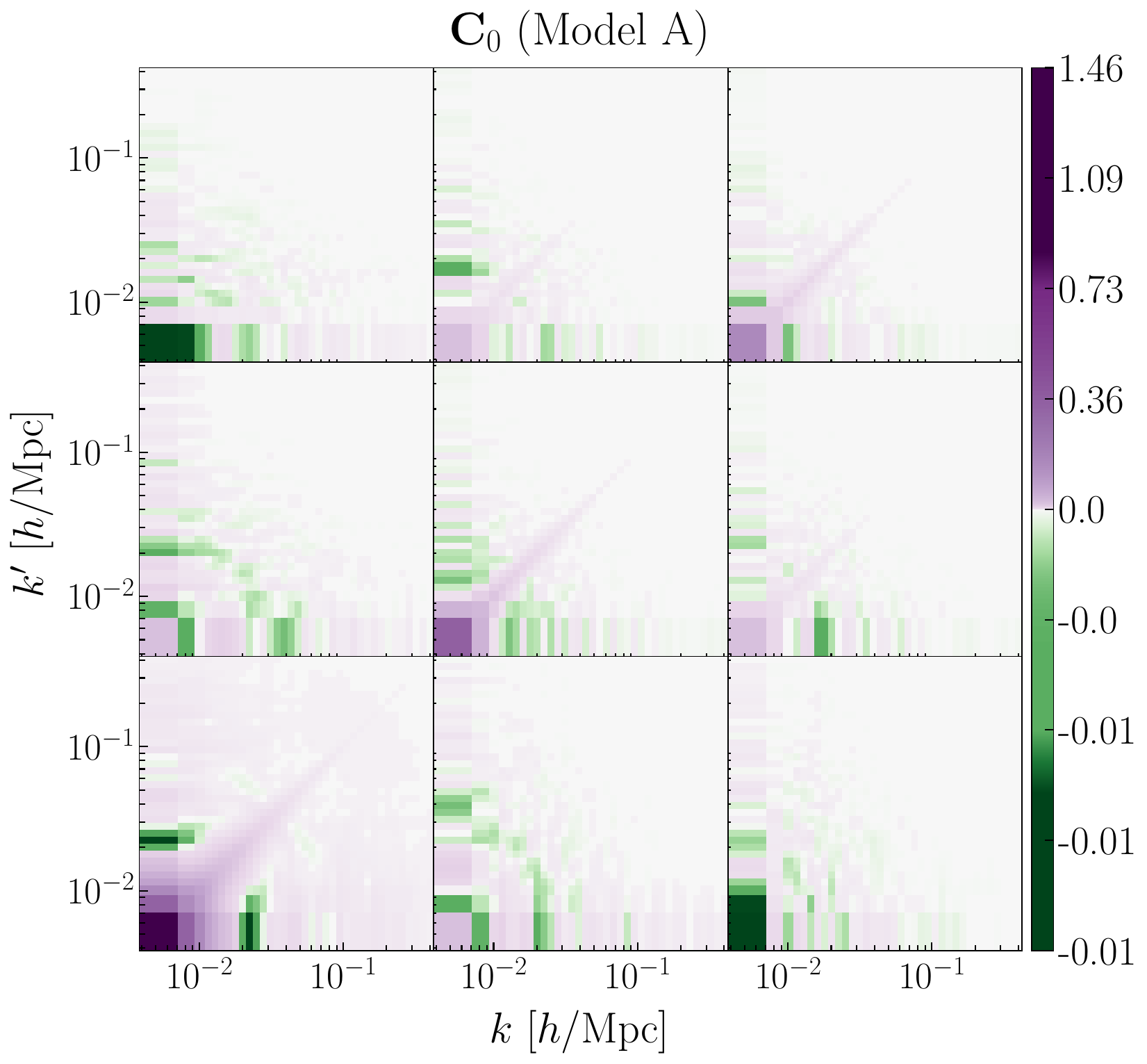}
\includegraphics[width=.53\textwidth]{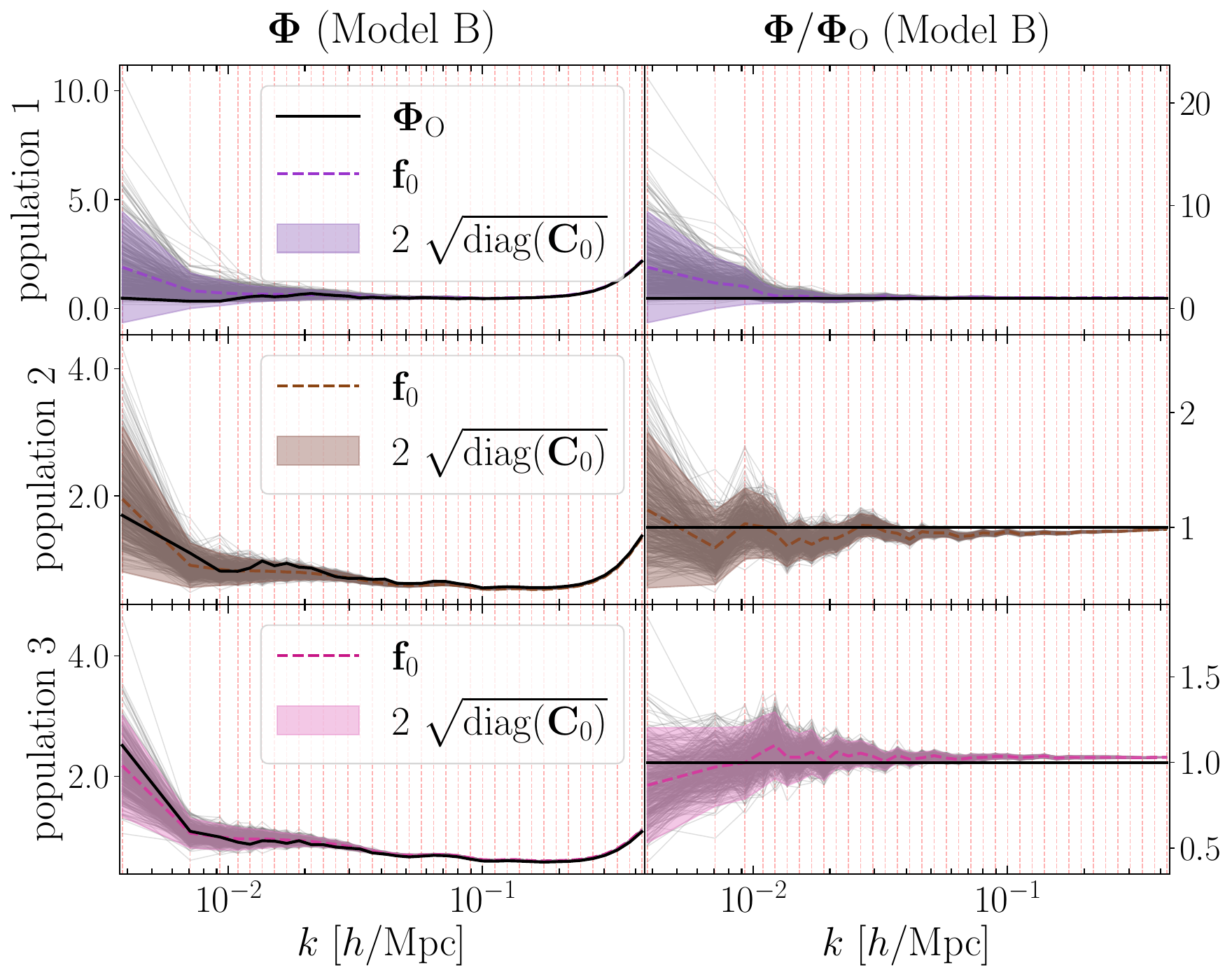}
\includegraphics[width=.451\textwidth]{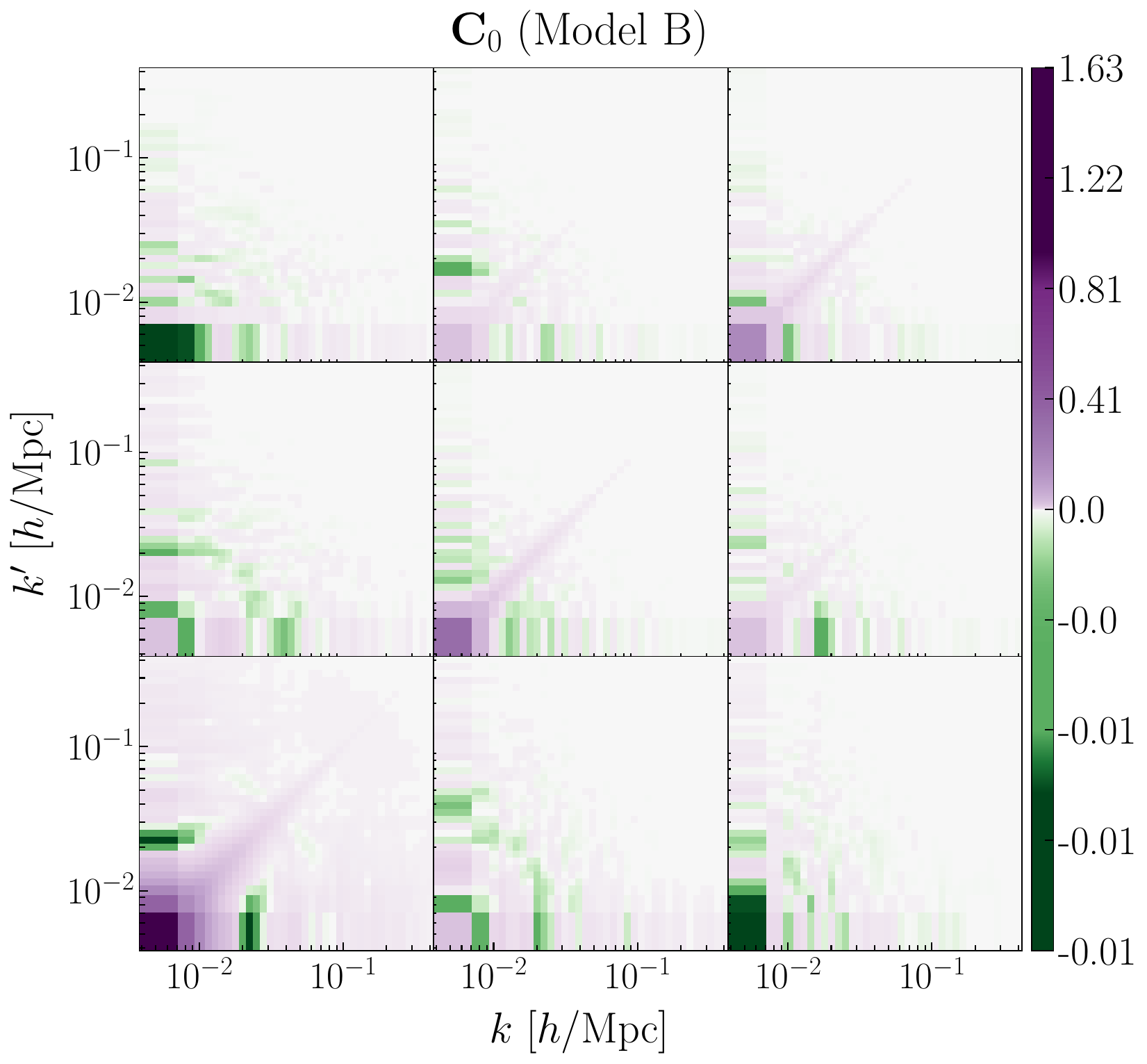}
\caption{\label{fig:selfi_sims}{\small{Data intervening in the computation of the \ac{selfi} posterior. (Left panels) Observed and simulated summary statistics for Model A (\textit{well-specified; upper panel}) and B (misspecified; lower panel). For each population, the simulated summaries are shown in grey and their means are represented as dotted coloured lines. The shaded areas correspond to $\pm 2\sigma$ around their mean. The solid black line corresponds to the observations $\boldsymbol{\Phi}_\textrm{O}$. The binning is indicated by the vertical dashed lines. (Right panels) Covariance matrices for Models A and B. For each $(k,k^\prime)$ entry, the colour scale represents the covariance between the $k$-th and $k'$-th modes. The diagonal blocs of the full covariance matrix correspond to the intra-population covariance; the extra-diagonal blocs correspond to the inter-populations covariances.}}}
\end{center}
\end{figure*}

We ran $N_0=500$ and $N_s \times S = 640$ simulations at and around the expansion point $\boldsymbol{\uptheta}_0$ defined by \Cref{eq:expansion_point,tab:cosmo_fiducial} to estimate $\textbf{f}_0$ and $\boldsymbol{\nabla}_{\uptheta}\textbf{f}_0$, respectively. The simulated summaries used for estimating $\textbf{f}_0$, along with their full covariance matrix, are displayed in \Cref{fig:selfi_sims}. Using these estimates, we computed the \ac{selfi} effective posterior $\p(\boldsymbol{\uptheta}|\boldsymbol{\Phi}_\mathrm{O})$ based on \Cref{eq:filter_mean,eq:filter_var}. From the simulated summaries and their covariance matrices alone, it is difficult to distinguish the well- and misspecified models, let alone identify the source of misspecification in Model B. More insight can be gained by examining the \ac{selfi} posteriors.

\begin{figure*}
\begin{center}
\includegraphics[width=\textwidth]{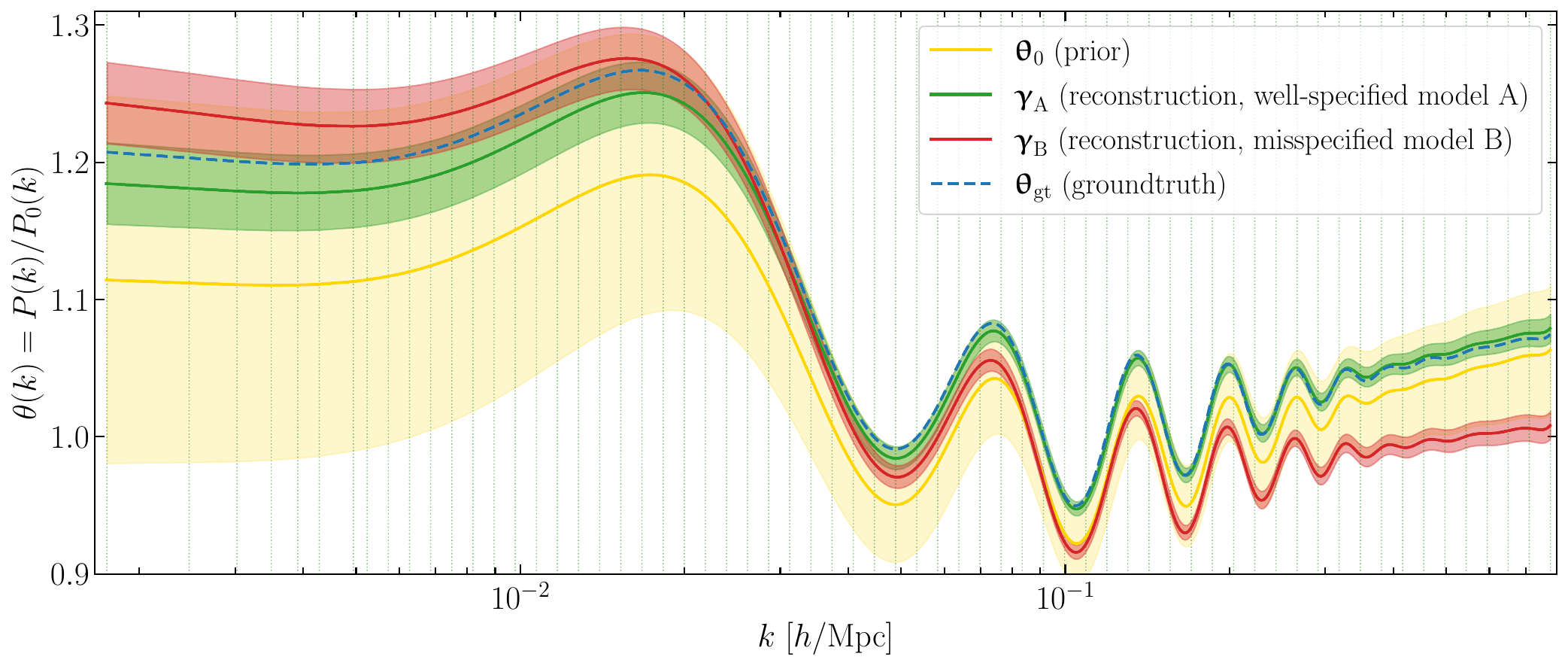}
\caption{\label{fig:posterior_selfi}\small{Prior and \ac{selfi} posteriors for the {\power} $\boldsymbol{\uptheta}$ given the observations $\boldsymbol{\Phi}_\mathrm{O}$. The ground truth $\boldsymbol{\uptheta}_\mathrm{gt}$ is indicated by the dashed blue line. The prior and posterior means, $\boldsymbol{\uptheta}_0$, $\boldsymbol{\upgamma}_{\mathrm{A}}$ and $\boldsymbol{\upgamma}_{\mathrm{B}}$, are represented respectively by the yellow, green and red lines, along with their $2\sigma$ credible regions (shaded yellow, green, and red areas). The vertical dashed lines indicate the support wave numbers for the {\power} representation. The posterior obtained with the well-specified Model A is unbiased over all scales, whilst the posterior obtained with the misspecified Model B exhibits an excess and a lack of power at the largest and smallest scales, respectively.}}
\end{center}
\end{figure*}

The posterior {\power} inferred with the well-specified Model A is represented in green in \Cref{fig:posterior_selfi}. It is compatible with the truth across all scales, demonstrating the self-consistency of the \ac{selfi} assumptions. We further verify that this holds true for a wide range of observed data vectors derived from different sets of cosmological parameters, confirming the unbiasedness of the method (\Cref{appendix:unbiasedness}). The posterior obtained with the misspecified Model B, in red in \Cref{fig:posterior_selfi}, is implausible: it exhibits an excess of power of about $\sim 2\,\sigma$ with respect to the prior mean at the largest scales and a lack of power of similar amplitude at the smallest scales, where non-linear physics dominate.

\begin{figure}
    \begin{center}
    \includegraphics[width=.5\textwidth]{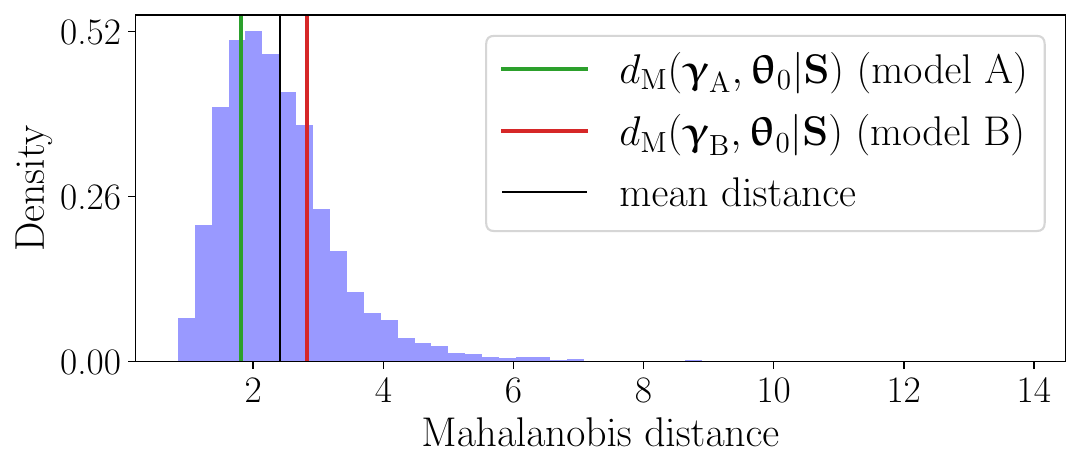}
    \caption{\small{Histogram of the Mahalanobis distances $d_\mathrm{M}(\boldsymbol{\uptheta}_n, \boldsymbol{\uptheta}_0 | \textbf{S})$ between the prior and $\num{5000}$ \psa $\boldsymbol{\uptheta}_n\eqdef\mathpzc{T}(\boldsymbol{\upomega}_n)$ sampled from \Cref{eq:natural_prior}. The mean value is indicated by the vertical solid black line. The green and red lines indicate the distances from $\boldsymbol{\upgamma}_\mathrm{A}$ and $\boldsymbol{\upgamma}_\mathrm{B}$ to the prior, respectively.
    }}\label{fig:Mahalanobis}
    \end{center}
\end{figure}

As a quantitative check, \Cref{fig:Mahalanobis} depicts the Mahalanobis distances between $\p(\boldsymbol{\uptheta})$ and the reconstructed mean {\psa} $\boldsymbol{\upgamma}$ for each model, computed using \Cref{eq:Mahalanobis_distance_def}, in comparison to those between $\p(\boldsymbol{\uptheta})$ and $\num{5000}$ random samples $\boldsymbol{\uptheta}_n=\mathpzc{T}(\boldsymbol{\upomega}_n)$. Each $\boldsymbol{\upomega}_n$ is drawn from the prior $\p(\boldsymbol{\upomega})$ on the cosmological parameters. The Mahalanobis distances are multivariate analogues of the standard scores, accounting for correlations between the components, and measure the deviation of the reconstructions from the prior distribution. Large deviations suggest significant disagreement, prompting further investigation. We found that the distance for Model A, $d_\mathrm{M}(\boldsymbol{\upgamma}_\textrm{A},\boldsymbol{\uptheta}_0|\textbf{S})=1.816$, is much smaller than the distance for Model B, $d_\mathrm{M}(\boldsymbol{\upgamma}_\textrm{B},\boldsymbol{\uptheta}_0|\textbf{S})=2.827$. A larger distance does not necessarily imply a worse model, but hints at the presence of systematic effects. For comparison, the mean Mahalanobis distance from $\p(\boldsymbol{\uptheta})$ to the $\num{5000}$ $\boldsymbol{\uptheta}_n$ samples is $\left\langle d_\mathrm{M}(\boldsymbol{\uptheta}_n, \boldsymbol{\uptheta}_0 | \textbf{S}) \right\rangle_n = 2.431$. The disagreement between $d_\mathrm{M}(\boldsymbol{\upgamma}_\textrm{B},\boldsymbol{\uptheta}_0|\textbf{S})$ and the value typically observed from random samples, $\left\langle d_\mathrm{M}(\boldsymbol{\uptheta}_n, \boldsymbol{\uptheta}_0 | \textbf{S}) \right\rangle_n$, highlights how percent-order variations in the modelling assumptions strongly influence the reconstruction of the {\power} $\boldsymbol{\uptheta}$, pulling it away from the prior. This motivates a careful investigation of systematic effects, which we undertake in the following sections.

Additionally, since the prior given by \Cref{eq:estimated_prior_covariance} embeds substantial information about the functional shape of the {\power}, one might question how the \ac{selfi} algorithm performs with a less informative prior. \citet{Leclercq2019SELFI} demonstrated that a wiggle-less prior centred on $\boldsymbol{\uptheta}_0 \eqdef \boldsymbol{1}_{\mathbb{R}^S}$ enables precise recovery of initial features such as the \ac{baos} up to mildly non-linear scales. Given the enhanced complexity of the data model employed here—incorporating extinction, point-source contamination, and radial selection effects, whilst using a lower mass-resolution—we successfully verified that the \ac{baos} are still retrievable in this set-up using a prior agnostic to the \ac{baos}, as is shown in \Cref{appendix:wiggleless}.

\subsubsection{Impact of galaxy biases on the initial power spectrum reconstruction}
\label{subsubsec:galaxy_biases}

Due to the non-linearity of the data model, the impact of the bias model on the posterior is in general out of reach of theoretical considerations. Linear galaxy biases, however, are expected to have a mostly scale-independent effect on the \ac{selfi} reconstruction $\boldsymbol{\upgamma}$ of the {\power}, as demonstrated in \Cref{appendix:linear_bias} based on qualitative considerations. This effect is observed on \Cref{fig:impact_systematics_a}, which illustrates the impact of the linear galaxy bias parameters on the \ac{selfi} posterior distribution for the {\power}. We considered a constant relative error in the $\pm 2\%$ range for the linear galaxy bias parameters of the three populations, as indicated by the colour scale. All scales experience a similar shift in standard deviations, which is consistent with the analytical expectations.

This demonstrates the ability of our framework to accurately diagnose the impact of systematic effects using the reconstructed {\power}. We retrieve that the effect of misspecified linear galaxy bias parameters closely resembles a simple rescaling of the {\power} by a constant factor, highlighting the well-known degeneracy between the linear galaxy bias parameters and the value of $\sigma_8$ \citep{beutler20126df, arnalte2016joint, Desjacques_2018, repp2020galaxy}.

\begin{figure*}[t]
\begin{center}
\subfloat[Impact of misspecified linear galaxy biases]{\includegraphics[width=.49\textwidth]{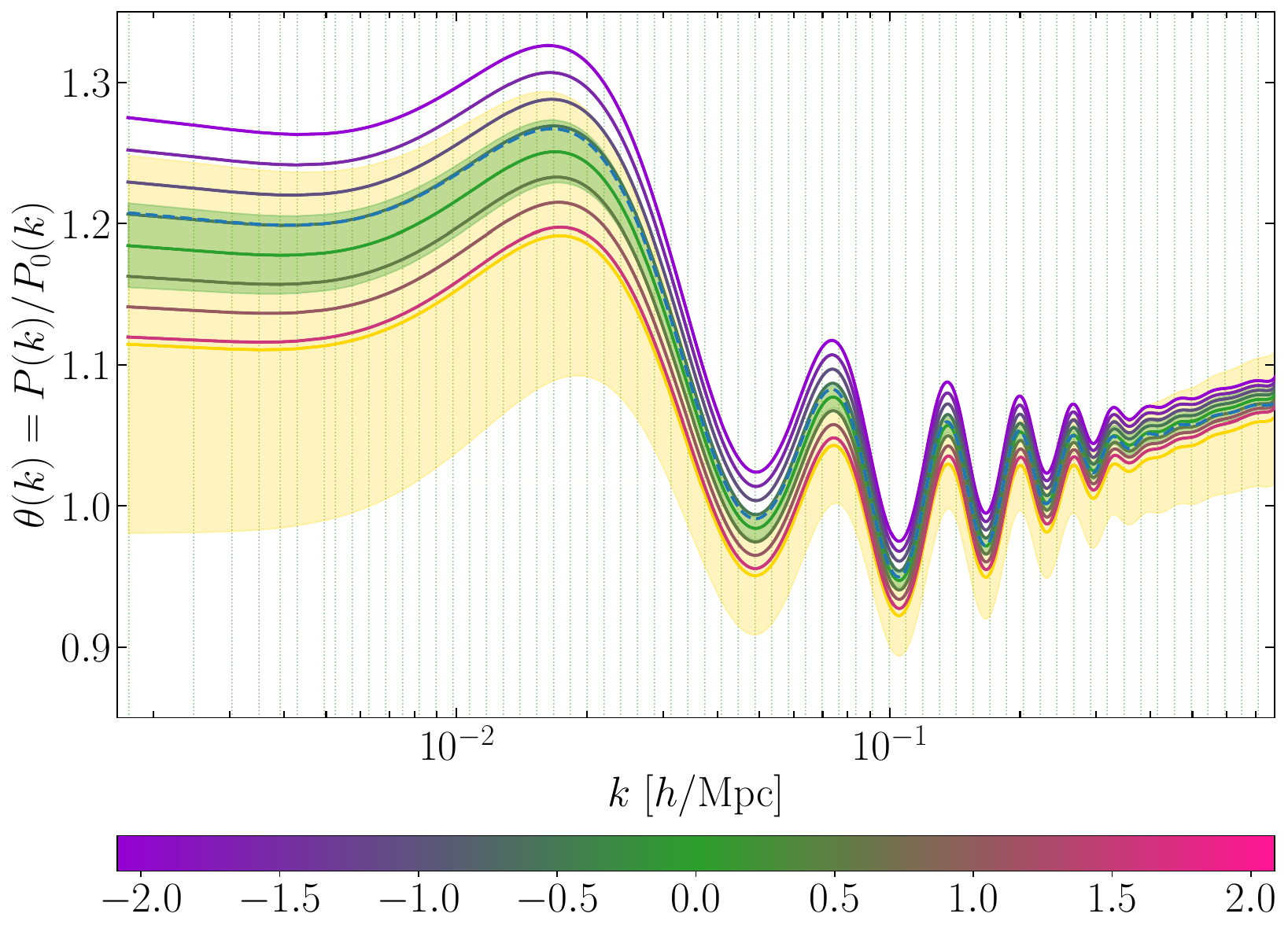}
\label{fig:impact_systematics_a}}%
\hfill
\subfloat[Impact of misspecified linear extinction]{\includegraphics[width=.49\textwidth]{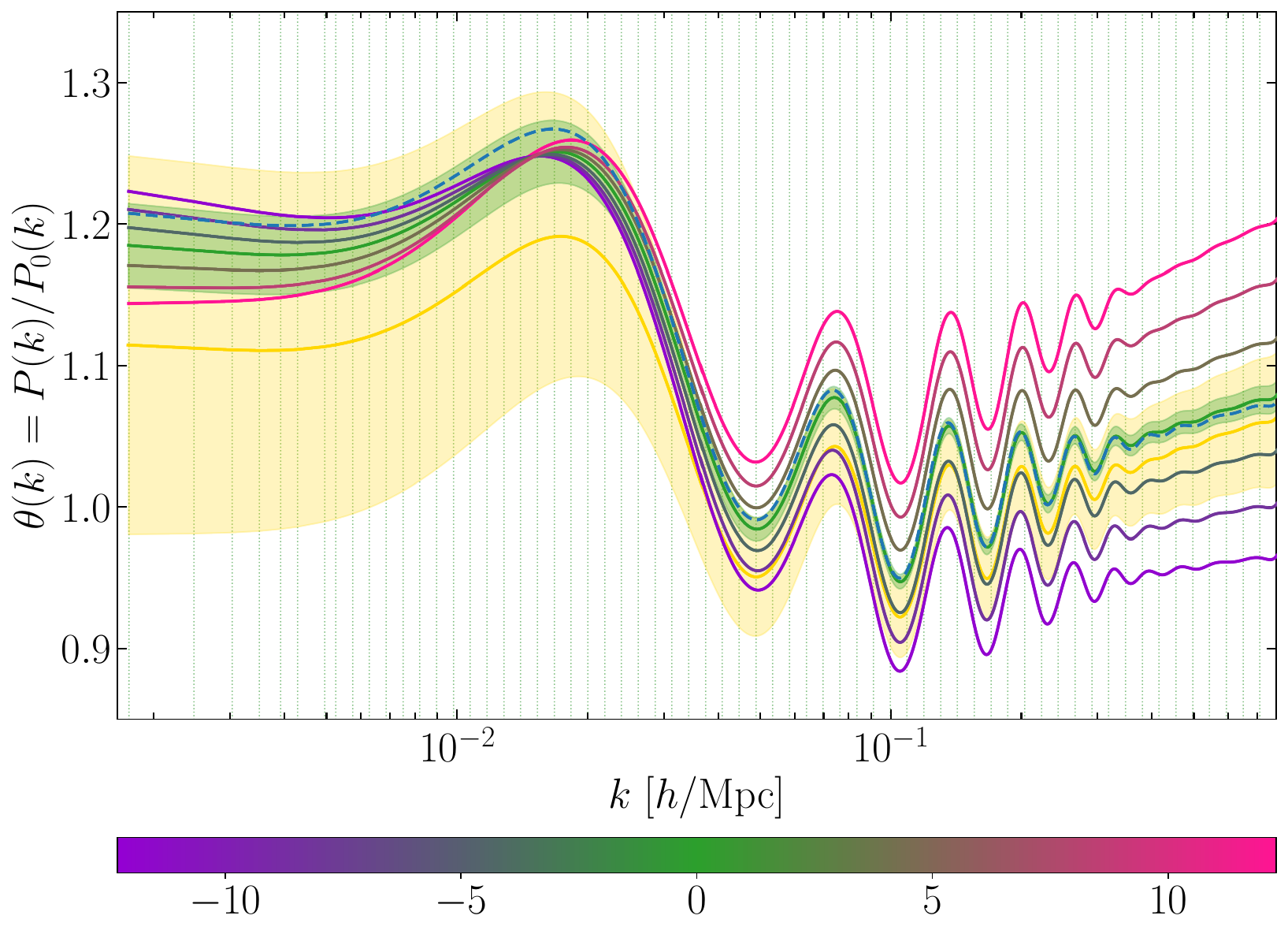}
\label{fig:impact_systematics_b}}%
\hfill
\subfloat[Impact of misspecified redshifts]{\includegraphics[width=.49\textwidth]{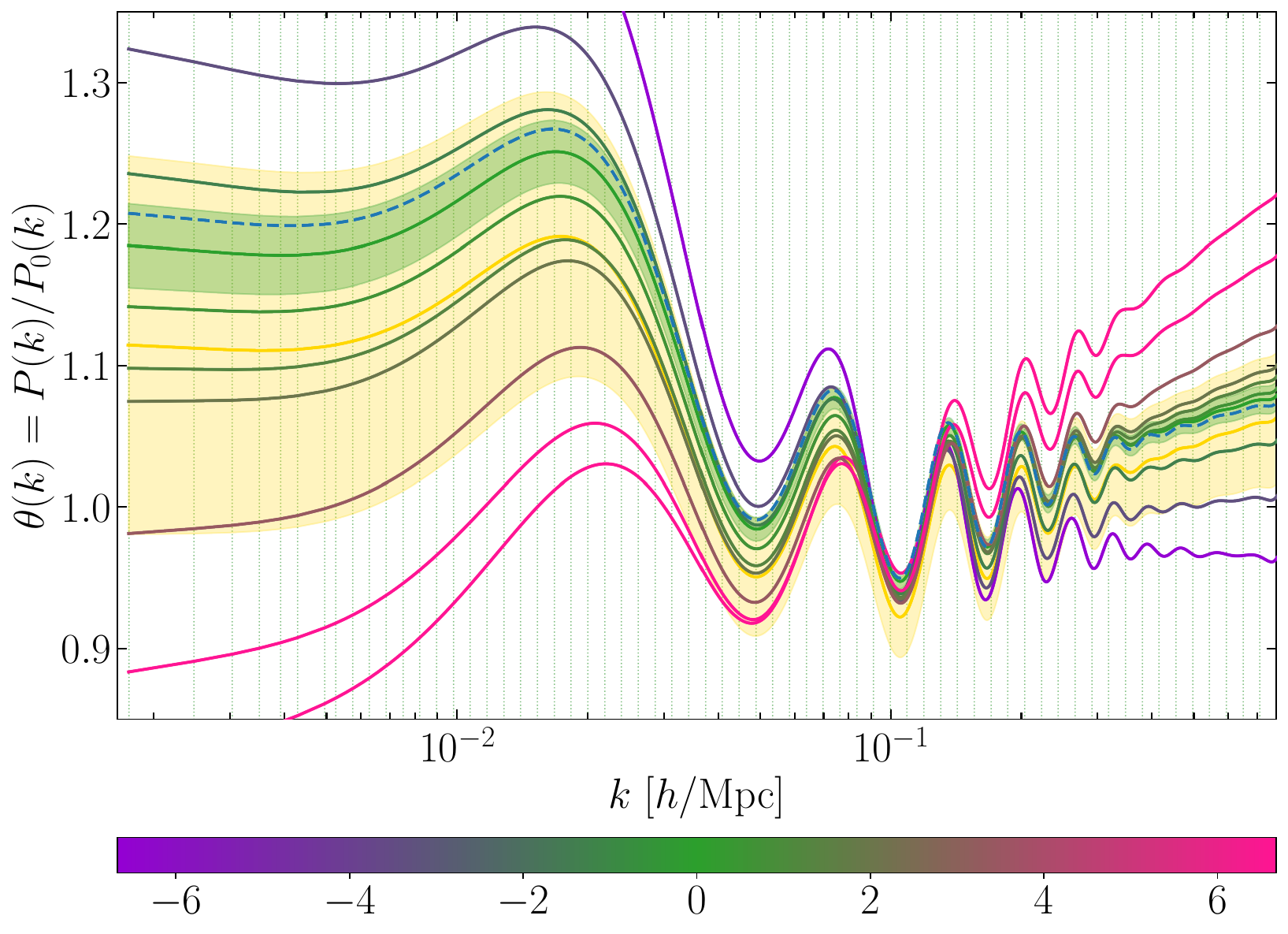}
\label{fig:impact_systematics_c}}%
\hfill
\subfloat[Impact of misspecified selection function variances]{\includegraphics[width=.49\textwidth]{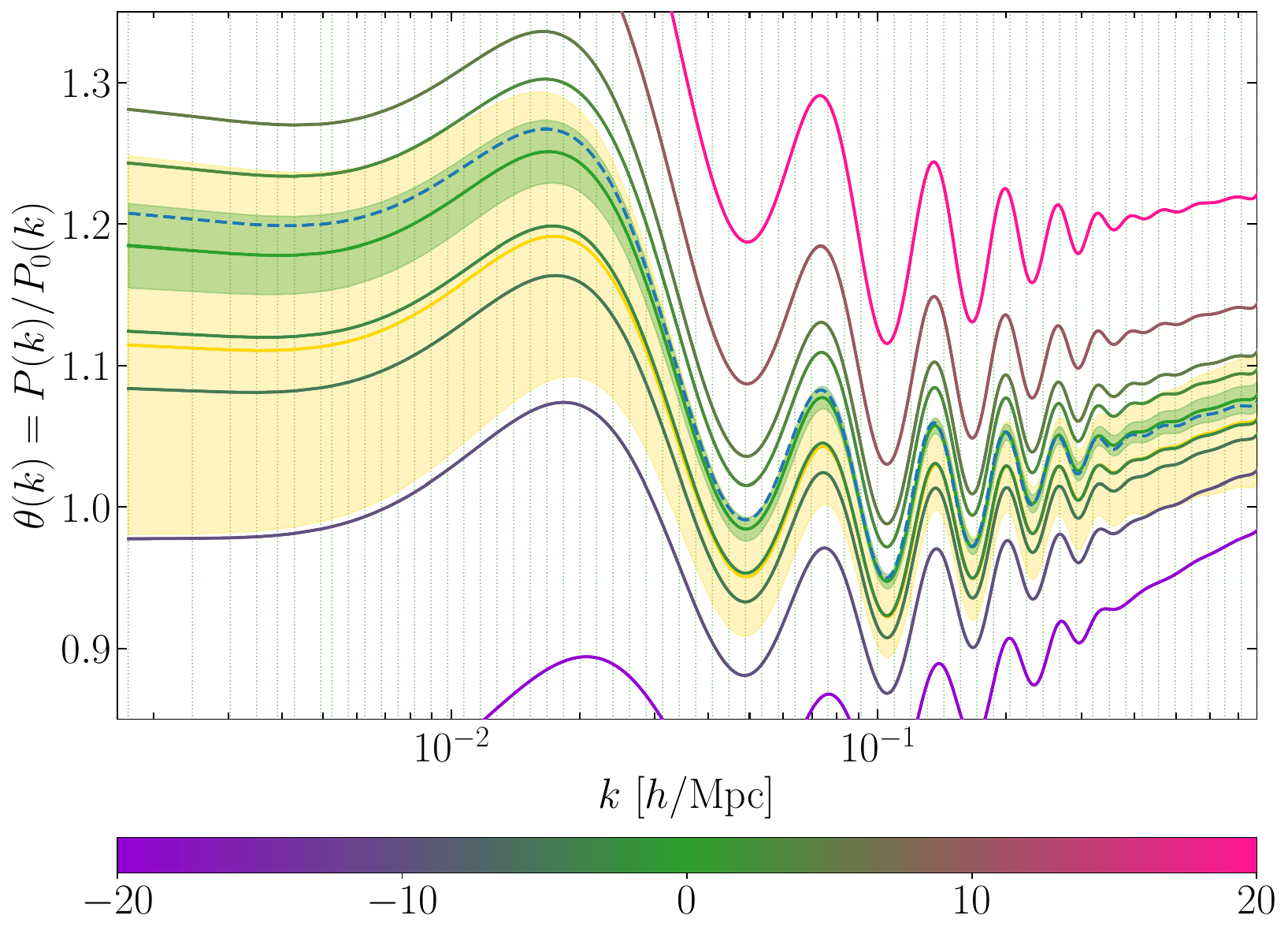}
\label{fig:impact_systematics_d}}%
\hfill
\subfloat[Impact of misspecified number of holes]{\includegraphics[width=.49\textwidth]{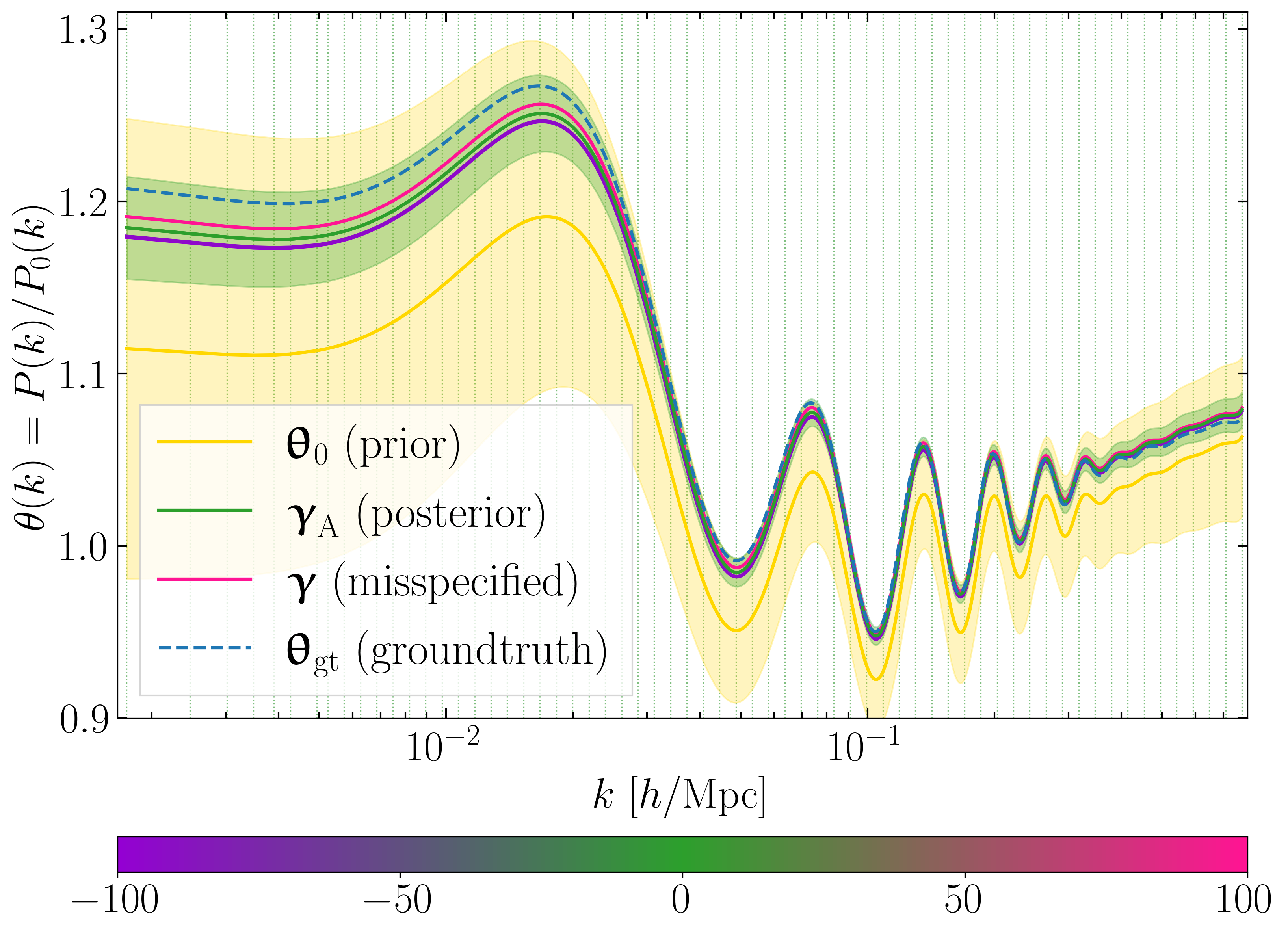}
\label{fig:impact_systematics_e}}%
\hfill
\subfloat[Impact of misspecified selection function variances (varying) under misspecified linear extinction (fixed)]{\includegraphics[width=.49\textwidth]{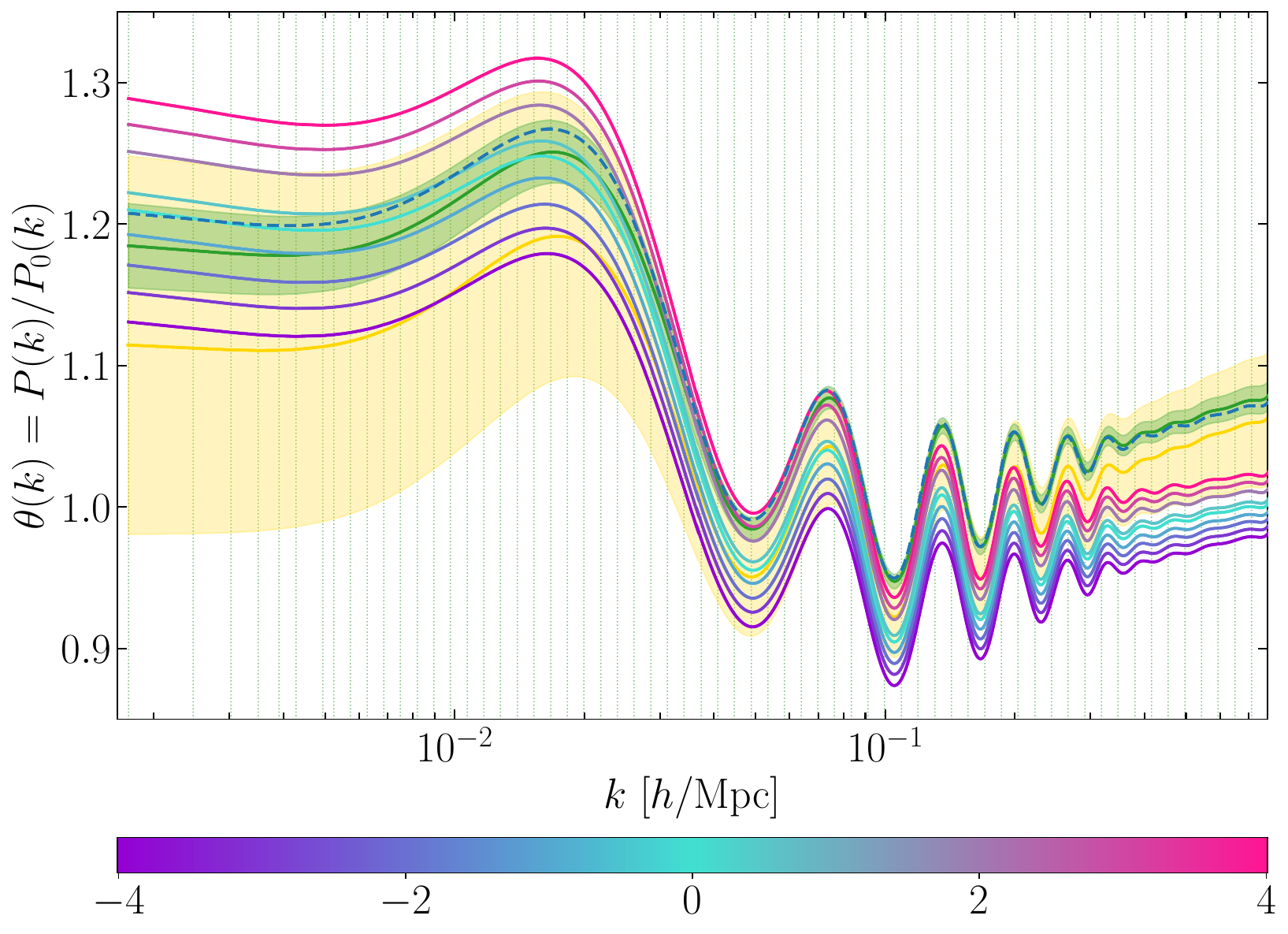}
\label{fig:impact_systematics_f}}
\caption{\small{Ensemble of \ac{selfi} posteriors for the {\power} $\boldsymbol{\uptheta}$ conditional on the observed data $\boldsymbol{\Phi}_\mathrm{O}$ for different observational models, used to diagnose systematic effects. All the posteriors are derived from a single, common set of $N$-body simulations. They make it possible to disentangle individual systematic contributions by comparing their differential impact}\label{fig:impact_systematics}}
\end{center}
\end{figure*}

\begin{figure*}[ht!]
    \ContinuedFloat
    \caption{\small{\textbf{Continued.} across the range of wave numbers spanned by the \psa. The colour scales indicate the relative error associated with the systematic effects considered in each sub-figure. The solid yellow line denotes the prior mean $\boldsymbol{\uptheta}_0$ and the ground truth $\boldsymbol{\uptheta}_\mathrm{gt}$ is indicated by the dashed blue line. The vertical dashed lines indicate the support wave numbers for the {\power} representation. The $2\sigma$ credible regions for the prior and the posterior with the well-specified Model A correspond to the shaded yellow and green areas, respectively. The posterior means $\boldsymbol{\upgamma}$ are represented by the coloured continuous lines for varying degrees of misspecification. For clarity, their corresponding credible regions have been omitted.
    }}
\end{figure*}

\subsubsection{Impact of linear extinction, selection functions, punctual contaminations, and inaccurate redshifts}
\label{subsubsec:variety_systematics}
Even when the internal mechanisms of the forward model are known and well understood, assessing how a particular systematic effect influences the reconstruction solely through analytical considerations is often impractical. If an analytical expression can be derived, it may not be accurate due to correlations with other systematic effects, and its validity is likely to depend on the specificities of the hidden-box forward model. In such cases, we show that our framework makes it possible to conduct a thorough numerical investigation by reconstructing $\boldsymbol{\upgamma}$ with \Cref{eq:filter_mean}, whilst varying the values of the parameters involved in the model of the systematic effect. As a demonstration, we examine the impact of misspecified extinction near the galactic plane, misspecified selection functions, inaccurate redshifts, and punctual contaminations. All these numerical investigations rely on the same, common set of $N$-body simulations previously computed to obtain the posterior with Model A and already used to evaluate the effect of linear galaxy bias parameters in \Cref{subsubsec:galaxy_biases}.

\Cref{fig:impact_systematics_b} illustrates the effect of over- or underestimating the amount of extinction near the galactic plane. The colour scale represents the relative mean visibility between the misspecified and well-specified models, averaged over the three-dimensional survey window function. The extinction reduces the amplitude of small scales correlations; thus, overestimating extinction in the misspecified mask leads to a power deficit at small scales in the simulated summaries, shifting the reconstructed spectrum upwards. Conversely, underestimating extinction shifts the reconstructed spectrum downwards at small scales. At larger scales, excessive extinction in the misspecified mask introduces spurious fluctuations which are not retrieved in the observations, adding power to the largest $k$-modes and pushing the reconstruction downwards. As is shown in \Cref{fig:impact_systematics_b}, the transition between these two regimes occurs at a constant scale, regardless of the degree of error in the extinction. This characteristic scale provides a means to distinguish the effects of extinction from other sources of systematic errors.

\begin{figure}
    \begin{center}
    \includegraphics[width=.5\textwidth]{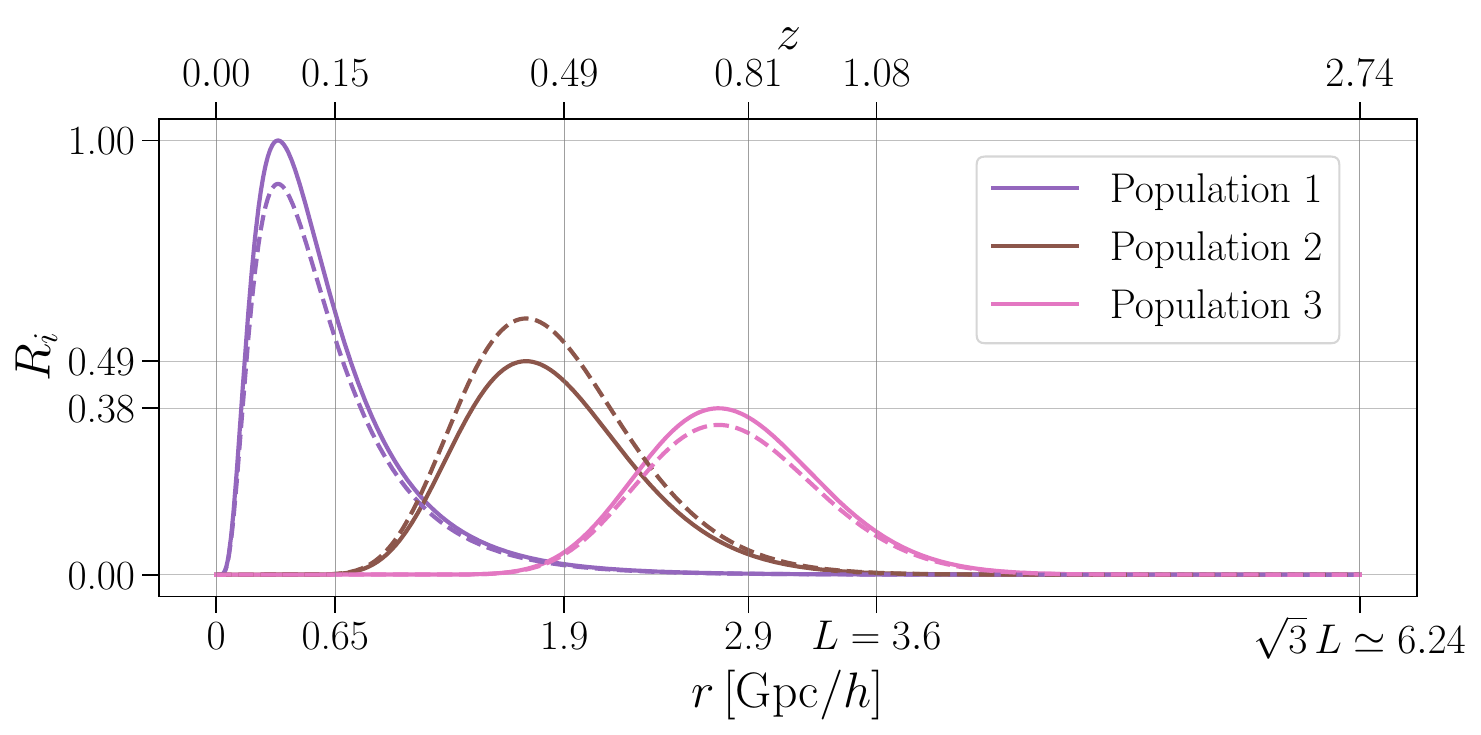}
    \caption{\small{Profiles of the well- and misspecified selection functions $R_i$ for the three mock galaxy populations. The continuous lines represent the selection functions used in Model A. The dashed lines are derived by underestimating $n_{1,2}(z)$ by $10\%$ whilst maintaining $n_\textrm{tot}(z)$ constant.}\label{fig:err_on_z}}
    \end{center}
\end{figure}

We emulated a systematic effect analogous to inaccurate redshifts due to line interlopers \citep[e.g.][]{pullen2016interloper,addison2019impact,massara2021line} by increasing (or decreasing) the observed density of the central spectroscopic galaxy bin whilst correspondingly decreasing (or increasing) the densities of the two other populations, ensuring that the total density remained unchanged. An example of the resulting misspecified radial selection functions is shown in \Cref{fig:err_on_z}. The resulting reconstructions, displayed in \Cref{fig:impact_systematics_c}, demonstrate that even small, percent-level variations significantly affect the posterior on the {\power}.

\Cref{fig:impact_systematics_d} shows the impact of another source of systematic error associated with selection functions, specifically the impact of misspecified variances in the radial log-normal distributions. The result is comparable to that of misspecified galaxy bias parameters, though it is slightly more pronounced at smaller scales.

In complex data models displaying multiple putative sources of systematic effects alongside potentially many corresponding model parameters, it is crucial to know which effects are most likely to have a significant impact on the {\power}, and should be prioritised for further investigation. Notably, some systematic effects may have a negligible impact on the {\power}. \Cref{fig:impact_systematics_e} illustrates the effect of misspecifying the number of holes in the survey mask in our set-up, using $0$, $256$, or all $512$ holes defined at the beginning of the study, which represent the masking of bright stars or other point sources of contamination expected in cosmological surveys. The reconstruction appears largely unaffected by the number of holes, as their small angular size causes them to be overwhelmed by small-scale noise.

\subsubsection{Joint impact of multiple systematic effects}
\label{subsec:joint_impact}

Naturally, one is not restricted to studying individual systematic effects in isolation, and the \ac{selfi} posterior can be employed to explore how jointly misspecified observational effects influence the reconstructed {\power}. This approach would be particularly beneficial for designing the data model of a real survey, enabling the assessment of the joint impact of different systematic effects on the cosmological inference whilst accounting for potential correlations between them. For example, \Cref{fig:impact_systematics_f} shows the combined impact on the {\power} of misspecified variances in the log-normal selection functions alongside incorrectly specified linear extinction.

Conversely, if a comprehensive study of multiple individual systematic effects has already been conducted, one can use it to disentangle their contributions when faced with a surprising posterior that warrants further investigation. For instance, if we obtain one of the reconstructions shown in \Cref{fig:impact_systematics_f}, assuming unknown additive contributions of the five effects described in \Cref{subsubsec:variety_systematics}, we can easily disentangle individual contributions as follows. Shifting the reconstruction until it aligns with one of the characteristic scales associated with misspecified extinction or redshifts, we deduce that the systematic error most consistent with the observed reconstruction is a $-10\%$ error in the extinction, along with the corresponding error in the variance of the selection functions. As the model complexity increases, greater caution is required when interpreting the results, since multiple systematic effects can have degenerate impacts on the reconstructed {\psa}. This underscores the importance of thoroughly investigating the influence of each individual systematic effect beforehand.

If known and unknown systematic effects are entangled, it remains possible to first identify the former using the characterisation presented in \Cref{fig:impact_systematics}. Running the framework while accounting for all known sources of systematic effects ensures that only unknown systematic effects and/or new physics remain to be characterised in the inferred {\power}. Consequently, our framework has significant potential for detecting previously unknown systematic effects. This approach is conceptually distinct from that introduced by \citet{porqueres2019explicit}, who identify and marginalise over unknown systematic effects within explicit likelihood inference.

\subsubsection{Impact of the gravity solver}
\label{subsec:gravity_model}

The gravity solver is a critical component of the data model. There is some flexibility in selecting the total number of time steps $n_\textrm{steps}$, dark matter particles $N_\textrm{p}$, and the resolution of the particle-mesh grid used to compute the gravitational forces. Often, a trade-off must be made between the numerical cost of the resolution and the precision required by the observations and scientific objectives. The standard approach is to specify a target precision for the first or second-order statistics of the final dark matter overdensity fields and adjust the gravity solver parameters accordingly, based on convergence tests or comparisons with higher resolution simulations.

Unlike the systematic effects discussed in \Cref{subsubsec:galaxy_biases,subsubsec:variety_systematics,subsec:joint_impact}, obtaining multiple posteriors on the {\power} for various configurations of the gravity solver implemented in the forward model requires re-running $N$-body simulations. However, the numerical cost of obtaining a single \ac{selfi} posterior for the {\power}—$\mathcal{O}(10^3)$ $N$-body simulations in this study—remains orders of magnitude lower than for the cosmological parameters—$\mathcal{O}(10^5)$ $N$-body simulations in this study. Consequently, our framework presents further value for investigating the error caused by misspecified gravity models on the posterior {\power}.
Similarly, our framework could be employed to diagnose the effect of baryons in the hidden-box model based on the inferred {\power}, provided one is willing to incur the numerical cost of incorporating baryonic physics into the simulator.

We refer to the error on the reconstructed {\power} as the inverse error, as opposed to the direct error on the measured galaxy {\psa}. We show that measuring this inverse error is particularly relevant for large-scale galaxy surveys: it highlights the sensitivity of the inference to the gravity solver parameters, which is not apparent from the direct error on the galaxy count fields.

\begin{figure*}
\begin{center}
\subfloat[Galaxy {\psa} with $n_\textrm{steps}=10$ compared to $n_\textrm{steps}=20$ for Model A ($P_\mathrm{A}$)]{\includegraphics[width=.45\textwidth]{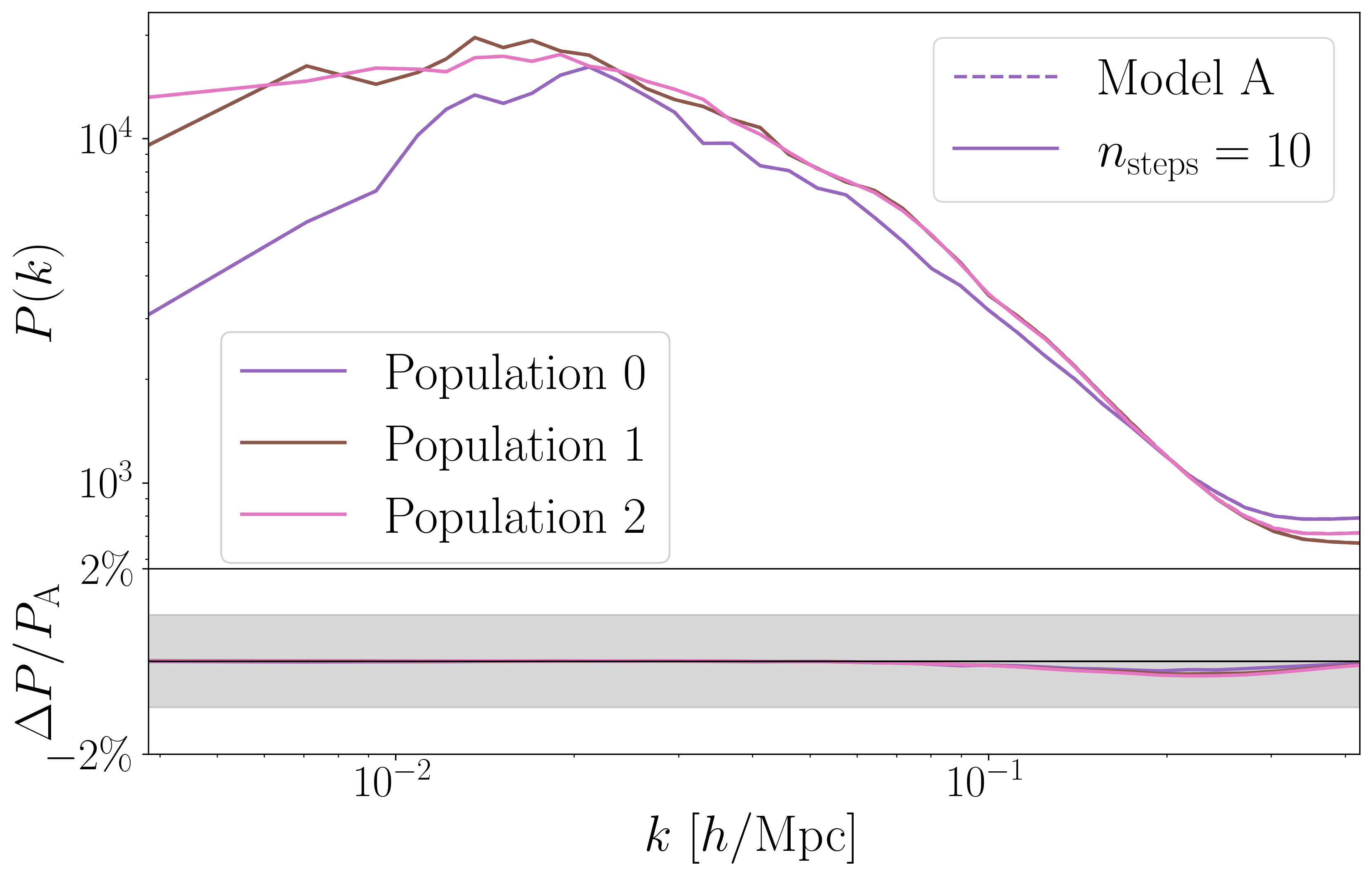}
\label{fig:misgrav_a}}%
\hfill
\subfloat[The \ac{selfi} posterior with $n_\textrm{steps}=10,20$]{
\includegraphics[width=.45\textwidth]{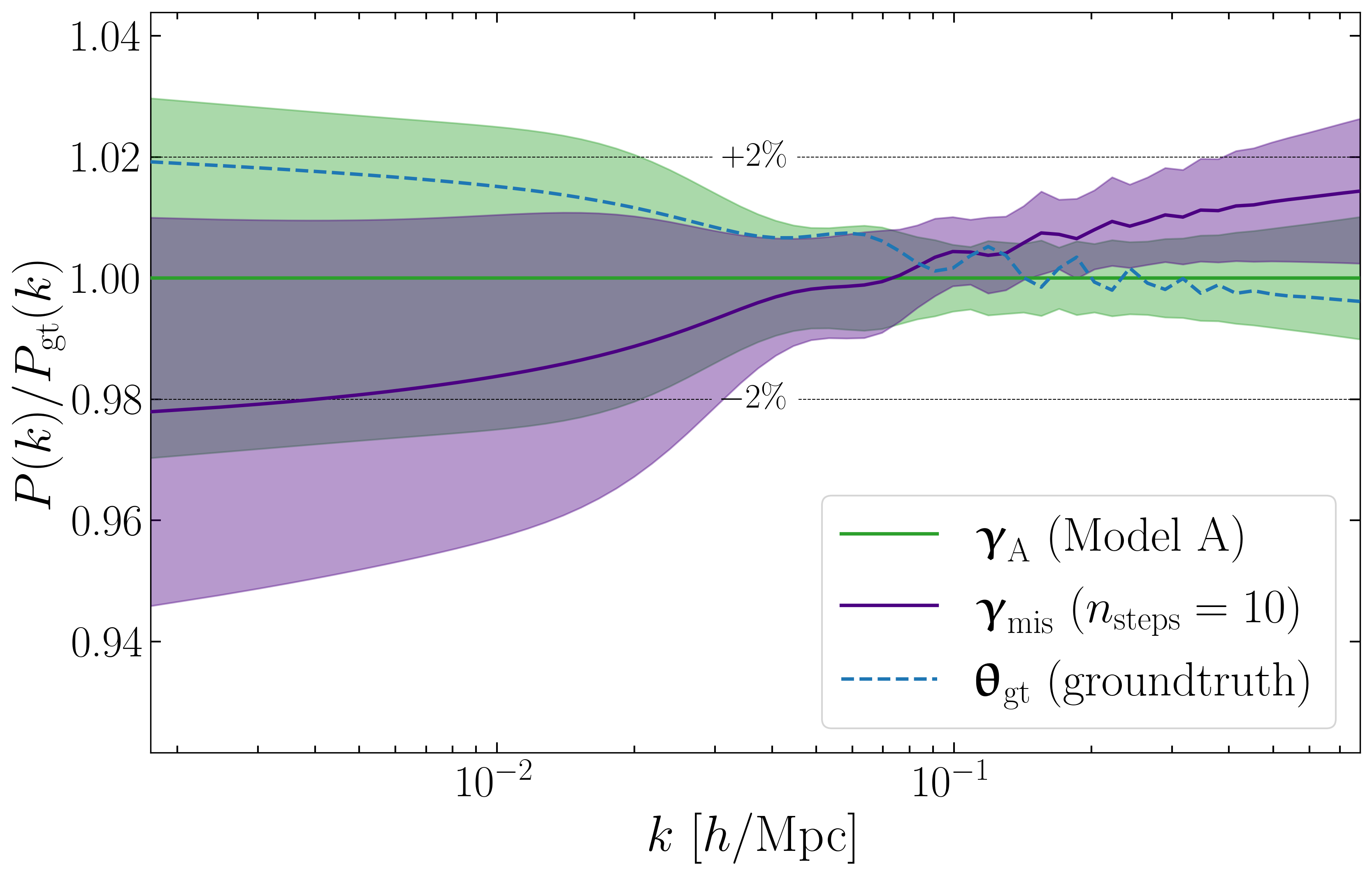}
\label{fig:misgrav_b}}%
\hfill
\subfloat[Galaxy {\psa} with ($P_\mathrm{A}$) and without light cone]{
\includegraphics[width=.45\textwidth]{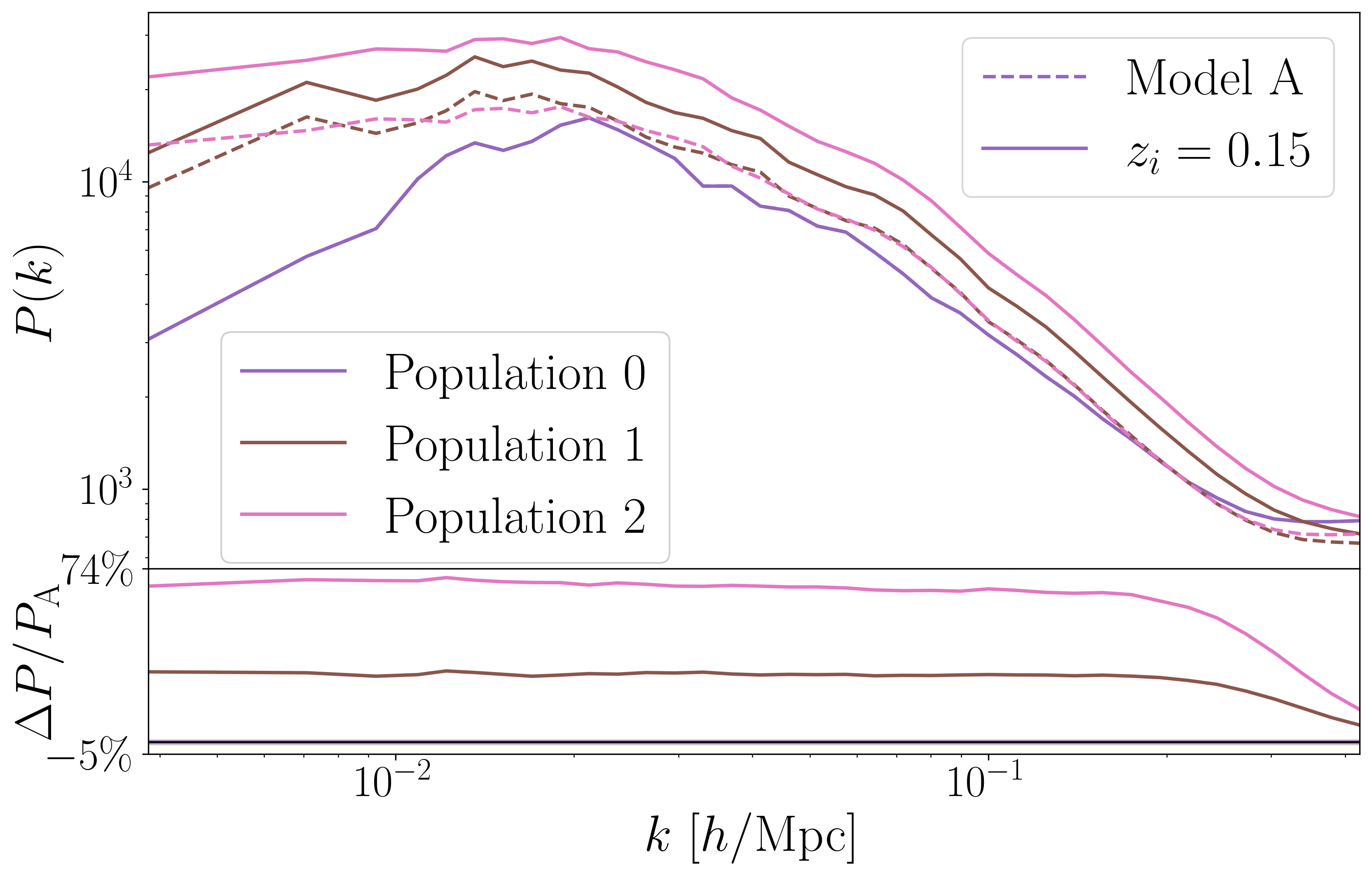}
\label{fig:misgrav_c}}%
\hfill
\subfloat[The \ac{selfi} posterior with or without light cone]{
\includegraphics[width=.45\textwidth]{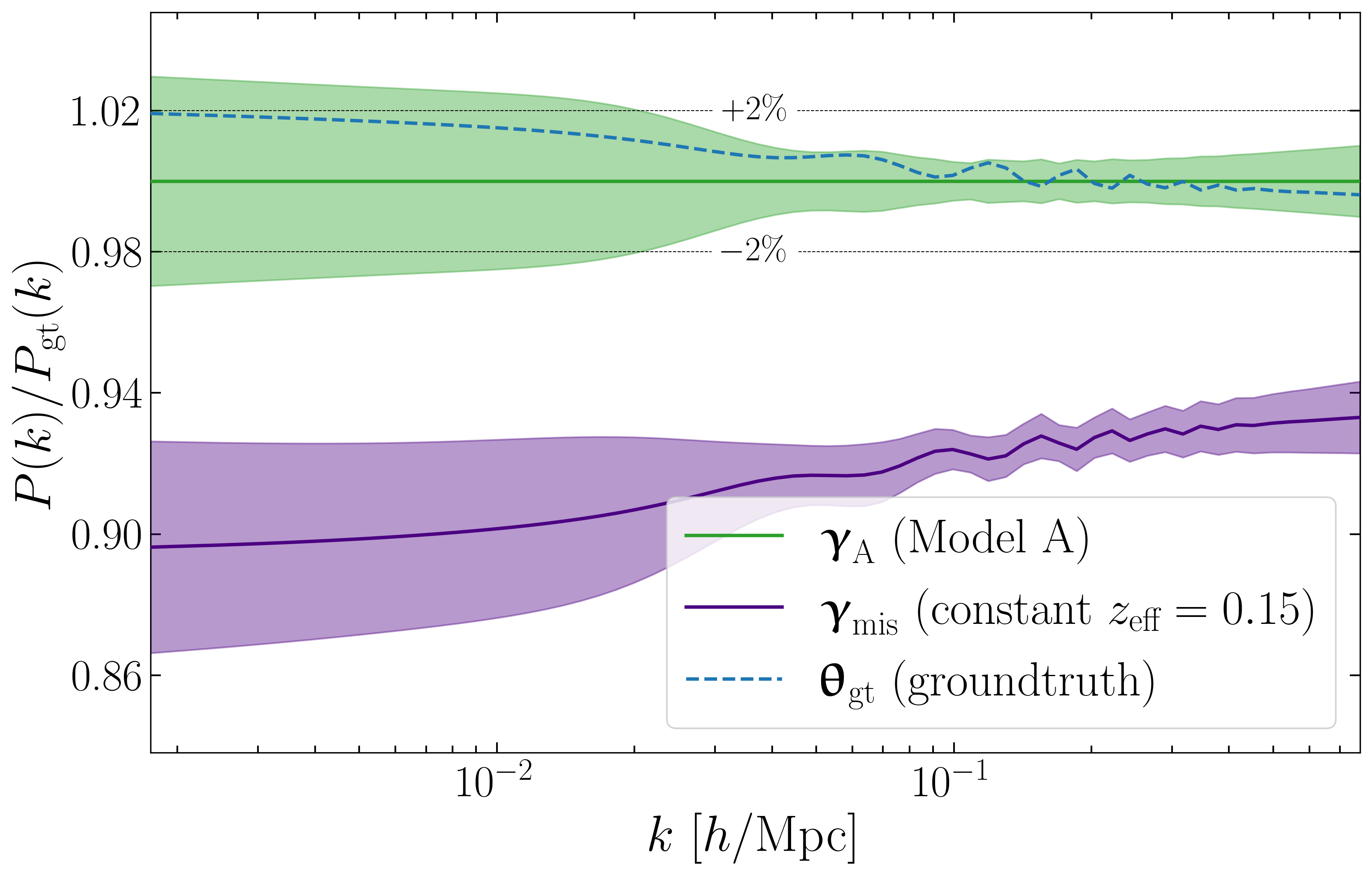}
\label{fig:misgrav_d}}%
\hfill
\subfloat[Galaxy with \ac{2lpt} instead of $N$-body ($P_\mathrm{A}$)]{
\includegraphics[width=.45\textwidth]{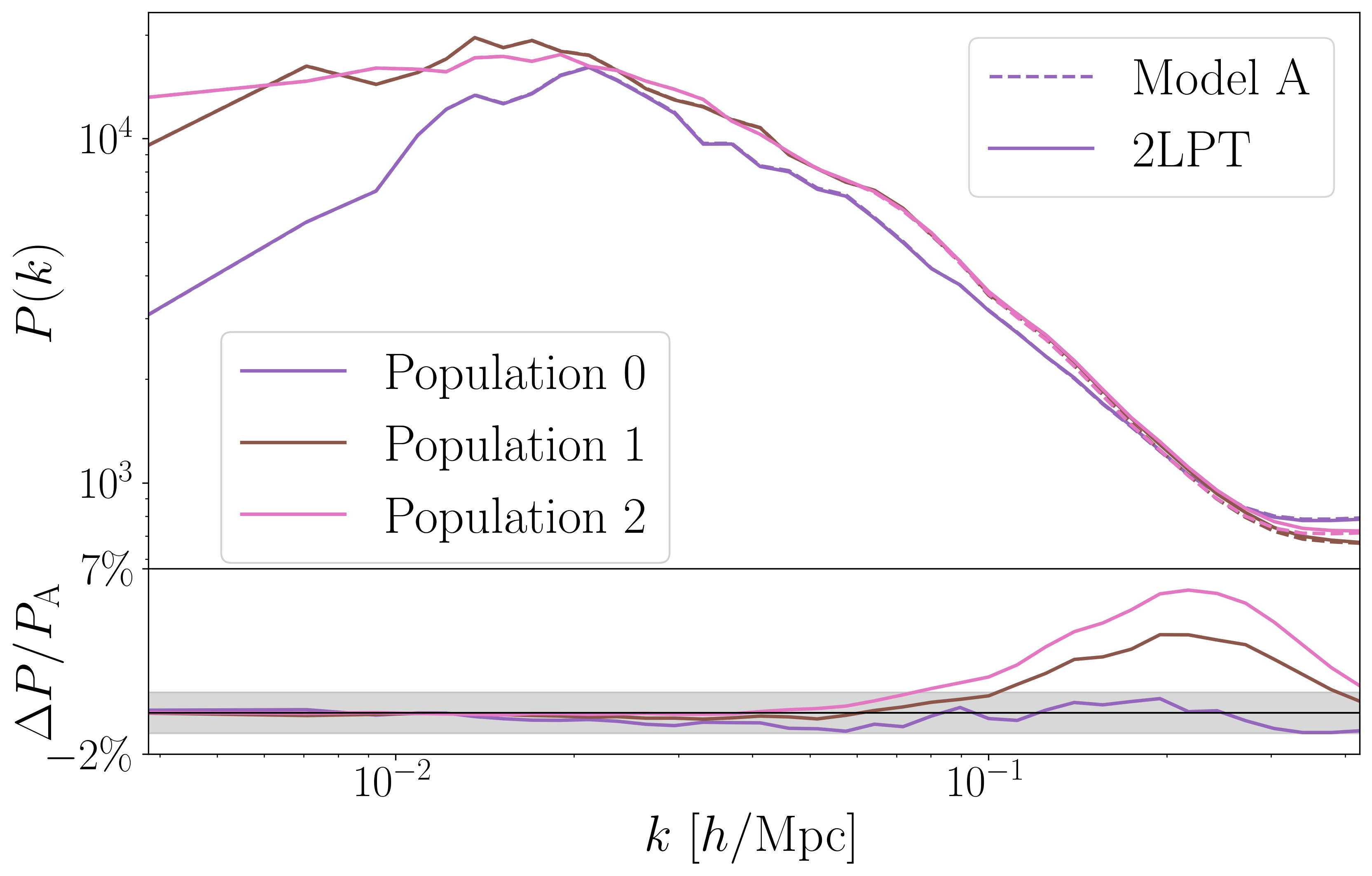}
\label{fig:misgrav_e}}%
\hfill
\subfloat[The \ac{selfi} posterior with \ac{2lpt} or $N$-body]{
\includegraphics[width=.45\textwidth]{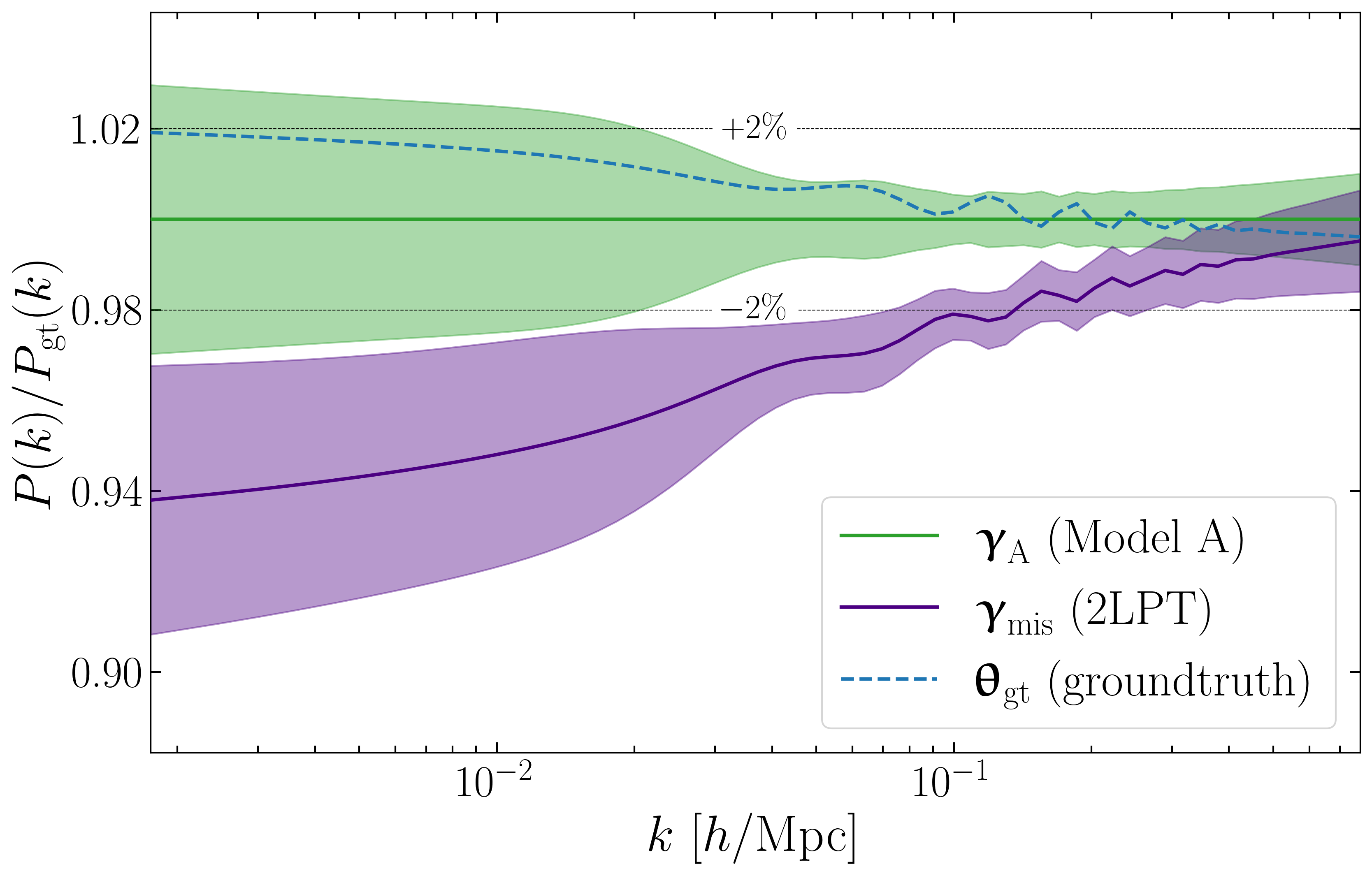}
\label{fig:misgrav_f}}%
\caption{\small{Impact of the gravity solver parameters on the \ac{selfi} posterior. Panels (a, c, e) display the summary statistics $\boldsymbol{\Phi}$ for the three galaxy count fields, comparing the well-specified Model A (dashed lines) with the misspecified model (solid lines). The lower parts of these panels show the relative error between the two models, with shaded regions corresponding to a $1\%$ error margin. Panels (b, d, f) show the corresponding \ac{selfi} posteriors for the {\power} $\boldsymbol{\uptheta}$, with the well-specified and misspecified models represented by green and indigo lines, respectively. (a) Example of summary statistics for a single realisation with $n_\textrm{steps}=10$ (solid lines) as opposed to $n_\textrm{steps}=20$ time steps for Model A (dashed lines) in the gravitational evolution with \ac{cola}. Although the direct error remains well below $1\%$ across all scales, the inverse error on the \ac{selfi} posterior (b) reaches $2\%$ at the largest scales, corresponding to a deviation greater than $1\sigma$ from the well-specified Model A. (c) Summary statistics using only the final snapshot at redshift $z_1=0.15$ for the three mock galaxy populations, thereby neglecting light cone effects (solid lines). The forward error reaches $74\%$ at the largest scales for the most distant galaxy population, and the corresponding \ac{selfi} posterior (d) significantly rejects the ground truth across all scales. (e) Summary statistics using a pure \ac{2lpt} gravity model instead of $N$-body simulations to define the galaxy populations (solid lines). The forward error reaches $\sim 7\%$ at the smallest scale, and the ground truth is significantly rejected by the \ac{selfi} posterior (f), which deviates by more than $2\sigma$ compared to Model A at the largest scales.
}
\label{fig:misgrav}}
\end{center}
\end{figure*}

\Cref{fig:misgrav_a,fig:misgrav_b} illustrate the effect of using $10$ instead of $20$ time steps for the gravitational evolution with \ac{cola}. Whilst the direct error in the measured galaxy {\psa} compared to Model A remains below $1\%$ across all scales, the inverse error in the \ac{selfi} posterior exceeds $2\%$ for $k\leq 4\cdot 10^{-3}$. This discrepancy arises due to the non-linearity of the inversion: small differences in the predicted galaxy power spectra between Models A and B can propagate into substantial biases in the posteriors after solving the inverse problem. As is shown in \Cref{fig:misgrav_b}, the posterior rejects the ground truth by more than $2\sigma$ across most scales. The \ac{selfi} posterior is therefore highly sensitive to the time-stepping in the gravity solver. To achieve, for example, a $1\%$ precision on the posterior, it is necessary to ensure an even higher precision in the galaxy count {\psa}.

Similarly, \Cref{fig:misgrav_c,fig:misgrav_d} highlight the importance of accounting for light cone effects in the forward model: the \ac{selfi} posterior consistently rejects the ground truth when all three galaxy populations are defined from the snapshot at redshift $z_1=0.15$, compared to using $z_1$, $z_2$, and $z_3$ in Model A. Since the cosmological structure is less evolved at higher redshifts, this results in overestimating the measured power spectrum amplitude for the two most distant galaxy populations, which pushes the \ac{selfi} posterior downwards at all scales. We note that this effect is highly degenerate with that of misspecified galaxy biases. Both risk biasing the value of $\sigma_8$ inferred from the data if not properly accounted for.

Strikingly, \Cref{fig:misgrav_e,fig:misgrav_f} show that using \ac{2lpt} instead of \ac{cola} in the forward model significantly biases the posterior on the {\power}. This arises because the range of scales considered in this study extends well beyond $k_{\mathrm{lin}} \simeq 0.15$~$h$/Mpc, commonly considered as the smallest scale of the linear regime \citep[e.g.][]{colless20012df}, up to $k_{\mathrm{max}} \simeq 0.4$~$h$/Mpc.
Within the SELFI framework, all scales are coupled by the non-linearity. Over-prediction of small-scale power by \ac{2lpt} (due to the mass resolution) results in under-prediction of large scale power to fit the data. Furthermore, since the number of small-scale modes vastly exceeds that of large-scale modes, the posterior computation sacrifices large scale accuracy to fit small scales.
The forward error on the galaxy count {\psa} reaches $\sim 7\%$ at the smallest scales, and the \ac{selfi} posterior rejects the ground truth by almost $2\sigma$ across most scales. This underscores the necessity of employing fully non-linear gravity solvers to accurately extract small-scale information from large-scale galaxy surveys.

\subsection{Second step: Cosmological parameters}
\label{sec:res_inference}

After checking for model misspecification, the second step of the framework consists of inferring the target cosmological parameters, $\boldsymbol{\upomega}$, given the compressed summaries $\boldsymbol{\widetilde{\upomega}}_\mathrm{O}$. For score compression, we re-used the estimates $\textbf{f}_0$ and $\boldsymbol{\nabla}_{\uptheta} \textbf{f}_0$ computed in the first step, and estimate the gradient $\boldsymbol{\nabla}\mathpzc{T}$ of \ac{class} at the fiducial cosmological parameters $\boldsymbol{\widetilde{\upomega}}_0$ using fifth order finite differences. To perform the inference, we use the \texttt{ELFI} package \citep{Lintusaari2018}, which implements the variant of \ac{pmc} proposed by \citet{Simola2021} and discussed in \Cref{subsubsec:PMC}.

\begin{figure*}
\begin{center}
\includegraphics[width=\textwidth]{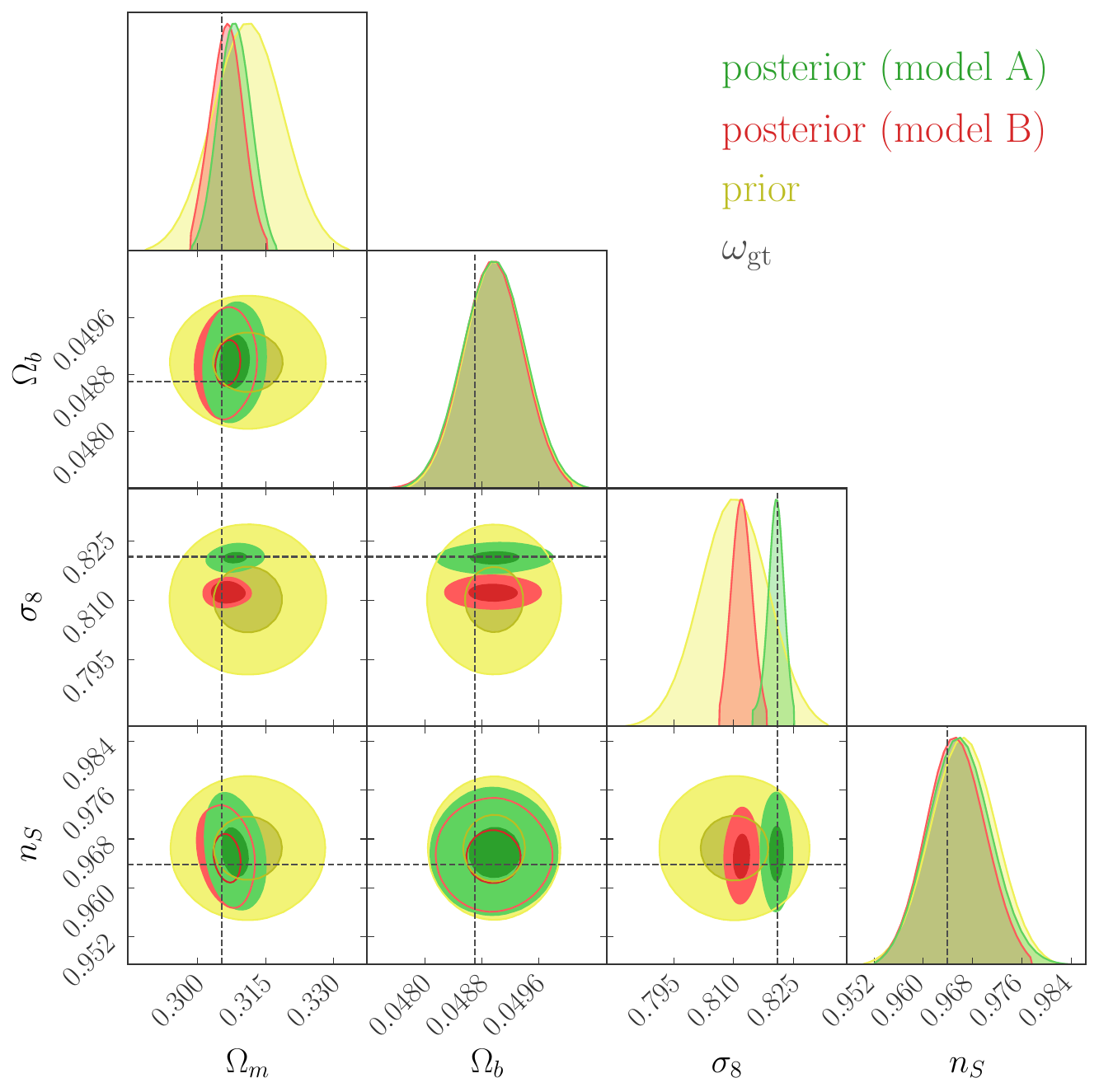}
\caption{\small{Prior $\p(\boldsymbol{\upomega})$ (yellow) and posterior $\p(\boldsymbol{\upomega}|\boldsymbol{\widetilde{\upomega}}_\mathrm{O})$ distributions of $(\Omega_m,\,\Omega_b,\,\sigma_8,\,n_S)$, estimated using Population Monte Carlo based on the Fisher-Rao distance $d_\mathrm{FR}(\boldsymbol{\widetilde{\upomega}}, \boldsymbol{\widetilde{\upomega}}_\mathrm{O})$ between simulated and observed summaries. The posterior with the well-specified Model A is represented by the green contours, whilst the red contours show the results with the misspecified Model B, both depicting the $1\sigma$ and $2\sigma$ credible regions. The percent-level misspecification in Model B introduces a bias exceeding $2\sigma$ in the $(\Omega_\mathrm{m}, \sigma_8)$ plane. The one-dimensional marginal posterior distributions of the parameters are displayed on the diagonal, following the same colour scheme. The dashed line indicates the ground truth parameters $\boldsymbol{\upomega}_\mathrm{gt}$.
}\label{fig:abc_posteriors}
}
\end{center}
\end{figure*}

To reduce computational costs, in this study, we inferred only $\Omega_\mathrm{m}$, $\Omega_\mathrm{b}$, $n_\mathrm{S}$, and $\sigma_8$ whilst fixing $h$ to the ground truth given in \Cref{tab:cosmo_gt}. This choice was further motivated by the fact that our data model carries no additional information beyond the prior concerning the value of $h$. For practical purposes, we used nine distinct \ac{pmc} populations of $64$ particles each, aggregating the $J=576$ final samples and corresponding weights to obtain a posterior distribution of $(\Omega_m,\,\Omega_b,\,\sigma_8,\,n_S)$. \Cref{fig:abc_posteriors} shows the posteriors obtained using a kernel density estimate with the $J$ weighted accepted samples. The shaded yellow area corresponds to $2\sigma$ credible contours on the prior. The posterior with the well-specified Model A is shown in green; it is unbiased for all cosmological parameters.

Replacing the accurate data Model A with the misspecified Model B for \ac{pmc} yields the posterior in red. This substitution results in over-concentrating the posterior distributions of $\sigma_8$, $\Omega_m$, and $n_S$, and introduces a bias exceeding $2\sigma$ in the $(\sigma_8,\Omega_m)$ plane. The bias in the marginal posterior of $\sigma_8$ is to be expected in light of the well-known degeneracy between $\sigma_8$ and galaxy bias parameters \citep{beutler20126df, arnalte2016joint, Desjacques_2018, repp2020galaxy}.

\section{Discussion and conclusion}
\label{sec:conclusion}

We have presented a method for thoroughly assessing individual or combined systematic effects in galaxy surveys using the {\power}. We provide a practical guide for evaluating the impact of various systematic effects, demonstrated using a spectroscopic galaxy survey model through a quantitative assessment of the effect of misspecified linear galaxy biases, extinction near the galactic plane, selection functions, punctual contaminations, inaccurate redshifts, and imprecise gravity solver.

The method introduced in this study offers a Bayesian framework for addressing model misspecification in field-based implicit-likelihood cosmological inference. Notably, we have shown that even small, percent-level errors can significantly skew cosmological inferences derived from galaxy clustering probes. This was illustrated by a bias exceeding $2\sigma$ in the $(\sigma_8,\Omega_m)$ plane (\Cref{fig:abc_posteriors}), which we were able to detect and correct (\Cref{fig:posterior_selfi,fig:impact_systematics}) prior to inferring the cosmological parameters. Correct identification of systematic effects is crucial for accurately measuring primordial features of the Universe, such as non-Gaussianities—a major scientific goal of next-generation galaxy surveys \citep{andrews2023bayesian}. Our method could serve as a critical tool for field-based \ac{ili} of these primordial features.

Our approach primarily centres on systematic effects that influence the {\power} reconstruction. While we have assumed that the inferred spectrum serves as a reliable proxy for cosmological parameters, there may be systematic effects that do not affect the {\power} yet still introduce biases in cosmological inferences. Such effects fall outside the current scope of this framework. In this paper, we assumed fixed deterministic linear galaxy bias parameters, but scale-dependent or non-linear bias models could be incorporated within the same framework. Addressing model misspecification for more sophisticated bias models would be particularly relevant for the inference of primordial non-Gaussianities. If a prior distribution of the bias values is available, they can be treated as additional nuisance parameters alongside $\boldsymbol{\uppsi}$ and marginalised over without altering the statistical framework presented here. Alternatively, they could be treated as additional input parameters alongside $\boldsymbol{\uptheta}$ and inferred jointly with the {\power}. These avenues remain for future investigation.

Potential extensions of the method include incorporating higher-order summary statistics from the full galaxy count fields into the inference \citep{gil2015power,philcox2022boss}, which provide valuable information, for instance for constraining dark energy \citep{fazolo2022skewness}. Other options include using wavelet scattering transform coefficients \citep{regaldo2024galaxy}, and additional cosmological probes such as peculiar velocity tracers \citep{prideaux2023field}. The compression step could be improved by using information-maximising neural networks \citep{charnock2018automatic,makinen2024hybrid,lanzieri2024optimal}, which complement physically motivated summary statistics. These beyond $2$-pt approaches are particularly relevant for extracting the maximum amount of information from upcoming galaxy surveys and could reveal additional sources of systematic effects which should be accounted for in cosmological inferences based on these summaries \citep{krause2024parameter}.

In conclusion, we have proposed a method to address the issue of model misspecification in implicit likelihood, field-based cosmological inferences. It enables a comprehensive investigation of both simulation- and observation-related systematic effects in galaxy surveys, leveraging both theoretical insights and prior knowledge of the {\power}. We demonstrated the effectiveness of this method by quantifying the impact of various systematic effects on the inferred {\power}. Furthermore, we emphasised the importance of using an accurate model for gravitational evolution within the hidden-box model: our results indicate that pure \ac{2lpt} is insufficient for the range of scales considered in this study, which is $k\in\left[ 1.75 \times 10^{-3},\, 4.47 \times 10^{-1}\right]$~$h$/Mpc. Our framework can be used with any hidden-box model of galaxy surveys developed for cosmological inference pipelines, making it a versatile tool for addressing model misspecification in cosmological data analysis. It has the potential to significantly enhance the robustness of cosmological inferences from upcoming galaxy surveys such as DESI, \textit{Euclid}, and LSST, for which the complexity of data models and the number of systematic effects will be substantial.

\section*{Code availability}

This study relies on the py{\selfi} implementation of the \ac{selfi} \citep{Leclercq2019SELFI} algorithm, which is publicly available at \href{https://pyselfi.florent-leclercq.eu}{pyselfi.florent-leclercq.eu}. The gravitational evolution is performed using a modified version of the publicly available \textsc{Simbelmyn\"e} cosmological solver \citep[\href{https://bitbucket.org/florent-leclercq/simbelmyne}{bitbucket.org/florent-leclercq/simbelmyne},][]{Leclercq2015ST}. The code and data underlying this paper, including the hidden-box model of galaxy surveys described in the article and routines to perform the inference using our framework in a cluster environment, are publicly available at \href{https://github.com/hoellin/selfisys\_public}{github.com/hoellin/selfisys\_public}.

\begin{appendix}

\section{Convergence of the posterior}
\label{appendix:convergence_covariance}

\begin{figure}[h!]
\begin{center}
\includegraphics[width=.5\textwidth]{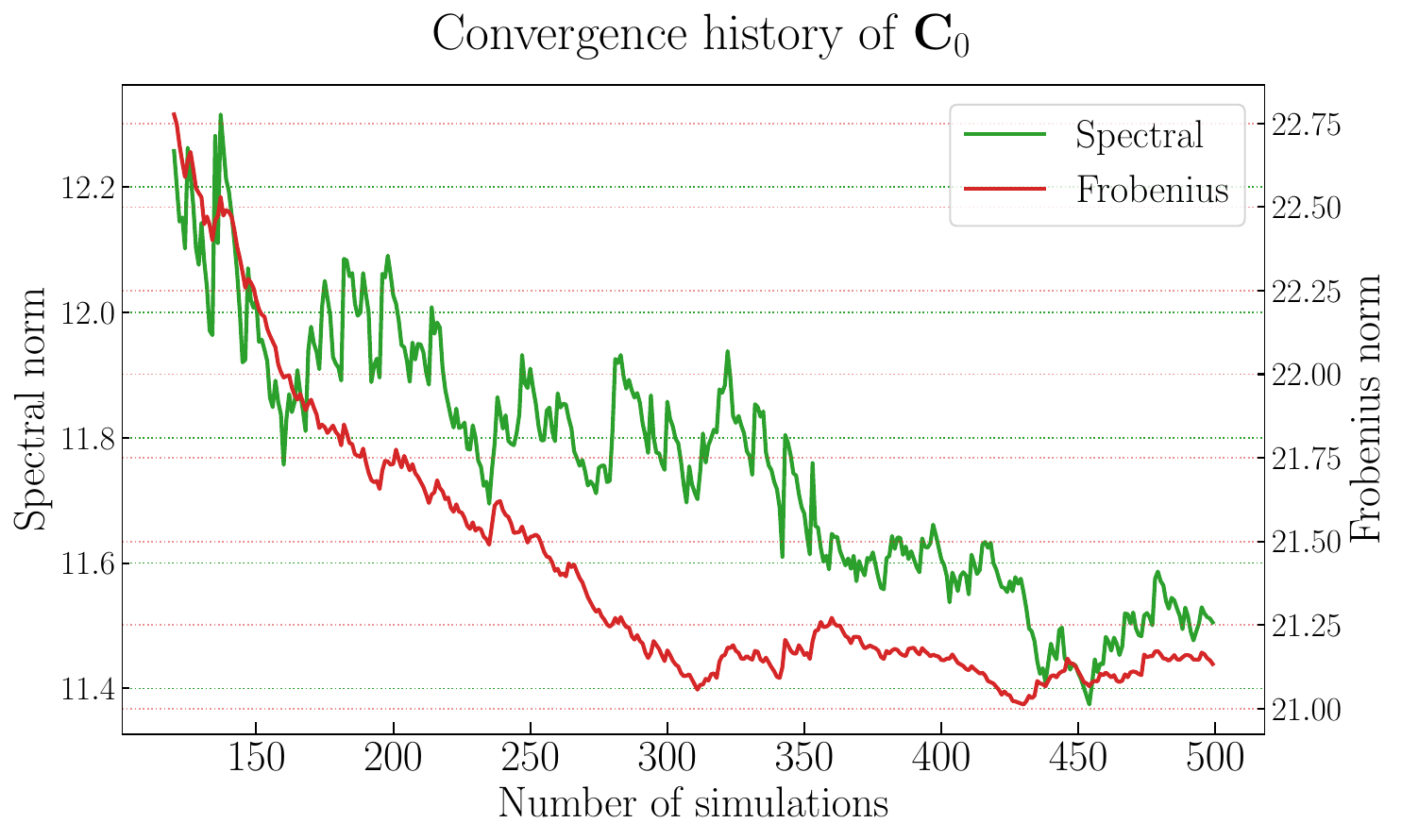}
\caption{\small{Convergence history of the covariance. We used the correlation matrix rather than $\textbf{C}_0$ to compute the norms.
}
\label{fig:cv_C0}}
\end{center}
\end{figure}

Considerations about the degrees of freedom in the estimated covariance matrix given by \Cref{eq:estimated_covariance_phi} impose the required number of simulations at the expansion point to be at least $N_0 \geq P+3=114$. However, the convergence may require more simulations depending for instance on the observational noise level. We checked based on the evolution of the spectral and Frobenius norms of $\textbf{C}_0$ that sufficient convergence was reached after $\simeq 450$ simulations (see \Cref{fig:cv_C0}), and we fixed the number of simulations at the expansion point to linearise the hidden-box to $N_0=500$. We also verified that using in between $\simeq 350$ and $\simeq 500$ simulations yielded vanishing differences in the corresponding {\power} posteriors.

Similarly, we ensured sufficient convergence for the gradients by looking at the evolution of $(\boldsymbol{\nabla}\textbf{f}_0)^\intercal \textbf{C}_0^{-1} \boldsymbol{\nabla} \textbf{f}_0$ where $\textbf{C}_0$ was estimated using all the $N_0=500$ simulations available at the expansion point. We fixed $N_S=10$ to obtain the \ac{selfi} posteriors presented in this article, and checked using Model A that restricting to any number of simulations between $2$ and $30$ per directional derivative yielded negligible difference in the posterior.

\section{Effective posterior for the initial matter power spectrum when the prior mean differs from the expansion point}
\label{appendix:effective_posterior}

We generalise the derivation of the \ac{selfi} equations from \citet{Leclercq2019SELFI}, to the case where the prior mean $\boldsymbol{\hat{\uptheta}}_{\boldsymbol{\upomega}}$ differs from the expansion point $\boldsymbol{\uptheta}_0$. The \ac{selfi} effective log-likelihood given by \Cref{eq:linearised_effective_likelihood} can be rewritten in canonical form as
\begin{equation}
    -2 \hat{\ell}^{N_0}_{\uptheta}(\boldsymbol{\uptheta}) = \log \left| 2\pi \textbf{C}_0 \right| + \boldsymbol{\upmu}_0^\intercal \textbf{N}_0 \boldsymbol{\upmu}_0 - 2\boldsymbol{\upmu}_0 \boldsymbol{\uptheta} + \boldsymbol{\uptheta}^\intercal \textbf{N}_0^{-1} \boldsymbol{\uptheta},
\end{equation}
with $\boldsymbol{\upmu}_0 \eqdef \textbf{N}_0^{-1} \textbf{y}_0 = \textbf{N}_0^{-1}\boldsymbol{\uptheta}_0 + (\nabla \textbf{f}_0)^\intercal \textbf{C}_0^{-1} (\boldsymbol{\Phi}_\mathrm{O}-\textbf{f}_0)$, where we have defined
\begin{eqnarray}
\textbf{N}_0 &\eqdef& \left[ (\nabla \textbf{f}_0)^\intercal \textbf{C}_0^{-1} \nabla \textbf{f}_0 \right]^{-1},\nonumber\\
\textbf{y}_0 &\eqdef& \boldsymbol{\uptheta}_0 + (\nabla \textbf{f}_0)^{-1} \cdot (\boldsymbol{\Phi}_\mathrm{O}-\textbf{f}_0).
\end{eqnarray}
Similarly, the prior given by \Cref{eq:Planck_prior} can be written
\begin{equation}
-2\log \p(\boldsymbol{\uptheta}) = \log\left| 2\pi\textbf{S} \right| + \boldsymbol{\upeta}^\intercal \textbf{S} \boldsymbol{\upeta} -2 \boldsymbol{\upeta}^\intercal \boldsymbol{\uptheta} + \boldsymbol{\uptheta}^\intercal \textbf{S}^{-1} \boldsymbol{\uptheta}
\end{equation}
in canonical form, where $\boldsymbol{\upeta} \eqdef \textbf{S}^{-1} \boldsymbol{\hat{\uptheta}}_{\boldsymbol{\upomega}}$. The effective posterior therefore verifies
\begin{eqnarray}
-2\log \p(\boldsymbol{\uptheta}|\boldsymbol{\Phi}_\mathrm{O}) &=& -2 \hat{\ell}^{N_0}_{\uptheta}(\boldsymbol{\uptheta}) -2\log \p(\boldsymbol{\uptheta}) \nonumber\\
&=& -2(\boldsymbol{\upmu}_0+\boldsymbol{\upeta})^\intercal \boldsymbol{\uptheta}
+ \boldsymbol{\uptheta}^\intercal (\textbf{N}_0^{-1} + \textbf{S}^{-1}) \boldsymbol{\uptheta}\nonumber\\
& & + \text{ constant terms},
\end{eqnarray}
where we recognise the canonical form of a Gaussian distribution with covariance matrix
\begin{equation}
\boldsymbol{\Gamma} \eqdef \left(\textbf{N}_0^{-1} + \textbf{S}^{-1}\right)^{-1}= \left[ (\nabla \textbf{f}_0)^\intercal \textbf{C}_0^{-1} \nabla \textbf{f}_0 + \textbf{S}^{-1} \right]^{-1},
\end{equation}
and mean
\begin{eqnarray}
\boldsymbol{\gamma} & \eqdef & \boldsymbol{\Gamma} (\boldsymbol{\upmu}_0+\boldsymbol{\upeta}) \nonumber\\
& = & \boldsymbol{\Gamma} \left[ (\nabla \textbf{f}_0)^\intercal \textbf{C}_0^{-1} \nabla \textbf{f}_0\boldsymbol{\uptheta}_0 + (\nabla \textbf{f}_0)^\intercal \textbf{C}_0^{-1} \nabla \textbf{f}_0 (\nabla \textbf{f}_0)^{-1}\right. \nonumber\\
& & \left. \cdot(\boldsymbol{\Phi}_\mathrm{O}-\textbf{f}_0) + \textbf{S}^{-1} \boldsymbol{\hat{\uptheta}}_{\boldsymbol{\upomega}} \right]\nonumber\\
& = & \boldsymbol{\Gamma} \left\{ \left[(\nabla \textbf{f}_0)^\intercal \textbf{C}_0^{-1} \nabla \textbf{f}_0 + \textbf{S}^{-1}\right] \boldsymbol{\uptheta}_0 - \textbf{S}^{-1} \boldsymbol{\uptheta}_0 \right. \nonumber\\
& & \left. + (\nabla \textbf{f}_0)^\intercal \textbf{C}_0^{-1} \cdot (\boldsymbol{\Phi}_\mathrm{O}-\textbf{f}_0) + \textbf{S}^{-1} \boldsymbol{\hat{\uptheta}}_{\boldsymbol{\upomega}} \right\} \nonumber\\
& = & \boldsymbol{\uptheta}_0 + \boldsymbol{\Gamma} \textbf{S}^{-1} \boldsymbol{\Delta} + \boldsymbol{\Gamma} (\nabla \textbf{f}_0)^\intercal \textbf{C}_0^{-1} \cdot (\boldsymbol{\Phi}_\mathrm{O}-\textbf{f}_0),
\end{eqnarray}
where $\boldsymbol{\Delta}\eqdef \left(\boldsymbol{\uptheta}_0 - \boldsymbol{\hat{\uptheta}}_{\boldsymbol{\upomega}}\right)$, giving \Cref{eq:filter_var}.

\section{Consistency checks}

\subsection{Unbiasedness of the posterior initial power spectrum}
\label{appendix:unbiasedness}
To consistently assess the unbiasedness of the method, we sampled ten different ground truth set of cosmological parameters from the prior, computed the corresponding observed vector $\boldsymbol\Phi_\mathrm{O}$ for each case, and confirmed both visually and quantitatively that the resulting \ac{selfi} posterior remained consistently unbiased. This demonstrates the robustness of the method with respect to the value of the ground truth cosmological parameters.

\subsection{Inference using a wiggle-less prior}
\label{appendix:wiggleless}

As a consistency check, we verified that \ac{selfi} successfully retrieves the \ac{baos} within the set-up described in this article, even when the prior does not provide any information about the oscillations. We postulate a Gaussian prior given by
\begin{equation}
    -2\log \p(\boldsymbol{\uptheta}) \eqdef \log\left| 2\pi \textbf{S} \right| + \left\lVert\boldsymbol{\uptheta}-\boldsymbol{\uptheta}_0\right\rVert^2_{\textbf{S}^{-1}},
    \label{eq:prior_wiggleless}
\end{equation}
where the expansion point $\boldsymbol{\uptheta}_0$ and the covariance matrix $\textbf{S}$ differ from those of \Cref{eq:Planck_prior}. Namely, we set the mean to $\boldsymbol{\uptheta}_0 \eqdef \boldsymbol{1}_{\mathbb{R}^S}$, corresponding to the BBKS power spectrum for the fiducial cosmological parameters, and, following the procedure described by \citet{Leclercq2019SELFI}, we define the prior covariance matrix as
\begin{equation}
    \textbf{S} \eqdef \theta_\mathrm{norm}^2 \, \textbf{u}\textbf{u}^\intercal \circ \textbf{K},
    \label{eq:prior_covariance_wiggleless}
\end{equation}
where $\circ$ denotes the Hadamard product and $\textbf{K}$ is a radial basis function defined by
\begin{equation}
\left(\textbf{K}\right)_{ss'} \eqdef \mathrm{exp} \left[ -\frac{1}{2} \left(\frac{k_s - k_{s'}}{k_\mathrm{corr}}\right)^2 \right],
\end{equation}
\begin{figure*}[ht!]
    \begin{center}
    \includegraphics[width=\textwidth]{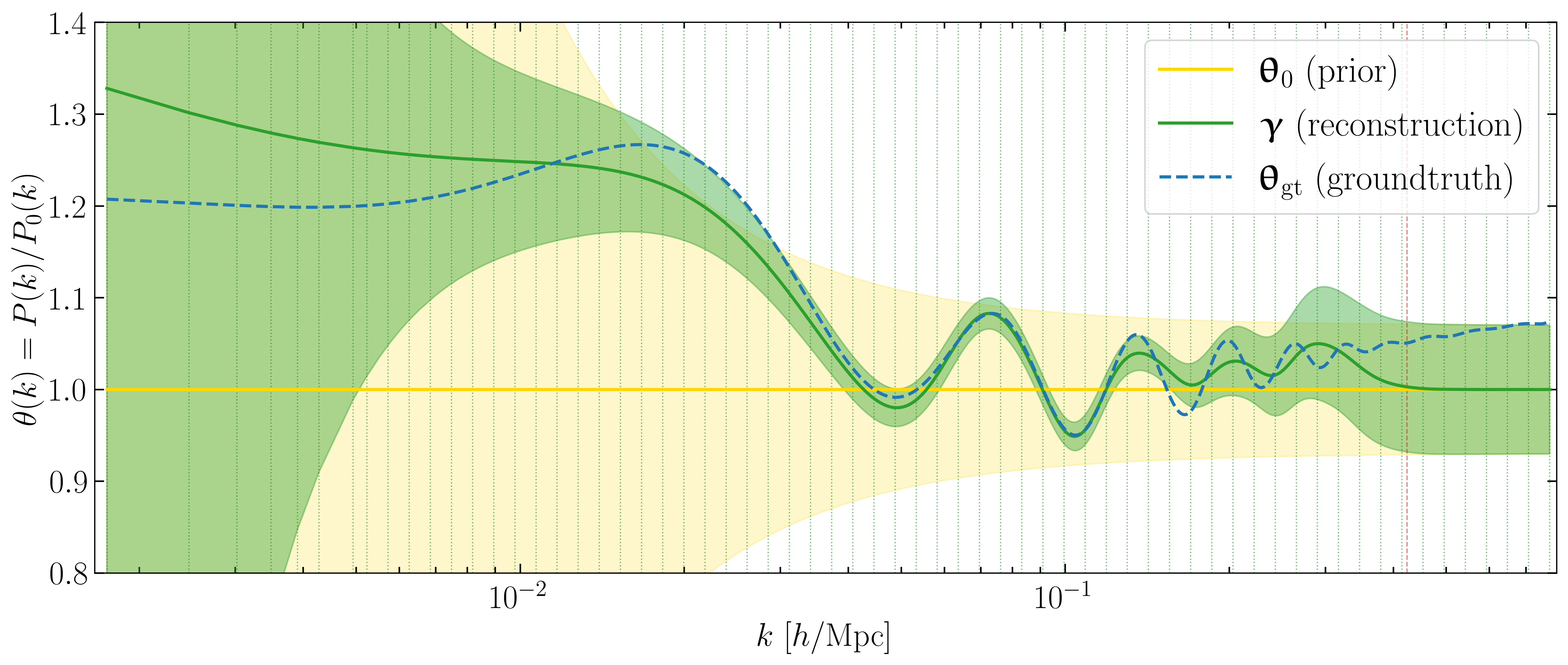}
    \caption{\small{\label{fig:posterior_selfi2019}\small{Wiggle-less prior and corresponding \ac{selfi} posterior on the {\power} $\boldsymbol{\uptheta}$. The prior mean $\boldsymbol{\uptheta}_0$, also used as expansion point, is indicated by the yellow line; its $2\sigma$ credible region corresponds to the yellow-shaded area. The posterior mean $\boldsymbol{\upgamma}$ and $2\sigma$ credible region correspond to the green line and shaded area. The ground truth $\boldsymbol{\uptheta}_\mathrm{gt}$ is indicated by the solid blue line. The dashed red line denotes the Nyquist frequency. Even though the prior contains no information about the \ac{baos}, the \ac{selfi} posterior} is consistently unbiased across all scales and accurately retrieves the oscillations up to $k\simeq 0.2$~$h$/Mpc.
    }}
    \end{center}
\end{figure*}
and $\textbf{u}$ is a correction accounting for cosmic variance,
\begin{equation}
(\textbf{u})_s \eqdef 1 + \sigma_s = 1+\frac{\alpha_\mathrm{cv}}{k_s^{3/2}}.
\end{equation}
$k_\mathrm{corr}$ determines the length scale on which power spectrum amplitudes of different wave number correlate with each other. $\alpha_\mathrm{cv}$ characterises the ``strength'' of cosmic variance given the effective volume considered and $\theta_\mathrm{norm}^2$ captures the overall amplitude of the covariance matrix $\textbf{S}$. The strength of cosmic variance within our simulation volume shall satisfy $\alpha_\mathrm{cv} = \sqrt{k^3/\Tilde{N_k}}$ at all scales $k$, where $\Tilde{N_k}$ is the effective number of modes at scale $k$, accounting for the window function. As in \citet{Leclercq2019SELFI}, the hyperparameters $k_\mathrm{corr}$ and $\theta_\mathrm{norm}$ are optimised to reproduce the shape of a fiducial wiggle vector $\boldsymbol{\uptheta}_\mathrm{fid}$, using the effective posterior distribution obtained in \Cref{subsubsec:selfi_posterior} as the likelihood; that is,
\begin{equation}
-2\log \p(k_\mathrm{corr},\theta_\mathrm{norm}\vert\boldsymbol{\uptheta}) \eqdef \log \left| 2\pi\boldsymbol{\Gamma} \right| + \left\lVert\boldsymbol{\uptheta}_\mathrm{fid}-\boldsymbol{\upgamma}\right\rVert _{\boldsymbol{\Gamma}^{-1}},
\end{equation}
where $\boldsymbol{\upgamma}$ and $\boldsymbol{\Gamma}$ inherit their dependence on $k_\mathrm{corr}$ and $\theta_\mathrm{norm}$ through $\textbf{S}$. We use Gaussian hyperpriors on $(k_\mathrm{corr},\theta_\mathrm{norm})$, namely $k_\mathrm{corr}\sim\mathcal{G}(0.020,0.015^2)$ $[h\,\mathrm{Mpc}^{-1}]$ and $\theta_\mathrm{norm}\sim\mathcal{G}(0.2,0.3^2)$. We estimated the maximum \textit{a posteriori} using L-BFGS \citep{liu1989limited} and found $k_\mathrm{corr}=0.012458\,h\textrm{Mpc}^{-1}$ and $\theta_\mathrm{norm}=0.034743$, values that were used to obtain the prior on $\boldsymbol{\uptheta}$ corresponding to the shaded yellow area in \Cref{fig:posterior_selfi2019}.

The posterior obtained with the wiggle-less prior is in green in \Cref{fig:posterior_selfi2019}. It is consistent with the ground truth. Notably, it accurately retrieves the oscillations up to $k\simeq 0.2$~$h$/Mpc, even though the prior is completely agnostic to the \ac{baos}, except through its hyperparameter optimisation.

\section{Impact of linear galaxy bias parameters}
\label{appendix:linear_bias}

The linear galaxy bias parameters have a quadratic scale-independent impact on the galaxy overdensity power spectrum. We show qualitatively that this is expected to hold true for the \ac{selfi} posterior, to some extent. Indeed, neglecting the term $\boldsymbol{\Gamma}\textbf{S}^{-1}\boldsymbol{\Delta}$, which we measured to be smaller than $10^{-3}$ at all scales in our set-up, the \ac{selfi} filter equation~\eqref{eq:filter_mean} for the mean of the posterior on the matter power spectrum can be rewritten:
\begin{equation}
    \boldsymbol{\upgamma} = \boldsymbol{\uptheta}_0 + \boldsymbol{\Gamma} \, (\boldsymbol{\nabla}_{\uptheta} \textbf{f}_0)^\intercal \, \textbf{C}_0^{-1}\boldsymbol{\Phi}_\mathrm{O}- \boldsymbol{\Gamma} \, (\boldsymbol{\nabla}_{\uptheta} \textbf{f}_0)^\intercal \, \textbf{C}_0^{-1}\textbf{f}_0.
    \label{eq:posterior_mis_bias}
\end{equation}
If we consider for simplicity the case where the overall noise level is $\sigma=0$, for a fixed phase, $P^{\mathrm{g}_i}(k_r)$ becomes a deterministic function of $b_{\mathrm{g}_i}$ given by
\begin{align*}
P^{\mathrm{g}_i}(k_r) &\propto b_{\mathrm{g}_i}^2\sum_{|\textbf{k}|=k_r} \left| \widehat{\textbf{W}_i\boldsymbol{\updelta}^{z_i}}(k) \right|^2,
\end{align*}
where $\widehat{\textbf{h}}$ denotes the discrete Fourier transform of $\textbf{h}$. Hence, in the absence of noise, a constant relative error on all the three squared linear galaxy biases has a scale-independent impact on any simulated summary $\boldsymbol{\Phi}$ obtained with the hidden-box simulator. By linearity of the $\mathbb{E}$ and $\partial_\theta$ operators, the relative impacts on $\textbf{f}_0$ and $\boldsymbol{\nabla}_{\uptheta} \textbf{f}_0$ are scale-independent as well and quadratic in $b_{\mathrm{g}_i}$. Similarly, from \Cref{eq:estimated_covariance_phi}, $\textbf{C}_0^{-1}$ scales as $b_{\mathrm{g}_i}^{-4}$, so that, from \Cref{eq:filter_var}, $\boldsymbol{\Gamma}$ and therefore $\boldsymbol{\Gamma} \, (\boldsymbol{\nabla}_{\uptheta} \textbf{f}_0)^\intercal \, \textbf{C}_0^{-1}\textbf{f}_0$ are left unaffected and only the $\boldsymbol{\Gamma} \, (\boldsymbol{\nabla}_{\uptheta} \textbf{f}_0)^\intercal \, \textbf{C}_0^{-1}\boldsymbol{\Phi}_\mathrm{O}$ term in \Cref{eq:posterior_mis_bias} is impacted by misspecified galaxy biases. If the misspecified biases are such that the relative error $(b^\textrm{mis}_i)^2/b_{\mathrm{g}_i}^2 = \epsilon$ is independent of the population $\textrm{g}_i$, then every component of the vector $(\boldsymbol{\nabla}_{\uptheta} \textbf{f}_0)^\intercal \, \textbf{C}_0^{-1}\boldsymbol{\Phi}_\mathrm{O}$ is affected by a factor $\epsilon$, and the overall effect inherits its scale-dependence solely through the covariance matrix $\boldsymbol{\Gamma}$. This is the behaviour observed on \Cref{fig:impact_systematics_a}, where, for a misspecified bias, all the scales undergo about the same shift in number of standard deviations.

\section{Implicit likelihood inference of cosmological parameters}

\subsection{Numerical set-up}

For the \ac{ili} of cosmological parameters, we used the variant of \ac{pmc} from \citet{Simola2021} implemented in the \texttt{ELFI} package \citep[\href{https://github.com/elfi-dev/elfi}{github.com/elfi-dev/elfi},][]{Lintusaari2018}, with minor technical modifications which do not affect the statistical method.

\subsection{Population Monte Carlo hyperparameters}

The Population Monte Carlo algorithm automatically selects a tolerance sequence based on the procedure described by \citet{Simola2021}. It is not completely $\epsilon$-free as one must choose a stopping rule and an initial acceptance threshold for the rejection sampler, the choice of the latter being crucial to prevent the algorithm from getting stuck in local overdense regions of the parameter space. In this study, we set the initial quantile used to determine the corresponding acceptance threshold for the rejection sampling stage to $q_0 = 0.2$.

\end{appendix}

\begin{acknowledgements}
We thank Chloé Barjou-Delayre, Guilhem Lavaux (\orcid{0000-0003-0143-8891}), Gary Mamon (\orcid{0000-0001-8956-5953}), Nhat-Minh Nguyen (\orcid{0000-0002-2542-7233}) and Fabian Schmidt (\orcid{0000-0002-9839-1614}) for useful discussions and feedback to enhance this manuscript.
The work has made use of the Infinity Cluster hosted by the Institut d'Astrophysique de Paris. We appreciate the efforts of Stéphane Rouberol for running the cluster smoothly. 
FL and TH acknowledge financial support from the Agence Nationale de la Recherche (ANR) through the grant INFOCW, under reference ANR-23-CE46-0006-01.
FL and TH acknowledge financial support from the Simons Collaboration on “Learning the Universe”.
This work was done within the Aquila Consortium (\url{https://www.aquila-consortium.org/}).
\end{acknowledgements}

\section*{References}
\bibliography{biblio_updated}

\begin{thebibliography}{132}%
\makeatletter
\providecommand \@ifxundefined [1]{%
 \@ifx{#1\undefined}
}%
\providecommand \@ifnum [1]{%
 \ifnum #1\expandafter \@firstoftwo
 \else \expandafter \@secondoftwo
 \fi
}%
\providecommand \@ifx [1]{%
 \ifx #1\expandafter \@firstoftwo
 \else \expandafter \@secondoftwo
 \fi
}%
\providecommand \natexlab [1]{#1}%
\providecommand \enquote  [1]{``#1''}%
\providecommand \bibnamefont  [1]{#1}%
\providecommand \bibfnamefont [1]{#1}%
\providecommand \citenamefont [1]{#1}%
\providecommand \href@noop [0]{\@secondoftwo}%
\providecommand \href [0]{\begingroup \@sanitize@url \@href}%
\providecommand \@href[1]{\@@startlink{#1}\@@href}%
\providecommand \@@href[1]{\endgroup#1\@@endlink}%
\providecommand \@sanitize@url [0]{\catcode `\\12\catcode `\$12\catcode `\&12\catcode `\#12\catcode `\^12\catcode `\_12\catcode `\%12\relax}%
\providecommand \@@startlink[1]{}%
\providecommand \@@endlink[0]{}%
\newcommand{\PineGreen}[1]{\textcolor{PineGreen}{#1}}%
\providecommand \url  [0]{\begingroup\@sanitize@url \@url }%
\providecommand \@url [1]{\endgroup\@href {#1}{\urlprefix }}%
\providecommand \urlprefix  [0]{URL }%
\providecommand \Eprint [0]{\href }%
\providecommand \doibase [0]{http://dx.doi.org/}%
\providecommand \selectlanguage [0]{\@gobble}%
\providecommand \bibinfo  [0]{\@secondoftwo}%
\providecommand \bibfield  [0]{\@secondoftwo}%
\providecommand \translation [1]{[#1]}%
\providecommand \BibitemOpen [0]{}%
\providecommand \bibitemStop [0]{}%
\providecommand \bibitemNoStop [0]{.\EOS\space}%
\providecommand \EOS [0]{\spacefactor3000\relax}%
\providecommand \BibitemShut  [1]{\csname bibitem#1\endcsname}%
\let\auto@bib@innerbib\@empty
\bibitem [{{Addison} {\textit{et~al}}\mbox{.}(2019)\citenamefont {{Addison}, {Bennett}, {Jeong}, {Komatsu},\ \&\ {Weiland}}}]{addison2019impact}%
{(\PineGreen{{Addison} {\textit{et~al}}\mbox{.}}, \PineGreen{2019})}  \BibitemOpen
  \bibfield  {author} {\bibinfo {author} {\bibfnamefont {G.~E.}\ \bibnamefont {{Addison}}}, \bibinfo {author} {\bibfnamefont {C.~L.}\ \bibnamefont {{Bennett}}}, \bibinfo {author} {\bibfnamefont {D.}~\bibnamefont {{Jeong}}}, \bibinfo {author} {\bibfnamefont {E.}~\bibnamefont {{Komatsu}}}, \bibinfo {author} {\bibfnamefont {J.~L.}\ \bibnamefont {{Weiland}}},\ }\emph {{The Impact of Line Misidentification on Cosmological Constraints from Euclid and Other Spectroscopic Galaxy Surveys}},\ \href {\doibase 10.3847/1538-4357/ab22a0} {\bibfield  {journal} {\bibinfo  {journal} {\apj}\ }\textbf {\bibinfo {volume} {879}},\ \bibinfo {eid} {15} (\bibinfo {year} {2019})},\ \Eprint {https://arxiv.org/abs/1811.10668} {arXiv:1811.10668 [astro-ph.CO]} \BibitemShut {NoStop}%
\bibitem [{{Albrecht} {\textit{et~al}}\mbox{.}(2006){Albrecht}, {Bernstein}, {Cahn}, {Freedman}, {Hewitt}, {Hu}, {Huth}, {Kamionkowski}, {Kolb}, {Knox} \emph {et~al.}}]{albrecht2006report}%
{(\PineGreen{{Albrecht} {\textit{et~al}}\mbox{.}}, \PineGreen{2006})}  \BibitemOpen
  \bibfield  {author} {\bibinfo {author} {\bibfnamefont {A.}~\bibnamefont {{Albrecht}}}, \bibinfo {author} {\bibfnamefont {G.}~\bibnamefont {{Bernstein}}}, \bibinfo {author} {\bibfnamefont {R.}~\bibnamefont {{Cahn}}}, \bibinfo {author} {\bibfnamefont {W.~L.}\ \bibnamefont {{Freedman}}}, \bibinfo {author} {\bibfnamefont {J.}~\bibnamefont {{Hewitt}}}, \bibinfo {author} {\bibfnamefont {W.}~\bibnamefont {{Hu}}}, \bibinfo {author} {\bibfnamefont {J.}~\bibnamefont {{Huth}}}, \bibinfo {author} {\bibfnamefont {M.}~\bibnamefont {{Kamionkowski}}}, \bibinfo {author} {\bibfnamefont {E.~W.}\ \bibnamefont {{Kolb}}}, \bibinfo {author} {\bibfnamefont {L.}~\bibnamefont {{Knox}}}, \emph {et~al.},\ }\emph {{Report of the Dark Energy Task Force}},\ \bibfield  {journal} {\bibinfo  {journal} {arXiv e-prints}\ }\href {\doibase 10.48550/arXiv.astro-ph/0609591} {10.48550/arXiv.astro-ph/0609591} (\bibinfo {year} {2006}),\ \bibinfo {note} {{arXiv:astro-ph/0609591}}\BibitemShut {NoStop}%
\bibitem [{{Alsing} \& {Wandelt}(2018)\citenamefont {{Alsing}\ \&\ {Wandelt}}}]{Alsing_Wandelt_2018}%
{(\PineGreen{{Alsing} \& {Wandelt}}, \PineGreen{2018})}  \BibitemOpen
  \bibfield  {author} {\bibinfo {author} {\bibfnamefont {J.}~\bibnamefont {{Alsing}}}, \bibinfo {author} {\bibfnamefont {B.}~\bibnamefont {{Wandelt}}},\ }\emph {{Generalized massive optimal data compression}},\ \href {\doibase 10.1093/mnrasl/sly029} {\bibfield  {journal} {\bibinfo  {journal} {\mnras}\ }\textbf {\bibinfo {volume} {476}},\ \bibinfo {pages} {L60} (\bibinfo {year} {2018})},\ \Eprint {https://arxiv.org/abs/1712.00012} {arXiv:1712.00012 [astro-ph.CO]} \BibitemShut {NoStop}%
\bibitem [{{Alsing}, {Wandelt} \& {Feeney}(2018)\citenamefont {{Alsing}, {Wandelt},\ \&\ {Feeney}}}]{alsing2018delfi}%
{(\PineGreen{{Alsing}, {Wandelt} \& {Feeney}}, \PineGreen{2018})}  \BibitemOpen
  \bibfield  {author} {\bibinfo {author} {\bibfnamefont {J.}~\bibnamefont {{Alsing}}}, \bibinfo {author} {\bibfnamefont {B.}~\bibnamefont {{Wandelt}}}, \bibinfo {author} {\bibfnamefont {S.}~\bibnamefont {{Feeney}}},\ }\emph {{Massive optimal data compression and density estimation for scalable, likelihood-free inference in cosmology}},\ \href {\doibase 10.1093/mnras/sty819} {\bibfield  {journal} {\bibinfo  {journal} {\mnras}\ }\textbf {\bibinfo {volume} {477}},\ \bibinfo {pages} {2874} (\bibinfo {year} {2018})},\ \Eprint {https://arxiv.org/abs/1801.01497} {arXiv:1801.01497 [astro-ph.CO]} \BibitemShut {NoStop}%
\bibitem [{{Alsing} {\textit{et~al}}\mbox{.}(2019)\citenamefont {{Alsing}, {Charnock}, {Feeney},\ \&\ {Wandelt}}}]{alsing2019fast}%
{(\PineGreen{{Alsing} {\textit{et~al}}\mbox{.}}, \PineGreen{2019})}  \BibitemOpen
  \bibfield  {author} {\bibinfo {author} {\bibfnamefont {J.}~\bibnamefont {{Alsing}}}, \bibinfo {author} {\bibfnamefont {T.}~\bibnamefont {{Charnock}}}, \bibinfo {author} {\bibfnamefont {S.}~\bibnamefont {{Feeney}}}, \bibinfo {author} {\bibfnamefont {B.}~\bibnamefont {{Wandelt}}},\ }\emph {{Fast likelihood-free cosmology with neural density estimators and active learning}},\ \href {\doibase 10.1093/mnras/stz1960} {\bibfield  {journal} {\bibinfo  {journal} {\mnras}\ }\textbf {\bibinfo {volume} {488}},\ \bibinfo {pages} {4440} (\bibinfo {year} {2019})},\ \Eprint {https://arxiv.org/abs/1903.00007} {arXiv:1903.00007 [astro-ph.CO]} \BibitemShut {NoStop}%
\bibitem [{{Alsing} {\textit{et~al}}\mbox{.}(2016)\citenamefont {{Alsing}, {Heavens}, {Jaffe}, {Kiessling}, {Wandelt},\ \&\ {Hoffmann}}}]{alsing2016hierarchical}%
{(\PineGreen{{Alsing} {\textit{et~al}}\mbox{.}}, \PineGreen{2016})}  \BibitemOpen
  \bibfield  {author} {\bibinfo {author} {\bibfnamefont {J.}~\bibnamefont {{Alsing}}}, \bibinfo {author} {\bibfnamefont {A.}~\bibnamefont {{Heavens}}}, \bibinfo {author} {\bibfnamefont {A.~H.}\ \bibnamefont {{Jaffe}}}, \bibinfo {author} {\bibfnamefont {A.}~\bibnamefont {{Kiessling}}}, \bibinfo {author} {\bibfnamefont {B.}~\bibnamefont {{Wandelt}}}, \bibinfo {author} {\bibfnamefont {T.}~\bibnamefont {{Hoffmann}}},\ }\emph {{Hierarchical cosmic shear power spectrum inference}},\ \href {\doibase 10.1093/mnras/stv2501} {\bibfield  {journal} {\bibinfo  {journal} {\mnras}\ }\textbf {\bibinfo {volume} {455}},\ \bibinfo {pages} {4452} (\bibinfo {year} {2016})},\ \Eprint {https://arxiv.org/abs/1505.07840} {arXiv:1505.07840 [astro-ph.CO]} \BibitemShut {NoStop}%
\bibitem [{{Anderson} {\textit{et~al}}\mbox{.}(2014){Anderson}, {Aubourg}, {Bailey}, {Beutler}, {Bhardwaj}, {Blanton}, {Bolton}, {Brinkmann}, {Brownstein}, {Burden} \emph {et~al.}}]{anderson2014clustering}%
{(\PineGreen{{Anderson} {\textit{et~al}}\mbox{.}}, \PineGreen{2014})}  \BibitemOpen
  \bibfield  {author} {\bibinfo {author} {\bibfnamefont {L.}~\bibnamefont {{Anderson}}}, \bibinfo {author} {\bibfnamefont {{\'E}.}~\bibnamefont {{Aubourg}}}, \bibinfo {author} {\bibfnamefont {S.}~\bibnamefont {{Bailey}}}, \bibinfo {author} {\bibfnamefont {F.}~\bibnamefont {{Beutler}}}, \bibinfo {author} {\bibfnamefont {V.}~\bibnamefont {{Bhardwaj}}}, \bibinfo {author} {\bibfnamefont {M.}~\bibnamefont {{Blanton}}}, \bibinfo {author} {\bibfnamefont {A.~S.}\ \bibnamefont {{Bolton}}}, \bibinfo {author} {\bibfnamefont {J.}~\bibnamefont {{Brinkmann}}}, \bibinfo {author} {\bibfnamefont {J.~R.}\ \bibnamefont {{Brownstein}}}, \bibinfo {author} {\bibfnamefont {A.}~\bibnamefont {{Burden}}}, \emph {et~al.},\ }\emph {{The clustering of galaxies in the SDSS-III Baryon Oscillation Spectroscopic Survey: baryon acoustic oscillations in the Data Releases 10 and 11 Galaxy samples}},\ \href {\doibase 10.1093/mnras/stu523} {\bibfield  {journal} {\bibinfo  {journal} {\mnras}\ }\textbf {\bibinfo {volume} {441}},\ \bibinfo {pages} {24} (\bibinfo {year} {2014})},\ \Eprint {https://arxiv.org/abs/1312.4877} {arXiv:1312.4877 [astro-ph.CO]} \BibitemShut {NoStop}%
\bibitem [{{Andrews} {\textit{et~al}}\mbox{.}(2023)\citenamefont {{Andrews}, {Jasche}, {Lavaux},\ \&\ {Schmidt}}}]{andrews2023bayesian}%
{(\PineGreen{{Andrews} {\textit{et~al}}\mbox{.}}, \PineGreen{2023})}  \BibitemOpen
  \bibfield  {author} {\bibinfo {author} {\bibfnamefont {A.}~\bibnamefont {{Andrews}}}, \bibinfo {author} {\bibfnamefont {J.}~\bibnamefont {{Jasche}}}, \bibinfo {author} {\bibfnamefont {G.}~\bibnamefont {{Lavaux}}}, \bibinfo {author} {\bibfnamefont {F.}~\bibnamefont {{Schmidt}}},\ }\emph {{Bayesian field-level inference of primordial non-Gaussianity using next-generation galaxy surveys}},\ \href {\doibase 10.1093/mnras/stad432} {\bibfield  {journal} {\bibinfo  {journal} {\mnras}\ }\textbf {\bibinfo {volume} {520}},\ \bibinfo {pages} {5746} (\bibinfo {year} {2023})},\ \Eprint {https://arxiv.org/abs/2203.08838} {arXiv:2203.08838 [astro-ph.CO]} \BibitemShut {NoStop}%
\bibitem [{{Arnalte-Mur} {\textit{et~al}}\mbox{.}(2016)\citenamefont {{Arnalte-Mur}, {Vielva}, {Mart{\'\i}nez}, {Sanz}, {Saar},\ \&\ {Paredes}}}]{arnalte2016joint}%
{(\PineGreen{{Arnalte-Mur} {\textit{et~al}}\mbox{.}}, \PineGreen{2016})}  \BibitemOpen
  \bibfield  {author} {\bibinfo {author} {\bibfnamefont {P.}~\bibnamefont {{Arnalte-Mur}}}, \bibinfo {author} {\bibfnamefont {P.}~\bibnamefont {{Vielva}}}, \bibinfo {author} {\bibfnamefont {V.~J.}\ \bibnamefont {{Mart{\'\i}nez}}}, \bibinfo {author} {\bibfnamefont {J.~L.}\ \bibnamefont {{Sanz}}}, \bibinfo {author} {\bibfnamefont {E.}~\bibnamefont {{Saar}}}, \bibinfo {author} {\bibfnamefont {S.}~\bibnamefont {{Paredes}}},\ }\emph {{Joint constraints on galaxy bias and {\ensuremath{\sigma}}$_{8}$ through the N-pdf of the galaxy number density}},\ \href {\doibase 10.1088/1475-7516/2016/03/005} {\bibfield  {journal} {\bibinfo  {journal} {\jcap}\ }\textbf {\bibinfo {volume} {2016}},\ \bibinfo {eid} {005} (\bibinfo {year} {2016})},\ \Eprint {https://arxiv.org/abs/1506.07794} {arXiv:1506.07794 [astro-ph.CO]} \BibitemShut {NoStop}%
\bibitem [{{Ay{\c{c}}oberry} {\textit{et~al}}\mbox{.}(2023)\citenamefont {{Ay{\c{c}}oberry}, {Ajani}, {Guinot}, {Kilbinger}, {Pettorino}, {Farrens}, {Starck}, {Gavazzi},\ \&\ {Hudson}}}]{ayccoberry2023unions}%
{(\PineGreen{{Ay{\c{c}}oberry} {\textit{et~al}}\mbox{.}}, \PineGreen{2023})}  \BibitemOpen
  \bibfield  {author} {\bibinfo {author} {\bibfnamefont {E.}~\bibnamefont {{Ay{\c{c}}oberry}}}, \bibinfo {author} {\bibfnamefont {V.}~\bibnamefont {{Ajani}}}, \bibinfo {author} {\bibfnamefont {A.}~\bibnamefont {{Guinot}}}, \bibinfo {author} {\bibfnamefont {M.}~\bibnamefont {{Kilbinger}}}, \bibinfo {author} {\bibfnamefont {V.}~\bibnamefont {{Pettorino}}}, \bibinfo {author} {\bibfnamefont {S.}~\bibnamefont {{Farrens}}}, \bibinfo {author} {\bibfnamefont {J.-L.}\ \bibnamefont {{Starck}}}, \bibinfo {author} {\bibfnamefont {R.}~\bibnamefont {{Gavazzi}}}, \bibinfo {author} {\bibfnamefont {M.~J.}\ \bibnamefont {{Hudson}}},\ }\emph {{UNIONS: The impact of systematic errors on weak-lensing peak counts}},\ \href {\doibase 10.1051/0004-6361/202243899} {\bibfield  {journal} {\bibinfo  {journal} {\aap}\ }\textbf {\bibinfo {volume} {671}},\ \bibinfo {eid} {A17} (\bibinfo {year} {2023})},\ \Eprint {https://arxiv.org/abs/2204.06280} {arXiv:2204.06280 [astro-ph.CO]} \BibitemShut {NoStop}%
\bibitem [{{Balkenhol} {\textit{et~al}}\mbox{.}(2023){Balkenhol}, {Dutcher}, {Spurio Mancini}, {Doussot}, {Benabed}, {Galli}, {Ade}, {Anderson}, {Ansarinejad}, {Archipley} \emph {et~al.}}]{balkenhol2023measurement}%
{(\PineGreen{{Balkenhol} {\textit{et~al}}\mbox{.}}, \PineGreen{2023})}  \BibitemOpen
  \bibfield  {author} {\bibinfo {author} {\bibfnamefont {L.}~\bibnamefont {{Balkenhol}}}, \bibinfo {author} {\bibfnamefont {D.}~\bibnamefont {{Dutcher}}}, \bibinfo {author} {\bibfnamefont {A.}~\bibnamefont {{Spurio Mancini}}}, \bibinfo {author} {\bibfnamefont {A.}~\bibnamefont {{Doussot}}}, \bibinfo {author} {\bibfnamefont {K.}~\bibnamefont {{Benabed}}}, \bibinfo {author} {\bibfnamefont {S.}~\bibnamefont {{Galli}}}, \bibinfo {author} {\bibfnamefont {P.~A.~R.}\ \bibnamefont {{Ade}}}, \bibinfo {author} {\bibfnamefont {A.~J.}\ \bibnamefont {{Anderson}}}, \bibinfo {author} {\bibfnamefont {B.}~\bibnamefont {{Ansarinejad}}}, \bibinfo {author} {\bibfnamefont {M.}~\bibnamefont {{Archipley}}}, \emph {et~al.},\ }\emph {{Measurement of the CMB temperature power spectrum and constraints on cosmology from the SPT-3G 2018 T T , T E , and E E dataset}},\ \href {\doibase 10.1103/PhysRevD.108.023510} {\bibfield  {journal} {\bibinfo  {journal} {\prd}\ }\textbf {\bibinfo {volume} {108}},\ \bibinfo {eid} {023510} (\bibinfo {year} {2023})},\ \Eprint {https://arxiv.org/abs/2212.05642} {arXiv:2212.05642 [astro-ph.CO]} \BibitemShut {NoStop}%
\bibitem [{{Bardeen} {\textit{et~al}}\mbox{.}(1986)\citenamefont {{Bardeen}, {Bond}, {Kaiser},\ \&\ {Szalay}}}]{bardeen1986statistics}%
{(\PineGreen{{Bardeen} {\textit{et~al}}\mbox{.}}, \PineGreen{1986})}  \BibitemOpen
  \bibfield  {author} {\bibinfo {author} {\bibfnamefont {J.~M.}\ \bibnamefont {{Bardeen}}}, \bibinfo {author} {\bibfnamefont {J.~R.}\ \bibnamefont {{Bond}}}, \bibinfo {author} {\bibfnamefont {N.}~\bibnamefont {{Kaiser}}}, \bibinfo {author} {\bibfnamefont {A.~S.}\ \bibnamefont {{Szalay}}},\ }\emph {{The Statistics of Peaks of Gaussian Random Fields}},\ \href {\doibase 10.1086/164143} {\bibfield  {journal} {\bibinfo  {journal} {\apj}\ }\textbf {\bibinfo {volume} {304}},\ \bibinfo {pages} {15} (\bibinfo {year} {1986})}\BibitemShut {NoStop}%
\bibitem [{{Barreira}(2020)\citenamefont {{Barreira}}}]{Barreira_2020_2}%
{(\PineGreen{{Barreira}}, \PineGreen{2020})}  \BibitemOpen
  \bibfield  {author} {\bibinfo {author} {\bibfnamefont {A.}~\bibnamefont {{Barreira}}},\ }\emph {{On the impact of galaxy bias uncertainties on primordial non-Gaussianity constraints}},\ \href {\doibase 10.1088/1475-7516/2020/12/031} {\bibfield  {journal} {\bibinfo  {journal} {\jcap}\ }\textbf {\bibinfo {volume} {2020}},\ \bibinfo {eid} {031} (\bibinfo {year} {2020})},\ \Eprint {https://arxiv.org/abs/2009.06622} {arXiv:2009.06622 [astro-ph.CO]} \BibitemShut {NoStop}%
\bibitem [{{Barreira}, {Lazeyras} \& {Schmidt}(2021)\citenamefont {{Barreira}, {Lazeyras},\ \&\ {Schmidt}}}]{Barreira_2021}%
{(\PineGreen{{Barreira}, {Lazeyras} \& {Schmidt}}, \PineGreen{2021})}  \BibitemOpen
  \bibfield  {author} {\bibinfo {author} {\bibfnamefont {A.}~\bibnamefont {{Barreira}}}, \bibinfo {author} {\bibfnamefont {T.}~\bibnamefont {{Lazeyras}}}, \bibinfo {author} {\bibfnamefont {F.}~\bibnamefont {{Schmidt}}},\ }\emph {{Galaxy bias from forward models: linear and second-order bias of IllustrisTNG galaxies}},\ \href {\doibase 10.1088/1475-7516/2021/08/029} {\bibfield  {journal} {\bibinfo  {journal} {\jcap}\ }\textbf {\bibinfo {volume} {2021}},\ \bibinfo {eid} {029} (\bibinfo {year} {2021})},\ \Eprint {https://arxiv.org/abs/2105.02876} {arXiv:2105.02876 [astro-ph.CO]} \BibitemShut {NoStop}%
\bibitem [{{Barreira} {\textit{et~al}}\mbox{.}(2020)\citenamefont {{Barreira}, {Cabass}, {Lozanov},\ \&\ {Schmidt}}}]{Barreira_2020_1}%
{(\PineGreen{{Barreira} {\textit{et~al}}\mbox{.}}, \PineGreen{2020})}  \BibitemOpen
  \bibfield  {author} {\bibinfo {author} {\bibfnamefont {A.}~\bibnamefont {{Barreira}}}, \bibinfo {author} {\bibfnamefont {G.}~\bibnamefont {{Cabass}}}, \bibinfo {author} {\bibfnamefont {K.~D.}\ \bibnamefont {{Lozanov}}}, \bibinfo {author} {\bibfnamefont {F.}~\bibnamefont {{Schmidt}}},\ }\emph {{Compensated isocurvature perturbations in the galaxy power spectrum}},\ \href {\doibase 10.1088/1475-7516/2020/07/049} {\bibfield  {journal} {\bibinfo  {journal} {\jcap}\ }\textbf {\bibinfo {volume} {2020}},\ \bibinfo {eid} {049} (\bibinfo {year} {2020})},\ \Eprint {https://arxiv.org/abs/2002.12931} {arXiv:2002.12931 [astro-ph.CO]} \BibitemShut {NoStop}%
\bibitem [{{Bartlett}, {Ho} \& {Wandelt}(2024)\citenamefont {{Bartlett}, {Ho},\ \&\ {Wandelt}}}]{bartlett2024bye}%
{(\PineGreen{{Bartlett}, {Ho} \& {Wandelt}}, \PineGreen{2024})}  \BibitemOpen
  \bibfield  {author} {\bibinfo {author} {\bibfnamefont {D.~J.}\ \bibnamefont {{Bartlett}}}, \bibinfo {author} {\bibfnamefont {M.}~\bibnamefont {{Ho}}}, \bibinfo {author} {\bibfnamefont {B.~D.}\ \bibnamefont {{Wandelt}}},\ }\emph {{Bye-bye, Local-in-matter-density Bias: The Statistics of the Halo Field Are Poorly Determined by the Local Mass Density}},\ \href {\doibase 10.3847/2041-8213/ad97b9} {\bibfield  {journal} {\bibinfo  {journal} {\apjl}\ }\textbf {\bibinfo {volume} {977}},\ \bibinfo {eid} {L44} (\bibinfo {year} {2024})},\ \Eprint {https://arxiv.org/abs/2405.00635} {arXiv:2405.00635 [astro-ph.CO]} \BibitemShut {NoStop}%
\bibitem [{Beaumont, Zhang \& Balding(2002)\citenamefont {Beaumont, Zhang,\ \&\ Balding}}]{beaumont2002approximate}%
{(\PineGreen{Beaumont, Zhang \& Balding}, \PineGreen{2002})}  \BibitemOpen
  \bibfield  {author} {\bibinfo {author} {\bibfnamefont {M.~A.}\ \bibnamefont {Beaumont}}, \bibinfo {author} {\bibfnamefont {W.}~\bibnamefont {Zhang}}, \bibinfo {author} {\bibfnamefont {D.~J.}\ \bibnamefont {Balding}},\ }\emph {Approximate Bayesian Computation in Population Genetics},\ \href {\doibase 10.1093/genetics/162.4.2025} {\bibfield  {journal} {\bibinfo  {journal} {Genetics}\ }\textbf {\bibinfo {volume} {162}},\ \bibinfo {pages} {2025} (\bibinfo {year} {2002})}\BibitemShut {NoStop}%
\bibitem [{Beaumont {\textit{et~al}}\mbox{.}(2009)\citenamefont {Beaumont, Cornuet, Marin,\ \&\ Robert}}]{beaumont2009adaptive}%
{(\PineGreen{Beaumont {\textit{et~al}}\mbox{.}}, \PineGreen{2009})}  \BibitemOpen
  \bibfield  {author} {\bibinfo {author} {\bibfnamefont {M.~A.}\ \bibnamefont {Beaumont}}, \bibinfo {author} {\bibfnamefont {J.-M.}\ \bibnamefont {Cornuet}}, \bibinfo {author} {\bibfnamefont {J.-M.}\ \bibnamefont {Marin}}, \bibinfo {author} {\bibfnamefont {C.~P.}\ \bibnamefont {Robert}},\ }\emph {Adaptive approximate Bayesian computation},\ \href {\doibase 10.1093/biomet/asp052} {\bibfield  {journal} {\bibinfo  {journal} {Biometrika}\ }\textbf {\bibinfo {volume} {96}},\ \bibinfo {pages} {983} (\bibinfo {year} {2009})},\ \Eprint {https://arxiv.org/abs/https://academic.oup.com/biomet/article-pdf/96/4/983/588237/asp052.pdf} {https://academic.oup.com/biomet/article-pdf/96/4/983/588237/asp052.pdf} \BibitemShut {NoStop}%
\bibitem [{{Berti}, {Dawson} \& {Dominguez}(2023)\citenamefont {{Berti}, {Dawson},\ \&\ {Dominguez}}}]{berti2023galaxy}%
{(\PineGreen{{Berti}, {Dawson} \& {Dominguez}}, \PineGreen{2023})}  \BibitemOpen
  \bibfield  {author} {\bibinfo {author} {\bibfnamefont {A.~M.}\ \bibnamefont {{Berti}}}, \bibinfo {author} {\bibfnamefont {K.~S.}\ \bibnamefont {{Dawson}}}, \bibinfo {author} {\bibfnamefont {W.}~\bibnamefont {{Dominguez}}},\ }\emph {{The Galaxy-Halo Connection of DESI Luminous Red Galaxies with Subhalo Abundance Matching}},\ \href {\doibase 10.3847/1538-4357/ace76e} {\bibfield  {journal} {\bibinfo  {journal} {\apj}\ }\textbf {\bibinfo {volume} {954}},\ \bibinfo {eid} {131} (\bibinfo {year} {2023})},\ \Eprint {https://arxiv.org/abs/2303.16096} {arXiv:2303.16096 [astro-ph.CO]} \BibitemShut {NoStop}%
\bibitem [{{Beutler} {\textit{et~al}}\mbox{.}(2012)\citenamefont {{Beutler}, {Blake}, {Colless}, {Jones}, {Staveley-Smith}, {Poole}, {Campbell}, {Parker}, {Saunders},\ \&\ {Watson}}}]{beutler20126df}%
{(\PineGreen{{Beutler} {\textit{et~al}}\mbox{.}}, \PineGreen{2012})}  \BibitemOpen
  \bibfield  {author} {\bibinfo {author} {\bibfnamefont {F.}~\bibnamefont {{Beutler}}}, \bibinfo {author} {\bibfnamefont {C.}~\bibnamefont {{Blake}}}, \bibinfo {author} {\bibfnamefont {M.}~\bibnamefont {{Colless}}}, \bibinfo {author} {\bibfnamefont {D.~H.}\ \bibnamefont {{Jones}}}, \bibinfo {author} {\bibfnamefont {L.}~\bibnamefont {{Staveley-Smith}}}, \bibinfo {author} {\bibfnamefont {G.~B.}\ \bibnamefont {{Poole}}}, \bibinfo {author} {\bibfnamefont {L.}~\bibnamefont {{Campbell}}}, \bibinfo {author} {\bibfnamefont {Q.}~\bibnamefont {{Parker}}}, \bibinfo {author} {\bibfnamefont {W.}~\bibnamefont {{Saunders}}}, \bibinfo {author} {\bibfnamefont {F.}~\bibnamefont {{Watson}}},\ }\emph {{The 6dF Galaxy Survey: z{\ensuremath{\approx}} 0 measurements of the growth rate and {\ensuremath{\sigma}}$_{8}$}},\ \href {\doibase 10.1111/j.1365-2966.2012.21136.x} {\bibfield  {journal} {\bibinfo  {journal} {\mnras}\ }\textbf {\bibinfo {volume} {423}},\ \bibinfo {pages} {3430} (\bibinfo {year} {2012})},\ \Eprint {https://arxiv.org/abs/1204.4725} {arXiv:1204.4725 [astro-ph.CO]} \BibitemShut {NoStop}%
\bibitem [{{Beyond-2pt Collaboration}(2024)\citenamefont {{Beyond-2pt Collaboration}}}]{krause2024parameter}%
{(\PineGreen{{Beyond-2pt Collaboration}}, \PineGreen{2024})}  \BibitemOpen
  \bibfield  {author} {\bibinfo {author} {\bibnamefont {{Beyond-2pt Collaboration}}},\ }\emph {{A Parameter-Masked Mock Data Challenge for Beyond-Two-Point Galaxy Clustering Statistics}},\ \href {\doibase 10.48550/arXiv.2405.02252} {\bibfield  {journal} {\bibinfo  {journal} {arXiv e-prints}\ ,\ \bibinfo {eid} {arXiv:2405.02252}} (\bibinfo {year} {2024})},\ \Eprint {https://arxiv.org/abs/2405.02252} {arXiv:2405.02252 [astro-ph.CO]} \BibitemShut {NoStop}%
\bibitem [{{Blas}, {Lesgourgues} \& {Tram}(2011)\citenamefont {{Blas}, {Lesgourgues},\ \&\ {Tram}}}]{Blas2011}%
{(\PineGreen{{Blas}, {Lesgourgues} \& {Tram}}, \PineGreen{2011})}  \BibitemOpen
  \bibfield  {author} {\bibinfo {author} {\bibfnamefont {D.}~\bibnamefont {{Blas}}}, \bibinfo {author} {\bibfnamefont {J.}~\bibnamefont {{Lesgourgues}}}, \bibinfo {author} {\bibfnamefont {T.}~\bibnamefont {{Tram}}},\ }\emph {{The Cosmic Linear Anisotropy Solving System (CLASS). Part II: Approximation schemes}},\ \href {\doibase 10.1088/1475-7516/2011/07/034} {\bibfield  {journal} {\bibinfo  {journal} {\jcap}\ }\textbf {\bibinfo {volume} {2011}},\ \bibinfo {eid} {034} (\bibinfo {year} {2011})},\ \Eprint {https://arxiv.org/abs/1104.2933} {arXiv:1104.2933 [astro-ph.CO]} \BibitemShut {NoStop}%
\bibitem [{{Blumenthal} {\textit{et~al}}\mbox{.}(1984)\citenamefont {{Blumenthal}, {Faber}, {Primack},\ \&\ {Rees}}}]{blumenthal1984formation}%
{(\PineGreen{{Blumenthal} {\textit{et~al}}\mbox{.}}, \PineGreen{1984})}  \BibitemOpen
  \bibfield  {author} {\bibinfo {author} {\bibfnamefont {G.~R.}\ \bibnamefont {{Blumenthal}}}, \bibinfo {author} {\bibfnamefont {S.~M.}\ \bibnamefont {{Faber}}}, \bibinfo {author} {\bibfnamefont {J.~R.}\ \bibnamefont {{Primack}}}, \bibinfo {author} {\bibfnamefont {M.~J.}\ \bibnamefont {{Rees}}},\ }\emph {{Formation of galaxies and large-scale structure with cold dark matter.}},\ \href {\doibase 10.1038/311517a0} {\bibfield  {journal} {\bibinfo  {journal} {\nat}\ }\textbf {\bibinfo {volume} {311}},\ \bibinfo {pages} {517} (\bibinfo {year} {1984})}\BibitemShut {NoStop}%
\bibitem [{{Bouchet} {\textit{et~al}}\mbox{.}(1995)\citenamefont {{Bouchet}, {Colombi}, {Hivon},\ \&\ {Juszkiewicz}}}]{bouchet1994perturbative}%
{(\PineGreen{{Bouchet} {\textit{et~al}}\mbox{.}}, \PineGreen{1995})}  \BibitemOpen
  \bibfield  {author} {\bibinfo {author} {\bibfnamefont {F.~R.}\ \bibnamefont {{Bouchet}}}, \bibinfo {author} {\bibfnamefont {S.}~\bibnamefont {{Colombi}}}, \bibinfo {author} {\bibfnamefont {E.}~\bibnamefont {{Hivon}}}, \bibinfo {author} {\bibfnamefont {R.}~\bibnamefont {{Juszkiewicz}}},\ }\emph {{Perturbative Lagrangian approach to gravitational instability.}},\ \href {\doibase 10.48550/arXiv.astro-ph/9406013} {\bibfield  {journal} {\bibinfo  {journal} {\aap}\ }\textbf {\bibinfo {volume} {296}},\ \bibinfo {pages} {575} (\bibinfo {year} {1995})},\ \Eprint {https://arxiv.org/abs/astro-ph/9406013} {arXiv:astro-ph/9406013 [astro-ph]} \BibitemShut {NoStop}%
\bibitem [{{Boulanger} {\textit{et~al}}\mbox{.}(1996)\citenamefont {{Boulanger}, {Abergel}, {Bernard}, {Burton}, {Desert}, {Hartmann}, {Lagache},\ \&\ {Puget}}}]{boulanger1996dust}%
{(\PineGreen{{Boulanger} {\textit{et~al}}\mbox{.}}, \PineGreen{1996})}  \BibitemOpen
  \bibfield  {author} {\bibinfo {author} {\bibfnamefont {F.}~\bibnamefont {{Boulanger}}}, \bibinfo {author} {\bibfnamefont {A.}~\bibnamefont {{Abergel}}}, \bibinfo {author} {\bibfnamefont {J.~P.}\ \bibnamefont {{Bernard}}}, \bibinfo {author} {\bibfnamefont {W.~B.}\ \bibnamefont {{Burton}}}, \bibinfo {author} {\bibfnamefont {F.~X.}\ \bibnamefont {{Desert}}}, \bibinfo {author} {\bibfnamefont {D.}~\bibnamefont {{Hartmann}}}, \bibinfo {author} {\bibfnamefont {G.}~\bibnamefont {{Lagache}}}, \bibinfo {author} {\bibfnamefont {J.~L.}\ \bibnamefont {{Puget}}},\ }\emph {{The dust/gas correlation at high Galactic latitude.}},\ \href {https://ui.adsabs.harvard.edu/abs/1996A&A...312..256B} {\bibfield  {journal} {\bibinfo  {journal} {\aap}\ }\textbf {\bibinfo {volume} {312}},\ \bibinfo {pages} {256} (\bibinfo {year} {1996})}\BibitemShut {NoStop}%
\bibitem [{{Bovy} {\textit{et~al}}\mbox{.}(2016)\citenamefont {{Bovy}, {Rix}, {Green}, {Schlafly},\ \&\ {Finkbeiner}}}]{bovy2016galactic}%
{(\PineGreen{{Bovy} {\textit{et~al}}\mbox{.}}, \PineGreen{2016})}  \BibitemOpen
  \bibfield  {author} {\bibinfo {author} {\bibfnamefont {J.}~\bibnamefont {{Bovy}}}, \bibinfo {author} {\bibfnamefont {H.-W.}\ \bibnamefont {{Rix}}}, \bibinfo {author} {\bibfnamefont {G.~M.}\ \bibnamefont {{Green}}}, \bibinfo {author} {\bibfnamefont {E.~F.}\ \bibnamefont {{Schlafly}}}, \bibinfo {author} {\bibfnamefont {D.~P.}\ \bibnamefont {{Finkbeiner}}},\ }\emph {{On Galactic Density Modeling in the Presence of Dust Extinction}},\ \href {\doibase 10.3847/0004-637X/818/2/130} {\bibfield  {journal} {\bibinfo  {journal} {\apj}\ }\textbf {\bibinfo {volume} {818}},\ \bibinfo {eid} {130} (\bibinfo {year} {2016})},\ \Eprint {https://arxiv.org/abs/1509.06751} {arXiv:1509.06751 [astro-ph.GA]} \BibitemShut {NoStop}%
\bibitem [{{Brehmer} {\textit{et~al}}\mbox{.}(2018)\citenamefont {{Brehmer}, {Louppe}, {Pavez},\ \&\ {Cranmer}}}]{brehmer2020mining}%
{(\PineGreen{{Brehmer} {\textit{et~al}}\mbox{.}}, \PineGreen{2018})}  \BibitemOpen
  \bibfield  {author} {\bibinfo {author} {\bibfnamefont {J.}~\bibnamefont {{Brehmer}}}, \bibinfo {author} {\bibfnamefont {G.}~\bibnamefont {{Louppe}}}, \bibinfo {author} {\bibfnamefont {J.}~\bibnamefont {{Pavez}}}, \bibinfo {author} {\bibfnamefont {K.}~\bibnamefont {{Cranmer}}},\ }\emph {{Mining gold from implicit models to improve likelihood-free inference}},\ \href {\doibase 10.48550/arXiv.1805.12244} {\bibfield  {journal} {\bibinfo  {journal} {arXiv e-prints}\ ,\ \bibinfo {eid} {arXiv:1805.12244}} (\bibinfo {year} {2018})},\ \Eprint {https://arxiv.org/abs/1805.12244} {arXiv:1805.12244 [stat.ML]} \BibitemShut {NoStop}%
\bibitem [{{Brout} \& {Riess}(2023)\citenamefont {{Brout}\ \&\ {Riess}}}]{brout2024impact}%
{(\PineGreen{{Brout} \& {Riess}}, \PineGreen{2023})}  \BibitemOpen
  \bibfield  {author} {\bibinfo {author} {\bibfnamefont {D.}~\bibnamefont {{Brout}}}, \bibinfo {author} {\bibfnamefont {A.}~\bibnamefont {{Riess}}},\ }\emph {{The Impact of Dust on Cepheid and Type Ia Supernova Distances}},\ \href {\doibase 10.48550/arXiv.2311.08253} {\bibfield  {journal} {\bibinfo  {journal} {arXiv e-prints}\ ,\ \bibinfo {eid} {arXiv:2311.08253}} (\bibinfo {year} {2023})},\ \Eprint {https://arxiv.org/abs/2311.08253} {arXiv:2311.08253 [astro-ph.CO]} \BibitemShut {NoStop}%
\bibitem [{{Campagne} {\textit{et~al}}\mbox{.}(2023)\citenamefont {{Campagne}, {Lanusse}, {Zuntz}, {Boucaud}, {Casas}, {Karamanis}, {Kirkby}, {Lanzieri}, {Peel},\ \&\ {Li}}}]{campagne2023jax}%
{(\PineGreen{{Campagne} {\textit{et~al}}\mbox{.}}, \PineGreen{2023})}  \BibitemOpen
  \bibfield  {author} {\bibinfo {author} {\bibfnamefont {J.-E.}\ \bibnamefont {{Campagne}}}, \bibinfo {author} {\bibfnamefont {F.}~\bibnamefont {{Lanusse}}}, \bibinfo {author} {\bibfnamefont {J.}~\bibnamefont {{Zuntz}}}, \bibinfo {author} {\bibfnamefont {A.}~\bibnamefont {{Boucaud}}}, \bibinfo {author} {\bibfnamefont {S.}~\bibnamefont {{Casas}}}, \bibinfo {author} {\bibfnamefont {M.}~\bibnamefont {{Karamanis}}}, \bibinfo {author} {\bibfnamefont {D.}~\bibnamefont {{Kirkby}}}, \bibinfo {author} {\bibfnamefont {D.}~\bibnamefont {{Lanzieri}}}, \bibinfo {author} {\bibfnamefont {A.}~\bibnamefont {{Peel}}}, \bibinfo {author} {\bibfnamefont {Y.}~\bibnamefont {{Li}}},\ }\emph {{JAX-COSMO: An End-to-End Differentiable and GPU Accelerated Cosmology Library}},\ \href {\doibase 10.21105/astro.2302.05163} {\bibfield  {journal} {\bibinfo  {journal} {The Open Journal of Astrophysics}\ }\textbf {\bibinfo {volume} {6}},\ \bibinfo {eid} {15} (\bibinfo {year} {2023})},\ \Eprint {https://arxiv.org/abs/2302.05163} {arXiv:2302.05163 [astro-ph.CO]} \BibitemShut {NoStop}%
\bibitem [{{Carreres} {\textit{et~al}}\mbox{.}(2023){Carreres}, {Bautista}, {Feinstein}, {Fouchez}, {Racine}, {Smith}, {Amenouche}, {Aubert}, {Dhawan}, {Ginolin} \emph {et~al.}}]{carreres2023growth}%
{(\PineGreen{{Carreres} {\textit{et~al}}\mbox{.}}, \PineGreen{2023})}  \BibitemOpen
  \bibfield  {author} {\bibinfo {author} {\bibfnamefont {B.}~\bibnamefont {{Carreres}}}, \bibinfo {author} {\bibfnamefont {J.~E.}\ \bibnamefont {{Bautista}}}, \bibinfo {author} {\bibfnamefont {F.}~\bibnamefont {{Feinstein}}}, \bibinfo {author} {\bibfnamefont {D.}~\bibnamefont {{Fouchez}}}, \bibinfo {author} {\bibfnamefont {B.}~\bibnamefont {{Racine}}}, \bibinfo {author} {\bibfnamefont {M.}~\bibnamefont {{Smith}}}, \bibinfo {author} {\bibfnamefont {M.}~\bibnamefont {{Amenouche}}}, \bibinfo {author} {\bibfnamefont {M.}~\bibnamefont {{Aubert}}}, \bibinfo {author} {\bibfnamefont {S.}~\bibnamefont {{Dhawan}}}, \bibinfo {author} {\bibfnamefont {M.}~\bibnamefont {{Ginolin}}}, \emph {et~al.},\ }\emph {{Growth-rate measurement with type-Ia supernovae using ZTF survey simulations}},\ \href {\doibase 10.1051/0004-6361/202346173} {\bibfield  {journal} {\bibinfo  {journal} {\aap}\ }\textbf {\bibinfo {volume} {674}},\ \bibinfo {eid} {A197} (\bibinfo {year} {2023})},\ \Eprint {https://arxiv.org/abs/2303.01198} {arXiv:2303.01198 [astro-ph.CO]} \BibitemShut {NoStop}%
\bibitem [{Charnock, Lavaux \& Wandelt(2018)\citenamefont {Charnock, Lavaux,\ \&\ Wandelt}}]{charnock2018automatic}%
{(\PineGreen{Charnock, Lavaux \& Wandelt}, \PineGreen{2018})}  \BibitemOpen
  \bibfield  {author} {\bibinfo {author} {\bibfnamefont {T.}~\bibnamefont {Charnock}}, \bibinfo {author} {\bibfnamefont {G.}~\bibnamefont {Lavaux}}, \bibinfo {author} {\bibfnamefont {B.~D.}\ \bibnamefont {Wandelt}},\ }\emph {{Automatic physical inference with information maximizing neural networks}},\ \href {\doibase 10.1103/PhysRevD.97.083004} {\bibfield  {journal} {\bibinfo  {journal} {\prd}\ }\textbf {\bibinfo {volume} {97}},\ \bibinfo {eid} {083004} (\bibinfo {year} {2018})},\ \Eprint {https://arxiv.org/abs/1802.03537} {arXiv:1802.03537 [astro-ph.IM]} \BibitemShut {NoStop}%
\bibitem [{Colless {\textit{et~al}}\mbox{.}(2001)Colless, Dalton, Maddox, Sutherland, Norberg, Cole, Bland-Hawthorn, Bridges, Cannon, Collins \emph {et~al.}}]{colless20012df}%
{(\PineGreen{Colless {\textit{et~al}}\mbox{.}}, \PineGreen{2001})}  \BibitemOpen
  \bibfield  {author} {\bibinfo {author} {\bibfnamefont {M.}~\bibnamefont {Colless}}, \bibinfo {author} {\bibfnamefont {G.}~\bibnamefont {Dalton}}, \bibinfo {author} {\bibfnamefont {S.}~\bibnamefont {Maddox}}, \bibinfo {author} {\bibfnamefont {W.}~\bibnamefont {Sutherland}}, \bibinfo {author} {\bibfnamefont {P.}~\bibnamefont {Norberg}}, \bibinfo {author} {\bibfnamefont {S.}~\bibnamefont {Cole}}, \bibinfo {author} {\bibfnamefont {J.}~\bibnamefont {Bland-Hawthorn}}, \bibinfo {author} {\bibfnamefont {T.}~\bibnamefont {Bridges}}, \bibinfo {author} {\bibfnamefont {R.}~\bibnamefont {Cannon}}, \bibinfo {author} {\bibfnamefont {C.}~\bibnamefont {Collins}}, \emph {et~al.},\ }\emph {The 2df galaxy redshift survey: spectra and redshifts},\ \href {https://academic.oup.com/mnras/article/328/4/1039/1082731} {\bibfield  {journal} {\bibinfo  {journal} {Monthly Notices of the Royal Astronomical Society}\ }\textbf {\bibinfo {volume} {328}},\ \bibinfo {pages} {1039} (\bibinfo {year} {2001})}\BibitemShut {NoStop}%
\bibitem [{{Corasaniti}(2006)\citenamefont {{Corasaniti}}}]{corasaniti2006impact}%
{(\PineGreen{{Corasaniti}}, \PineGreen{2006})}  \BibitemOpen
  \bibfield  {author} {\bibinfo {author} {\bibfnamefont {P.~S.}\ \bibnamefont {{Corasaniti}}},\ }\emph {{The impact of cosmic dust on supernova cosmology}},\ \href {\doibase 10.1111/j.1365-2966.2006.10825.x} {\bibfield  {journal} {\bibinfo  {journal} {\mnras}\ }\textbf {\bibinfo {volume} {372}},\ \bibinfo {pages} {191} (\bibinfo {year} {2006})},\ \Eprint {https://arxiv.org/abs/astro-ph/0603833} {arXiv:astro-ph/0603833 [astro-ph]} \BibitemShut {NoStop}%
\bibitem [{{Cranmer}, {Brehmer} \& {Louppe}(2020)\citenamefont {{Cranmer}, {Brehmer},\ \&\ {Louppe}}}]{cranmer2020frontier}%
{(\PineGreen{{Cranmer}, {Brehmer} \& {Louppe}}, \PineGreen{2020})}  \BibitemOpen
  \bibfield  {author} {\bibinfo {author} {\bibfnamefont {K.}~\bibnamefont {{Cranmer}}}, \bibinfo {author} {\bibfnamefont {J.}~\bibnamefont {{Brehmer}}}, \bibinfo {author} {\bibfnamefont {G.}~\bibnamefont {{Louppe}}},\ }\emph {{The frontier of simulation-based inference}},\ \href {\doibase 10.1073/pnas.1912789117} {\bibfield  {journal} {\bibinfo  {journal} {Proceedings of the National Academy of Science}\ }\textbf {\bibinfo {volume} {117}},\ \bibinfo {pages} {30055} (\bibinfo {year} {2020})},\ \Eprint {https://arxiv.org/abs/1911.01429} {arXiv:1911.01429 [stat.ML]} \BibitemShut {NoStop}%
\bibitem [{{Davis} {\textit{et~al}}\mbox{.}(2011){Davis}, {Hui}, {Frieman}, {Haugb{\o}lle}, {Kessler}, {Sinclair}, {Sollerman}, {Bassett}, {Marriner}, {M{\"o}rtsell} \emph {et~al.}}]{davis2011effect}%
{(\PineGreen{{Davis} {\textit{et~al}}\mbox{.}}, \PineGreen{2011})}  \BibitemOpen
  \bibfield  {author} {\bibinfo {author} {\bibfnamefont {T.~M.}\ \bibnamefont {{Davis}}}, \bibinfo {author} {\bibfnamefont {L.}~\bibnamefont {{Hui}}}, \bibinfo {author} {\bibfnamefont {J.~A.}\ \bibnamefont {{Frieman}}}, \bibinfo {author} {\bibfnamefont {T.}~\bibnamefont {{Haugb{\o}lle}}}, \bibinfo {author} {\bibfnamefont {R.}~\bibnamefont {{Kessler}}}, \bibinfo {author} {\bibfnamefont {B.}~\bibnamefont {{Sinclair}}}, \bibinfo {author} {\bibfnamefont {J.}~\bibnamefont {{Sollerman}}}, \bibinfo {author} {\bibfnamefont {B.}~\bibnamefont {{Bassett}}}, \bibinfo {author} {\bibfnamefont {J.}~\bibnamefont {{Marriner}}}, \bibinfo {author} {\bibfnamefont {E.}~\bibnamefont {{M{\"o}rtsell}}}, \emph {et~al.},\ }\emph {{The Effect of Peculiar Velocities on Supernova Cosmology}},\ \href {\doibase 10.1088/0004-637X/741/1/67} {\bibfield  {journal} {\bibinfo  {journal} {\apj}\ }\textbf {\bibinfo {volume} {741}},\ \bibinfo {eid} {67} (\bibinfo {year} {2011})},\ \Eprint {https://arxiv.org/abs/1012.2912} {arXiv:1012.2912 [astro-ph.CO]} \BibitemShut {NoStop}%
\bibitem [{{DESI Collaboration}(2016)\citenamefont {{DESI Collaboration}}}]{aghamousa2016desi}%
{(\PineGreen{{DESI Collaboration}}, \PineGreen{2016})}  \BibitemOpen
  \bibfield  {author} {\bibinfo {author} {\bibnamefont {{DESI Collaboration}}},\ }\emph {{The DESI Experiment Part I: Science,Targeting, and Survey Design}},\ \href {\doibase 10.48550/arXiv.1611.00036} {\bibfield  {journal} {\bibinfo  {journal} {arXiv e-prints}\ ,\ \bibinfo {eid} {arXiv:1611.00036}} (\bibinfo {year} {2016})},\ \Eprint {https://arxiv.org/abs/1611.00036} {arXiv:1611.00036 [astro-ph.IM]} \BibitemShut {NoStop}%
\bibitem [{{Desjacques}, {Jeong} \& {Schmidt}(2018)\citenamefont {{Desjacques}, {Jeong},\ \&\ {Schmidt}}}]{Desjacques_2018}%
{(\PineGreen{{Desjacques}, {Jeong} \& {Schmidt}}, \PineGreen{2018})}  \BibitemOpen
  \bibfield  {author} {\bibinfo {author} {\bibfnamefont {V.}~\bibnamefont {{Desjacques}}}, \bibinfo {author} {\bibfnamefont {D.}~\bibnamefont {{Jeong}}}, \bibinfo {author} {\bibfnamefont {F.}~\bibnamefont {{Schmidt}}},\ }\emph {{Large-scale galaxy bias}},\ \href {\doibase 10.1016/j.physrep.2017.12.002} {\bibfield  {journal} {\bibinfo  {journal} {\physrep}\ }\textbf {\bibinfo {volume} {733}},\ \bibinfo {pages} {1} (\bibinfo {year} {2018})},\ \Eprint {https://arxiv.org/abs/1611.09787} {arXiv:1611.09787 [astro-ph.CO]} \BibitemShut {NoStop}%
\bibitem [{{Ding}, {Lavaux} \& {Jasche}(2024)\citenamefont {{Ding}, {Lavaux},\ \&\ {Jasche}}}]{ding2024pinetree}%
{(\PineGreen{{Ding}, {Lavaux} \& {Jasche}}, \PineGreen{2024})}  \BibitemOpen
  \bibfield  {author} {\bibinfo {author} {\bibfnamefont {S.}~\bibnamefont {{Ding}}}, \bibinfo {author} {\bibfnamefont {G.}~\bibnamefont {{Lavaux}}}, \bibinfo {author} {\bibfnamefont {J.}~\bibnamefont {{Jasche}}},\ }\emph {{PineTree: A generative, fast, and differentiable halo model for wide-field galaxy surveys}},\ \href {\doibase 10.1051/0004-6361/202451343} {\bibfield  {journal} {\bibinfo  {journal} {\aap}\ }\textbf {\bibinfo {volume} {690}},\ \bibinfo {eid} {A236} (\bibinfo {year} {2024})},\ \Eprint {https://arxiv.org/abs/2407.01391} {arXiv:2407.01391 [astro-ph.CO]} \BibitemShut {NoStop}%
\bibitem [{{Doeser} {\textit{et~al}}\mbox{.}(2024)\citenamefont {{Doeser}, {Jamieson}, {Stopyra}, {Lavaux}, {Leclercq},\ \&\ {Jasche}}}]{doeser2023bayesian}%
{(\PineGreen{{Doeser} {\textit{et~al}}\mbox{.}}, \PineGreen{2024})}  \BibitemOpen
  \bibfield  {author} {\bibinfo {author} {\bibfnamefont {L.}~\bibnamefont {{Doeser}}}, \bibinfo {author} {\bibfnamefont {D.}~\bibnamefont {{Jamieson}}}, \bibinfo {author} {\bibfnamefont {S.}~\bibnamefont {{Stopyra}}}, \bibinfo {author} {\bibfnamefont {G.}~\bibnamefont {{Lavaux}}}, \bibinfo {author} {\bibfnamefont {F.}~\bibnamefont {{Leclercq}}}, \bibinfo {author} {\bibfnamefont {J.}~\bibnamefont {{Jasche}}},\ }\emph {{Bayesian inference of initial conditions from non-linear cosmic structures using field-level emulators}},\ \href {\doibase 10.1093/mnras/stae2429} {\bibfield  {journal} {\bibinfo  {journal} {\mnras}\ }\textbf {\bibinfo {volume} {535}},\ \bibinfo {pages} {1258} (\bibinfo {year} {2024})},\ \Eprint {https://arxiv.org/abs/2312.09271} {arXiv:2312.09271 [astro-ph.CO]} \BibitemShut {NoStop}%
\bibitem [{{Duane} {\textit{et~al}}\mbox{.}(1987)\citenamefont {{Duane}, {Kennedy}, {Pendleton},\ \&\ {Roweth}}}]{duane1987hybrid}%
{(\PineGreen{{Duane} {\textit{et~al}}\mbox{.}}, \PineGreen{1987})}  \BibitemOpen
  \bibfield  {author} {\bibinfo {author} {\bibfnamefont {S.}~\bibnamefont {{Duane}}}, \bibinfo {author} {\bibfnamefont {A.~D.}\ \bibnamefont {{Kennedy}}}, \bibinfo {author} {\bibfnamefont {B.~J.}\ \bibnamefont {{Pendleton}}}, \bibinfo {author} {\bibfnamefont {D.}~\bibnamefont {{Roweth}}},\ }\emph {{Hybrid Monte Carlo}},\ \href {\doibase 10.1016/0370-2693(87)91197-X} {\bibfield  {journal} {\bibinfo  {journal} {Physics Letters B}\ }\textbf {\bibinfo {volume} {195}},\ \bibinfo {pages} {216} (\bibinfo {year} {1987})}\BibitemShut {NoStop}%
\bibitem [{{Eisenstein}(2015)\citenamefont {{Eisenstein}}}]{eisenstein2015baryon}%
{(\PineGreen{{Eisenstein}}, \PineGreen{2015})}  \BibitemOpen
  \bibfield  {author} {\bibinfo {author} {\bibfnamefont {D.}~\bibnamefont {{Eisenstein}}},\ }\emph {{The Baryon Oscillation Spectroscopic Survey (BOSS): Dark Energy from the World's Largest Redshift Survey}},\ in\ \href {https://ui.adsabs.harvard.edu/abs/2015APS..APR.Z2001E} {\emph {\bibinfo {booktitle} {APS April Meeting Abstracts}}},\ \bibinfo {series} {APS Meeting Abstracts}, Vol.\ \bibinfo {volume} {2015}\ (\bibinfo {year} {2015})\ p.\ \bibinfo {pages} {Z2.001}\BibitemShut {NoStop}%
\bibitem [{{Eisenstein} \& {Hu}(1998)\citenamefont {{Eisenstein}\ \&\ {Hu}}}]{Eisenstein1998}%
{(\PineGreen{{Eisenstein} \& {Hu}}, \PineGreen{1998})}  \BibitemOpen
  \bibfield  {author} {\bibinfo {author} {\bibfnamefont {D.~J.}\ \bibnamefont {{Eisenstein}}}, \bibinfo {author} {\bibfnamefont {W.}~\bibnamefont {{Hu}}},\ }\emph {{Baryonic Features in the Matter Transfer Function}},\ \href {\doibase 10.1086/305424} {\bibfield  {journal} {\bibinfo  {journal} {\apj}\ }\textbf {\bibinfo {volume} {496}},\ \bibinfo {pages} {605} (\bibinfo {year} {1998})},\ \Eprint {https://arxiv.org/abs/astro-ph/9709112} {arXiv:astro-ph/9709112 [astro-ph]} \BibitemShut {NoStop}%
\bibitem [{{Eisenstein} {\textit{et~al}}\mbox{.}(2001){Eisenstein}, {Annis}, {Gunn}, {Szalay}, {Connolly}, {Nichol}, {Bahcall}, {Bernardi}, {Burles}, {Castander} \emph {et~al.}}]{eisenstein2001spectroscopic}%
{(\PineGreen{{Eisenstein} {\textit{et~al}}\mbox{.}}, \PineGreen{2001})}  \BibitemOpen
  \bibfield  {author} {\bibinfo {author} {\bibfnamefont {D.~J.}\ \bibnamefont {{Eisenstein}}}, \bibinfo {author} {\bibfnamefont {J.}~\bibnamefont {{Annis}}}, \bibinfo {author} {\bibfnamefont {J.~E.}\ \bibnamefont {{Gunn}}}, \bibinfo {author} {\bibfnamefont {A.~S.}\ \bibnamefont {{Szalay}}}, \bibinfo {author} {\bibfnamefont {A.~J.}\ \bibnamefont {{Connolly}}}, \bibinfo {author} {\bibfnamefont {R.~C.}\ \bibnamefont {{Nichol}}}, \bibinfo {author} {\bibfnamefont {N.~A.}\ \bibnamefont {{Bahcall}}}, \bibinfo {author} {\bibfnamefont {M.}~\bibnamefont {{Bernardi}}}, \bibinfo {author} {\bibfnamefont {S.}~\bibnamefont {{Burles}}}, \bibinfo {author} {\bibfnamefont {F.~J.}\ \bibnamefont {{Castander}}}, \emph {et~al.},\ }\emph {{Spectroscopic Target Selection for the Sloan Digital Sky Survey: The Luminous Red Galaxy Sample}},\ \href {\doibase 10.1086/323717} {\bibfield  {journal} {\bibinfo  {journal} {\aj}\ }\textbf {\bibinfo {volume} {122}},\ \bibinfo {pages} {2267} (\bibinfo {year} {2001})},\ \Eprint {https://arxiv.org/abs/astro-ph/0108153} {arXiv:astro-ph/0108153 [astro-ph]} \BibitemShut {NoStop}%
\bibitem [{{Eisenstein} {\textit{et~al}}\mbox{.}(2005){Eisenstein}, {Zehavi}, {Hogg}, {Scoccimarro}, {Blanton}, {Nichol}, {Scranton}, {Seo}, {Tegmark}, {Zheng} \emph {et~al.}}]{eisenstein2005detection}%
{(\PineGreen{{Eisenstein} {\textit{et~al}}\mbox{.}}, \PineGreen{2005})}  \BibitemOpen
  \bibfield  {author} {\bibinfo {author} {\bibfnamefont {D.~J.}\ \bibnamefont {{Eisenstein}}}, \bibinfo {author} {\bibfnamefont {I.}~\bibnamefont {{Zehavi}}}, \bibinfo {author} {\bibfnamefont {D.~W.}\ \bibnamefont {{Hogg}}}, \bibinfo {author} {\bibfnamefont {R.}~\bibnamefont {{Scoccimarro}}}, \bibinfo {author} {\bibfnamefont {M.~R.}\ \bibnamefont {{Blanton}}}, \bibinfo {author} {\bibfnamefont {R.~C.}\ \bibnamefont {{Nichol}}}, \bibinfo {author} {\bibfnamefont {R.}~\bibnamefont {{Scranton}}}, \bibinfo {author} {\bibfnamefont {H.-J.}\ \bibnamefont {{Seo}}}, \bibinfo {author} {\bibfnamefont {M.}~\bibnamefont {{Tegmark}}}, \bibinfo {author} {\bibfnamefont {Z.}~\bibnamefont {{Zheng}}}, \emph {et~al.},\ }\emph {{Detection of the Baryon Acoustic Peak in the Large-Scale Correlation Function of SDSS Luminous Red Galaxies}},\ \href {\doibase 10.1086/466512} {\bibfield  {journal} {\bibinfo  {journal} {\apj}\ }\textbf {\bibinfo {volume} {633}},\ \bibinfo {pages} {560} (\bibinfo {year} {2005})},\ \Eprint {https://arxiv.org/abs/astro-ph/0501171} {arXiv:astro-ph/0501171 [astro-ph]} \BibitemShut {NoStop}%
\bibitem [{{Euclid Collaboration}(2024)\citenamefont {{Euclid Collaboration}}}]{mellier2024euclid}%
{(\PineGreen{{Euclid Collaboration}}, \PineGreen{2024})}  \BibitemOpen
  \bibfield  {author} {\bibinfo {author} {\bibnamefont {{Euclid Collaboration}}},\ }\emph {{Euclid. I. Overview of the Euclid mission}},\ \href {\doibase 10.48550/arXiv.2405.13491} {\bibfield  {journal} {\bibinfo  {journal} {arXiv e-prints}\ ,\ \bibinfo {eid} {arXiv:2405.13491}} (\bibinfo {year} {2024})},\ \Eprint {https://arxiv.org/abs/2405.13491} {arXiv:2405.13491 [astro-ph.CO]} \BibitemShut {NoStop}%
\bibitem [{{Euclid Collaboration}(2020)\citenamefont {{Euclid Collaboration}}}]{euclid2020euclid}%
{(\PineGreen{{Euclid Collaboration}}, \PineGreen{2020})}  \BibitemOpen
  \bibfield  {author} {\bibinfo {author} {\bibnamefont {{Euclid Collaboration}}},\ }\emph {{Euclid preparation. VII. Forecast validation for Euclid cosmological probes}},\ \href {\doibase 10.1051/0004-6361/202038071} {\bibfield  {journal} {\bibinfo  {journal} {\aap}\ }\textbf {\bibinfo {volume} {642}},\ \bibinfo {eid} {A191} (\bibinfo {year} {2020})},\ \Eprint {https://arxiv.org/abs/1910.09273} {arXiv:1910.09273 [astro-ph.CO]} \BibitemShut {NoStop}%
\bibitem [{{Fazolo}, {Amendola} \& {Velten}(2022)\citenamefont {{Fazolo}, {Amendola},\ \&\ {Velten}}}]{fazolo2022skewness}%
{(\PineGreen{{Fazolo}, {Amendola} \& {Velten}}, \PineGreen{2022})}  \BibitemOpen
  \bibfield  {author} {\bibinfo {author} {\bibfnamefont {R.~E.}\ \bibnamefont {{Fazolo}}}, \bibinfo {author} {\bibfnamefont {L.}~\bibnamefont {{Amendola}}}, \bibinfo {author} {\bibfnamefont {H.}~\bibnamefont {{Velten}}},\ }\emph {{Skewness as a test of dark energy perturbations}},\ \href {\doibase 10.1103/PhysRevD.105.103521} {\bibfield  {journal} {\bibinfo  {journal} {\prd}\ }\textbf {\bibinfo {volume} {105}},\ \bibinfo {eid} {103521} (\bibinfo {year} {2022})}\BibitemShut {NoStop}%
\bibitem [{{Frazier}, {Robert} \& {Rousseau}(2017)\citenamefont {{Frazier}, {Robert},\ \&\ {Rousseau}}}]{frazier2017model}%
{(\PineGreen{{Frazier}, {Robert} \& {Rousseau}}, \PineGreen{2017})}  \BibitemOpen
  \bibfield  {author} {\bibinfo {author} {\bibfnamefont {D.~T.}\ \bibnamefont {{Frazier}}}, \bibinfo {author} {\bibfnamefont {C.~P.}\ \bibnamefont {{Robert}}}, \bibinfo {author} {\bibfnamefont {J.}~\bibnamefont {{Rousseau}}},\ }\emph {{Model Misspecification in ABC: Consequences and Diagnostics}},\ \href {\doibase 10.48550/arXiv.1708.01974} {\bibfield  {journal} {\bibinfo  {journal} {arXiv e-prints}\ ,\ \bibinfo {eid} {arXiv:1708.01974}} (\bibinfo {year} {2017})},\ \Eprint {https://arxiv.org/abs/1708.01974} {arXiv:1708.01974 [math.ST]} \BibitemShut {NoStop}%
\bibitem [{{Galametz} {\textit{et~al}}\mbox{.}(2017)\citenamefont {{Galametz}, {Saglia}, {Paltani}, {Apostolakos},\ \&\ {Dubath}}}]{galametz2017sed}%
{(\PineGreen{{Galametz} {\textit{et~al}}\mbox{.}}, \PineGreen{2017})}  \BibitemOpen
  \bibfield  {author} {\bibinfo {author} {\bibfnamefont {A.}~\bibnamefont {{Galametz}}}, \bibinfo {author} {\bibfnamefont {R.}~\bibnamefont {{Saglia}}}, \bibinfo {author} {\bibfnamefont {S.}~\bibnamefont {{Paltani}}}, \bibinfo {author} {\bibfnamefont {N.}~\bibnamefont {{Apostolakos}}}, \bibinfo {author} {\bibfnamefont {P.}~\bibnamefont {{Dubath}}},\ }\emph {{SED-dependent galactic extinction prescription for Euclid and future cosmological surveys}},\ \href {\doibase 10.1051/0004-6361/201629333} {\bibfield  {journal} {\bibinfo  {journal} {\aap}\ }\textbf {\bibinfo {volume} {598}},\ \bibinfo {eid} {A20} (\bibinfo {year} {2017})},\ \Eprint {https://arxiv.org/abs/1609.08624} {arXiv:1609.08624 [astro-ph.CO]} \BibitemShut {NoStop}%
\bibitem [{{Gavazzi} \& {Jaffe}(1986)\citenamefont {{Gavazzi}\ \&\ {Jaffe}}}]{gavazzi1986radio}%
{(\PineGreen{{Gavazzi} \& {Jaffe}}, \PineGreen{1986})}  \BibitemOpen
  \bibfield  {author} {\bibinfo {author} {\bibfnamefont {G.}~\bibnamefont {{Gavazzi}}}, \bibinfo {author} {\bibfnamefont {W.}~\bibnamefont {{Jaffe}}},\ }\emph {{Radio Continuum Survey of the Coma/A1367 Supercluster. III. Radio Properties of Galaxies in Different Density Environments}},\ \href {\doibase 10.1086/164664} {\bibfield  {journal} {\bibinfo  {journal} {\apj}\ }\textbf {\bibinfo {volume} {310}},\ \bibinfo {pages} {53} (\bibinfo {year} {1986})}\BibitemShut {NoStop}%
\bibitem [{{Gil-Marín} {\textit{et~al}}\mbox{.}(2015)\citenamefont {{Gil-Marín}, {Nore{\~n}a}, {Verde}, {Percival}, {Wagner}, {Manera},\ \&\ {Schneider}}}]{gil2015power}%
{(\PineGreen{{Gil-Marín} {\textit{et~al}}\mbox{.}}, \PineGreen{2015})}  \BibitemOpen
  \bibfield  {author} {\bibinfo {author} {\bibfnamefont {H.}~\bibnamefont {{Gil-Marín}}}, \bibinfo {author} {\bibfnamefont {J.}~\bibnamefont {{Nore{\~n}a}}}, \bibinfo {author} {\bibfnamefont {L.}~\bibnamefont {{Verde}}}, \bibinfo {author} {\bibfnamefont {W.~J.}\ \bibnamefont {{Percival}}}, \bibinfo {author} {\bibfnamefont {C.}~\bibnamefont {{Wagner}}}, \bibinfo {author} {\bibfnamefont {M.}~\bibnamefont {{Manera}}}, \bibinfo {author} {\bibfnamefont {D.~P.}\ \bibnamefont {{Schneider}}},\ }\emph {{The power spectrum and bispectrum of SDSS DR11 BOSS galaxies - I. Bias and gravity}},\ \href {\doibase 10.1093/mnras/stv961} {\bibfield  {journal} {\bibinfo  {journal} {\mnras}\ }\textbf {\bibinfo {volume} {451}},\ \bibinfo {pages} {539} (\bibinfo {year} {2015})},\ \Eprint {https://arxiv.org/abs/1407.5668} {arXiv:1407.5668 [astro-ph.CO]} \BibitemShut {NoStop}%
\bibitem [{{Glanville}, {Howlett} \& {Davis}(2021)\citenamefont {{Glanville}, {Howlett},\ \&\ {Davis}}}]{glanville2021effect}%
{(\PineGreen{{Glanville}, {Howlett} \& {Davis}}, \PineGreen{2021})}  \BibitemOpen
  \bibfield  {author} {\bibinfo {author} {\bibfnamefont {A.}~\bibnamefont {{Glanville}}}, \bibinfo {author} {\bibfnamefont {C.}~\bibnamefont {{Howlett}}}, \bibinfo {author} {\bibfnamefont {T.~M.}\ \bibnamefont {{Davis}}},\ }\emph {{The effect of systematic redshift biases in BAO cosmology}},\ \href {\doibase 10.1093/mnras/stab657} {\bibfield  {journal} {\bibinfo  {journal} {\mnras}\ }\textbf {\bibinfo {volume} {503}},\ \bibinfo {pages} {3510} (\bibinfo {year} {2021})},\ \Eprint {https://arxiv.org/abs/2011.04210} {arXiv:2011.04210 [astro-ph.CO]} \BibitemShut {NoStop}%
\bibitem [{{Goldstein} {\textit{et~al}}\mbox{.}(2023)\citenamefont {{Goldstein}, {Park}, {Raveri}, {Jain},\ \&\ {Samushia}}}]{goldstein2023beyond}%
{(\PineGreen{{Goldstein} {\textit{et~al}}\mbox{.}}, \PineGreen{2023})}  \BibitemOpen
  \bibfield  {author} {\bibinfo {author} {\bibfnamefont {S.}~\bibnamefont {{Goldstein}}}, \bibinfo {author} {\bibfnamefont {M.}~\bibnamefont {{Park}}}, \bibinfo {author} {\bibfnamefont {M.}~\bibnamefont {{Raveri}}}, \bibinfo {author} {\bibfnamefont {B.}~\bibnamefont {{Jain}}}, \bibinfo {author} {\bibfnamefont {L.}~\bibnamefont {{Samushia}}},\ }\emph {{Beyond dark energy Fisher forecasts: How the Dark Energy Spectroscopic Instrument will constrain LCDM and quintessence models}},\ \href {\doibase 10.1103/PhysRevD.107.063530} {\bibfield  {journal} {\bibinfo  {journal} {\prd}\ }\textbf {\bibinfo {volume} {107}},\ \bibinfo {eid} {063530} (\bibinfo {year} {2023})},\ \Eprint {https://arxiv.org/abs/2207.01612} {arXiv:2207.01612 [astro-ph.CO]} \BibitemShut {NoStop}%
\bibitem [{{Hahn}, {List} \& {Porqueres}(2024)\citenamefont {{Hahn}, {List},\ \&\ {Porqueres}}}]{hahn2024disco}%
{(\PineGreen{{Hahn}, {List} \& {Porqueres}}, \PineGreen{2024})}  \BibitemOpen
  \bibfield  {author} {\bibinfo {author} {\bibfnamefont {O.}~\bibnamefont {{Hahn}}}, \bibinfo {author} {\bibfnamefont {F.}~\bibnamefont {{List}}}, \bibinfo {author} {\bibfnamefont {N.}~\bibnamefont {{Porqueres}}},\ }\emph {{DISCO-DJ I: a differentiable Einstein-Boltzmann solver for cosmology}},\ \href {\doibase 10.1088/1475-7516/2024/06/063} {\bibfield  {journal} {\bibinfo  {journal} {\jcap}\ }\textbf {\bibinfo {volume} {2024}},\ \bibinfo {eid} {063} (\bibinfo {year} {2024})},\ \Eprint {https://arxiv.org/abs/2311.03291} {arXiv:2311.03291 [astro-ph.CO]} \BibitemShut {NoStop}%
\bibitem [{{Hartlap}, {Simon} \& {Schneider}(2007)\citenamefont {{Hartlap}, {Simon},\ \&\ {Schneider}}}]{hartlap2007your}%
{(\PineGreen{{Hartlap}, {Simon} \& {Schneider}}, \PineGreen{2007})}  \BibitemOpen
  \bibfield  {author} {\bibinfo {author} {\bibfnamefont {J.}~\bibnamefont {{Hartlap}}}, \bibinfo {author} {\bibfnamefont {P.}~\bibnamefont {{Simon}}}, \bibinfo {author} {\bibfnamefont {P.}~\bibnamefont {{Schneider}}},\ }\emph {{Why your model parameter confidences might be too optimistic. Unbiased estimation of the inverse covariance matrix}},\ \href {\doibase 10.1051/0004-6361:20066170} {\bibfield  {journal} {\bibinfo  {journal} {\aap}\ }\textbf {\bibinfo {volume} {464}},\ \bibinfo {pages} {399} (\bibinfo {year} {2007})},\ \Eprint {https://arxiv.org/abs/astro-ph/0608064} {arXiv:astro-ph/0608064 [astro-ph]} \BibitemShut {NoStop}%
\bibitem [{{Heavens}, {Jimenez} \& {Lahav}(2000)\citenamefont {{Heavens}, {Jimenez},\ \&\ {Lahav}}}]{heavens2000massive}%
{(\PineGreen{{Heavens}, {Jimenez} \& {Lahav}}, \PineGreen{2000})}  \BibitemOpen
  \bibfield  {author} {\bibinfo {author} {\bibfnamefont {A.~F.}\ \bibnamefont {{Heavens}}}, \bibinfo {author} {\bibfnamefont {R.}~\bibnamefont {{Jimenez}}}, \bibinfo {author} {\bibfnamefont {O.}~\bibnamefont {{Lahav}}},\ }\emph {{Massive lossless data compression and multiple parameter estimation from galaxy spectra}},\ \href {\doibase 10.1046/j.1365-8711.2000.03692.x} {\bibfield  {journal} {\bibinfo  {journal} {\mnras}\ }\textbf {\bibinfo {volume} {317}},\ \bibinfo {pages} {965} (\bibinfo {year} {2000})},\ \Eprint {https://arxiv.org/abs/astro-ph/9911102} {arXiv:astro-ph/9911102 [astro-ph]} \BibitemShut {NoStop}%
\bibitem [{{Ho} {\textit{et~al}}\mbox{.}(2024){Ho}, {Bartlett}, {Chartier}, {Cuesta-Lazaro}, {Ding}, {Lapel}, {Lemos}, {Lovell}, {Makinen}, {Modi} \emph {et~al.}}]{ho2024ltu}%
{(\PineGreen{{Ho} {\textit{et~al}}\mbox{.}}, \PineGreen{2024})}  \BibitemOpen
  \bibfield  {author} {\bibinfo {author} {\bibfnamefont {M.}~\bibnamefont {{Ho}}}, \bibinfo {author} {\bibfnamefont {D.~J.}\ \bibnamefont {{Bartlett}}}, \bibinfo {author} {\bibfnamefont {N.}~\bibnamefont {{Chartier}}}, \bibinfo {author} {\bibfnamefont {C.}~\bibnamefont {{Cuesta-Lazaro}}}, \bibinfo {author} {\bibfnamefont {S.}~\bibnamefont {{Ding}}}, \bibinfo {author} {\bibfnamefont {A.}~\bibnamefont {{Lapel}}}, \bibinfo {author} {\bibfnamefont {P.}~\bibnamefont {{Lemos}}}, \bibinfo {author} {\bibfnamefont {C.~C.}\ \bibnamefont {{Lovell}}}, \bibinfo {author} {\bibfnamefont {T.~L.}\ \bibnamefont {{Makinen}}}, \bibinfo {author} {\bibfnamefont {C.}~\bibnamefont {{Modi}}}, \emph {et~al.},\ }\emph {{LtU-ILI: An All-in-One Framework for Implicit Inference in Astrophysics and Cosmology}},\ \href {\doibase 10.33232/001c.120559} {\bibfield  {journal} {\bibinfo  {journal} {The Open Journal of Astrophysics}\ }\textbf {\bibinfo {volume} {7}},\ \bibinfo {eid} {54} (\bibinfo {year} {2024})},\ \Eprint {https://arxiv.org/abs/2402.05137} {arXiv:2402.05137 [astro-ph.IM]} \BibitemShut {NoStop}%
\bibitem [{{Ho} {\textit{et~al}}\mbox{.}(2015){Ho}, {Agarwal}, {Myers}, {Lyons}, {Disbrow}, {Seo}, {Ross}, {Hirata}, {Padmanabhan}, {O'Connell} \emph {et~al.}}]{ho2015sloan}%
{(\PineGreen{{Ho} {\textit{et~al}}\mbox{.}}, \PineGreen{2015})}  \BibitemOpen
  \bibfield  {author} {\bibinfo {author} {\bibfnamefont {S.}~\bibnamefont {{Ho}}}, \bibinfo {author} {\bibfnamefont {N.}~\bibnamefont {{Agarwal}}}, \bibinfo {author} {\bibfnamefont {A.~D.}\ \bibnamefont {{Myers}}}, \bibinfo {author} {\bibfnamefont {R.}~\bibnamefont {{Lyons}}}, \bibinfo {author} {\bibfnamefont {A.}~\bibnamefont {{Disbrow}}}, \bibinfo {author} {\bibfnamefont {H.-J.}\ \bibnamefont {{Seo}}}, \bibinfo {author} {\bibfnamefont {A.}~\bibnamefont {{Ross}}}, \bibinfo {author} {\bibfnamefont {C.}~\bibnamefont {{Hirata}}}, \bibinfo {author} {\bibfnamefont {N.}~\bibnamefont {{Padmanabhan}}}, \bibinfo {author} {\bibfnamefont {R.}~\bibnamefont {{O'Connell}}}, \emph {et~al.},\ }\emph {{Sloan Digital Sky Survey III photometric quasar clustering: probing the initial conditions of the Universe}},\ \href {\doibase 10.1088/1475-7516/2015/05/040} {\bibfield  {journal} {\bibinfo  {journal} {\jcap}\ }\textbf {\bibinfo {volume} {2015}},\ \bibinfo {pages} {040} (\bibinfo {year} {2015})},\ \Eprint {https://arxiv.org/abs/1311.2597} {arXiv:1311.2597 [astro-ph.CO]} \BibitemShut {NoStop}%
\bibitem [{{Hockney} \& {Eastwood}(1988)\citenamefont {{Hockney}\ \&\ {Eastwood}}}]{Hockney1981}%
{(\PineGreen{{Hockney} \& {Eastwood}}, \PineGreen{1988})}  \BibitemOpen
  \bibfield  {author} {\bibinfo {author} {\bibfnamefont {R.~W.}\ \bibnamefont {{Hockney}}}, \bibinfo {author} {\bibfnamefont {J.~W.}\ \bibnamefont {{Eastwood}}},\ }\href {\doibase 10.1201/9780367806934} {\emph {{Computer simulation using particles}}}\ (\bibinfo  {publisher} {Taylor \& Francis Group},\ \bibinfo {year} {1988})\BibitemShut {NoStop}%
\bibitem [{{Howlett} {\textit{et~al}}\mbox{.}(2015)\citenamefont {{Howlett}, {Ross}, {Samushia}, {Percival},\ \&\ {Manera}}}]{HowlettRoss2015}%
{(\PineGreen{{Howlett} {\textit{et~al}}\mbox{.}}, \PineGreen{2015})}  \BibitemOpen
  \bibfield  {author} {\bibinfo {author} {\bibfnamefont {C.}~\bibnamefont {{Howlett}}}, \bibinfo {author} {\bibfnamefont {A.~J.}\ \bibnamefont {{Ross}}}, \bibinfo {author} {\bibfnamefont {L.}~\bibnamefont {{Samushia}}}, \bibinfo {author} {\bibfnamefont {W.~J.}\ \bibnamefont {{Percival}}}, \bibinfo {author} {\bibfnamefont {M.}~\bibnamefont {{Manera}}},\ }\emph {{The clustering of the SDSS main galaxy sample - II. Mock galaxy catalogues and a measurement of the growth of structure from redshift space distortions at z = 0.15}},\ \href {\doibase 10.1093/mnras/stu2693} {\bibfield  {journal} {\bibinfo  {journal} {\mnras}\ }\textbf {\bibinfo {volume} {449}},\ \bibinfo {pages} {848} (\bibinfo {year} {2015})},\ \Eprint {https://arxiv.org/abs/1409.3238} {arXiv:1409.3238 [astro-ph.CO]} \BibitemShut {NoStop}%
\bibitem [{{Huterer}, {Cunha} \& {Fang}(2013)\citenamefont {{Huterer}, {Cunha},\ \&\ {Fang}}}]{huterer2013calibration}%
{(\PineGreen{{Huterer}, {Cunha} \& {Fang}}, \PineGreen{2013})}  \BibitemOpen
  \bibfield  {author} {\bibinfo {author} {\bibfnamefont {D.}~\bibnamefont {{Huterer}}}, \bibinfo {author} {\bibfnamefont {C.~E.}\ \bibnamefont {{Cunha}}}, \bibinfo {author} {\bibfnamefont {W.}~\bibnamefont {{Fang}}},\ }\emph {{Calibration errors unleashed: effects on cosmological parameters and requirements for large-scale structure surveys}},\ \href {\doibase 10.1093/mnras/stt653} {\bibfield  {journal} {\bibinfo  {journal} {\mnras}\ }\textbf {\bibinfo {volume} {432}},\ \bibinfo {pages} {2945} (\bibinfo {year} {2013})},\ \Eprint {https://arxiv.org/abs/1211.1015} {arXiv:1211.1015 [astro-ph.CO]} \BibitemShut {NoStop}%
\bibitem [{Ishida {\textit{et~al}}\mbox{.}(2015)Ishida, Vitenti, Penna-Lima, Cisewski, de~Souza, Trindade, Cameron, Busti, collaboration \emph {et~al.}}]{ishida2015cosmoabc}%
{(\PineGreen{Ishida {\textit{et~al}}\mbox{.}}, \PineGreen{2015})}  \BibitemOpen
  \bibfield  {author} {\bibinfo {author} {\bibfnamefont {E.~E.}\ \bibnamefont {Ishida}}, \bibinfo {author} {\bibfnamefont {S.~D.}\ \bibnamefont {Vitenti}}, \bibinfo {author} {\bibfnamefont {M.}~\bibnamefont {Penna-Lima}}, \bibinfo {author} {\bibfnamefont {J.}~\bibnamefont {Cisewski}}, \bibinfo {author} {\bibfnamefont {R.~S.}\ \bibnamefont {de~Souza}}, \bibinfo {author} {\bibfnamefont {A.~M.}\ \bibnamefont {Trindade}}, \bibinfo {author} {\bibfnamefont {E.}~\bibnamefont {Cameron}}, \bibinfo {author} {\bibfnamefont {V.~C.}\ \bibnamefont {Busti}}, \bibinfo {author} {\bibfnamefont {C.}~\bibnamefont {collaboration}}, \bibinfo {author} {\bibnamefont {others}},\ }\emph {Cosmoabc: likelihood-free inference via population Monte Carlo approximate Bayesian computation},\ \href {\doibase 10.1016/j.ascom.2015.09.001} {\bibfield  {journal} {\bibinfo  {journal} {Astronomy and Computing}\ }\textbf {\bibinfo {volume} {13}},\ \bibinfo {pages} {1} (\bibinfo {year} {2015})}\BibitemShut {NoStop}%
\bibitem [{Ivezi{\'c} {\textit{et~al}}\mbox{.}(2019)Ivezi{\'c}, Kahn, Tyson, Abel, Acosta, Allsman, Alonso, AlSayyad, Anderson, Andrew \emph {et~al.}}]{ivezic2019lsst}%
{(\PineGreen{Ivezi{\'c} {\textit{et~al}}\mbox{.}}, \PineGreen{2019})}  \BibitemOpen
  \bibfield  {author} {\bibinfo {author} {\bibfnamefont {{\v{Z}}.}~\bibnamefont {Ivezi{\'c}}}, \bibinfo {author} {\bibfnamefont {S.~M.}\ \bibnamefont {Kahn}}, \bibinfo {author} {\bibfnamefont {J.~A.}\ \bibnamefont {Tyson}}, \bibinfo {author} {\bibfnamefont {B.}~\bibnamefont {Abel}}, \bibinfo {author} {\bibfnamefont {E.}~\bibnamefont {Acosta}}, \bibinfo {author} {\bibfnamefont {R.}~\bibnamefont {Allsman}}, \bibinfo {author} {\bibfnamefont {D.}~\bibnamefont {Alonso}}, \bibinfo {author} {\bibfnamefont {Y.}~\bibnamefont {AlSayyad}}, \bibinfo {author} {\bibfnamefont {S.~F.}\ \bibnamefont {Anderson}}, \bibinfo {author} {\bibfnamefont {J.}~\bibnamefont {Andrew}}, \emph {et~al.},\ }\emph {LSST: from science drivers to reference design and anticipated data products},\ \href {https://iopscience.iop.org/article/10.3847/1538-4357/ab042c/pdf} {\bibfield  {journal} {\bibinfo  {journal} {The Astrophysical Journal}\ }\textbf {\bibinfo {volume} {873}},\ \bibinfo {pages} {111} (\bibinfo {year} {2019})}\BibitemShut {NoStop}%
\bibitem [{{Jasche} \& {Lavaux}(2019)\citenamefont {{Jasche}\ \&\ {Lavaux}}}]{jasche2019physical}%
{(\PineGreen{{Jasche} \& {Lavaux}}, \PineGreen{2019})}  \BibitemOpen
  \bibfield  {author} {\bibinfo {author} {\bibfnamefont {J.}~\bibnamefont {{Jasche}}}, \bibinfo {author} {\bibfnamefont {G.}~\bibnamefont {{Lavaux}}},\ }\emph {{Physical Bayesian modelling of the non-linear matter distribution: New insights into the nearby universe}},\ \href {\doibase 10.1051/0004-6361/201833710} {\bibfield  {journal} {\bibinfo  {journal} {\aap}\ }\textbf {\bibinfo {volume} {625}},\ \bibinfo {eid} {A64} (\bibinfo {year} {2019})},\ \Eprint {https://arxiv.org/abs/1806.11117} {arXiv:1806.11117 [astro-ph.CO]} \BibitemShut {NoStop}%
\bibitem [{{Jasche} \& {Wandelt}(2013)\citenamefont {{Jasche}\ \&\ {Wandelt}}}]{Jasche2013BORG}%
{(\PineGreen{{Jasche} \& {Wandelt}}, \PineGreen{2013})}  \BibitemOpen
  \bibfield  {author} {\bibinfo {author} {\bibfnamefont {J.}~\bibnamefont {{Jasche}}}, \bibinfo {author} {\bibfnamefont {B.~D.}\ \bibnamefont {{Wandelt}}},\ }\emph {{Bayesian physical reconstruction of initial conditions from large-scale structure surveys}},\ \href {\doibase 10.1093/mnras/stt449} {\bibfield  {journal} {\bibinfo  {journal} {\mnras}\ }\textbf {\bibinfo {volume} {432}},\ \bibinfo {pages} {894} (\bibinfo {year} {2013})},\ \Eprint {https://arxiv.org/abs/1203.3639} {arXiv:1203.3639 [astro-ph.CO]} \BibitemShut {NoStop}%
\bibitem [{{Jasche} \& {Lavaux}(2017)\citenamefont {{Jasche}\ \&\ {Lavaux}}}]{jasche2017bayesian}%
{(\PineGreen{{Jasche} \& {Lavaux}}, \PineGreen{2017})}  \BibitemOpen
  \bibfield  {author} {\bibinfo {author} {\bibfnamefont {J.}~\bibnamefont {{Jasche}}}, \bibinfo {author} {\bibfnamefont {G.}~\bibnamefont {{Lavaux}}},\ }\emph {{Bayesian power spectrum inference with foreground and target contamination treatment}},\ \href {\doibase 10.1051/0004-6361/201730909} {\bibfield  {journal} {\bibinfo  {journal} {\aap}\ }\textbf {\bibinfo {volume} {606}},\ \bibinfo {eid} {A37} (\bibinfo {year} {2017})},\ \Eprint {https://arxiv.org/abs/1706.08971} {arXiv:1706.08971 [astro-ph.CO]} \BibitemShut {NoStop}%
\bibitem [{{Jasche} {\textit{et~al}}\mbox{.}(2010)\citenamefont {{Jasche}, {Kitaura}, {Wandelt},\ \&\ {En{\ss}lin}}}]{Jasche2010b}%
{(\PineGreen{{Jasche} {\textit{et~al}}\mbox{.}}, \PineGreen{2010})}  \BibitemOpen
  \bibfield  {author} {\bibinfo {author} {\bibfnamefont {J.}~\bibnamefont {{Jasche}}}, \bibinfo {author} {\bibfnamefont {F.~S.}\ \bibnamefont {{Kitaura}}}, \bibinfo {author} {\bibfnamefont {B.~D.}\ \bibnamefont {{Wandelt}}}, \bibinfo {author} {\bibfnamefont {T.~A.}\ \bibnamefont {{En{\ss}lin}}},\ }\emph {{Bayesian power-spectrum inference for large-scale structure data}},\ \href {\doibase 10.1111/j.1365-2966.2010.16610.x} {\bibfield  {journal} {\bibinfo  {journal} {\mnras}\ }\textbf {\bibinfo {volume} {406}},\ \bibinfo {pages} {60} (\bibinfo {year} {2010})},\ \Eprint {https://arxiv.org/abs/0911.2493} {arXiv:0911.2493 [astro-ph.CO]} \BibitemShut {NoStop}%
\bibitem [{{Kaiser}(1984)\citenamefont {{Kaiser}}}]{kaiser1984spatial}%
{(\PineGreen{{Kaiser}}, \PineGreen{1984})}  \BibitemOpen
  \bibfield  {author} {\bibinfo {author} {\bibfnamefont {N.}~\bibnamefont {{Kaiser}}},\ }\emph {{On the spatial correlations of Abell clusters.}},\ \href {\doibase 10.1086/184341} {\bibfield  {journal} {\bibinfo  {journal} {\apjl}\ }\textbf {\bibinfo {volume} {284}},\ \bibinfo {pages} {L9} (\bibinfo {year} {1984})}\BibitemShut {NoStop}%
\bibitem [{{Karchev} {\textit{et~al}}\mbox{.}(2024)\citenamefont {{Karchev}, {Grayling}, {Boyd}, {Trotta}, {Mandel},\ \&\ {Weniger}}}]{karchev2024side}%
{(\PineGreen{{Karchev} {\textit{et~al}}\mbox{.}}, \PineGreen{2024})}  \BibitemOpen
  \bibfield  {author} {\bibinfo {author} {\bibfnamefont {K.}~\bibnamefont {{Karchev}}}, \bibinfo {author} {\bibfnamefont {M.}~\bibnamefont {{Grayling}}}, \bibinfo {author} {\bibfnamefont {B.~M.}\ \bibnamefont {{Boyd}}}, \bibinfo {author} {\bibfnamefont {R.}~\bibnamefont {{Trotta}}}, \bibinfo {author} {\bibfnamefont {K.~S.}\ \bibnamefont {{Mandel}}}, \bibinfo {author} {\bibfnamefont {C.}~\bibnamefont {{Weniger}}},\ }\emph {{SIDE-real: Supernova Ia Dust Extinction with truncated marginal neural ratio estimation applied to real data}},\ \href {\doibase 10.1093/mnras/stae995} {\bibfield  {journal} {\bibinfo  {journal} {\mnras}\ }\textbf {\bibinfo {volume} {530}},\ \bibinfo {pages} {3881} (\bibinfo {year} {2024})},\ \Eprint {https://arxiv.org/abs/2403.07871} {arXiv:2403.07871 [astro-ph.CO]} \BibitemShut {NoStop}%
\bibitem [{{Kim} {\textit{et~al}}\mbox{.}(2004)\citenamefont {{Kim}, {Linder}, {Miquel},\ \&\ {Mostek}}}]{kim2004effects}%
{(\PineGreen{{Kim} {\textit{et~al}}\mbox{.}}, \PineGreen{2004})}  \BibitemOpen
  \bibfield  {author} {\bibinfo {author} {\bibfnamefont {A.~G.}\ \bibnamefont {{Kim}}}, \bibinfo {author} {\bibfnamefont {E.~V.}\ \bibnamefont {{Linder}}}, \bibinfo {author} {\bibfnamefont {R.}~\bibnamefont {{Miquel}}}, \bibinfo {author} {\bibfnamefont {N.}~\bibnamefont {{Mostek}}},\ }\emph {{Effects of systematic uncertainties on the supernova determination of cosmological parameters}},\ \href {\doibase 10.1111/j.1365-2966.2004.07260.x} {\bibfield  {journal} {\bibinfo  {journal} {\mnras}\ }\textbf {\bibinfo {volume} {347}},\ \bibinfo {pages} {909} (\bibinfo {year} {2004})},\ \Eprint {https://arxiv.org/abs/astro-ph/0304509} {arXiv:astro-ph/0304509 [astro-ph]} \BibitemShut {NoStop}%
\bibitem [{{Kitching}, {Taylor} \& {Heavens}(2008)\citenamefont {{Kitching}, {Taylor},\ \&\ {Heavens}}}]{kitching2008systematic}%
{(\PineGreen{{Kitching}, {Taylor} \& {Heavens}}, \PineGreen{2008})}  \BibitemOpen
  \bibfield  {author} {\bibinfo {author} {\bibfnamefont {T.~D.}\ \bibnamefont {{Kitching}}}, \bibinfo {author} {\bibfnamefont {A.~N.}\ \bibnamefont {{Taylor}}}, \bibinfo {author} {\bibfnamefont {A.~F.}\ \bibnamefont {{Heavens}}},\ }\emph {{Systematic effects on dark energy from 3D weak shear}},\ \href {\doibase 10.1111/j.1365-2966.2008.13419.x} {\bibfield  {journal} {\bibinfo  {journal} {\mnras}\ }\textbf {\bibinfo {volume} {389}},\ \bibinfo {pages} {173} (\bibinfo {year} {2008})},\ \Eprint {https://arxiv.org/abs/0801.3270} {arXiv:0801.3270 [astro-ph]} \BibitemShut {NoStop}%
\bibitem [{{Kitching} {\textit{et~al}}\mbox{.}(2016)\citenamefont {{Kitching}, {Verde}, {Heavens},\ \&\ {Jimenez}}}]{kitching2016discrepancies}%
{(\PineGreen{{Kitching} {\textit{et~al}}\mbox{.}}, \PineGreen{2016})}  \BibitemOpen
  \bibfield  {author} {\bibinfo {author} {\bibfnamefont {T.~D.}\ \bibnamefont {{Kitching}}}, \bibinfo {author} {\bibfnamefont {L.}~\bibnamefont {{Verde}}}, \bibinfo {author} {\bibfnamefont {A.~F.}\ \bibnamefont {{Heavens}}}, \bibinfo {author} {\bibfnamefont {R.}~\bibnamefont {{Jimenez}}},\ }\emph {{Discrepancies between CFHTLenS cosmic shear and Planck: new physics or systematic effects?}},\ \href {\doibase 10.1093/mnras/stw707} {\bibfield  {journal} {\bibinfo  {journal} {\mnras}\ }\textbf {\bibinfo {volume} {459}},\ \bibinfo {pages} {971} (\bibinfo {year} {2016})},\ \Eprint {https://arxiv.org/abs/1602.02960} {arXiv:1602.02960 [astro-ph.CO]} \BibitemShut {NoStop}%
\bibitem [{{Kosti{\'c}} {\textit{et~al}}\mbox{.}(2023)\citenamefont {{Kosti{\'c}}, {Nguyen}, {Schmidt},\ \&\ {Reinecke}}}]{kostic2023consistency}%
{(\PineGreen{{Kosti{\'c}} {\textit{et~al}}\mbox{.}}, \PineGreen{2023})}  \BibitemOpen
  \bibfield  {author} {\bibinfo {author} {\bibfnamefont {A.}~\bibnamefont {{Kosti{\'c}}}}, \bibinfo {author} {\bibfnamefont {N.-M.}\ \bibnamefont {{Nguyen}}}, \bibinfo {author} {\bibfnamefont {F.}~\bibnamefont {{Schmidt}}}, \bibinfo {author} {\bibfnamefont {M.}~\bibnamefont {{Reinecke}}},\ }\emph {{Consistency tests of field level inference with the EFT likelihood}},\ \href {\doibase 10.1088/1475-7516/2023/07/063} {\bibfield  {journal} {\bibinfo  {journal} {\jcap}\ }\textbf {\bibinfo {volume} {2023}},\ \bibinfo {eid} {063} (\bibinfo {year} {2023})},\ \Eprint {https://arxiv.org/abs/2212.07875} {arXiv:2212.07875 [astro-ph.CO]} \BibitemShut {NoStop}%
\bibitem [{Kullback \& Leibler(1951)\citenamefont {Kullback\ \&\ Leibler}}]{kullback1951information}%
{(\PineGreen{Kullback \& Leibler}, \PineGreen{1951})}  \BibitemOpen
  \bibfield  {author} {\bibinfo {author} {\bibfnamefont {S.}~\bibnamefont {Kullback}}, \bibinfo {author} {\bibfnamefont {R.~A.}\ \bibnamefont {Leibler}},\ }\emph {On information and sufficiency},\ \href {\doibase 10.1214/aoms/1177729694} {\bibfield  {journal} {\bibinfo  {journal} {The annals of mathematical statistics}\ }\textbf {\bibinfo {volume} {22}},\ \bibinfo {pages} {79} (\bibinfo {year} {1951})}\BibitemShut {NoStop}%
\bibitem [{{Lanzieri} {\textit{et~al}}\mbox{.}(2024)\citenamefont {{Lanzieri}, {Zeghal}, {Makinen}, {Boucaud}, {Starck},\ \&\ {Lanusse}}}]{lanzieri2024optimal}%
{(\PineGreen{{Lanzieri} {\textit{et~al}}\mbox{.}}, \PineGreen{2024})}  \BibitemOpen
  \bibfield  {author} {\bibinfo {author} {\bibfnamefont {D.}~\bibnamefont {{Lanzieri}}}, \bibinfo {author} {\bibfnamefont {J.}~\bibnamefont {{Zeghal}}}, \bibinfo {author} {\bibfnamefont {T.~L.}\ \bibnamefont {{Makinen}}}, \bibinfo {author} {\bibfnamefont {A.}~\bibnamefont {{Boucaud}}}, \bibinfo {author} {\bibfnamefont {J.-L.}\ \bibnamefont {{Starck}}}, \bibinfo {author} {\bibfnamefont {F.}~\bibnamefont {{Lanusse}}},\ }\emph {{Optimal Neural Summarisation for Full-Field Weak Lensing Cosmological Implicit Inference}},\ \href {\doibase 10.48550/arXiv.2407.10877} {\bibfield  {journal} {\bibinfo  {journal} {arXiv e-prints}\ ,\ \bibinfo {eid} {arXiv:2407.10877}} (\bibinfo {year} {2024})},\ \Eprint {https://arxiv.org/abs/2407.10877} {arXiv:2407.10877 [astro-ph.CO]} \BibitemShut {NoStop}%
\bibitem [{{Laureijs} {\textit{et~al}}\mbox{.}(2011){Laureijs}, {Amiaux}, {Arduini}, {Augu{\`e}res}, {Brinchmann}, {Cole}, {Cropper}, {Dabin}, {Duvet}, {Ealet} \emph {et~al.}}]{laureijs2011euclid}%
{(\PineGreen{{Laureijs} {\textit{et~al}}\mbox{.}}, \PineGreen{2011})}  \BibitemOpen
  \bibfield  {author} {\bibinfo {author} {\bibfnamefont {R.}~\bibnamefont {{Laureijs}}}, \bibinfo {author} {\bibfnamefont {J.}~\bibnamefont {{Amiaux}}}, \bibinfo {author} {\bibfnamefont {S.}~\bibnamefont {{Arduini}}}, \bibinfo {author} {\bibfnamefont {J.~L.}\ \bibnamefont {{Augu{\`e}res}}}, \bibinfo {author} {\bibfnamefont {J.}~\bibnamefont {{Brinchmann}}}, \bibinfo {author} {\bibfnamefont {R.}~\bibnamefont {{Cole}}}, \bibinfo {author} {\bibfnamefont {M.}~\bibnamefont {{Cropper}}}, \bibinfo {author} {\bibfnamefont {C.}~\bibnamefont {{Dabin}}}, \bibinfo {author} {\bibfnamefont {L.}~\bibnamefont {{Duvet}}}, \bibinfo {author} {\bibfnamefont {A.}~\bibnamefont {{Ealet}}}, \emph {et~al.},\ }\emph {{Euclid Definition Study Report}},\ \href {\doibase 10.48550/arXiv.1110.3193} {\bibfield  {journal} {\bibinfo  {journal} {arXiv e-prints}\ ,\ \bibinfo {eid} {arXiv:1110.3193}} (\bibinfo {year} {2011})},\ \Eprint {https://arxiv.org/abs/1110.3193} {arXiv:1110.3193 [astro-ph.CO]} \BibitemShut {NoStop}%
\bibitem [{{Leclercq}(2018)\citenamefont {{Leclercq}}}]{Leclercq2018BOLFI}%
{(\PineGreen{{Leclercq}}, \PineGreen{2018})}  \BibitemOpen
  \bibfield  {author} {\bibinfo {author} {\bibfnamefont {F.}~\bibnamefont {{Leclercq}}},\ }\emph {{Bayesian optimization for likelihood-free cosmological inference}},\ \href {\doibase 10.1103/PhysRevD.98.063511} {\bibfield  {journal} {\bibinfo  {journal} {\prd}\ }\textbf {\bibinfo {volume} {98}},\ \bibinfo {eid} {063511} (\bibinfo {year} {2018})},\ \Eprint {https://arxiv.org/abs/1805.07152} {arXiv:1805.07152 [astro-ph.CO]} \BibitemShut {NoStop}%
\bibitem [{Leclercq(2022)\citenamefont {Leclercq}}]{Leclercq_2022LV}%
{(\PineGreen{Leclercq}, \PineGreen{2022})}  \BibitemOpen
  \bibfield  {author} {\bibinfo {author} {\bibfnamefont {F.}~\bibnamefont {Leclercq}},\ }\emph {Simulation-Based Inference of Bayesian Hierarchical Models While Checking for Model Misspecification},\ in\ \href {\doibase 10.3390/psf2022005004} {\emph {\bibinfo {booktitle} {{MaxEnt} 2022}}}\ (\bibinfo  {publisher} {{MDPI}},\ \bibinfo {year} {2022})\BibitemShut {NoStop}%
\bibitem [{{Leclercq}, {Jasche} \& {Wandelt}(2015)\citenamefont {{Leclercq}, {Jasche},\ \&\ {Wandelt}}}]{Leclercq2015ST}%
{(\PineGreen{{Leclercq}, {Jasche} \& {Wandelt}}, \PineGreen{2015})}  \BibitemOpen
  \bibfield  {author} {\bibinfo {author} {\bibfnamefont {F.}~\bibnamefont {{Leclercq}}}, \bibinfo {author} {\bibfnamefont {J.}~\bibnamefont {{Jasche}}}, \bibinfo {author} {\bibfnamefont {B.}~\bibnamefont {{Wandelt}}},\ }\emph {{Bayesian analysis of the dynamic cosmic web in the SDSS galaxy survey}},\ \href {\doibase 10.1088/1475-7516/2015/06/015} {\bibfield  {journal} {\bibinfo  {journal} {\jcap}\ }\textbf {\bibinfo {volume} {2015}},\ \bibinfo {pages} {015} (\bibinfo {year} {2015})},\ \Eprint {https://arxiv.org/abs/1502.02690} {arXiv:1502.02690 [astro-ph.CO]} \BibitemShut {NoStop}%
\bibitem [{{Leclercq} {\textit{et~al}}\mbox{.}(2019)\citenamefont {{Leclercq}, {Enzi}, {Jasche},\ \&\ {Heavens}}}]{Leclercq2019SELFI}%
{(\PineGreen{{Leclercq} {\textit{et~al}}\mbox{.}}, \PineGreen{2019})}  \BibitemOpen
  \bibfield  {author} {\bibinfo {author} {\bibfnamefont {F.}~\bibnamefont {{Leclercq}}}, \bibinfo {author} {\bibfnamefont {W.}~\bibnamefont {{Enzi}}}, \bibinfo {author} {\bibfnamefont {J.}~\bibnamefont {{Jasche}}}, \bibinfo {author} {\bibfnamefont {A.}~\bibnamefont {{Heavens}}},\ }\emph {{Primordial power spectrum and cosmology from black-box galaxy surveys}},\ \href {\doibase 10.1093/mnras/stz2718} {\bibfield  {journal} {\bibinfo  {journal} {\mnras}\ }\textbf {\bibinfo {volume} {490}},\ \bibinfo {pages} {4237} (\bibinfo {year} {2019})},\ \Eprint {https://arxiv.org/abs/1902.10149} {arXiv:1902.10149 [astro-ph.CO]} \BibitemShut {NoStop}%
\bibitem [{{Leistedt} \& {Peiris}(2014)\citenamefont {{Leistedt}\ \&\ {Peiris}}}]{leistedt2014exploiting}%
{(\PineGreen{{Leistedt} \& {Peiris}}, \PineGreen{2014})}  \BibitemOpen
  \bibfield  {author} {\bibinfo {author} {\bibfnamefont {B.}~\bibnamefont {{Leistedt}}}, \bibinfo {author} {\bibfnamefont {H.~V.}\ \bibnamefont {{Peiris}}},\ }\emph {{Exploiting the full potential of photometric quasar surveys: optimal power spectra through blind mitigation of systematics}},\ \href {\doibase 10.1093/mnras/stu1439} {\bibfield  {journal} {\bibinfo  {journal} {\mnras}\ }\textbf {\bibinfo {volume} {444}},\ \bibinfo {pages} {2} (\bibinfo {year} {2014})},\ \Eprint {https://arxiv.org/abs/1404.6530} {arXiv:1404.6530 [astro-ph.CO]} \BibitemShut {NoStop}%
\bibitem [{{Lewis} \& {Challinor}(2011)\citenamefont {{Lewis}\ \&\ {Challinor}}}]{lewis2011camb}%
{(\PineGreen{{Lewis} \& {Challinor}}, \PineGreen{2011})}  \BibitemOpen
  \bibfield  {author} {\bibinfo {author} {\bibfnamefont {A.}~\bibnamefont {{Lewis}}}, \bibinfo {author} {\bibfnamefont {A.}~\bibnamefont {{Challinor}}},\ }\emph {{CAMB: Code for Anisotropies in the Microwave Background}},\ \href {https://ui.adsabs.harvard.edu/abs/2011ascl.soft02026L} {\bibfield  {journal} {\bibinfo  {journal} {Astrophysics Source Code Library}\ ,\ \bibinfo {eid} {ascl:1102.026}} (\bibinfo {year} {2011})}\BibitemShut {NoStop}%
\bibitem [{Lintusaari {\textit{et~al}}\mbox{.}(2017)\citenamefont {Lintusaari, Gutmann, Dutta, Kaski,\ \&\ Corander}}]{lintusaari2017fundamentals}%
{(\PineGreen{Lintusaari {\textit{et~al}}\mbox{.}}, \PineGreen{2017})}  \BibitemOpen
  \bibfield  {author} {\bibinfo {author} {\bibfnamefont {J.}~\bibnamefont {Lintusaari}}, \bibinfo {author} {\bibfnamefont {M.~U.}\ \bibnamefont {Gutmann}}, \bibinfo {author} {\bibfnamefont {R.}~\bibnamefont {Dutta}}, \bibinfo {author} {\bibfnamefont {S.}~\bibnamefont {Kaski}}, \bibinfo {author} {\bibfnamefont {J.}~\bibnamefont {Corander}},\ }\emph {Fundamentals and recent developments in approximate Bayesian computation},\ \href {https://academic.oup.com/sysbio/article/66/1/e66/2420817} {\bibfield  {journal} {\bibinfo  {journal} {Systematic biology}\ }\textbf {\bibinfo {volume} {66}},\ \bibinfo {pages} {e66} (\bibinfo {year} {2017})}\BibitemShut {NoStop}%
\bibitem [{{Lintusaari} {\textit{et~al}}\mbox{.}(2018)\citenamefont {{Lintusaari}, {Vuollekoski}, {Kangasr{\"a}{\"a}si{\"o}}, {Skyt{\'e}n}, {J{\"a}rvenp{\"a}{\"a}}, {Marttinen}, {Gutmann}, {Vehtari}, {Corander},\ \&\ {Kaski}}}]{Lintusaari2018}%
{(\PineGreen{{Lintusaari} {\textit{et~al}}\mbox{.}}, \PineGreen{2018})}  \BibitemOpen
  \bibfield  {author} {\bibinfo {author} {\bibfnamefont {J.}~\bibnamefont {{Lintusaari}}}, \bibinfo {author} {\bibfnamefont {H.}~\bibnamefont {{Vuollekoski}}}, \bibinfo {author} {\bibfnamefont {A.}~\bibnamefont {{Kangasr{\"a}{\"a}si{\"o}}}}, \bibinfo {author} {\bibfnamefont {K.}~\bibnamefont {{Skyt{\'e}n}}}, \bibinfo {author} {\bibfnamefont {M.}~\bibnamefont {{J{\"a}rvenp{\"a}{\"a}}}}, \bibinfo {author} {\bibfnamefont {P.}~\bibnamefont {{Marttinen}}}, \bibinfo {author} {\bibfnamefont {M.~U.}\ \bibnamefont {{Gutmann}}}, \bibinfo {author} {\bibfnamefont {A.}~\bibnamefont {{Vehtari}}}, \bibinfo {author} {\bibfnamefont {J.}~\bibnamefont {{Corander}}}, \bibinfo {author} {\bibfnamefont {S.}~\bibnamefont {{Kaski}}},\ }\emph {{ELFI: Engine for Likelihood-Free Inference}},\ \href {\doibase 10.48550/arXiv.1708.00707} {\bibfield  {journal} {\bibinfo  {journal} {Journal of Machine Learning Research}\ }\textbf {\bibinfo {volume} {19}},\ \bibinfo {pages} {1} (\bibinfo {year} {2018})}\BibitemShut {NoStop}%
\bibitem [{Liu \& Nocedal(1989)\citenamefont {Liu\ \&\ Nocedal}}]{liu1989limited}%
{(\PineGreen{Liu \& Nocedal}, \PineGreen{1989})}  \BibitemOpen
  \bibfield  {author} {\bibinfo {author} {\bibfnamefont {D.~C.}\ \bibnamefont {Liu}}, \bibinfo {author} {\bibfnamefont {J.}~\bibnamefont {Nocedal}},\ }\emph {On the limited memory BFGS method for large scale optimization},\ \href {\doibase 10.1007/BF01589116} {\bibfield  {journal} {\bibinfo  {journal} {Mathematical programming}\ }\textbf {\bibinfo {volume} {45}},\ \bibinfo {pages} {503} (\bibinfo {year} {1989})}\BibitemShut {NoStop}%
\bibitem [{{Loureiro} {\textit{et~al}}\mbox{.}(2023)\citenamefont {{Loureiro}, {Whiteway}, {Sellentin}, {Silva Lafaurie}, {Jaffe},\ \&\ {Heavens}}}]{loureiro2022almanac}%
{(\PineGreen{{Loureiro} {\textit{et~al}}\mbox{.}}, \PineGreen{2023})}  \BibitemOpen
  \bibfield  {author} {\bibinfo {author} {\bibfnamefont {A.}~\bibnamefont {{Loureiro}}}, \bibinfo {author} {\bibfnamefont {L.}~\bibnamefont {{Whiteway}}}, \bibinfo {author} {\bibfnamefont {E.}~\bibnamefont {{Sellentin}}}, \bibinfo {author} {\bibfnamefont {J.}~\bibnamefont {{Silva Lafaurie}}}, \bibinfo {author} {\bibfnamefont {A.~H.}\ \bibnamefont {{Jaffe}}}, \bibinfo {author} {\bibfnamefont {A.~F.}\ \bibnamefont {{Heavens}}},\ }\emph {{Almanac: Weak Lensing power spectra and map inference on the masked sphere}},\ \href {\doibase 10.21105/astro.2210.13260} {\bibfield  {journal} {\bibinfo  {journal} {The Open Journal of Astrophysics}\ }\textbf {\bibinfo {volume} {6}},\ \bibinfo {eid} {6} (\bibinfo {year} {2023})},\ \Eprint {https://arxiv.org/abs/2210.13260} {arXiv:2210.13260 [astro-ph.CO]} \BibitemShut {NoStop}%
\bibitem [{{LSST Dark Energy Science Collaboration}(2012)\citenamefont {{LSST Dark Energy Science Collaboration}}}]{lsst2012large}%
{(\PineGreen{{LSST Dark Energy Science Collaboration}}, \PineGreen{2012})}  \BibitemOpen
  \bibfield  {author} {\bibinfo {author} {\bibnamefont {{LSST Dark Energy Science Collaboration}}},\ }\emph {Large synoptic survey telescope: dark energy science collaboration},\ \href {https://arxiv.org/pdf/1211.0310} {\bibfield  {journal} {\bibinfo  {journal} {arXiv preprint arXiv:1211.0310}\ } (\bibinfo {year} {2012})}\BibitemShut {NoStop}%
\bibitem [{{Makinen} {\textit{et~al}}\mbox{.}(2021)\citenamefont {{Makinen}, {Charnock}, {Alsing},\ \&\ {Wandelt}}}]{Makinen_2021}%
{(\PineGreen{{Makinen} {\textit{et~al}}\mbox{.}}, \PineGreen{2021})}  \BibitemOpen
  \bibfield  {author} {\bibinfo {author} {\bibfnamefont {T.~L.}\ \bibnamefont {{Makinen}}}, \bibinfo {author} {\bibfnamefont {T.}~\bibnamefont {{Charnock}}}, \bibinfo {author} {\bibfnamefont {J.}~\bibnamefont {{Alsing}}}, \bibinfo {author} {\bibfnamefont {B.~D.}\ \bibnamefont {{Wandelt}}},\ }\emph {{Lossless, scalable implicit likelihood inference for cosmological fields}},\ \href {\doibase 10.1088/1475-7516/2021/11/049} {\bibfield  {journal} {\bibinfo  {journal} {\jcap}\ }\textbf {\bibinfo {volume} {2021}},\ \bibinfo {eid} {049} (\bibinfo {year} {2021})},\ \Eprint {https://arxiv.org/abs/2107.07405} {arXiv:2107.07405 [astro-ph.CO]} \BibitemShut {NoStop}%
\bibitem [{{Makinen} {\textit{et~al}}\mbox{.}(2024)\citenamefont {{Makinen}, {Heavens}, {Porqueres}, {Charnock}, {Lapel},\ \&\ {Wandelt}}}]{makinen2024hybrid}%
{(\PineGreen{{Makinen} {\textit{et~al}}\mbox{.}}, \PineGreen{2024})}  \BibitemOpen
  \bibfield  {author} {\bibinfo {author} {\bibfnamefont {T.~L.}\ \bibnamefont {{Makinen}}}, \bibinfo {author} {\bibfnamefont {A.}~\bibnamefont {{Heavens}}}, \bibinfo {author} {\bibfnamefont {N.}~\bibnamefont {{Porqueres}}}, \bibinfo {author} {\bibfnamefont {T.}~\bibnamefont {{Charnock}}}, \bibinfo {author} {\bibfnamefont {A.}~\bibnamefont {{Lapel}}}, \bibinfo {author} {\bibfnamefont {B.~D.}\ \bibnamefont {{Wandelt}}},\ }\emph {{Hybrid summary statistics: neural weak lensing inference beyond the power spectrum}},\ \href {\doibase 10.48550/arXiv.2407.18909} {\bibfield  {journal} {\bibinfo  {journal} {arXiv e-prints}\ ,\ \bibinfo {eid} {arXiv:2407.18909}} (\bibinfo {year} {2024})},\ \Eprint {https://arxiv.org/abs/2407.18909} {arXiv:2407.18909 [astro-ph.CO]} \BibitemShut {NoStop}%
\bibitem [{{Massara} {\textit{et~al}}\mbox{.}(2021)\citenamefont {{Massara}, {Ho}, {Hirata}, {DeRose}, {Wechsler},\ \&\ {Fang}}}]{massara2021line}%
{(\PineGreen{{Massara} {\textit{et~al}}\mbox{.}}, \PineGreen{2021})}  \BibitemOpen
  \bibfield  {author} {\bibinfo {author} {\bibfnamefont {E.}~\bibnamefont {{Massara}}}, \bibinfo {author} {\bibfnamefont {S.}~\bibnamefont {{Ho}}}, \bibinfo {author} {\bibfnamefont {C.~M.}\ \bibnamefont {{Hirata}}}, \bibinfo {author} {\bibfnamefont {J.}~\bibnamefont {{DeRose}}}, \bibinfo {author} {\bibfnamefont {R.~H.}\ \bibnamefont {{Wechsler}}}, \bibinfo {author} {\bibfnamefont {X.}~\bibnamefont {{Fang}}},\ }\emph {{Line confusion in spectroscopic surveys and its possible effects: shifts in Baryon Acoustic Oscillations position}},\ \href {\doibase 10.1093/mnras/stab2628} {\bibfield  {journal} {\bibinfo  {journal} {\mnras}\ }\textbf {\bibinfo {volume} {508}},\ \bibinfo {pages} {4193} (\bibinfo {year} {2021})},\ \Eprint {https://arxiv.org/abs/2010.00047} {arXiv:2010.00047 [astro-ph.CO]} \BibitemShut {NoStop}%
\bibitem [{{Meiksin}, {White} \& {Peacock}(1999)\citenamefont {{Meiksin}, {White},\ \&\ {Peacock}}}]{Meiksin1999}%
{(\PineGreen{{Meiksin}, {White} \& {Peacock}}, \PineGreen{1999})}  \BibitemOpen
  \bibfield  {author} {\bibinfo {author} {\bibfnamefont {A.}~\bibnamefont {{Meiksin}}}, \bibinfo {author} {\bibfnamefont {M.}~\bibnamefont {{White}}}, \bibinfo {author} {\bibfnamefont {J.~A.}\ \bibnamefont {{Peacock}}},\ }\emph {{Baryonic signatures in large-scale structure}},\ \href {\doibase 10.1046/j.1365-8711.1999.02369.x} {\bibfield  {journal} {\bibinfo  {journal} {\mnras}\ }\textbf {\bibinfo {volume} {304}},\ \bibinfo {pages} {851} (\bibinfo {year} {1999})},\ \Eprint {https://arxiv.org/abs/astro-ph/9812214} {arXiv:astro-ph/9812214 [astro-ph]} \BibitemShut {NoStop}%
\bibitem [{M{\'e}nard {\textit{et~al}}\mbox{.}(2008)\citenamefont {M{\'e}nard, Nestor, Turnshek, Quider, Richards, Chelouche,\ \&\ Rao}}]{menard2008lensing}%
{(\PineGreen{M{\'e}nard {\textit{et~al}}\mbox{.}}, \PineGreen{2008})}  \BibitemOpen
  \bibfield  {author} {\bibinfo {author} {\bibfnamefont {B.}~\bibnamefont {M{\'e}nard}}, \bibinfo {author} {\bibfnamefont {D.}~\bibnamefont {Nestor}}, \bibinfo {author} {\bibfnamefont {D.}~\bibnamefont {Turnshek}}, \bibinfo {author} {\bibfnamefont {A.}~\bibnamefont {Quider}}, \bibinfo {author} {\bibfnamefont {G.}~\bibnamefont {Richards}}, \bibinfo {author} {\bibfnamefont {D.}~\bibnamefont {Chelouche}}, \bibinfo {author} {\bibfnamefont {S.}~\bibnamefont {Rao}},\ }\emph {Lensing, reddening and extinction effects of Mg ii absorbers from z= 0.4 to 2},\ \href {\doibase 10.1111/j.1365-2966.2008.12909.x} {\bibfield  {journal} {\bibinfo  {journal} {Monthly Notices of the Royal Astronomical Society}\ }\textbf {\bibinfo {volume} {385}},\ \bibinfo {pages} {1053} (\bibinfo {year} {2008})}\BibitemShut {NoStop}%
\bibitem [{Metropolis {\textit{et~al}}\mbox{.}(1953)\citenamefont {Metropolis, Rosenbluth, Rosenbluth, Teller,\ \&\ Teller}}]{metropolis1953equation}%
{(\PineGreen{Metropolis {\textit{et~al}}\mbox{.}}, \PineGreen{1953})}  \BibitemOpen
  \bibfield  {author} {\bibinfo {author} {\bibfnamefont {N.}~\bibnamefont {Metropolis}}, \bibinfo {author} {\bibfnamefont {A.~W.}\ \bibnamefont {Rosenbluth}}, \bibinfo {author} {\bibfnamefont {M.~N.}\ \bibnamefont {Rosenbluth}}, \bibinfo {author} {\bibfnamefont {A.~H.}\ \bibnamefont {Teller}}, \bibinfo {author} {\bibfnamefont {E.}~\bibnamefont {Teller}},\ }\emph {Equation of state calculations by fast computing machines},\ \href {https://doi.org/10.2172/4390578} {\bibfield  {journal} {\bibinfo  {journal} {The journal of chemical physics}\ }\textbf {\bibinfo {volume} {21}},\ \bibinfo {pages} {1087} (\bibinfo {year} {1953})}\BibitemShut {NoStop}%
\bibitem [{{Mishra-Sharma}, {Alonso} \& {Dunkley}(2018)\citenamefont {{Mishra-Sharma}, {Alonso},\ \&\ {Dunkley}}}]{mishra2018neutrino}%
{(\PineGreen{{Mishra-Sharma}, {Alonso} \& {Dunkley}}, \PineGreen{2018})}  \BibitemOpen
  \bibfield  {author} {\bibinfo {author} {\bibfnamefont {S.}~\bibnamefont {{Mishra-Sharma}}}, \bibinfo {author} {\bibfnamefont {D.}~\bibnamefont {{Alonso}}}, \bibinfo {author} {\bibfnamefont {J.}~\bibnamefont {{Dunkley}}},\ }\emph {{Neutrino masses and beyond-$\Lambda$ CDM cosmology with LSST and future CMB experiments}},\ \href {\doibase 10.1103/PhysRevD.97.123544} {\bibfield  {journal} {\bibinfo  {journal} {\prd}\ }\textbf {\bibinfo {volume} {97}},\ \bibinfo {eid} {123544} (\bibinfo {year} {2018})},\ \Eprint {https://arxiv.org/abs/1803.07561} {arXiv:1803.07561 [astro-ph.CO]} \BibitemShut {NoStop}%
\bibitem [{{Modi}, {Lanusse} \& {Seljak}(2021)\citenamefont {{Modi}, {Lanusse},\ \&\ {Seljak}}}]{modi2021flowpm}%
{(\PineGreen{{Modi}, {Lanusse} \& {Seljak}}, \PineGreen{2021})}  \BibitemOpen
  \bibfield  {author} {\bibinfo {author} {\bibfnamefont {C.}~\bibnamefont {{Modi}}}, \bibinfo {author} {\bibfnamefont {F.}~\bibnamefont {{Lanusse}}}, \bibinfo {author} {\bibfnamefont {U.}~\bibnamefont {{Seljak}}},\ }\emph {{FlowPM: Distributed TensorFlow implementation of the FastPM cosmological N-body solver}},\ \href {\doibase 10.1016/j.ascom.2021.100505} {\bibfield  {journal} {\bibinfo  {journal} {Astronomy and Computing}\ }\textbf {\bibinfo {volume} {37}},\ \bibinfo {eid} {100505} (\bibinfo {year} {2021})},\ \Eprint {https://arxiv.org/abs/2010.11847} {arXiv:2010.11847 [astro-ph.CO]} \BibitemShut {NoStop}%
\bibitem [{{Mootoovaloo} {\textit{et~al}}\mbox{.}(2022)\citenamefont {{Mootoovaloo}, {Jaffe}, {Heavens},\ \&\ {Leclercq}}}]{mootoovaloo2022kernel}%
{(\PineGreen{{Mootoovaloo} {\textit{et~al}}\mbox{.}}, \PineGreen{2022})}  \BibitemOpen
  \bibfield  {author} {\bibinfo {author} {\bibfnamefont {A.}~\bibnamefont {{Mootoovaloo}}}, \bibinfo {author} {\bibfnamefont {A.~H.}\ \bibnamefont {{Jaffe}}}, \bibinfo {author} {\bibfnamefont {A.~F.}\ \bibnamefont {{Heavens}}}, \bibinfo {author} {\bibfnamefont {F.}~\bibnamefont {{Leclercq}}},\ }\emph {{Kernel-based emulator for the 3D matter power spectrum from CLASS}},\ \href {\doibase 10.1016/j.ascom.2021.100508} {\bibfield  {journal} {\bibinfo  {journal} {Astronomy and Computing}\ }\textbf {\bibinfo {volume} {38}},\ \bibinfo {eid} {100508} (\bibinfo {year} {2022})},\ \Eprint {https://arxiv.org/abs/2105.02256} {arXiv:2105.02256 [astro-ph.CO]} \BibitemShut {NoStop}%
\bibitem [{{More}, {Bovy} \& {Hogg}(2009)\citenamefont {{More}, {Bovy},\ \&\ {Hogg}}}]{more2009cosmic}%
{(\PineGreen{{More}, {Bovy} \& {Hogg}}, \PineGreen{2009})}  \BibitemOpen
  \bibfield  {author} {\bibinfo {author} {\bibfnamefont {S.}~\bibnamefont {{More}}}, \bibinfo {author} {\bibfnamefont {J.}~\bibnamefont {{Bovy}}}, \bibinfo {author} {\bibfnamefont {D.~W.}\ \bibnamefont {{Hogg}}},\ }\emph {{Cosmic Transparency: A Test with the Baryon Acoustic Feature and Type Ia Supernovae}},\ \href {\doibase 10.1088/0004-637X/696/2/1727} {\bibfield  {journal} {\bibinfo  {journal} {\apj}\ }\textbf {\bibinfo {volume} {696}},\ \bibinfo {pages} {1727} (\bibinfo {year} {2009})},\ \Eprint {https://arxiv.org/abs/0810.5553} {arXiv:0810.5553 [astro-ph]} \BibitemShut {NoStop}%
\bibitem [{{Moutarde} {\textit{et~al}}\mbox{.}(1991)\citenamefont {{Moutarde}, {Alimi}, {Bouchet}, {Pellat},\ \&\ {Ramani}}}]{Moutarde1991}%
{(\PineGreen{{Moutarde} {\textit{et~al}}\mbox{.}}, \PineGreen{1991})}  \BibitemOpen
  \bibfield  {author} {\bibinfo {author} {\bibfnamefont {F.}~\bibnamefont {{Moutarde}}}, \bibinfo {author} {\bibfnamefont {J.~M.}\ \bibnamefont {{Alimi}}}, \bibinfo {author} {\bibfnamefont {F.~R.}\ \bibnamefont {{Bouchet}}}, \bibinfo {author} {\bibfnamefont {R.}~\bibnamefont {{Pellat}}}, \bibinfo {author} {\bibfnamefont {A.}~\bibnamefont {{Ramani}}},\ }\emph {{Precollapse Scale Invariance in Gravitational Instability}},\ \href {\doibase 10.1086/170728} {\bibfield  {journal} {\bibinfo  {journal} {\apj}\ }\textbf {\bibinfo {volume} {382}},\ \bibinfo {pages} {377} (\bibinfo {year} {1991})}\BibitemShut {NoStop}%
\bibitem [{Neumann(1951)\citenamefont {Neumann}}]{neumann1951various}%
{(\PineGreen{Neumann}, \PineGreen{1951})}  \BibitemOpen
  \bibfield  {author} {\bibinfo {author} {\bibfnamefont {V.}~\bibnamefont {Neumann}},\ }\emph {Various techniques used in connection with random digits},\ \href@noop {} {\bibfield  {journal} {\bibinfo  {journal} {Notes by GE Forsythe}\ ,\ \bibinfo {pages} {36}} (\bibinfo {year} {1951})}\BibitemShut {NoStop}%
\bibitem [{{Nguyen} {\textit{et~al}}\mbox{.}(2021)\citenamefont {{Nguyen}, {Schmidt}, {Lavaux},\ \&\ {Jasche}}}]{nguyen2021impacts}%
{(\PineGreen{{Nguyen} {\textit{et~al}}\mbox{.}}, \PineGreen{2021})}  \BibitemOpen
  \bibfield  {author} {\bibinfo {author} {\bibfnamefont {N.-M.}\ \bibnamefont {{Nguyen}}}, \bibinfo {author} {\bibfnamefont {F.}~\bibnamefont {{Schmidt}}}, \bibinfo {author} {\bibfnamefont {G.}~\bibnamefont {{Lavaux}}}, \bibinfo {author} {\bibfnamefont {J.}~\bibnamefont {{Jasche}}},\ }\emph {{Impacts of the physical data model on the forward inference of initial conditions from biased tracers}},\ \href {\doibase 10.1088/1475-7516/2021/03/058} {\bibfield  {journal} {\bibinfo  {journal} {\jcap}\ }\textbf {\bibinfo {volume} {2021}},\ \bibinfo {eid} {058} (\bibinfo {year} {2021})},\ \Eprint {https://arxiv.org/abs/2011.06587} {arXiv:2011.06587 [astro-ph.CO]} \BibitemShut {NoStop}%
\bibitem [{{Nguyen} {\textit{et~al}}\mbox{.}(2024)\citenamefont {{Nguyen}, {Schmidt}, {Tucci}, {Reinecke},\ \&\ {Kosti{\'c}}}}]{nguyen2024much}%
{(\PineGreen{{Nguyen} {\textit{et~al}}\mbox{.}}, \PineGreen{2024})}  \BibitemOpen
  \bibfield  {author} {\bibinfo {author} {\bibfnamefont {N.-M.}\ \bibnamefont {{Nguyen}}}, \bibinfo {author} {\bibfnamefont {F.}~\bibnamefont {{Schmidt}}}, \bibinfo {author} {\bibfnamefont {B.}~\bibnamefont {{Tucci}}}, \bibinfo {author} {\bibfnamefont {M.}~\bibnamefont {{Reinecke}}}, \bibinfo {author} {\bibfnamefont {A.}~\bibnamefont {{Kosti{\'c}}}},\ }\emph {{How Much Information Can Be Extracted from Galaxy Clustering at the Field Level?}},\ \href {\doibase 10.1103/PhysRevLett.133.221006} {\bibfield  {journal} {\bibinfo  {journal} {\prl}\ }\textbf {\bibinfo {volume} {133}},\ \bibinfo {eid} {221006} (\bibinfo {year} {2024})},\ \Eprint {https://arxiv.org/abs/2403.03220} {arXiv:2403.03220 [astro-ph.CO]} \BibitemShut {NoStop}%
\bibitem [{Parzen(1962)\citenamefont {Parzen}}]{parzen1962estimation}%
{(\PineGreen{Parzen}, \PineGreen{1962})}  \BibitemOpen
  \bibfield  {author} {\bibinfo {author} {\bibfnamefont {E.}~\bibnamefont {Parzen}},\ }\emph {On estimation of a probability density function and mode},\ \href {\doibase 10.1214/aoms/1177704472} {\bibfield  {journal} {\bibinfo  {journal} {The annals of mathematical statistics}\ }\textbf {\bibinfo {volume} {33}},\ \bibinfo {pages} {1065} (\bibinfo {year} {1962})}\BibitemShut {NoStop}%
\bibitem [{{Peebles}(1980)\citenamefont {{Peebles}}}]{peebles1980large}%
{(\PineGreen{{Peebles}}, \PineGreen{1980})}  \BibitemOpen
  \bibfield  {author} {\bibinfo {author} {\bibfnamefont {P.~J.~E.}\ \bibnamefont {{Peebles}}},\ }\href {\doibase 10.2307/j.ctvxrpz4n} {\emph {{The large-scale structure of the universe}}}\ (\bibinfo  {publisher} {Princeton University Press},\ \bibinfo {year} {1980})\BibitemShut {NoStop}%
\bibitem [{{Philcox} \& {Ivanov}(2022)\citenamefont {{Philcox}\ \&\ {Ivanov}}}]{philcox2022boss}%
{(\PineGreen{{Philcox} \& {Ivanov}}, \PineGreen{2022})}  \BibitemOpen
  \bibfield  {author} {\bibinfo {author} {\bibfnamefont {O.~H.~E.}\ \bibnamefont {{Philcox}}}, \bibinfo {author} {\bibfnamefont {M.~M.}\ \bibnamefont {{Ivanov}}},\ }\emph {{BOSS DR12 full-shape cosmology: {\ensuremath{\Lambda}} CDM constraints from the large-scale galaxy power spectrum and bispectrum monopole}},\ \href {\doibase 10.1103/PhysRevD.105.043517} {\bibfield  {journal} {\bibinfo  {journal} {\prd}\ }\textbf {\bibinfo {volume} {105}},\ \bibinfo {eid} {043517} (\bibinfo {year} {2022})},\ \Eprint {https://arxiv.org/abs/2112.04515} {arXiv:2112.04515 [astro-ph.CO]} \BibitemShut {NoStop}%
\bibitem [{{Planck Collaboration}(2020)\citenamefont {{Planck Collaboration}}}]{aghanim2020planck}%
{(\PineGreen{{Planck Collaboration}}, \PineGreen{2020})}  \BibitemOpen
  \bibfield  {author} {\bibinfo {author} {\bibnamefont {{Planck Collaboration}}},\ }\emph {{Planck 2018 results. VI. Cosmological parameters}},\ \href {\doibase 10.1051/0004-6361/201833910} {\bibfield  {journal} {\bibinfo  {journal} {\aap}\ }\textbf {\bibinfo {volume} {641}},\ \bibinfo {eid} {A6} (\bibinfo {year} {2020})},\ \Eprint {https://arxiv.org/abs/1807.06209} {arXiv:1807.06209 [astro-ph.CO]} \BibitemShut {NoStop}%
\bibitem [{{Porqueres} {\textit{et~al}}\mbox{.}(2022)\citenamefont {{Porqueres}, {Heavens}, {Mortlock},\ \&\ {Lavaux}}}]{porqueres2022lifting}%
{(\PineGreen{{Porqueres} {\textit{et~al}}\mbox{.}}, \PineGreen{2022})}  \BibitemOpen
  \bibfield  {author} {\bibinfo {author} {\bibfnamefont {N.}~\bibnamefont {{Porqueres}}}, \bibinfo {author} {\bibfnamefont {A.}~\bibnamefont {{Heavens}}}, \bibinfo {author} {\bibfnamefont {D.}~\bibnamefont {{Mortlock}}}, \bibinfo {author} {\bibfnamefont {G.}~\bibnamefont {{Lavaux}}},\ }\emph {{Lifting weak lensing degeneracies with a field-based likelihood}},\ \href {\doibase 10.1093/mnras/stab3234} {\bibfield  {journal} {\bibinfo  {journal} {\mnras}\ }\textbf {\bibinfo {volume} {509}},\ \bibinfo {pages} {3194} (\bibinfo {year} {2022})},\ \Eprint {https://arxiv.org/abs/2108.04825} {arXiv:2108.04825 [astro-ph.CO]} \BibitemShut {NoStop}%
\bibitem [{{Porqueres} {\textit{et~al}}\mbox{.}(2019)\citenamefont {{Porqueres}, {Kodi Ramanah}, {Jasche},\ \&\ {Lavaux}}}]{porqueres2019explicit}%
{(\PineGreen{{Porqueres} {\textit{et~al}}\mbox{.}}, \PineGreen{2019})}  \BibitemOpen
  \bibfield  {author} {\bibinfo {author} {\bibfnamefont {N.}~\bibnamefont {{Porqueres}}}, \bibinfo {author} {\bibfnamefont {D.}~\bibnamefont {{Kodi Ramanah}}}, \bibinfo {author} {\bibfnamefont {J.}~\bibnamefont {{Jasche}}}, \bibinfo {author} {\bibfnamefont {G.}~\bibnamefont {{Lavaux}}},\ }\emph {{Explicit Bayesian treatment of unknown foreground contaminations in galaxy surveys}},\ \href {\doibase 10.1051/0004-6361/201834844} {\bibfield  {journal} {\bibinfo  {journal} {\aap}\ }\textbf {\bibinfo {volume} {624}},\ \bibinfo {eid} {A115} (\bibinfo {year} {2019})},\ \Eprint {https://arxiv.org/abs/1812.05113} {arXiv:1812.05113 [astro-ph.CO]} \BibitemShut {NoStop}%
\bibitem [{{Prideaux-Ghee} {\textit{et~al}}\mbox{.}(2023)\citenamefont {{Prideaux-Ghee}, {Leclercq}, {Lavaux}, {Heavens},\ \&\ {Jasche}}}]{prideaux2023field}%
{(\PineGreen{{Prideaux-Ghee} {\textit{et~al}}\mbox{.}}, \PineGreen{2023})}  \BibitemOpen
  \bibfield  {author} {\bibinfo {author} {\bibfnamefont {J.}~\bibnamefont {{Prideaux-Ghee}}}, \bibinfo {author} {\bibfnamefont {F.}~\bibnamefont {{Leclercq}}}, \bibinfo {author} {\bibfnamefont {G.}~\bibnamefont {{Lavaux}}}, \bibinfo {author} {\bibfnamefont {A.}~\bibnamefont {{Heavens}}}, \bibinfo {author} {\bibfnamefont {J.}~\bibnamefont {{Jasche}}},\ }\emph {{Field-based physical inference from peculiar velocity tracers}},\ \href {\doibase 10.1093/mnras/stac3346} {\bibfield  {journal} {\bibinfo  {journal} {\mnras}\ }\textbf {\bibinfo {volume} {518}},\ \bibinfo {pages} {4191} (\bibinfo {year} {2023})},\ \Eprint {https://arxiv.org/abs/2204.00023} {arXiv:2204.00023 [astro-ph.CO]} \BibitemShut {NoStop}%
\bibitem [{{Pullen} {\textit{et~al}}\mbox{.}(2016)\citenamefont {{Pullen}, {Hirata}, {Dor{\'e}},\ \&\ {Raccanelli}}}]{pullen2016interloper}%
{(\PineGreen{{Pullen} {\textit{et~al}}\mbox{.}}, \PineGreen{2016})}  \BibitemOpen
  \bibfield  {author} {\bibinfo {author} {\bibfnamefont {A.~R.}\ \bibnamefont {{Pullen}}}, \bibinfo {author} {\bibfnamefont {C.~M.}\ \bibnamefont {{Hirata}}}, \bibinfo {author} {\bibfnamefont {O.}~\bibnamefont {{Dor{\'e}}}}, \bibinfo {author} {\bibfnamefont {A.}~\bibnamefont {{Raccanelli}}},\ }\emph {{Interloper bias in future large-scale structure surveys}},\ \href {\doibase 10.1093/pasj/psv118} {\bibfield  {journal} {\bibinfo  {journal} {\pasj}\ }\textbf {\bibinfo {volume} {68}},\ \bibinfo {eid} {12} (\bibinfo {year} {2016})},\ \Eprint {https://arxiv.org/abs/1507.05092} {arXiv:1507.05092 [astro-ph.CO]} \BibitemShut {NoStop}%
\bibitem [{{Ramanah} {\textit{et~al}}\mbox{.}(2019)\citenamefont {{Ramanah}, {Lavaux}, {Jasche},\ \&\ {Wandelt}}}]{ramanah2019cosmological}%
{(\PineGreen{{Ramanah} {\textit{et~al}}\mbox{.}}, \PineGreen{2019})}  \BibitemOpen
  \bibfield  {author} {\bibinfo {author} {\bibfnamefont {D.~K.}\ \bibnamefont {{Ramanah}}}, \bibinfo {author} {\bibfnamefont {G.}~\bibnamefont {{Lavaux}}}, \bibinfo {author} {\bibfnamefont {J.}~\bibnamefont {{Jasche}}}, \bibinfo {author} {\bibfnamefont {B.~D.}\ \bibnamefont {{Wandelt}}},\ }\emph {{Cosmological inference from Bayesian forward modelling of deep galaxy redshift surveys}},\ \href {\doibase 10.1051/0004-6361/201834117} {\bibfield  {journal} {\bibinfo  {journal} {\aap}\ }\textbf {\bibinfo {volume} {621}},\ \bibinfo {eid} {A69} (\bibinfo {year} {2019})},\ \Eprint {https://arxiv.org/abs/1808.07496} {arXiv:1808.07496 [astro-ph.CO]} \BibitemShut {NoStop}%
\bibitem [{R{\'e}galdo-Saint~Blancard {\textit{et~al}}\mbox{.}(2024)R{\'e}galdo-Saint~Blancard, Hahn, Ho, Hou, Lemos, Massara, Modi, Dizgah, Parker, Yao \emph {et~al.}}]{regaldo2024galaxy}%
{(\PineGreen{R{\'e}galdo-Saint~Blancard {\textit{et~al}}\mbox{.}}, \PineGreen{2024})}  \BibitemOpen
  \bibfield  {author} {\bibinfo {author} {\bibfnamefont {B.}~\bibnamefont {R{\'e}galdo-Saint~Blancard}}, \bibinfo {author} {\bibfnamefont {C.}~\bibnamefont {Hahn}}, \bibinfo {author} {\bibfnamefont {S.}~\bibnamefont {Ho}}, \bibinfo {author} {\bibfnamefont {J.}~\bibnamefont {Hou}}, \bibinfo {author} {\bibfnamefont {P.}~\bibnamefont {Lemos}}, \bibinfo {author} {\bibfnamefont {E.}~\bibnamefont {Massara}}, \bibinfo {author} {\bibfnamefont {C.}~\bibnamefont {Modi}}, \bibinfo {author} {\bibfnamefont {A.~M.}\ \bibnamefont {Dizgah}}, \bibinfo {author} {\bibfnamefont {L.}~\bibnamefont {Parker}}, \bibinfo {author} {\bibfnamefont {Y.}~\bibnamefont {Yao}}, \emph {et~al.},\ }\emph {Galaxy clustering analysis with SimBIG and the wavelet scattering transform},\ \href {\doibase 10.1103/PhysRevD.109.083535} {\bibfield  {journal} {\bibinfo  {journal} {Physical Review D}\ }\textbf {\bibinfo {volume} {109}},\ \bibinfo {pages} {083535} (\bibinfo {year} {2024})}\BibitemShut {NoStop}%
\bibitem [{{Repp} \& {Szapudi}(2020)\citenamefont {{Repp}\ \&\ {Szapudi}}}]{repp2020galaxy}%
{(\PineGreen{{Repp} \& {Szapudi}}, \PineGreen{2020})}  \BibitemOpen
  \bibfield  {author} {\bibinfo {author} {\bibfnamefont {A.}~\bibnamefont {{Repp}}}, \bibinfo {author} {\bibfnamefont {I.}~\bibnamefont {{Szapudi}}},\ }\emph {{Galaxy bias and {\ensuremath{\sigma}}$_{8}$ from counts in cells from the SDSS main sample}},\ \href {\doibase 10.1093/mnrasl/slaa139} {\bibfield  {journal} {\bibinfo  {journal} {\mnras}\ }\textbf {\bibinfo {volume} {498}},\ \bibinfo {pages} {L125} (\bibinfo {year} {2020})},\ \Eprint {https://arxiv.org/abs/2006.01146} {arXiv:2006.01146 [astro-ph.CO]} \BibitemShut {NoStop}%
\bibitem [{Rosenblatt(1956)\citenamefont {Rosenblatt}}]{rosenblatt1956remarks}%
{(\PineGreen{Rosenblatt}, \PineGreen{1956})}  \BibitemOpen
  \bibfield  {author} {\bibinfo {author} {\bibfnamefont {M.}~\bibnamefont {Rosenblatt}},\ }\emph {Remarks on Some Nonparametric Estimates of a Density Function},\ \href {\doibase 10.1214/aoms/1177728190} {\bibfield  {journal} {\bibinfo  {journal} {The Annals of Mathematical Statistics}\ }\textbf {\bibinfo {volume} {27}},\ \bibinfo {pages} {832 } (\bibinfo {year} {1956})}\BibitemShut {NoStop}%
\bibitem [{{Ross} {\textit{et~al}}\mbox{.}(2013){Ross}, {Percival}, {Carnero}, {Zhao}, {Manera}, {Raccanelli}, {Aubourg}, {Bizyaev}, {Brewington}, {Brinkmann} \emph {et~al.}}]{ross2013clustering}%
{(\PineGreen{{Ross} {\textit{et~al}}\mbox{.}}, \PineGreen{2013})}  \BibitemOpen
  \bibfield  {author} {\bibinfo {author} {\bibfnamefont {A.~J.}\ \bibnamefont {{Ross}}}, \bibinfo {author} {\bibfnamefont {W.~J.}\ \bibnamefont {{Percival}}}, \bibinfo {author} {\bibfnamefont {A.}~\bibnamefont {{Carnero}}}, \bibinfo {author} {\bibfnamefont {G.-b.}\ \bibnamefont {{Zhao}}}, \bibinfo {author} {\bibfnamefont {M.}~\bibnamefont {{Manera}}}, \bibinfo {author} {\bibfnamefont {A.}~\bibnamefont {{Raccanelli}}}, \bibinfo {author} {\bibfnamefont {E.}~\bibnamefont {{Aubourg}}}, \bibinfo {author} {\bibfnamefont {D.}~\bibnamefont {{Bizyaev}}}, \bibinfo {author} {\bibfnamefont {H.}~\bibnamefont {{Brewington}}}, \bibinfo {author} {\bibfnamefont {J.}~\bibnamefont {{Brinkmann}}}, \emph {et~al.},\ }\emph {{The clustering of galaxies in the SDSS-III DR9 Baryon Oscillation Spectroscopic Survey: constraints on primordial non-Gaussianity}},\ \href {\doibase 10.1093/mnras/sts094} {\bibfield  {journal} {\bibinfo  {journal} {\mnras}\ }\textbf {\bibinfo {volume} {428}},\ \bibinfo {pages} {1116} (\bibinfo {year} {2013})},\ \Eprint {https://arxiv.org/abs/1208.1491} {arXiv:1208.1491 [astro-ph.CO]} \BibitemShut {NoStop}%
\bibitem [{{Said} {\textit{et~al}}\mbox{.}(2020)\citenamefont {{Said}, {Colless}, {Magoulas}, {Lucey},\ \&\ {Hudson}}}]{said2020joint}%
{(\PineGreen{{Said} {\textit{et~al}}\mbox{.}}, \PineGreen{2020})}  \BibitemOpen
  \bibfield  {author} {\bibinfo {author} {\bibfnamefont {K.}~\bibnamefont {{Said}}}, \bibinfo {author} {\bibfnamefont {M.}~\bibnamefont {{Colless}}}, \bibinfo {author} {\bibfnamefont {C.}~\bibnamefont {{Magoulas}}}, \bibinfo {author} {\bibfnamefont {J.~R.}\ \bibnamefont {{Lucey}}}, \bibinfo {author} {\bibfnamefont {M.~J.}\ \bibnamefont {{Hudson}}},\ }\emph {{Joint analysis of 6dFGS and SDSS peculiar velocities for the growth rate of cosmic structure and tests of gravity}},\ \href {\doibase 10.1093/mnras/staa2032} {\bibfield  {journal} {\bibinfo  {journal} {\mnras}\ }\textbf {\bibinfo {volume} {497}},\ \bibinfo {pages} {1275} (\bibinfo {year} {2020})},\ \Eprint {https://arxiv.org/abs/2007.04993} {arXiv:2007.04993 [astro-ph.CO]} \BibitemShut {NoStop}%
\bibitem [{{Salvati}, {Douspis} \& {Aghanim}(2020)\citenamefont {{Salvati}, {Douspis},\ \&\ {Aghanim}}}]{salvati2020impact}%
{(\PineGreen{{Salvati}, {Douspis} \& {Aghanim}}, \PineGreen{2020})}  \BibitemOpen
  \bibfield  {author} {\bibinfo {author} {\bibfnamefont {L.}~\bibnamefont {{Salvati}}}, \bibinfo {author} {\bibfnamefont {M.}~\bibnamefont {{Douspis}}}, \bibinfo {author} {\bibfnamefont {N.}~\bibnamefont {{Aghanim}}},\ }\emph {{Impact of systematics on cosmological parameters from future galaxy cluster surveys}},\ \href {\doibase 10.1051/0004-6361/202038465} {\bibfield  {journal} {\bibinfo  {journal} {\aap}\ }\textbf {\bibinfo {volume} {643}},\ \bibinfo {eid} {A20} (\bibinfo {year} {2020})},\ \Eprint {https://arxiv.org/abs/2005.10204} {arXiv:2005.10204 [astro-ph.CO]} \BibitemShut {NoStop}%
\bibitem [{{Sellentin} {\textit{et~al}}\mbox{.}(2023)\citenamefont {{Sellentin}, {Loureiro}, {Whiteway}, {Lafaurie}, {Balan}, {Olamaie}, {Jaffe},\ \&\ {Heavens}}}]{sellentin2023almanac}%
{(\PineGreen{{Sellentin} {\textit{et~al}}\mbox{.}}, \PineGreen{2023})}  \BibitemOpen
  \bibfield  {author} {\bibinfo {author} {\bibfnamefont {E.}~\bibnamefont {{Sellentin}}}, \bibinfo {author} {\bibfnamefont {A.}~\bibnamefont {{Loureiro}}}, \bibinfo {author} {\bibfnamefont {L.}~\bibnamefont {{Whiteway}}}, \bibinfo {author} {\bibfnamefont {J.~S.}\ \bibnamefont {{Lafaurie}}}, \bibinfo {author} {\bibfnamefont {S.~T.}\ \bibnamefont {{Balan}}}, \bibinfo {author} {\bibfnamefont {M.}~\bibnamefont {{Olamaie}}}, \bibinfo {author} {\bibfnamefont {A.~H.}\ \bibnamefont {{Jaffe}}}, \bibinfo {author} {\bibfnamefont {A.~F.}\ \bibnamefont {{Heavens}}},\ }\emph {{Almanac: MCMC-based signal extraction of power spectra and maps on the sphere}},\ \href {\doibase 10.21105/astro.2305.16134} {\bibfield  {journal} {\bibinfo  {journal} {The Open Journal of Astrophysics}\ }\textbf {\bibinfo {volume} {6}},\ \bibinfo {eid} {31} (\bibinfo {year} {2023})},\ \Eprint {https://arxiv.org/abs/2305.16134} {arXiv:2305.16134 [astro-ph.CO]} \BibitemShut {NoStop}%
\bibitem [{Simola {\textit{et~al}}\mbox{.}(2021)\citenamefont {Simola, Cisewski-Kehe, Gutmann,\ \&\ Corander}}]{Simola2021}%
{(\PineGreen{Simola {\textit{et~al}}\mbox{.}}, \PineGreen{2021})}  \BibitemOpen
  \bibfield  {author} {\bibinfo {author} {\bibfnamefont {U.}~\bibnamefont {Simola}}, \bibinfo {author} {\bibfnamefont {J.}~\bibnamefont {Cisewski-Kehe}}, \bibinfo {author} {\bibfnamefont {M.~U.}\ \bibnamefont {Gutmann}}, \bibinfo {author} {\bibfnamefont {J.}~\bibnamefont {Corander}},\ }\emph {{Adaptive Approximate Bayesian Computation Tolerance Selection}},\ \href {\doibase 10.1214/20-BA1211} {\bibfield  {journal} {\bibinfo  {journal} {Bayesian Analysis}\ }\textbf {\bibinfo {volume} {16}},\ \bibinfo {pages} {397 } (\bibinfo {year} {2021})}\BibitemShut {NoStop}%
\bibitem [{{Spurio Mancini} {\textit{et~al}}\mbox{.}(2022)\citenamefont {{Spurio Mancini}, {Piras}, {Alsing}, {Joachimi},\ \&\ {Hobson}}}]{spurio2022cosmopower}%
{(\PineGreen{{Spurio Mancini} {\textit{et~al}}\mbox{.}}, \PineGreen{2022})}  \BibitemOpen
  \bibfield  {author} {\bibinfo {author} {\bibfnamefont {A.}~\bibnamefont {{Spurio Mancini}}}, \bibinfo {author} {\bibfnamefont {D.}~\bibnamefont {{Piras}}}, \bibinfo {author} {\bibfnamefont {J.}~\bibnamefont {{Alsing}}}, \bibinfo {author} {\bibfnamefont {B.}~\bibnamefont {{Joachimi}}}, \bibinfo {author} {\bibfnamefont {M.~P.}\ \bibnamefont {{Hobson}}},\ }\emph {{COSMOPOWER: emulating cosmological power spectra for accelerated Bayesian inference from next-generation surveys}},\ \href {\doibase 10.1093/mnras/stac064} {\bibfield  {journal} {\bibinfo  {journal} {\mnras}\ }\textbf {\bibinfo {volume} {511}},\ \bibinfo {pages} {1771} (\bibinfo {year} {2022})},\ \Eprint {https://arxiv.org/abs/2106.03846} {arXiv:2106.03846 [astro-ph.CO]} \BibitemShut {NoStop}%
\bibitem [{{Sunn{\r{a}}ker} {\textit{et~al}}\mbox{.}(2013)\citenamefont {{Sunn{\r{a}}ker}, {Busetto}, {Numminen}, {Corander}, {Foll},\ \&\ {Dessimoz}}}]{beaumont2019approximate}%
{(\PineGreen{{Sunn{\r{a}}ker} {\textit{et~al}}\mbox{.}}, \PineGreen{2013})}  \BibitemOpen
  \bibfield  {author} {\bibinfo {author} {\bibfnamefont {M.}~\bibnamefont {{Sunn{\r{a}}ker}}}, \bibinfo {author} {\bibfnamefont {A.~G.}\ \bibnamefont {{Busetto}}}, \bibinfo {author} {\bibfnamefont {E.}~\bibnamefont {{Numminen}}}, \bibinfo {author} {\bibfnamefont {J.}~\bibnamefont {{Corander}}}, \bibinfo {author} {\bibfnamefont {M.}~\bibnamefont {{Foll}}}, \bibinfo {author} {\bibfnamefont {C.}~\bibnamefont {{Dessimoz}}},\ }\emph {{Approximate Bayesian Computation}},\ \href {\doibase 10.1371/journal.pcbi.1002803} {\bibfield  {journal} {\bibinfo  {journal} {PLoS Computational Biology}\ }\textbf {\bibinfo {volume} {9}},\ \bibinfo {pages} {e1002803} (\bibinfo {year} {2013})}\BibitemShut {NoStop}%
\bibitem [{{Tassev}, {Zaldarriaga} \& {Eisenstein}(2013)\citenamefont {{Tassev}, {Zaldarriaga},\ \&\ {Eisenstein}}}]{Tassev2013}%
{(\PineGreen{{Tassev}, {Zaldarriaga} \& {Eisenstein}}, \PineGreen{2013})}  \BibitemOpen
  \bibfield  {author} {\bibinfo {author} {\bibfnamefont {S.}~\bibnamefont {{Tassev}}}, \bibinfo {author} {\bibfnamefont {M.}~\bibnamefont {{Zaldarriaga}}}, \bibinfo {author} {\bibfnamefont {D.~J.}\ \bibnamefont {{Eisenstein}}},\ }\emph {{Solving large scale structure in ten easy steps with COLA}},\ \href {\doibase 10.1088/1475-7516/2013/06/036} {\bibfield  {journal} {\bibinfo  {journal} {\jcap}\ }\textbf {\bibinfo {volume} {2013}},\ \bibinfo {eid} {036} (\bibinfo {year} {2013})},\ \Eprint {https://arxiv.org/abs/1301.0322} {arXiv:1301.0322 [astro-ph.CO]} \BibitemShut {NoStop}%
\bibitem [{{Taylor} {\textit{et~al}}\mbox{.}(2019)\citenamefont {{Taylor}, {Kitching}, {Alsing}, {Wandelt}, {Feeney},\ \&\ {McEwen}}}]{taylor2019cosmic}%
{(\PineGreen{{Taylor} {\textit{et~al}}\mbox{.}}, \PineGreen{2019})}  \BibitemOpen
  \bibfield  {author} {\bibinfo {author} {\bibfnamefont {P.~L.}\ \bibnamefont {{Taylor}}}, \bibinfo {author} {\bibfnamefont {T.~D.}\ \bibnamefont {{Kitching}}}, \bibinfo {author} {\bibfnamefont {J.}~\bibnamefont {{Alsing}}}, \bibinfo {author} {\bibfnamefont {B.~D.}\ \bibnamefont {{Wandelt}}}, \bibinfo {author} {\bibfnamefont {S.~M.}\ \bibnamefont {{Feeney}}}, \bibinfo {author} {\bibfnamefont {J.~D.}\ \bibnamefont {{McEwen}}},\ }\emph {{Cosmic shear: Inference from forward models}},\ \href {\doibase 10.1103/PhysRevD.100.023519} {\bibfield  {journal} {\bibinfo  {journal} {\prd}\ }\textbf {\bibinfo {volume} {100}},\ \bibinfo {eid} {023519} (\bibinfo {year} {2019})},\ \Eprint {https://arxiv.org/abs/1904.05364} {arXiv:1904.05364 [astro-ph.CO]} \BibitemShut {NoStop}%
\bibitem [{{Tegmark} {\textit{et~al}}\mbox{.}(2004){Tegmark}, {Strauss}, {Blanton}, {Abazajian}, {Dodelson}, {Sandvik}, {Wang}, {Weinberg}, {Zehavi}, {Bahcall} \emph {et~al.}}]{tegmark2004cosmological}%
{(\PineGreen{{Tegmark} {\textit{et~al}}\mbox{.}}, \PineGreen{2004})}  \BibitemOpen
  \bibfield  {author} {\bibinfo {author} {\bibfnamefont {M.}~\bibnamefont {{Tegmark}}}, \bibinfo {author} {\bibfnamefont {M.~A.}\ \bibnamefont {{Strauss}}}, \bibinfo {author} {\bibfnamefont {M.~R.}\ \bibnamefont {{Blanton}}}, \bibinfo {author} {\bibfnamefont {K.}~\bibnamefont {{Abazajian}}}, \bibinfo {author} {\bibfnamefont {S.}~\bibnamefont {{Dodelson}}}, \bibinfo {author} {\bibfnamefont {H.}~\bibnamefont {{Sandvik}}}, \bibinfo {author} {\bibfnamefont {X.}~\bibnamefont {{Wang}}}, \bibinfo {author} {\bibfnamefont {D.~H.}\ \bibnamefont {{Weinberg}}}, \bibinfo {author} {\bibfnamefont {I.}~\bibnamefont {{Zehavi}}}, \bibinfo {author} {\bibfnamefont {N.~A.}\ \bibnamefont {{Bahcall}}}, \emph {et~al.},\ }\emph {{Cosmological parameters from SDSS and WMAP}},\ \href {\doibase 10.1103/PhysRevD.69.103501} {\bibfield  {journal} {\bibinfo  {journal} {\prd}\ }\textbf {\bibinfo {volume} {69}},\ \bibinfo {eid} {103501} (\bibinfo {year} {2004})},\ \Eprint {https://arxiv.org/abs/astro-ph/0310723} {arXiv:astro-ph/0310723 [astro-ph]} \BibitemShut {NoStop}%
\bibitem [{Thomas {\textit{et~al}}\mbox{.}(2020)\citenamefont {Thomas, S{\'a}-Le{\~a}o, de~Lencastre, Kaski, Corander,\ \&\ Pesonen}}]{thomas2020misspecification}%
{(\PineGreen{Thomas {\textit{et~al}}\mbox{.}}, \PineGreen{2020})}  \BibitemOpen
  \bibfield  {author} {\bibinfo {author} {\bibfnamefont {O.}~\bibnamefont {Thomas}}, \bibinfo {author} {\bibfnamefont {R.}~\bibnamefont {S{\'a}-Le{\~a}o}}, \bibinfo {author} {\bibfnamefont {H.}~\bibnamefont {de~Lencastre}}, \bibinfo {author} {\bibfnamefont {S.}~\bibnamefont {Kaski}}, \bibinfo {author} {\bibfnamefont {J.}~\bibnamefont {Corander}}, \bibinfo {author} {\bibfnamefont {H.}~\bibnamefont {Pesonen}},\ }\emph {Misspecification-robust likelihood-free inference in high dimensions},\ \href {https://arxiv.org/abs/2002.09377} {\bibfield  {journal} {\bibinfo  {journal} {arXiv preprint arXiv:2002.09377}\ } (\bibinfo {year} {2020})}\BibitemShut {NoStop}%
\bibitem [{{Tucci} \& {Schmidt}(2024)\citenamefont {{Tucci}\ \&\ {Schmidt}}}]{tucci2024eftoflss}%
{(\PineGreen{{Tucci} \& {Schmidt}}, \PineGreen{2024})}  \BibitemOpen
  \bibfield  {author} {\bibinfo {author} {\bibfnamefont {B.}~\bibnamefont {{Tucci}}}, \bibinfo {author} {\bibfnamefont {F.}~\bibnamefont {{Schmidt}}},\ }\emph {{EFTofLSS meets simulation-based inference: {\ensuremath{\sigma}} $_{8}$ from biased tracers}},\ \href {\doibase 10.1088/1475-7516/2024/05/063} {\bibfield  {journal} {\bibinfo  {journal} {\jcap}\ }\textbf {\bibinfo {volume} {2024}},\ \bibinfo {eid} {063} (\bibinfo {year} {2024})},\ \Eprint {https://arxiv.org/abs/2310.03741} {arXiv:2310.03741 [astro-ph.CO]} \BibitemShut {NoStop}%
\bibitem [{{Wandelt}, {Larson} \& {Lakshminarayanan}(2004)\citenamefont {{Wandelt}, {Larson},\ \&\ {Lakshminarayanan}}}]{wandelt2004global}%
{(\PineGreen{{Wandelt}, {Larson} \& {Lakshminarayanan}}, \PineGreen{2004})}  \BibitemOpen
  \bibfield  {author} {\bibinfo {author} {\bibfnamefont {B.~D.}\ \bibnamefont {{Wandelt}}}, \bibinfo {author} {\bibfnamefont {D.~L.}\ \bibnamefont {{Larson}}}, \bibinfo {author} {\bibfnamefont {A.}~\bibnamefont {{Lakshminarayanan}}},\ }\emph {{Global, exact cosmic microwave background data analysis using Gibbs sampling}},\ \href {\doibase 10.1103/PhysRevD.70.083511} {\bibfield  {journal} {\bibinfo  {journal} {\prd}\ }\textbf {\bibinfo {volume} {70}},\ \bibinfo {eid} {083511} (\bibinfo {year} {2004})},\ \Eprint {https://arxiv.org/abs/astro-ph/0310080} {arXiv:astro-ph/0310080 [astro-ph]} \BibitemShut {NoStop}%
\bibitem [{{Wang} {\textit{et~al}}\mbox{.}(2014)\citenamefont {{Wang}, {Mo}, {Yang}, {Jing},\ \&\ {Lin}}}]{wang2014elucid}%
{(\PineGreen{{Wang} {\textit{et~al}}\mbox{.}}, \PineGreen{2014})}  \BibitemOpen
  \bibfield  {author} {\bibinfo {author} {\bibfnamefont {H.}~\bibnamefont {{Wang}}}, \bibinfo {author} {\bibfnamefont {H.~J.}\ \bibnamefont {{Mo}}}, \bibinfo {author} {\bibfnamefont {X.}~\bibnamefont {{Yang}}}, \bibinfo {author} {\bibfnamefont {Y.~P.}\ \bibnamefont {{Jing}}}, \bibinfo {author} {\bibfnamefont {W.~P.}\ \bibnamefont {{Lin}}},\ }\emph {{ELUCID{\textemdash}Exploring the Local Universe with the Reconstructed Initial Density Field. I. Hamiltonian Markov Chain Monte Carlo Method with Particle Mesh Dynamics}},\ \href {\doibase 10.1088/0004-637X/794/1/94} {\bibfield  {journal} {\bibinfo  {journal} {\apj}\ }\textbf {\bibinfo {volume} {794}},\ \bibinfo {eid} {94} (\bibinfo {year} {2014})},\ \Eprint {https://arxiv.org/abs/1407.3451} {arXiv:1407.3451 [astro-ph.CO]} \BibitemShut {NoStop}%
\bibitem [{{Wang} {\textit{et~al}}\mbox{.}(2016)\citenamefont {{Wang}, {Mo}, {Yang}, {Zhang}, {Shi}, {Jing}, {Liu}, {Li}, {Kang},\ \&\ {Gao}}}]{wang2016elucid}%
{(\PineGreen{{Wang} {\textit{et~al}}\mbox{.}}, \PineGreen{2016})}  \BibitemOpen
  \bibfield  {author} {\bibinfo {author} {\bibfnamefont {H.}~\bibnamefont {{Wang}}}, \bibinfo {author} {\bibfnamefont {H.~J.}\ \bibnamefont {{Mo}}}, \bibinfo {author} {\bibfnamefont {X.}~\bibnamefont {{Yang}}}, \bibinfo {author} {\bibfnamefont {Y.}~\bibnamefont {{Zhang}}}, \bibinfo {author} {\bibfnamefont {J.}~\bibnamefont {{Shi}}}, \bibinfo {author} {\bibfnamefont {Y.~P.}\ \bibnamefont {{Jing}}}, \bibinfo {author} {\bibfnamefont {C.}~\bibnamefont {{Liu}}}, \bibinfo {author} {\bibfnamefont {S.}~\bibnamefont {{Li}}}, \bibinfo {author} {\bibfnamefont {X.}~\bibnamefont {{Kang}}}, \bibinfo {author} {\bibfnamefont {Y.}~\bibnamefont {{Gao}}},\ }\emph {{ELUCID - Exploring the Local Universe with ReConstructed Initial Density Field III: Constrained Simulation in the SDSS Volume}},\ \href {\doibase 10.3847/0004-637X/831/2/164} {\bibfield  {journal} {\bibinfo  {journal} {\apj}\ }\textbf {\bibinfo {volume} {831}},\ \bibinfo {eid} {164} (\bibinfo {year} {2016})},\ \Eprint {https://arxiv.org/abs/1608.01763} {arXiv:1608.01763 [astro-ph.CO]} \BibitemShut {NoStop}%
\bibitem [{{Weyant}, {Schafer} \& {Wood-Vasey}(2013)\citenamefont {{Weyant}, {Schafer},\ \&\ {Wood-Vasey}}}]{weyant2013likelihood}%
{(\PineGreen{{Weyant}, {Schafer} \& {Wood-Vasey}}, \PineGreen{2013})}  \BibitemOpen
  \bibfield  {author} {\bibinfo {author} {\bibfnamefont {A.}~\bibnamefont {{Weyant}}}, \bibinfo {author} {\bibfnamefont {C.}~\bibnamefont {{Schafer}}}, \bibinfo {author} {\bibfnamefont {W.~M.}\ \bibnamefont {{Wood-Vasey}}},\ }\emph {{Likelihood-free Cosmological Inference with Type Ia Supernovae: Approximate Bayesian Computation for a Complete Treatment of Uncertainty}},\ \href {\doibase 10.1088/0004-637X/764/2/116} {\bibfield  {journal} {\bibinfo  {journal} {\apj}\ }\textbf {\bibinfo {volume} {764}},\ \bibinfo {eid} {116} (\bibinfo {year} {2013})},\ \Eprint {https://arxiv.org/abs/1206.2563} {arXiv:1206.2563 [astro-ph.CO]} \BibitemShut {NoStop}%
\bibitem [{{Zeghal} {\textit{et~al}}\mbox{.}(2024)\citenamefont {{Zeghal}, {Lanzieri}, {Lanusse}, {Boucaud}, {Louppe}, {Aubourg}, {Bayer},\ \&\ {The LSST Dark Energy Science Collaboration}}}]{zeghal2024simulation}%
{(\PineGreen{{Zeghal} {\textit{et~al}}\mbox{.}}, \PineGreen{2024})}  \BibitemOpen
  \bibfield  {author} {\bibinfo {author} {\bibfnamefont {J.}~\bibnamefont {{Zeghal}}}, \bibinfo {author} {\bibfnamefont {D.}~\bibnamefont {{Lanzieri}}}, \bibinfo {author} {\bibfnamefont {F.}~\bibnamefont {{Lanusse}}}, \bibinfo {author} {\bibfnamefont {A.}~\bibnamefont {{Boucaud}}}, \bibinfo {author} {\bibfnamefont {G.}~\bibnamefont {{Louppe}}}, \bibinfo {author} {\bibfnamefont {E.}~\bibnamefont {{Aubourg}}}, \bibinfo {author} {\bibfnamefont {A.~E.}\ \bibnamefont {{Bayer}}}, \bibinfo {author} {\bibnamefont {{The LSST Dark Energy Science Collaboration}}},\ }\emph {{Simulation-Based Inference Benchmark for LSST Weak Lensing Cosmology}},\ \href {\doibase 10.48550/arXiv.2409.17975} {\bibfield  {journal} {\bibinfo  {journal} {arXiv e-prints}\ ,\ \bibinfo {eid} {arXiv:2409.17975}} (\bibinfo {year} {2024})},\ \Eprint {https://arxiv.org/abs/2409.17975} {arXiv:2409.17975 [astro-ph.CO]} \BibitemShut {NoStop}%
\bibitem [{{Zhou} {\textit{et~al}}\mbox{.}(2024)\citenamefont {{Zhou}, {Li}, {Dodelson},\ \&\ {Mandelbaum}}}]{zhou2024accurate}%
{(\PineGreen{{Zhou} {\textit{et~al}}\mbox{.}}, \PineGreen{2024})}  \BibitemOpen
  \bibfield  {author} {\bibinfo {author} {\bibfnamefont {A.~J.}\ \bibnamefont {{Zhou}}}, \bibinfo {author} {\bibfnamefont {X.}~\bibnamefont {{Li}}}, \bibinfo {author} {\bibfnamefont {S.}~\bibnamefont {{Dodelson}}}, \bibinfo {author} {\bibfnamefont {R.}~\bibnamefont {{Mandelbaum}}},\ }\emph {{Accurate field-level weak lensing inference for precision cosmology}},\ \href {\doibase 10.1103/PhysRevD.110.023539} {\bibfield  {journal} {\bibinfo  {journal} {\prd}\ }\textbf {\bibinfo {volume} {110}},\ \bibinfo {eid} {023539} (\bibinfo {year} {2024})},\ \Eprint {https://arxiv.org/abs/2312.08934} {arXiv:2312.08934 [astro-ph.CO]} \BibitemShut {NoStop}%
\bibitem [{{Zhou} {\textit{et~al}}\mbox{.}(2023){Zhou}, {Dey}, {Newman}, {Eisenstein}, {Dawson}, {Bailey}, {Berti}, {Guy}, {Lan}, {Zou} \emph {et~al.}}]{zhou2023target}%
{(\PineGreen{{Zhou} {\textit{et~al}}\mbox{.}}, \PineGreen{2023})}  \BibitemOpen
  \bibfield  {author} {\bibinfo {author} {\bibfnamefont {R.}~\bibnamefont {{Zhou}}}, \bibinfo {author} {\bibfnamefont {B.}~\bibnamefont {{Dey}}}, \bibinfo {author} {\bibfnamefont {J.~A.}\ \bibnamefont {{Newman}}}, \bibinfo {author} {\bibfnamefont {D.~J.}\ \bibnamefont {{Eisenstein}}}, \bibinfo {author} {\bibfnamefont {K.}~\bibnamefont {{Dawson}}}, \bibinfo {author} {\bibfnamefont {S.}~\bibnamefont {{Bailey}}}, \bibinfo {author} {\bibfnamefont {A.}~\bibnamefont {{Berti}}}, \bibinfo {author} {\bibfnamefont {J.}~\bibnamefont {{Guy}}}, \bibinfo {author} {\bibfnamefont {T.-W.}\ \bibnamefont {{Lan}}}, \bibinfo {author} {\bibfnamefont {H.}~\bibnamefont {{Zou}}}, \emph {et~al.},\ }\emph {{Target Selection and Validation of DESI Luminous Red Galaxies}},\ \href {\doibase 10.3847/1538-3881/aca5fb} {\bibfield  {journal} {\bibinfo  {journal} {\aj}\ }\textbf {\bibinfo {volume} {165}},\ \bibinfo {eid} {58} (\bibinfo {year} {2023})},\ \Eprint {https://arxiv.org/abs/2208.08515} {arXiv:2208.08515 [astro-ph.CO]} \BibitemShut {NoStop}%
\end{thebibliography}%

\end{document}